\documentclass[fleqn,usenatbib]{mnras}

\usepackage{newtxtext,newtxmath}

\usepackage[T1]{fontenc}
\usepackage{ae,aecompl}

\usepackage{multicol}
\usepackage{graphicx}    
\usepackage{amsmath}    
\usepackage{xcolor}

\usepackage{etoolbox}

\title[Chemical evolution of the solar neighbourhood]{The chemical evolution of the solar neighbourhood for planet-hosting stars}

\author[M. Pignatari et al.]{Marco Pignatari$^{1,2,3,8,9}$,
Thomas C.~L. Trueman$^{1,3,8}$,
Kate A. Womack$^{3}$,
Brad K. Gibson$^{3,9}$,\newauthor
Benoit C\^ot\'e$^{1,8,9,10}$,
Diego Turrini$^{4,5,6}$,
Christopher Sneden$^{7}$,
Stephen J. Mojzsis$^{1,11}$,\newauthor
Richard J. Stancliffe$^{8,12}$,
Paul Fong$^{3,8}$,
Thomas V. Lawson$^{3,8,13}$,
James D. Keegans$^{14,8}$,\newauthor
Kate Pilkington$^{15}$,
Jean-Claude Passy$^{16}$,
Timothy C. Beers$^{17,9}$,
Maria Lugaro$^{1,2,18,19}$
\\
$^{1}$Konkoly Observatory, Research Centre for Astronomy and Earth Sciences (CSFK), ELKH, H-1121, Budapest, Konkoly Thege M. \'ut 15–17., Hungary\\
$^{2}$CSFK, MTA Centre of Excellence, Budapest, Konkoly Thege Miklós út 15-17., H-1121, Hungary\\
$^{3}$E.~A.~Milne Centre for Astrophysics, Department of Physics and Mathematics, University of Hull, HU6 7RX, United Kingdom\\
$^{4}$ INAF - Turin Astrophysical Observatory, Via Osservatorio 20, 10025, Pino Torinese, Italy\\
$^{5}$ INAF - Institute of Space Astrophysics and Planetology, Via Fosso del Cavaliere 100, 00133, Rome, Italy\\
$^{6}$ ICSC National Research Centre for High Performance Computing, Big Data and Quantum Computing, Via Magnanelli 2, 40033, Casalecchio di Reno, Italy\\
$^{7}$Department of Astronomy and McDonald Observatory, The University of Texas, Austin, TX 78712, USA\\
$^{8}$NuGrid Collaboration, \url{http://nugridstars.org}\\
$^{9}$Joint Institute for Nuclear Astrophysics - Center for the Evolution of the Elements, USA\\
$^{10}$Department of Physics and Astronomy, University of Victoria, Victoria, BC V8P 5C2, Canada\\
$^{11}$Department of Petrology and Geochemistry, Eötvös Loránd University (ELTE), Pázmány Péter sétány, 1/c, Budapest H-1117, Hungary\\
$^{12}$H.H. Wills Physics Laboratory, University of Bristol, Tyndall Avenue, Bristol, BS8 1TL, United Kingdom\\
$^{13}$Centre of Excellence for Data Science, Artificial Intelligence and Modelling, University of Hull, HU6 7RX, United Kingdom\\
$^{14}$Astrophysics Group, Lennard-Jones Laboratories, Keele University, Keele ST5 5BG, United Kingdom\\
$^{15}$School of Physical Sciences, The Open University, Milton Keynes, MK7~6AA, United Kingdom\\
$^{16}$Max Planck Institute for Intelligent Systems, Tübingen, Germany \\
$^{17}$Department of Physics and Astronomy, University of Notre Dame, Notre Dame, IN, 46556, USA\\
$^{18}$ELTE Eötvös Loránd University, Institute of Physics, Budapest 1117, Pázmány Péter sétány 1/A, Hungary\\
$^{19}$School of Physics and Astronomy, Monash University, VIC 3800, Australia
}
\date{Accepted XXX. Received YYY; in original form ZZZ}

\pubyear{2023}

\begin{document}

\label{firstpage}
\pagerange{\pageref{firstpage}--\pageref{lastpage}}
\maketitle
\begin{abstract}
Theoretical physical-chemical models for the formation of planetary systems depend on data quality for the Sun's composition, that of stars in the solar neighbourhood, and 
of the estimated "pristine" compositions 
for stellar systems. 
The effective scatter and the observational uncertainties of 
elements within a few hundred parsecs from the Sun,
even for the most abundant metals like carbon, oxygen and silicon, are still controversial. 
Here we analyse the stellar production and the chemical evolution of 
key elements that underpin the formation of rocky (C, O, Mg, Si) and gas/ice giant planets (C, N, O, S). 
We calculate 198 galactic chemical evolution (GCE) models of the solar neighbourhood to analyse the impact of different sets of stellar yields, of the upper mass limit for massive stars contributing to GCE ($M_{\rm up}$) and of 
supernovae from massive-star progenitors which do not eject the bulk of the 
iron-peak elements (faint supernovae).
Even considering 
the GCE variation 
produced via different sets of stellar yields, 
the observed dispersion of 
elements reported for stars in 
the Milky Way disk is not reproduced. Among others, the observed range of super-solar [Mg/Si] ratios, sub-solar [S/N], and the dispersion of 
up to 0.5 dex for [S/Si] challenge our models. 
The impact of varying $M_{\rm up}$ depends on the adopted 
supernova yields. Thus, 
observations do not provide a 
constraint on the M$_{\rm up}$ parametrization. When including the impact of 
faint supernova models in GCE calculations, elemental ratios vary by up to 0.1-0.2 dex in the Milky Way disk; this modification better reproduces observations. 
\end{abstract}




\section{Introduction}
\label{sec: intro}

The chemical enrichment history of the elements observed in the Sun and in other stars in the solar neighbourhood 
serves as the basis for our information about the formation and the chemical evolution of the Milky Way (MW) disk \citep[e.g.,][]{truran:71, tinsley:78, timmes:95, goswami:00, matteucci:21}. Galactic chemical evolution (GCE) simulations attempt to model the change with time of the chemical elements by taking into account the formation of the MW disk and including theoretical stellar yields from different generations of stars. The GCE models are then compared to stellar abundance trends with metallicity or age of the MW disk, and to the solar abundance pattern \citep[e.g.,][]{matteucci:86,gibson:03, kobayashi:11, molla:15, mishenina:17,prantzos:18, kobayashi:20, prantzos:23}. The composition for all the elements can be measured in the Sun and in meteorites \citep[][and references therein]{lodders:19}, whereas a more limited number of elements are available for other stars. Nevertheless, elemental abundances are preserved with limited modification over time at the stellar surfaces, and are therefore taken to be indicative of the pristine stellar abundances \citep[e.g.,][]{piersanti:07}.

Analyses become rather more complex for discussions of likely compositions of the planets that may have formed around these stars. Stars typically represent $>$98$\%$ of the mass of a star+planet(s) system. So, the original abundances of stellar systems are recapitulated in the composition of the host star, which in turn mirrors the composition at the start of the planetary-formation process. The bulk composition of the stellar system is the ultimate arbiter for the properties of the planets that will form. This holds not only for stable elements, but may also be true for the short-lived radioactive isotopes relevant for the
heating of planetesimals \citep[mostly $^{26}$Al in the case of the early Solar System, e.g.,][]{kleine:05,lichtenberg:16,lugaro:18} and the radiogenic heating of planet interiors via the long-lived radionuclides $^{40}$K, $^{232}$Th, $^{235}$U and $^{238}$U \citep[e.g., ][]{frank:14,unterborn:15,wang:20}.

Nevertheless, depending on where and how the planets formed, their migration history, the global dynamical history of their system, the planetary formation process can modify or even erase the signatures of the initial chemical abundances of the system for ultra-volatile elements like H, C, N, O and S at different distances from the host stars  
\citep[see later in the Introduction, and e.g.,][and references therein]{madhusudhan:16,madhusudhan:19,cridland:19,cridland:20,turrini:21a,turrini:22,adibekyan:21,drazkowska:22,pacetti:22}. 

Iron is also problematic, since it is by definition siderophile and along with other such elements (Ni), tends to be sequestered into the metallic\footnote{Note that in this context metallic cores are made mostly by Fe and Ni. In the rest of the paper, we refer as "metals" all the nuclides heavier than H and He.} cores of rocky planets
during their differentiation process. On the other hand, refractory lithophile elements like Ca and the Rare Earth Elements are unaffected by these processes. Moderately refractory lithophile elements such as Li, Mg and Si, and some other moderately volatile lithophile elements such as K and Na, follow a devolatilization trend based on 
the different condensation temperatures
of the elements \citep[][]{yoshizaki:20,wang:22,spaargaren:23}. 


A criterion often invoked in arguments for the geodynamic predisposition of a planet to host life (so-called "habitability") is the metal enrichment \citep[][]{lineweaver:04,spitoni:14,spitoni:17}. Consequently, initial major elemental ratios such as C/O and Mg/Si are regarded as especially crucial in modulating the chemistry of early condensates and the mineralogy of rocky planets that are conducive for biological activity to take hold \citep[][]{Mojzsis22}. 

Based on observational results, it has been proposed that these ratios also modulate the types of planet formed. For instance, \cite{adibekyan:15} found that low-mass planets are more prevalent around stars with Mg/Si higher than solar, and in general for stellar hosts with high [Mg/Si] ratios after removing the GCE trend of the two elements. From the theoretical point of view this is expected \citep[e.g.,][]{frank:14}, but a broader analysis of Mg/Si with respect to exoplanet populations \citep[][]{spaargaren:23} is warranted.

Given the above criteria, simulations of planetary formation and evolution demand a better understanding of the connection between GCE models for the composition of stars in the solar neighbourhood, and the particular compositional characteristics observed for the planet-hosting stars and for their planets \citep[e.g.,][]{santos:17,turrini:21a,turrini:22,adibekyan:21,khorshid:21,reggiani:22,jorge:22,pacetti:22,fonte:23}. 
Consequently, GCE models can be used then as a theoretical source for the initial abundances at planet formation for all elements (observed with different uncertainties or not available in the stellar spectra) at different times and locations in the Galaxy, and as a benchmark for the results of planet formation obtained from simulations \citep[][]{frank:14,mojzsis:22}. 

With respect to the origin of the gas and ice giant planets in our Solar System and beyond, their present C/O-ratio has been proposed as a diagnostic to distinguish between different formation processes where gas accretion or capture of planetary material may dominate, with following modifications of the initial C/O-ratio \citep[e.g.,][]{oberg:11,madhusudhan:16,madhusudhan:19}. In particular, \cite{turrini:21a} show that when the capture of planetary material is the dominant source of planetary metallicity, the C/O-ratio of giant planets is close to the stellar C/O-ratio \citep{turrini:22}. 
For giant planets where the accretion of disc gas is the dominant source of the planetary metallicity, the C/O-ratio can be both super-stellar and sub-stellar depending on the chemical structure of the circumstellar disc where the giant planet was born \citep[][]{pacetti:22}. The precise determination of the stellar C/O-ratio therefore may provide information on the planet-formation history and the native circumstellar disc if the C and O abundances of giant planets can be determined \citep[][]{turrini:22,pacetti:22}. For observational validations see also \cite{carleo:22}, \cite{guilluy:22} and \cite{biazzo:22}.  
Recent studies further expanded the range of elements that can be used to investigate the formation history of gas giant planets to N \citep[][]{oberg:19,bosman:19,cridland:20,turrini:21a,turrini:22,pacetti:22} and S \citep[][]{turrini:21a,turrini:22,pacetti:22}. In particular, \cite{turrini:21a} and \cite{turrini:22} argue 
that the combined use of the abundance ratios of elements with different volatility like C, O, N, and S provides more unequivocal constraints on the planet-formation history than C/O alone. As an example, the C/N ratio will monotonically grow with migration for solid-enriched giant planets and decrease for gas-dominated giant planets, also in those cases where the C/O-ratio remains close to stellar \citep{turrini:21a,turrini:22}. Furthermore, \cite{turrini:21a} and \cite{turrini:22} showed that the information provided by these elemental ratios becomes immediately accessible once the planetary abundances are normalised to the stellar abundances and that the use of this normalised scale allows for the straightforward comparison between giant planets formed around different stars, as later supported by observational studies \citep[][]{kolecki:21,biazzo:22}. For a recent application with C and S on JWST data, see \cite{crossfield:23}.

While the observed and/or inferred elemental ratios from other nearby planetary systems constrains our knowledge about them, the Solar System will still remain a fundamental benchmark for theoretical planetary models. In this case, physical properties and isotopic anomalies found in meteoritic material provide the data to constrain the main features and structures in the earliest stages of the proto-solar disk \citep[e.g.,][]{burkhardt:19, brasser:20}, and/or the following core formation and accretion timescale of giant planets, in particular of Jupiter \citep[e.g.,][]{kruijer:17, nanne:19}.
The Solar System also shows us the importance of using the information on the stellar composition to validate exoplanetary observations. The comparison between Jupiter's elemental abundances and the Solar ones, in particular, may allow us to infer the Jovian Mg/O, Fe/O e Si/O ratios and quantity how the formation of refractory oxides alters the atmospheric C/O ratio of the giant planet \citep[][]{fonte:23}.

Whereas the collection and improvement of abundance data for stars within a few hundred parsecs has long been a priority (e.g., the GAIA-ESO and the GALAH surveys, see \citealt{gilmore:12} and \citealt{desilva:15}, respectively), such a capability for planets is in its infancy. In the next two decades, observatories like JWST \citep[James Webb Space Telescope,][]{beichman:14} and ARIEL \citep[Atmospheric Remote-sensing Infrared Exoplanet Large-survey,][]{tinetti:18,turrini:18,edwards:19} will expand and deepen the amount of planetary abundance data mostly from retrieved spectra from (hot) planetary atmospheres, gathering data in alignment with other existing and future facilities like e.g., TESS \citep[Transiting Exoplanet Survey Satellite,][]{ricker:15}, CHEOPS \citep[Characterizing ExoPlanets Satellite,][]{broeg:13} and PLATO \citep[Planetary Transits and Oscillations of stars,][]{rauer:14}. For a comprehensive list of present and future facilities and observatories, we refer to \cite{tinetti:18}. This will be the framework in the coming years where GCE, planet formation, nuclear astrophysics and stellar and planetary observations will be parallelized to provide a new comprehensive picture of stellar and planet systems formation and evolution.   
 
This work begins with an analysis of the stellar production (\S~\ref{sec: stars}) and the galactic chemical evolution (\S~\ref{sec: tk_plots/models}) of the elements that are crucial for the formation of planets, as discussed above: C, N, O, Mg, Si, and S. We present the main uncertainties associated with stellar observations and solar abundances in \S~\ref{sec: solar_wasp_nest}, followed by the main results in \S~\ref{sec: results}, and conclusions in \S~\ref{sec: summary}.

\section{Production of elements in stars}
\label{sec: stars}

It is well established that multiple stellar sources contributed to the chemical enrichment of the Milky Way disk. In particular, the main source of metals are Core-Collapse Supernovae \citep[CCSNe, see e.g.,][]{woosley:02,nomoto:13}, Asymptotic Giant Branch stars \citep[AGB stars e.g.,][]{herwig:05,karakas:14} and Thermonuclear Supernovae \citep[SNIa, e.g.,][]{hillebrandt:13}. We discuss each of these stellar metal sources that ultimately go into planet formation and present some examples below. 

\begin{figure}
    \centering
    \includegraphics[width=\columnwidth]{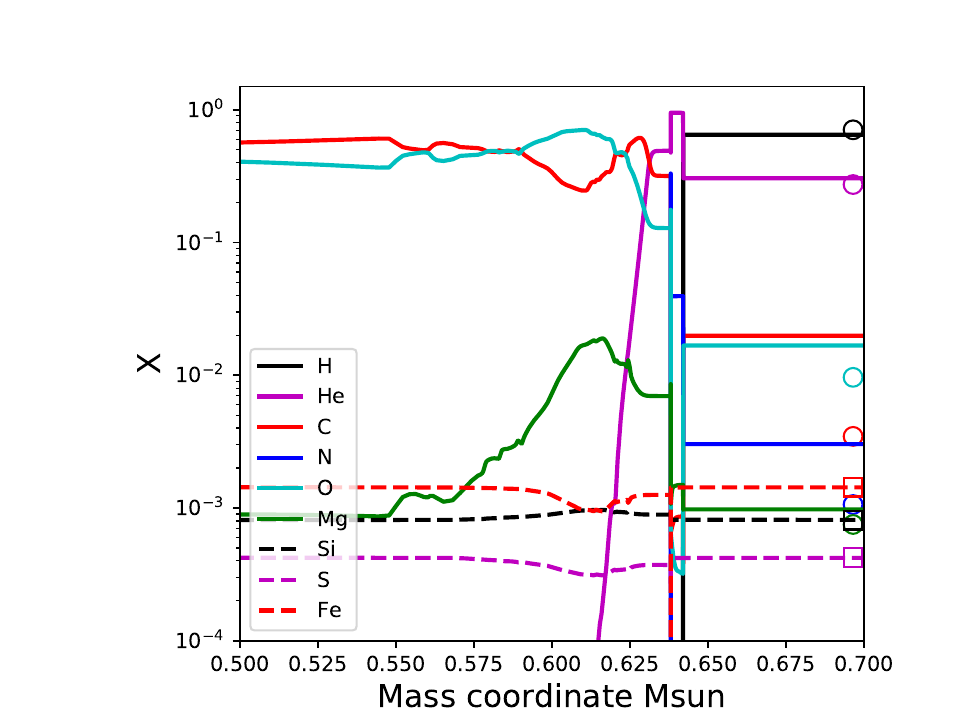}
    \includegraphics[width=\columnwidth]{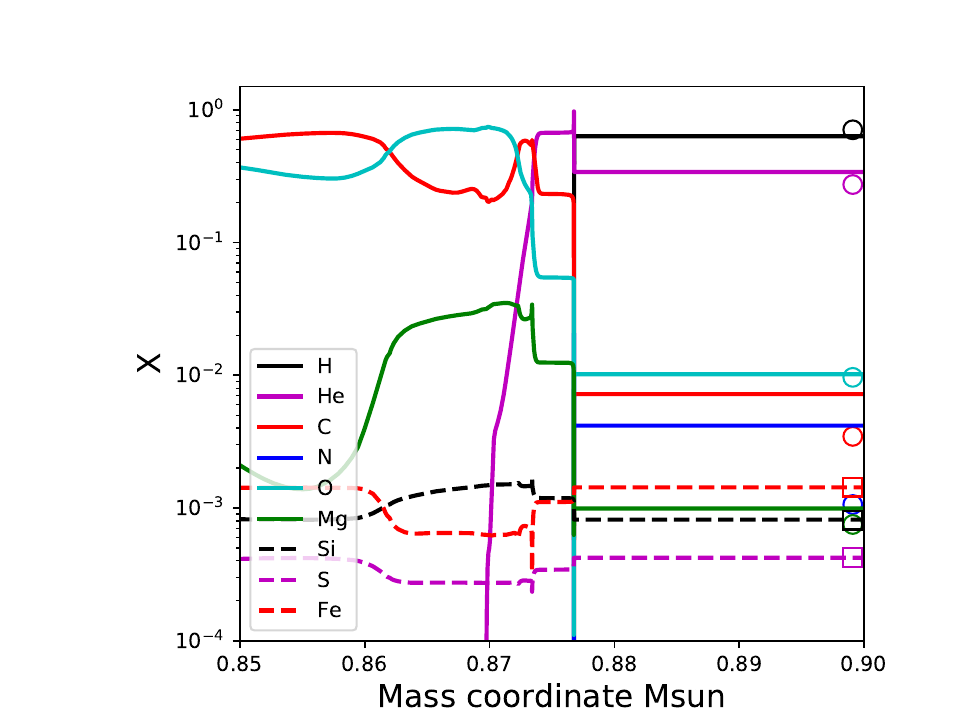}
    \caption{Model abundance (in mass fraction X) profile for the elements H, He, C, N, O, Mg, Si, S and Fe for two AGB stars with initial mass $M$=3M$_{\odot}$ (upper panel) and $M$=5M$_{\odot}$ (lower panel), and Z=0.02 \citep[][]{ritter:18}. The boundary of the AGB envelope is at about M=0.64M$_{\odot}$ and M=0.876M$_{\odot}$ for the 3M$_{\odot}$ and 5M$_{\odot}$ models, respectively. As a reference, the initial elemental abundances are reported with empty symbols of the same color of the reference lines (circles and squares are used for elements plotted with continuous lines, and squares for dashed lines).
    }
    \label{fig: agb_elements}
\end{figure}

For the metallicity range typical of stars in the solar neighbourhood, 
the bulk inventory of N can be explained by production from low-mass and intermediate-mass AGB stars (M$\lesssim$8M$_{\odot}$). AGB stars also make an amount of C comparable to the CCSNe contribution \citep[][]{kobayashi:20} or larger \citep[e.g., this work and][]{goswami:00, chiappini:03}, with the relative relevance of the two sources that is still matter of debate \citep[e.g.,][]{prantzos:94, romano:17}. 
The AGB phase is the last evolutionary stage before these low-mass and intermediate-mass stars eject their entire envelope into the Interstellar Medium (ISM) and evolve as planetary nebula, then develop into white dwarfs. With respect to the production of the elements, the AGB phase is crucial since this is when the bulk of new metals under consideration here are made and ejected. For recent studies of the nucleosynthesis in AGB stars, we refer to e.g., \cite{cristallo:15}, \cite{karakas:16},  \cite{jones:16}, \cite{bisterzo:17}, \cite{battino:19} and \cite{denhartogh:19}.
In Figure~\ref{fig: agb_elements}, the abundance profiles for 3M$_{\odot}$ and 5M$_{\odot}$ AGB stellar models by \cite{ritter:18} are shown close to the end of the evolution of these stars. The stellar envelope (to the right in the plot, where H is present) contributes to the enrichment of the ISM, while the interior part is the remnant that will form the future White Dwarf. 
In the 3M$_{\odot}$ model (top panel, Figure~\ref{fig: agb_elements}), the largest enhancement appears for the species C and N, with much smaller increase for O, Mg and He. In particular, the envelope becomes C-rich, with C/O $>$ 1. Conversely, the initial Si, S, and Fe show no modification. 
Only small variations are triggered by the activation of shell He-burning and shell H-burning in the He-intershell just below the envelope, which are not sufficient to modify the pristine elemental abundances. 
In the bottom panel of Figure~\ref{fig: agb_elements}, the AGB envelope in the 5M$_{\odot}$ model shows a relevant enrichment in N and C, with smaller increases of Mg, He, and O. 

Moreover, massive stars and CCSNe produced the bulk of O and Mg in stars in the MW disk, as well as relevant quantities 
of C, N, Si, S, and Fe. For these stellar objects, a substantial amount of the initial mass is lost by stellar winds, but most of the metals are ejected in the final CCSN explosion \citep[e.g.,][]{woosley:02}. A contemporary view, however, is that not all massive stars will successfully explode as CCSN. Depending on the progenitor structure and on details associated with the as yet poorly understood explosion mechanism, not all the material ejected by the SN explosion manages to escape and fall back to the compact central object \citep[e.g.,][]{fryer:12, ugliano:12, ott:18, fryer:18}. In the most extreme cases, all the ejecta fall back and only the winds contribute to the ISM enrichment. Present uncertainties of the CCSN mechanism, however, continue to undermine the efficacy of theoretical simulations \citep[e.g.,][]{fryer:18,mueller:16,janka:16,burrows:21}. Direct observations of recent CCSN remnants appear to confirm that a large variety of ejecta and energetic configurations are indeed possible \citep[e.g.,][]{nomoto:13,smartt:15,martinez:22}. It is unlikely that these different possibilities observed in recent CCSNe are mostly due to physics mechanisms seldom included in stellar model sets like rotation \citep[e.g.,][]{heger:00, hirschi:05, limongi:18} and/or more exotic types of supernova explosion like Pair-Instability Supernovae \citep[e.g.,][]{heger:02, kozyreva:14, takahashi:18, goswami:22}, which are expected to be more relevant for stars at low metallicity. The start of a successful or weaker CCSN explosion shown by the observations is instead most likely affected by the details of the stellar progenitor structure and by its interaction with the forming SN shock \citep[][]{wongwathanarat:13, burrows:21, varma:23}. For recent studies of the nucleosynthesis in massive stars and CCSNe, we refer to e.g., \cite{pignatari:16}, \cite{sukhbold:16}, \cite{ritter:18}, \cite{limongi:18}, \cite{curtis:19}, \cite{ebinger:20}, \cite{andrews:20}.

\begin{figure}
    \centering
    \includegraphics[width=\columnwidth]{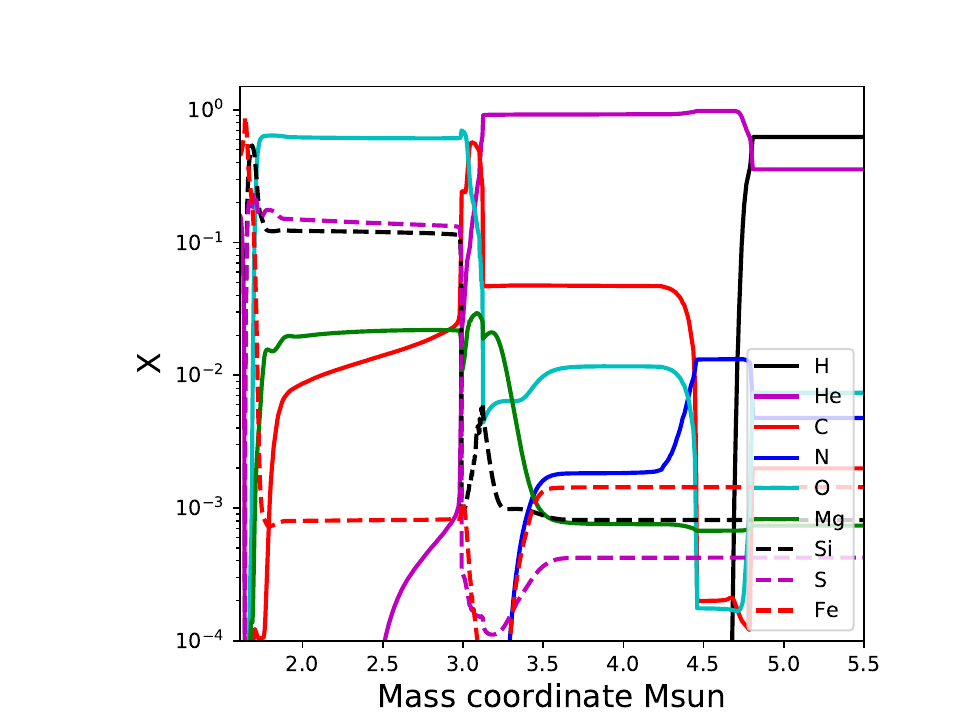}
    \includegraphics[width=\columnwidth]{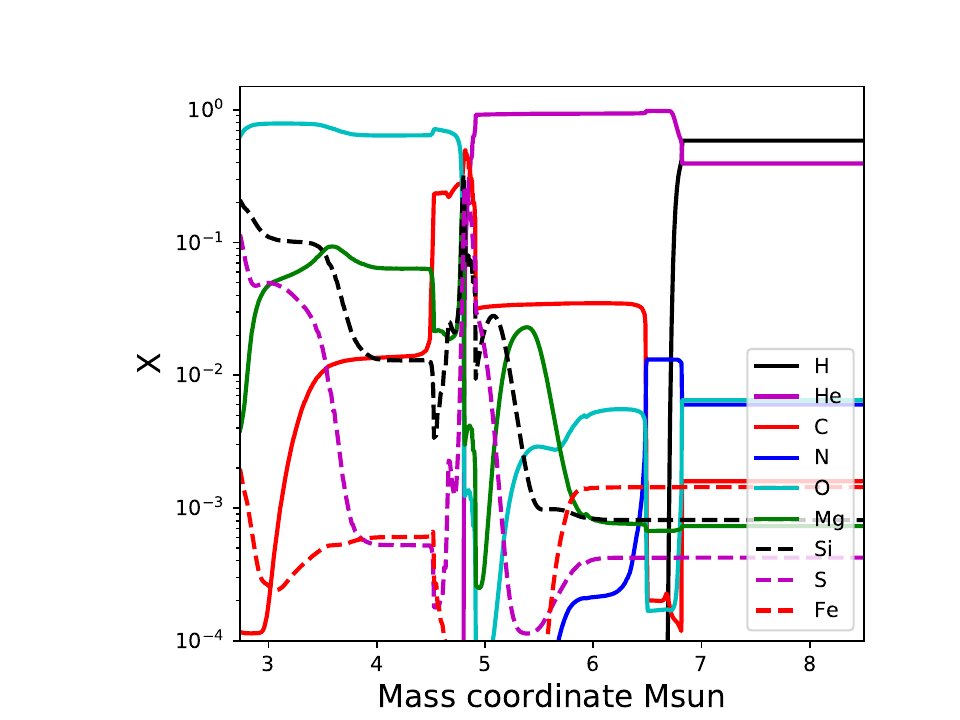}
    \includegraphics[width=\columnwidth]{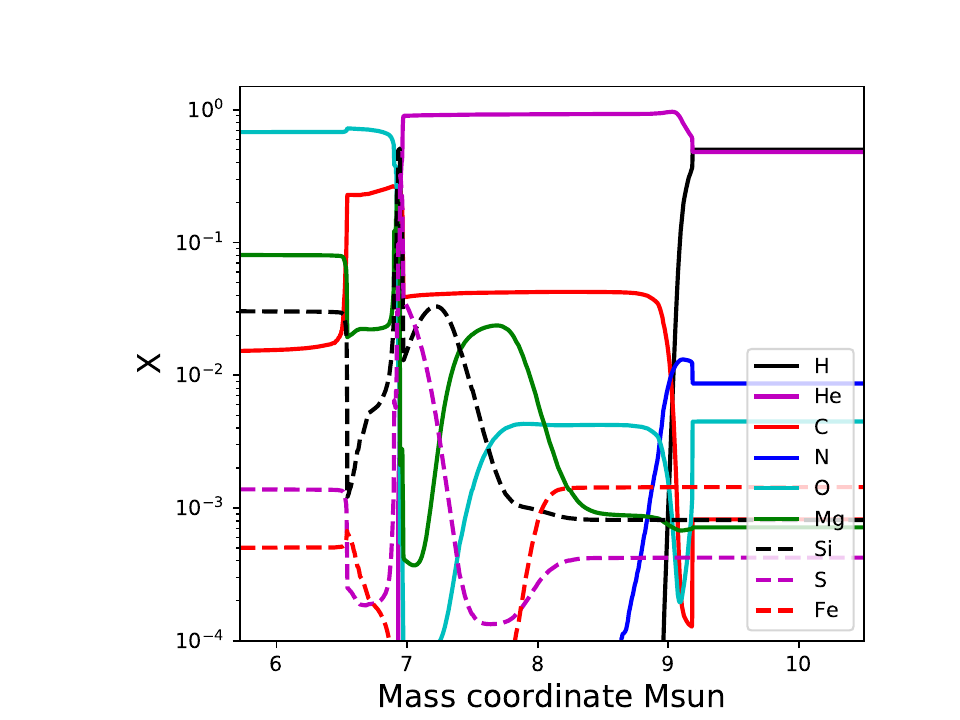}
    \caption{Abundance (in mass fraction X) profiles for the elements H, He, C, N, O, Mg, Si, S, and Fe for CCSN models from massive-star progenitors with initial mass M=15 M$_{\odot}$ (upper panel), M=20 M$_{\odot}$ (central panel) and M=25 M$_{\odot}$ (lower panel) and Z=0.02 \citep[][]{ritter:18}. 
The deepest ejecta are shown to the left, while the H-rich envelope is located to the right of the plots. Material that forms the compact central neutron star directly, or that will afterwards fall back onto it, is not considered.   
    }
    \label{fig: ccsn_elements}
\end{figure}

Figure~\ref{fig: ccsn_elements} shows the abundance profile of the CCSN ejecta of 15M$_{\odot}$, 20M$_{\odot}$ and 25M$_{\odot}$ models by \cite{ritter:18}. 
The 15M$_{\odot}$ model shows the most classical onion-layer structure (top panel, Figure~\ref{fig: ccsn_elements}). The explosive Si-burning and explosive O-burning ejecta are squeezed within the inner mass range 0.2-0.3 M$_{\odot}$, producing a peak of Fe (originally made as $^{56}$Ni, which will decay to $^{56}$Fe powering the peak of the CCSN lightcurve) and a peak of Si and S, respectively. 
Moving outward (toward the right in the plot), we find the O-rich and Mg-rich extended ashes of pre-supernova C fusion. The next large He-rich region represents the remains of the He-shell, with enrichment in C and O and the signature of explosive He-burning at the bottom (i.e., the Mg peak and a small Si peak at M=3.1M$_{\odot}$). At the top of the He-ashes are the remains of the pre-supernova H-shell, where C and O are consumed to make N. Finally, at the right edge of the plot is the H-rich envelope of the star. Such a structure is common and is shared across several sets of one-dimensional CCSN models \citep[e.g.,][]{woosley:95,thielemann:96}. 

The 20M$_{\odot}$ and 25M$_{\odot}$ models show a fundamental difference when compared to the 15M$_{\odot}$ model. In the 20M$_{\odot}$ model case the explosive Si-burning layers are not ejected, and only a fraction of the explosive O-burning escapes the gravitational bounds of the central compact object. Such a model represents what is dubbed a faint supernova \citep[e.g.,][]{heger:03, nomoto:13}, where the same nucleosynthesis as that in the 15M$_{\odot}$ star produces very different Si/Fe or Mg/Si ratios in the ejecta, due to the different outcome of the CCSN explosion. 

Faint CCSNe are considered potential sources of the peculiar stellar abundances observed in a number of old metal-poor stars, such as the so-called CEMP-no stars, i,e. carbon-enhanced, metal-poor stars with no enrichment of heavy elements \citep[e.g.,][and references therein]{beers:05,ishigaki:14, bonifacio:15, maeder:15, lee:19, zepeda:23}.
In fact, the most Fe-poor star known, SMSS J031300.36-670839.3, was proposed to carry the unique abundance signature of an Fe-poor faint CCSNe \citep[][]{keller:14, bessell:15, nordlander:17}. 
\cite{wehmeyer:19} discussed the contribution of faint CCSNe to explain the observed scatter of heavy element r-process enrichments with respect to iron in the early Galaxy. Those GCE simulations assume as main r-process sources neutron-star mergers and neutron star- black hole mergers, where black holes are considered as the remnants of faint CCSNe. Nevertheless, the contribution of faint CCSNe to GCE is not well defined. This applies also to the MW disk, 
although SN lightcurves and remnants of recent faint CCSNe explosions are observed \citep[e.g.,][]{nomoto:13}.

In Figure~\ref{fig: ccsn_elements}, the 25M$_{\odot}$ model exhibits an even more extreme case of faint supernovae, where the whole explosive O-burning layers are not ejected.
Even so, depending on the CCSN explosion energy and the progenitor structure, a more or less effective explosive He-burning can produce Mg, Si, and even S in relevant quantities (see mass coordinates M = 5M$_{\odot}$ and M = 7-7.5M$_{\odot}$ for the 20M$_{\odot}$ and 25M$_{\odot}$ models, in the central and bottom panels in Figure~\ref{fig: ccsn_elements}, respectively). 
Analysis of the presence of a C/Si zone at the bottom of the He shell during the CCSN explosion explains the anomalous abundance signature measured in C-rich presolar grains made in CCSNe \citep[][]{pignatari:13}. In terms of contribution to the total CCSN yields, the explosive He-burning contribution to Mg, Si and S is typically small compared to that of the explosive Si- and O-burning ejecta. Yet, in the case of faint CCSNe, the contribution of the external layers to the total ejecta of these elements can be relevant. The relative contribution between standard CCSNe and faint CCSNe to the GCE of the Milky Way disk and of the solar neighbourhood is not known. We will return to this point in the next section. 

There are additional major uncertainties that need to be considered. According to basic stellar evolution principles, and considering the nuclear reactions involved, O and Mg are produced and ejected in the same CCSN layers and therefore ought to scale nearly perfectly to each other in their GCE history. Pre-supernova He-burning makes O via the $^{12}$C($\alpha$,$\gamma$)$^{16}$O reaction, together with a small amount of Mg. During C-fusion, O is left mostly unchanged whereas Mg is made efficiently in the form of $^{24}$Mg via the $^{20}$Ne($\alpha$,$\gamma$)$^{24}$Mg reaction. In the following evolutionary stage, O-fusion destroys both O and Mg. Therefore, the Mg/O ratio should be quite similar in the C-burning ashes of all CCSNe \citep[e.g., ][]{arnett:85,thielemann:85,chieffi:98}. Then again, explosive He-burning decouples O and Mg, where O feeds the production of Mg (and eventually Si and S) via a sequence of $\alpha$-capture reactions \citep[][]{pignatari:13}.  

In a similar way, the production of Si and S is generally expected to be connected, since these elements are produced together by the two main O-burning fusion channels in the form of their stable isotopes $^{28}$Si and $^{32}$S, respectively \citep[e.g.,][]{thielemann:85}. It is apparent that abundance profiles of Si and S, however, change significantly in the models shown in Figure~\ref{fig: ccsn_elements}. In the C-burning ashes of the 20 M$_{\odot}$ and 25 M$_{\odot}$ models, nuclear reactions have already started to make Si, while S is only marginally modified. In these CCSN explosions, it is evident that the C-ashes and eventually explosive He-burning products shape the S/Si ratio in the yields. In the 15M$_{\odot}$ model, we find instead that the ratio S/Si$<$1 typical of O-burning is shown only in the small region shaped by the explosive O-burning (M$\sim$1.6M$_{\odot}$). The C-ashes instead show a ratio S/Si$>$1, with both Si and S being more than 10$\%$ in mass fraction. This is the signature of the C-O shell merger, which occurs during the pre-SN evolution of the star and allows for the pollution of the C shell with O-burning products, with a signature quite different compared to pure O-burning material. 

The study of the interaction between the convective C-shell and O shell up to a complete C-O shell mergers was considered in previous nucleosynthesis studies \citep[][]{rauscher:02,ritter:18a,clarkson:18}. In these conditions, the predictive power of one-dimensional models is limited and multi-dimensional hydrodynamics simulations are required \citep[e.g.,][]{meakin:06, cristini:17, andrassy:20, clarkson:20}. The potential relevance of these events in triggering the asymmetries in the progenitor structure favouring successful CCSN explosions is also a matter of debate \citep[e.g.,][]{janka:17, ott:18}. 
For the purpose of our analysis, this implies that some scatter can be expected for the S/Si ratio in CCSN ejecta. 
The same scatter could be possibly visible in stellar observations, if observational uncertainties are small enough \citep[][]{chen:02,reddy:03,reddy:06}. Therefore, the common assumption made in forward simulations of planetary formation, where the initial Si and S abundances scale together with respect to the solar abundances \citep[e.g.,][]{bitsch:20} needs to be carefully checked against the spectroscopic data from the stellar host, or with GCE simulations, when S observations are not available. 
 
The last stellar source we consider here are Supernovae Type-Ia (SNIa) which are responsible for the synthesis of the bulk of Fe measured in stars in the Milky Way disk, as well as a significant fraction of Si and S. Uncertainties surround the relative importance of the single-degenerate scenario (where the SNIa explosion is triggered by accretion of material on a CO-WD reaching the Chandrasekhar mass) with respect to the double-degenerate scenario (where the SNIa explosion is triggered by the merger of two CO-WDs) in the SNIa population of galaxies at present. \citep[][]{hillebrandt:13}. A complementary matter of discussion is the relative contributions of SNIa explosions from Chandrasekhar-mass progenitors and sub-Chandrasekhar-mass explosions, where also in the single-degenerate scenario a SNIa may explode before reaching 1.44M$_{\odot}$. To explain the GCE of the ratio [Mn/Fe]\footnote{With spectroscopic square-bracket notation we refer to the ratio of two elements X and Y, represented as the logarithm of the ratio relative to the same ratio in the Sun: [X/Y] = log$_{10}$(X/Y)$_{\rm star}$ $-$ log$_{10}$(X/Y)$_{\odot}$.}
in the MW disk, \cite{kobayashi:20a}, \cite{seitenzahl:13} and \cite{eitner:20} concluded that only 75$\%$, 50$\%$ and 25$\%$ of all the SNIa population should be from Chandrasekhar-mass progenitors, respectively. On the other hand, based on observational surveys of early-type galaxies \cite{woods:13} and \cite{johansson:16} determined that only a few per cent of all SNIa should be from Chandrasekhar-mass progenitors. These two conclusions are obviously at odds with one another and warrant further study. Finally, it is also important to note that from a study of 407 SNIa in older massive Red-Sequence galaxies and younger less massive Blue-Cloud galaxies, \cite{hakobyan:20} showed that about one-third of all SNIa events are peculiar, possibly related to the contribution from double-degenerate WD mergers, and that the diversity of SNIa progenitors may also be due to the age of the progenitor. Such a phenomenological diversity is difficult to capture within GCE simulations, but it will need to be considered in the future. 

\begin{figure}
    \centering
    \includegraphics[width=\columnwidth]{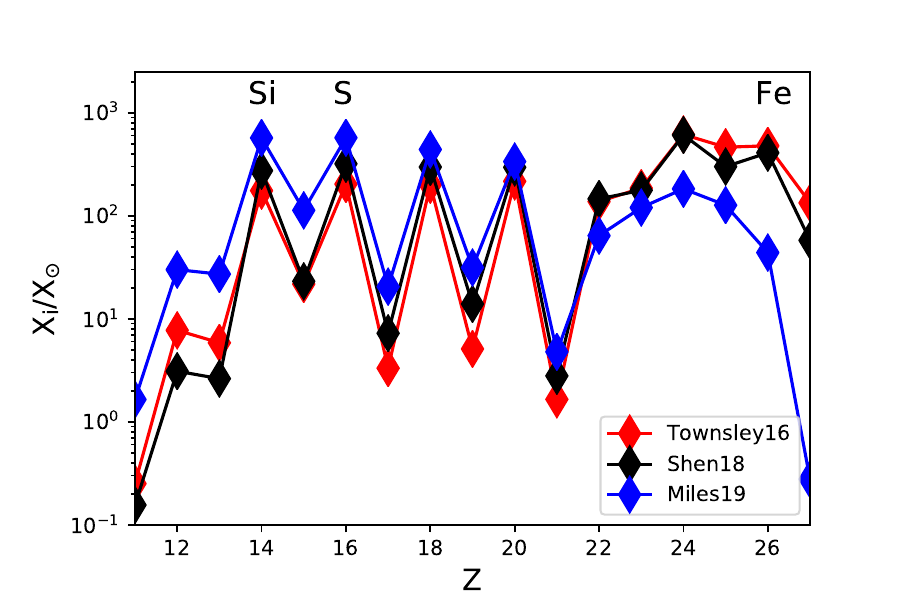}
    \caption{Elemental abundances (normalized to the solar values) in the mass region between Si and Fe for the SNIa models by \protect\cite{townsley:16}, \protect\cite{shen:18} and \protect\cite{miles:19}, with progenitor masses 1.37M$_{\odot}$, 1M$_{\odot}$ and 0.8M$_{\odot}$, respectively. Yields are calculated by \protect\cite{keegans:23}.}
    \label{fig: snia_prodfac}
\end{figure}

Figure~\ref{fig: snia_prodfac} shows the abundances normalized to their solar values for the elements between Mg and Fe for SNIa models computed with the same initial $^{22}$Ne abundance equivalent to a metallicity of Z=0.014 \citep[][]{keegans:23}. The three models correspond to explosions of different masses of WD: 1.37 M$_{\odot}$ \citep[][]{townsley:16}, 1 M$_{\odot}$ \citep{shen:18} and 0.8 M$_{\odot}$ \citep{miles:19}. For the elements around Fe, the 1.37 M$_{\odot}$ and 1 M$_{\odot}$ progenitors produce very similar distributions, while the lowest mass progenitor produces far less of these in absolute abundances. For Fe itself, the production factor in the low-mass case is almost an order of magnitude lower than in the other two models. On the other hand, for the elements of interest in our discussion here, Si and S production factors are similar in the figure, and show only minor variations between the different models. This is because Si and S are typically produced in the same explosive conditions, and they are ejected together in the SNIa ejecta. This makes their production less sensitive to the relevant stellar uncertainties. 
Note that the contribution timescale (or delay-time) to GCE from different types of SNIa explosions may change depending on how the explosion was triggered in the stellar progenitors. Standard Chandrasekhar-mass SNIa accreting H will have a long delay-time \citep[in the order of 1 Gyr, see e.g.,][]{ruiter:09}. This would be the case for the models by \cite{townsley:16}, shown in Figure~\ref{fig: snia_prodfac}. Sub-Chandrasekhar-mass SNIa accreting He from a WD companion have a comparably long delay-time to the standard Chandrasekhar-mass SNIa \citep[e.g.,][]{gronow:21}. This would be the scenario compatible with the \cite{shen:18} and the \cite{miles:19} models shown in Figure~\ref{fig: snia_prodfac}. On the other hand, in case He would have been accreted by a He-burning star, the delay-time of Sub-Chandrasekhar-mass SNIa would be much shorter \citep[in the order of few hundred million years, see e.g.,][]{ruiter:14}. However, in the present context where different types of SNIa produce yields with similar S/Si abundance ratios, the GCE impact of varying the delay-time for different SNIa progenitors would be marginal.

\section{GCE Models and simulations}
\label{sec: tk_plots/models}

To explore the GCE of the solar neighbourhood we produced a set of 198 models; 
we use the \texttt{OMEGA} GCE code \citep[One-zone Model for the Evolution of GAlaxies][]{cote17} to calculate elemental abundance ratios in the ISM for several choices of stellar-yield sets and of the contribution from faint CCSNe yields. 
\texttt{OMEGA} has also an option for including any number of extra enrichment sources, in addition to the mass- and metallicity-dependent yields from AGB stars, massive stars, CCSNe, and SNe Ia described above which are included by default in the code. The \texttt{OMEGA} code, the simple stellar population (SSP) model \texttt{SYGMA} \citep[Stellar Yields for Galactic Modeling Applications][]{rit18}, and the \texttt{STELLAB} (STELLar ABundances) module used to plot the observational data are all part of the publicly available NuGrid chemical evolution framework\footnote{\url{http://nugrid.github.io/NuPyCEE}}. Here we give a brief description of the \texttt{OMEGA} code. 

The evolution of the mass of the gas-reservoir in the Galaxy as a function of time can be expressed in terms of the gas inflow rate $\dot{M}_{\text{inflow}}(t)$, the rate at which gas is ejected from stars $\dot{M}_{\text{ej}}(t)$, the star-formation rate (SFR; $\dot{M}_{\star}(t)$), and the outflow rate of gas from the galaxy $\dot{M}_{\text{outflow}}(t)$ (e.g., \citealt{tinsley:80,gibson:03,matteucci:21}), where 

\begin{equation}
    \dot{M}_{\text{gas}}(t) = \dot{M}_{\text{inflow}}(t) + \dot{M}_{\text{ej}}(t) - \dot{M}_{\star}(t) - \dot{M}_{\text{outflow}}(t).
\end{equation}
The gas ejected by stars is assumed to mix instantaneously with the ISM, so that the metallicity of the gas-reservoir is always homogeneously distributed, as is the case for all one-zone GCE models. 
\texttt{OMEGA} also includes a simple treatment of galactic inflows and outflows. Since stellar feedback is assumed to drive the gas outflows, then


\begin{equation}
    \dot{M}_{\text{outflow}}(t)=\eta \dot{M}_{\star}(t),
\end{equation}
where the mass-loading factor $\eta$ is a free parameter controlling the magnitude of the gas outflow (e.g., \citealt{murray:05,muratov:15}). The inflow of primordial matter is a catalyst for star-formation in the Galaxy, however \texttt{OMEGA} uses the star-formation history to determine the SFR rather than the inflow rate. For further details regarding the treatment of galactic inflows in \texttt{OMEGA} we refer the reader to \cite{cote17}.  

At each timestep, \texttt{OMEGA} creates a single stellar population (SSP) with a mass proportional to the SFR at that time, which is proportional to the total mass of gas in the Galaxy following the Kennicutt-Schmidt law \citep{schmidt59, kennicutt}:

\begin{equation}
    \dot{M}_{\star}(t)=f_\star \dot{M}_{\text{gas}}(t),
\end{equation}
where $f_{\star}=\epsilon_{\star}/\tau_{\star}$ is a combination of the dimensionless star-formation efficiency $\epsilon_{\star}$ and the star-formation timescale $\tau_{\star}$. All stars in a given SSP have the same initial metallicity - that of the gas-reservoir - since they are all assumed to have formed from the same parent gas cloud. At each timestep, the \texttt{SYGMA} code calculates the combined integrated yields from all the SSPs that are currently in the Galaxy. However, each star in an SSP will eject material at different times according to the delay-time distribution (DTD) function of the specific progenitor. For a galaxy with $j$ SSPs at time $t$, the combined integrated yield returned by \texttt{SYGMA} is given by

\begin{equation}
\dot{M}_{\mathrm{ej}}(t)=\sum_{j} \dot{M}_{\mathrm{ej}}\left(M_{j}, Z_{j}, t-t_{j}\right),
\end{equation}
where $Z_j$ and $M_j$ are the initial mass and metallicity of the $j^{\text{th}}$ SSP that was born at time $t_j$. Significantly for the present study, if an extra source (e.g., faint CCSNe) is added to \texttt{SYGMA}, then a DTD must also be assigned to it, in addition to specifying the yields, the number of events per M$_{\odot}$, and the mass ejected per event. 

Here, we assume that a fraction of massive stars $f_{\text{faint}}$ with initial mass $M$ above a given mass threshold $M_{\text{min}}$ will explode as faint CCSNe, rather than wholly as regular CCSNe. 
At the moment, it is still unclear if faint CCSNe have a significant impact on the GCE. Indeed, their presence may be hidden in the variations due to the uncertainties affecting the yields of CCSNe. Although previous GCE studies adopted various CCSN yields from the literature, they might have mitigated faint CCSN uncertainties by varying other parameters such as the slope of the IMF, the range of stellar masses contributing to the nucleosynthesis, the star-formation efficiency, and the strength of large-scale gas flows \citep[e.g.,][]{gibson:97,romano:10,molla:15,cote:17,philcox:18}. 
Given the lack of constraints for the value of $f_{\text{faint}}$, we consider values between 0 and 1 in order to fully explore the potential impact of faint CCSNe.

Since both faint and regular CCSNe share the same type of progenitors, the only defining factor between the two types of explosion mechanisms are their yields. Therefore, for M>M$_{\text{min}}$ we make the simplification that faint and regular CCSNe occur at the same frequency in the Galaxy, but we apply a $f_{\text{faint}}$ correction factor to the yields of faint CCSNe, and a $1-f_{\text{faint}}$ correction factor to yields of the latter. We make no modifications to the yields of CCSNe that result from massive-star progenitors with M<$M_{\text{min}}$. The models considered for this work are summarized in Table~\ref{tab: list_tk_plots/models}. The adopted name scheme is o<yield$\_$identifier><faintSN$\_$model><faintSN$\_$weight>. The term $M_{\rm up}$ represents the mass of the most massive stars that can contribute to the GCE. All stars with an initial mass above $M_{\rm up}$ are assumed to directly collapse into a black hole without any ejecta.

With regards to stellar yields, for AGB stars the oK06, oR18, oR18d and oR18h sets use \cite{ritter:18}, while the oK10 and oL18 models use the yields by \cite{karakas:10}. 
The CCSN yields by \cite{kobayashi:06} are adopted in the oK06 and oK10 models, and non-rotating massive-star yields by \cite{limongi:18} are adopted in the oL18 models. The remaining sets of models use different yield setups by \cite{ritter:18}. In particular, oR18d, oR18 and oR18h use the same AGB stellar yields, but oR18d use the CCSN models with a delayed explosion setup \citep[][]{fryer:12}, oR18h is the same as oR18d but the 12M$_{\odot}$ yields are not included, while for oR18 CCSN models adopt a classical mass cut defined by the electron fraction ($Y_e$) jump in the progenitor structure \citep[][]{cote:17}. 
For all GCE models, the \texttt{OMEGA} default W7 SNIa yields by \cite{iwamoto:99} are used for all metallicities. Notice that within our GCE platform there are multiple sets of SNIa yields available \citep[e.g.,][]{lach:20, grunov:21}. However, since for the elements considered in this study the impact of using different SNIa yields is much smaller compared to CCSN yields, we decided to not modify the default setup for the models discussed here.  

For each GCE model setup described above, 
a number of models are generated with five 
faint CCSN weighting factors of 10$\%$, 25$\%$, 50$\%$, 75$\%$ 
and 100$\%$ respectively, and two types of faint CCSNe: the stellar model of 20 M$_{\odot}$ by \cite{ritter:18} (model m20, Figure~\ref{fig: ccsn_elements}, central panel), and 25 M$_{\odot}$ (model m25, Figure~\ref{fig: ccsn_elements}, lower panel). The additional contribution of faint CCSNe is considered for stellar progenitor masses larger than $M_{\rm min}$ = 15M$_{\odot}$. The weighting factor mentioned above provides the relative contribution of faint CCSNe compared to default SNe yields for the mass range M > $M_{\rm min}$. 

Analogous GCE model sets are generated considering three different 
$M_{\rm up}$: 20 M$_{\odot}$, 40 M$_{\odot}$ and 100 M$_{\odot}$ as the value of M$_{\rm up}$ is uncertain and it is still a matter of debate. While $M_{\rm up}$ = 40 M$_{\odot}$ and 100 M$_{\odot}$ are more typical choices, we also considered in our calculations the lower value at 20M$_{\odot}$, which would be more consistent with observations from CCSN remnants and their progenitors \citep[e.g.,][]{smartt:15,davies:18}.
Such a lack of CCSNe from massive stars with initial mass M $\gtrsim$ 20 M$_{\odot}$ also seems to be plausible for stellar simulations, where a relevant population of massive stars with initial mass larger than 20 M$_{\odot}$ may fail to explode \citep[e.g.,][]{sukhbold:16,fryer:18}, and it is indepently confirmed by direct element observations of late-time supernova spectra \citep[e.g.,][and references therein]{jerkstrand:14, jerkstrand:15, silverman:17}.

\begin{table}
	\centering
	\caption{Summary of the GCE models used in this work with their properties: name (used in the text), stellar yields set identifier (see the details in the text), faint CCSN model if included, and its frequency with respect to the default yields. The extra sources "m20" and "m25" correspond to the M=20 M$_{\odot}$ and the M=25 M$_{\odot}$ faint CCSN models by \protect\cite{ritter:18}, respectively (Figure~\ref{fig: ccsn_elements}, central panel and lower panel). The last column provides the upper limit of stellar masses contributing to the GCE (in solar masses). The models name are designed as follows: o<yield$\_$identifier><faintSN$\_$model><faintSN$\_$weight>, where the initial o stands for \texttt{OMEGA}. 
}
	\label{tab: list_tk_plots/models}
	\begin{tabular}{lccc} 
		\hline
		yield$\_$identifier & faintSN$\_$model & faintSN$\_$weight  & M$_{\rm up}$ (M$_{\odot}$) \\
		\hline
		K06    &    -     &  no,                  & 20, 40, 100 \\
                       & m20, m25 &  f0p10, f0p25, f0p50, & 20, 40, 100 \\
                       & m20, m25 &  f0p75, f1p00  & 20, 40, 100 \\
		\hline
		K10    &    -     &  no,                  & 20, 40, 100 \\
                       & m20, m25 &  f0p10, f0p25, f0p50, & 20, 40, 100 \\
                       & m20, m25 &  f0p75, f1p00  & 20, 40, 100 \\
		\hline
		R18    &    -     &  no,                  & 20, 40, 100 \\
                       & m20, m25 &  f0p10, f0p25, f0p50, & 20, 40, 100 \\
                       & m20, m25 &  f0p75, f1p00  & 20, 40, 100 \\
		\hline
		R18d   &    -     &  no,                  & 20, 40, 100 \\
                       & m20, m25 &  f0p10, f0p25, f0p50, & 20, 40, 100 \\
                       & m20, m25 &  f0p75, f1p00  & 20, 40, 100 \\
		\hline
		R18h   &    -     &  no,                  & 20, 40, 100 \\
                       & m20, m25 &  f0p10, f0p25, f0p50, & 20, 40, 100 \\
                       & m20, m25 &  f0p75, f1p00  & 20, 40, 100 \\
		\hline  
		L18   &    -     &  no,                  & 20, 40, 100 \\
                       & m20, m25 &  f0p10, f0p25, f0p50, & 20, 40, 100 \\
                       & m20, m25 &  f0p75, f1p00  & 20, 40, 100 \\
        \hline
	\end{tabular}
\end{table}

That the chemical evolution of the solar neighbourhood is complex and a challenging task for GCE, is an understatement \citep[e.g.,][]{goswami:00, kobayashi:11, molla:15, prantzos:18, kobayashi:20, prantzos:23}. This is because stars that are observed within a few hundred parsecs from the Sun may have formed from material with radically different chemical evolution histories from one another and from our star. In fact, the observed [Fe/H] varies by about an order of magnitude and some stars may have formed after the Sun, or billions of years earlier and shortly after the formation of the Galaxy \citep[e.g., HD140283][]{siqueira-mello:15}.
This variety should be taken into account, 
because the relevance of different stellar sources varies during the galactic evolution timescale \citep[e.g.,][]{matteucci:86}. Stellar ages of nearby stars can be derived with a precision of about 1 billion years \citep[e.g.,][]{nissen:20}. We emphasize that the age of the star needs to be considered together with the stellar abundances in order to fully understand the elemental composition directly observed using spectroscopic data, and that can only be inferred \citep[e.g.,][]{spina:16}. 

\begin{figure}
    \centering
    \includegraphics[width=\columnwidth]{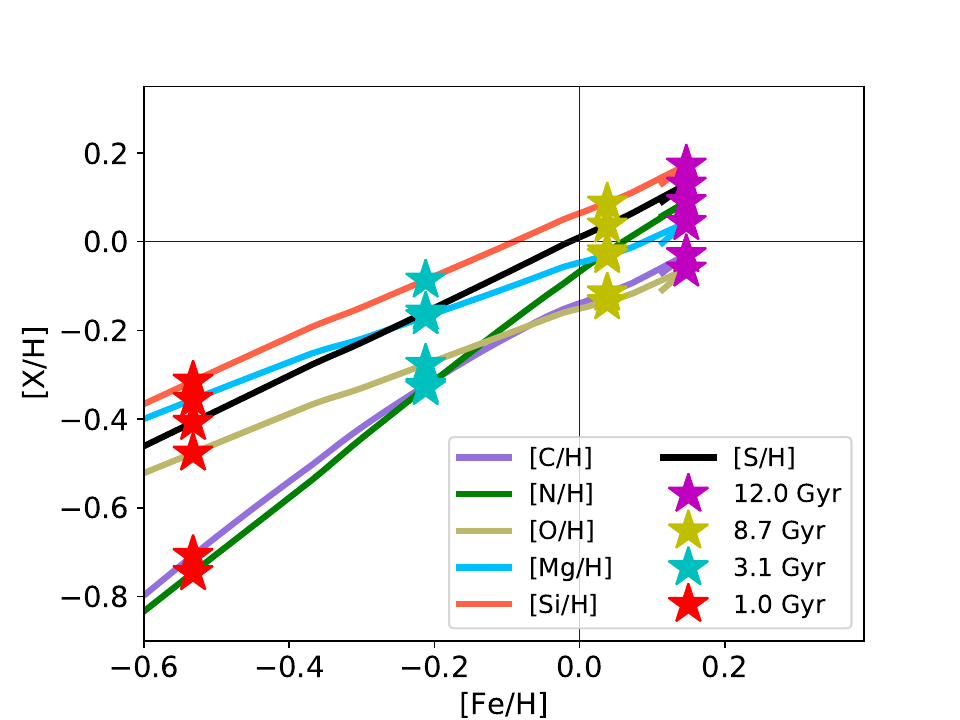}
    \caption{The evolution of 
    the elements of interest with [Fe/H] for the model oK06no (see Table~\ref{tab: list_tk_plots/models} for the model name explanation). Stars of different colors represent evolution times from the beginning of the simulation. The time 8.7 Gyr corresponds to the formation of the Sun. }
    \label{fig: tk_plots/ok06no_xh}
\end{figure}

As an example, Figure~\ref{fig: tk_plots/ok06no_xh} shows the evolution with [Fe/H] of the elements in model oK06no along with some reference evolution timescales. Model oK06no provides a good match to the solar abundances for the elements considered at the time when the Sun formed (8.7 Gyr). The predicted [Fe/H] is about 10$\%$ higher than solar. Carbon (mostly made by AGB stars) and O (mostly made by massive stars) are both about 60$\%$ too low. The elements N, Mg, and S are reproduced within 10$\%$ for solar material, while Si (made by both massive stars and SNIa) is about 20$\%$ higher. However, this same model, when considering both the chemical enrichment and evolution timescale, may not be appropriate to use for another star, even one with solar metallicity. 
We will use the same four reference GCE timescales shown in Figure~\ref{fig: tk_plots/ok06no_xh} (1.0, 3.1, 8.7, and 12.0 Gyr) in the next section, to also analyze the evolution curves of elemental ratios with respect to time. 


\section{Stellar observation samples and solar abundances}
\label{sec: solar_wasp_nest}

Figure~\ref{fig: solar_ratios} shows the abundances from two different stellar samples from the galactic disk by \cite{reddy:03} and \cite{reddy:06} (R03, R06) and \cite{suarez-andres:18} (S18), together with their reference solar ratios. No scaling or normalization has been applied to the C/O and Mg/Si ratios. 
The two observed distributions exhibit clear discrepancies, where the S18 data have on average both higher C/O and Mg/Si ratios. 
This difference also appears in the solar abundances used in these surveys, with the S18 C/O and Mg/Si being factors of 1.38 and 1.29 higher than those of R03 and R06, respectively.
Figure~\ref{fig: solar_ratios} also plots C/O and Mg/Si ratios from several other solar chemical composition studies. 
The C/O values range by a factor of two, from 0.83 \citep[][D10 HARPS]{delgadomena:10} to 0.43 \citep[][AG89]{anders:89}. 
The Mg/Si ratios are also scattered, varying between 0.83 \citep[][based on their own solar analysis]{reddy:03}, and 1.23 \citep[][A09]{asplund:09}.
These variations indicate that solar abundance differences are not limited to the two stellar surveys employed in our investigation.
The details of these abundance determinations can be found in the survey papers. We report below some general remarks about observational uncertainties.

\subsection{Comparing Results from the Abundance Surveys Considered Here}
\label{sec: observed_abundance_details}

\begin{figure}
    \centering
    \includegraphics[width=\columnwidth]{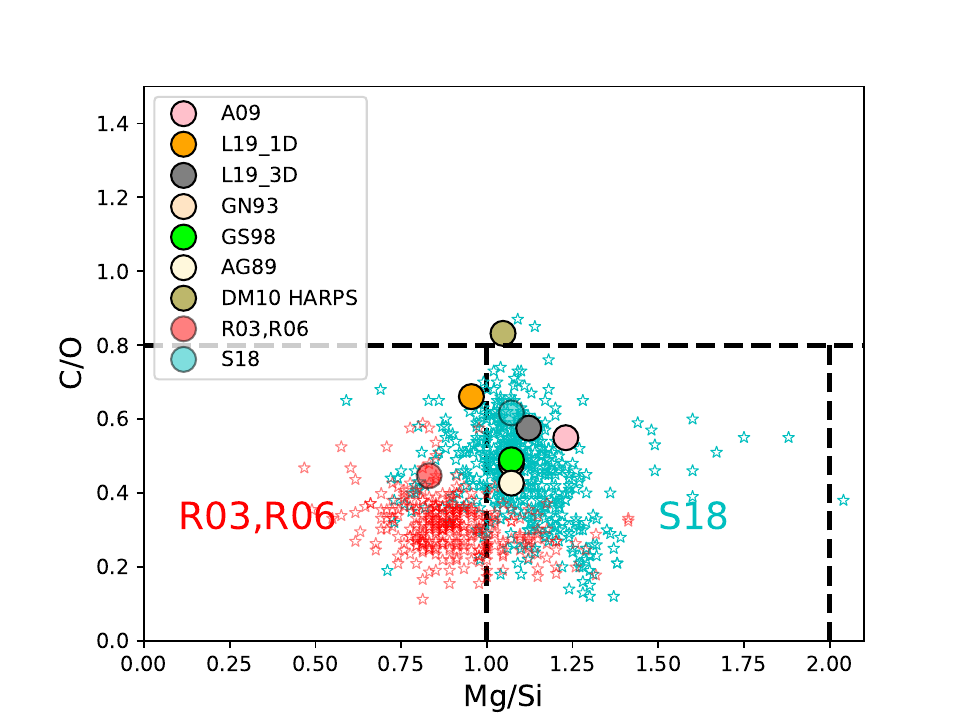}
    \caption{The C/O and Mg/Si ratios for stars by \protect\cite{reddy:03} and \protect\cite{reddy:06} (R03 and R06, red stars) and \protect\cite{suarez-andres:18} (S18, cyan stars). The solar ratios used as reference in the two papers are reported, together with a collection of other solar abundances from: \protect\cite{asplund:09} (A09), \protect\cite{lodders:19} using 1D and 3D models for the solar atmosphere (L19$\_1$D and L19$\_$3D, respectively), \protect\cite{grevesse:93} (GN93), \protect\cite{grevesse:98} (GS98), \protect\cite{anders:89} (AG89) and the solar abundances measured using HARPS \citep[][]{delgadomena:10} (D10 HARPS). Ratios discussed by \protect\cite{bond:10} as relevant to the chemistry and dynamics of rocky planets are also reported as black-dashed lines.}
    \label{fig: solar_ratios}
\end{figure}

\begin{figure}
    \centering
    \includegraphics[width=\columnwidth]{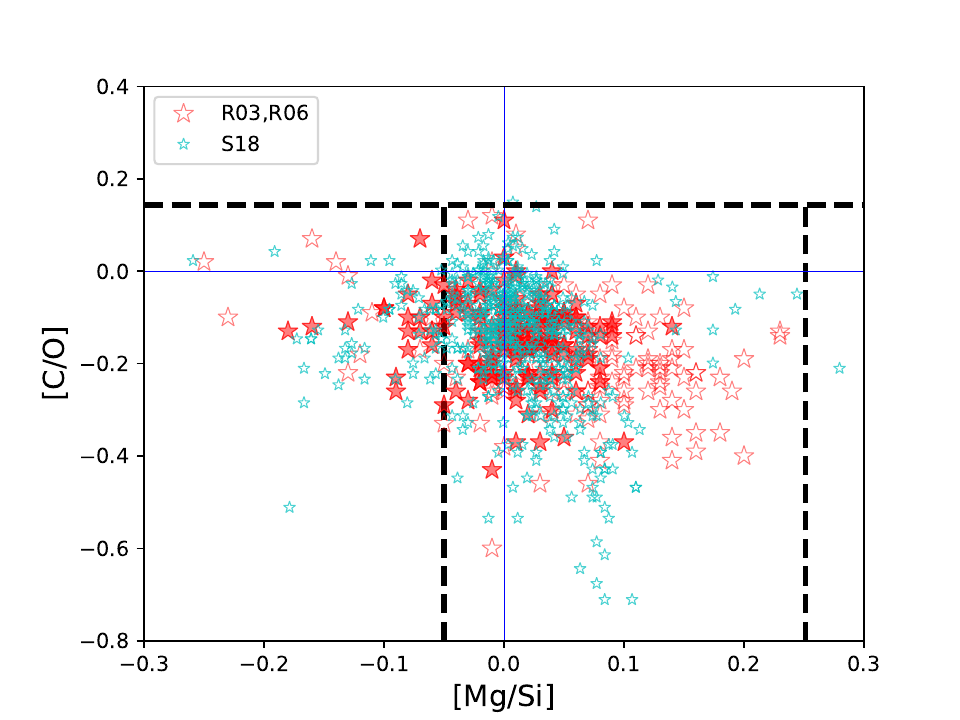}
    \caption{The [C/O] and [Mg/Si] ratios for stellar data by R03 and R06 (full and open red stars, respectively) and S18 (cyan stars). The indicative abundance errors reported by R03 and R06 are in the order of $\pm$0.2 dex for [C/O] and $\pm$0.1 dex for [Mg/Si]. The average errors reported by S18 are about a factor of two smaller. The same reference lines of Figure~\ref{fig: solar_ratios} are reported, normalized to solar (L19$\_$3D by \protect\cite{lodders:19}). The reference solar L19$\_$3D is very close to the solar of S18 (see Figure~\ref{fig: solar_ratios} for comparison).}
    \label{fig: ratios_s18_r03}
\end{figure}

Some of the apparent clashes in Figure~\ref{fig: solar_ratios} between the C/O and Mg/Si ratios of R03, R06 and S18 can be alleviated by a normalisation to the solar abundances derived with the same analysis setup.
In this way, it is possible to remove at least some of systematic uncertainties. 
In Figure~\ref{fig: ratios_s18_r03}, we report the same stellar ratios shown in Figure~\ref{fig: solar_ratios}, but in logarithmic notation and normalized to their respective (and different) solar reference ratios. 
Because of these, the two stellar samples show a much better overlap compared to Figure~\ref{fig: solar_ratios}. 
The [C/O] ranges are similar, and the [Mg/Si] is also consistent (albeit with significantly more scatter), when excluding outliers with [C/O] $\lesssim$ $-$0.5 and [Mg/Si]$\lesssim$ $-$0.2. 
The set by S18 is concentrated around the solar values or slightly higher, while R03 and R06 data are more scattered toward larger Mg/Si values, up to about 0.2 dex.
The larger ranges of both C/O and Mg/Si in the R03 and R06 sample combined reveals that the two surveys may not draw their targets from the same Galactic metallicity/kinematic samples. Indeed, from Figure~\ref{fig: ratios_s18_r03} we can see that the high Mg/Si-signature is mostly given by R06 stars, which are mostly thick-disk stars. The R03 stars are instead in better agreement with the S18 scatter.

The abundance surveys considered here were conducted using similar methods.
They both used model atmospheres from the ATLAS grid \citep{kurucz:11,kurucz:18} and performed equivalent width and synthetic spectrum analyses with the current versions of the same LTE plane-parallel code \citep{sneden:73}. On the other hand, R03, R06 and S18 surveys have different selection functions. The HARPS S18 sample is a subset of $\sim$500 stars from the HARPS \cite{adibekyan:12} sample, who chose their objects based on suitability for radial velocity surveys (slowly rotating FGK stars
without chromospheric activity); the R06 sample was primarily selected from kinematically thick Galactic disk but within a given distance and R03 mostly includes selected stars from the kinematically thin Galactic disk.
Different software was employed to measure equivalent widths, but the basic procedure is straightforward and accurate for unblended spectral lines.
The two surveys differ either in the choice of species for study or in the selection of the individual spectral features.  
Here we briefly comment on each element.
The cited papers describe the details of the analyses and discuss the uncertainties.

\textit{Carbon:}
R03 and R06 employed six high-excitation C~{\sc i} lines, using transition probabilities in agreement with those currently recommended by the curated NIST Atomic Spectra Database\footnote{https://www.nist.gov/pml/atomic-spectra-database}.
S18 adopted C abundances derived from the CH molecular G-band by \cite{suarez-andres:17}.
Both C~{\sc i} and CH can yield reliable C abundances in solar-type stars, but the response of these two species to variations in atmospheric temperature and gravity parameters is dissimilar. This could justify the existence of relevant differences in the derived abundances when using these two observational sources.
For additional comments on transitions and references to solar C studies see \cite{asplund:21}. Note that \cite{delgado-mena:21} re-derived the C abundance for the HARPS sample (S18) using atomic lines C~{\sc i} (like R03 and R06): they derived typically larger C abundances compared to S18, especially for cool and metal-poor stars. Therefore, the same [C/O] variation seen in e.g., Figure~\ref{fig: solar_ratios} for S18 with respect to R03 and R06 would have been comparable or larger if we would have used \cite{delgado-mena:21} data instead of S18.

\textit{Oxygen:} No molecular species are available in the optical spectral regime, and there are very few detectable O~{\sc{i}} lines.
R03 and R06 derived O abundances exclusively from the high-excitation O~I 7700~\AA\ triplet lines.
S18 adopted instead their abundances from \cite{bertrandelis:15}, who employed a unique combination of the [O~{\sc i}] 6300~\AA\ ground-state line and a high-excitation O~{\sc i} line at 6158~\AA.
The high-excitation lines of this species have long been known to exhibit departures from local thermodynamic equilibrium \citep[][]{caffau:08,asplund:21}.  
The 6300~\AA [O~{\sc i}] line is very weak in solar-type stars, and is significantly blended with a Ni~{\sc i} transition \citep{allendeprieto:01}.
Such concerns, along with different transition choices in the two surveys, serve as cautionary notes.

\textit{Magnesium:}
Mg~{\sc i} lines are the only reliable Mg abundance sources.
MgH lines are detectable near 5000~\AA, but they are weak and very blended with strong atomic lines and C$_2$ molecular features.
There are relatively few available Mg~{\sc i} transitions, and those well known ones are often very strong.  
Many of the usually employed lines (4730.00~\AA, 4730.30~\AA, the Fraunhofer "b" triplet, 5528~\AA, and 5711.10~\AA) are saturated in the solar spectrum:  log($EW/\lambda$) > -4.8 \citep[][]{moore:66}.
Therefore, the derived Mg abundances depend on atomic damping parameters and on the adopted atmosphere conditions in the outer photospheric model.
R03 and R06 selected three Mg~{\sc i} lines, two of which are weak enough to be relatively sensitive to Mg abundances.
S18 adopted the abundances from \cite{adibekyan:12}, who in turn used the line lists of \cite{neves:09} for their study of three Mg~{\sc i} lines, with just one of them being in common with R03 and R06.
The log($gf$) values are generally in accord with the values recommended by NIST, however we note that the NIST laboratory sources are decades old and would benefit from modern re-analysis.  
Finally, a carefully developed line list from 4750-8950~\AA~has been constructed by the Gaia-ESO consortium \citep{heiter:21}.  
Their transition probabilities for Mg~{\sc i} are in accord with those used the the two surveys of interest here.

\textit{Silicon:}
A rich Si~{\sc i} spectrum is available in the optical spectra of solar-type stars, but transition probabilities have not been subjected to comprehensive laboratory analyses in recent decades.  
The R03, R06 and S18 (again, based on the earlier papers by \cite{adibekyan:12} and \cite{neves:09}) used 7 and 18 lines; their log($gf$) scales agree reasonably well within 
0.0 $\pm$ 0.07~dex, for the 5 lines in common. However, the line-to-line scatter between \cite{neves:09} and NIST (0.21~dex) and between \cite{neves:09} and \cite{heiter:21} (0.14 dex) would be eminently more useful if it was not so large, and deserves to be re-investigated.

Our brief summary of line list issues in the two surveys should be viewed as illustrative; such questions need to be kept in mind for all abundance data sets.
Another fundamental problem that needs to be addressed in the near future is the lack of recent comprehensive investigations by the atomic physics community of transition probabilities.  
With current data it is reasonable to hope for survey-to-survey agreement to the $\simeq$0.05~dex level.
Deriving abundance uncertainties to $\simeq$0.05~dex remains a future goal.

\subsection{Brief Comments on other Surveys}
\label{sec: observed_other}

\begin{figure}
    \centering
    \includegraphics[width=\columnwidth]{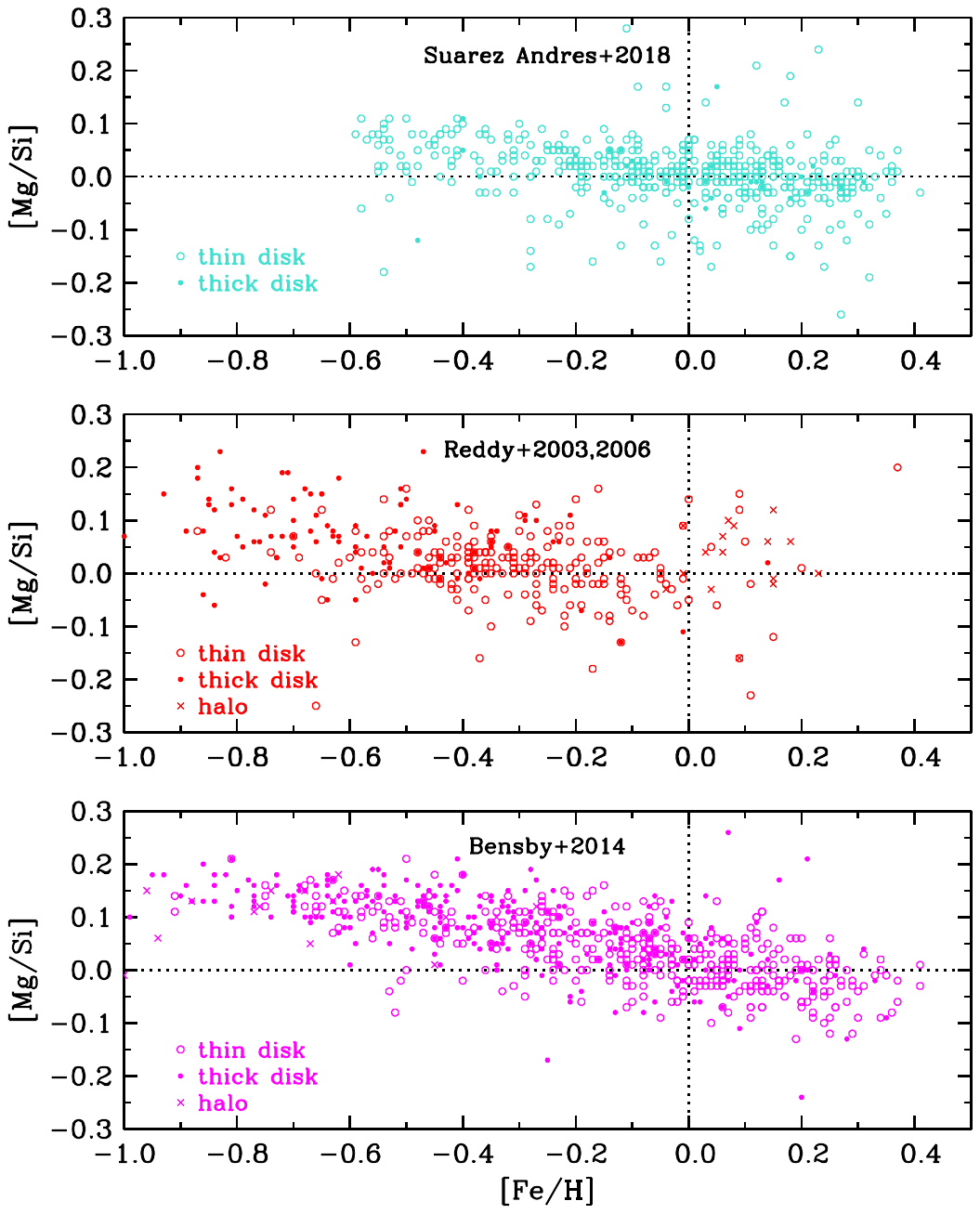}
    \vspace*{-15mm}
    \caption{[Mg/Si] abundance ratios as functions of [Fe/H], for S18 (top panel), R03 and R06 (middle panel), and \protect\cite{bensby:14} (bottom panel). Thin disk, thick disk, and halo stellar populations as defined in the individual papers are shown with different symbols, that are identified in the figure legends.  }
    \label{fig:mgsi2}
\end{figure}

We have concentrated on the R03, R06 and S18 studies because they have extensive abundance data on all four elements of interest for understanding gross planetary characteristics, and used similar analytical procedures.
Other groups have made significant contributions to Galactic disk abundance surveys.
The $\alpha$ elements as well as C and O have been studied in various surveys at different spectral resolutions in different spectral regions, such as by GALAH (e.g,, \citealt{clark:22}, \citealt{sharma:22}), and Gaia-ESO \citep[e.g.,][]{kordopatis:15}.
Here we call attention to a noteworthy contribution by T. Bensby and collaborators.
Figure~\ref{fig:mgsi2} shows the [Mg/Si] ratios versus [Fe/H] for our main surveys and the 714 star sample of \cite{bensby:14}. These authors used extensive line lists of Mg~{\sc i} and Si~{\sc i}, and transition probabilities from laboratory work and reverse solar analyses, as discussed in \cite{bensby:03}.
Inspection of Figure~\ref{fig:mgsi2} reveals a drift to larger [Mg/Si] values with decreasing [Fe/H]. 
The addition of the \cite{bensby:14} sample highlights this trend, which is weaker in R03, R06 and S18.
Note that it appears to be independent of Galactic thin-disk, thick-disk, and halo-population memberships.

\cite{bensby:14} reported O but not C abundances, so their data could not be added into Figure~\ref{fig: solar_ratios}.
It is worth noting that \cite{bensby:06} derived [C/O] ratios from a unique forbidden-line combination:  [C~{\sc i}] at 8727~\AA\ and [O~{\sc i}] 6300~\AA\ transitions.
In general, their [C/O] ratios are $\simeq$0.15~dex higher than those of R03, R06 and S18.  
The \citeauthor{bensby:06} sample size is relatively small compared to the other samples discussed here, $\simeq$50 stars, and it overlaps with the \cite{bensby:14} list only for a few stars. 
In general, a large sample investigation of [C/Fe] ratios from the combination of CH, C$_2$, C~{\sc i}, [C~{\sc i}], and possibly CO in unevolved disk stars would be an important anchor for 
all future C abundance studies.

\subsection{Astrophysical Implications of C/O and Mg/Si in Local Samples}
\label{sec: other_surveys}

The stellar samples considered in Figure \ref{fig: solar_ratios} and Figure~\ref{fig:mgsi2} show significant star-to-star scatter in [C/O] and [Mg/Si], $\sigma$~$\simeq$~0.2~dex.
As there are systematic dependencies on [Fe/H] and stellar populations that lie beyond observational uncertainties, the origin of the internal dispersion is a matter of debate.
Some of the scatter must be observational, as outlined in \ref{sec: observed_abundance_details}  
but some of this effect may be an intrinsic property of stars in the solar neighbourhood, which is difficult to quantify with the present observational errors. 
While, e.g., R03, R06, \cite{bond:10} and S18 find a significant dispersion of stellar abundances, \cite{bedell:18} obtain a consistency within 10\% for Sun-like stars (the stars in their sample are solar twins with very similar metallicities) within 100 pc. 
It will be paramount in future work to definitively solve these discrepancies and clarify the diversity in composition of the solar neighbourhood.

In section \ref{sec: results}, we will compare GCE simulations with solar-scaled observations, so that the systematic uncertainties discussed here will not directly affect our analysis. However, since the absolute elemental abundances are needed for the simulations of planetary systems, we are highlighting the uncertainties discussed in this section as an issue that will need to be addressed. The variations seen in Figure \ref{fig: solar_ratios} are a clear source of degeneracy for future planet formation and evolution studies. 

For the comparison with GCE simulations, we will only use stellar data by R03 and R06, where all the elemental abundances are provided for 
the elements discussed in the following sections (i.e., C, N, O, Mg, Si and S). Additionally, this allows us to compare GCE simulations with the abundance dispersion observed by R03 and R06, which is larger and more conservative compared to the results by, e.g., \cite{bedell:18}. We have seen that R06 includes indeed a large number of thick-disk stars.
While we can expect that the planet-host stellar samples from e.g., TESS and ARIEL will be biased towards more metal-rich, thin-disk stars, the first observational efforts to homogeneously characterize the physical parameters of planet-host stars in the Ariel Reference Sample \citep[][]{edwards:19} show the presence of both thick-disk stars and stars that could dynamically belong to either the thin and thick disks \citep[][]{magrini:22}.  Furthermore, it is expected that a number of thick-disk stars will be discovered hosting planetary systems (e.g., from TESS Kepler-444, \citealt{campante:15}, and TOI-561, \citealt{weiss:21}). The study of these stars and their planetary systems will be a fundamental benchmark for planet formation and the evolution of planetary systems.

Finally, we note that among the elements considered in this work N is not available in the HARPS data used by S18 and \cite{bedell:18}. However, for a sample of 74 stars \cite{suarez-andres:16} also combined HARPS data for relevant elements to derive the N abundance from the NH band in UVES spectra.
 
%

\section{Results of the GCE simulations}
\label{sec: results}

We now summarize our analysis for the elements of interest discussed in Section \S~\ref{sec: stars}, using the GCE models introduced in the previous section. All the plots showing the comparison between GCE model predictions and observations are provided in this section or in Appendix \S~\ref{app_1: plots extra}.

In Figure~\ref{fig: tk_plots/ratios_mup40}, selected element ratios are plotted against each other and compared to abundances observed in stars within 150 parsecs by R03 (which comprises mostly thin-disk stars) and R06 (which instead comprises mostly thick-disk stars). 
This is a similar range of distance to consider for stellar hosts of TESS and ARIEL planetary targets \citep[][]{edwards:19,magrini:22}. Therefore, we may assume that
this observational sample is consistent with the abundance variations that merit exploration.   

The upper-left panel of Figure~\ref{fig: tk_plots/ratios_mup40} is the same diagram shown by \cite{bond:10} but in logarithmic notation, and including GCE models. In general, only the theoretical GCE curves between the red and the magenta stars (which correspond to a GCE evolution age between 1 Gyr and today, Figure~\ref{fig: tk_plots/ok06no_xh}), should be considered as representative of the evolution over time of the Milky Way disk. Indeed, the observed properties of the old stellar population of the Milky Way halo are consistent with the first Gyr of active star formation, while to reproduce the age and metallicity distribution of the stars in the Milky Way disk much longer times are required (see e.g, \citealt{fenner:03}). 

The observational scatter of about a factor of 2.5 is obtained for both [C/O] and [Mg/Si]. At time of Sun formation, the models oK06no and oK10no produce [C/O] ratios 0.2 dex and 0.3 dex lower than the solar abundances, respectively. However, they reproduce the ratio observed in the majority of stars, with a [C/O] ratio increasing over time (or with metallicity) until about 5 Gyr ago \citep[][]{bitsch:20}. After Sun formation, the calculated [C/O] is almost constant. The oR18dno model produces a final solar [C/O] ratio. Between the three GCE models using \cite{ritter:18} CCSN sets, the different CCSN explosion parametrizations affect the O production with a variation of the [C/O] ratio by about 0.2 dex. At evolution timescales representative of the MW disk, oL18 models show a solar [C/O] ratio with only marginal variation. 

The [Mg/Si] is about 30$\%$ lower than the Sun in oK06no and oK10no, between a factor of 1.8 and 2.2 lower for the R18 models, and sub-solar by a factor of $\sim1.8$ for oL18no. While the bulk of the stars in the solar neighbourhood have a solar-like or super-solar ratio up to [Mg/Si]$\sim$0.2, the GCE models predict a ratio lower than solar for all the combinations of stellar yields considered. In particular, the largest departure seen in oR18no is mostly due to the contribution of energetic CCSN explosions for the 12M$_{\odot}$ and 15M$_{\odot}$ models by \cite{ritter:18}. Note that these results would not have changed by considering a different observational stellar sample, e.g. R03 only without the R06 or the S18 sample (see Figure~\ref{fig: ratios_s18_r03}).
We also cannot expect one-zone GCE simulations to reproduce the observed [Mg/Si] scatter, since at a given evolution time of the model the result is given by a single data point, not by some statistical distribution. However, the predictions should still be compatible with the bulk of the observed stars. Even by varying the stellar yields - one of the main uncertainty sources of GCE - we do not achieve this result. It is true that GCE uncertainties could play a relevant role in the abundance analysis, and in our single-zone GCE simulations we do not take into account relevant processes like stellar migration and past infall of fresh material in the galactic disk \citep[][and reference therein]{matteucci:21,prantzos:23}. However, the impact of such processes should be significantly reduced by studying the evolution of primary elements sharing a similar stellar production. For instance, in the specific case of Mg and Si they are both mostly produced by short-lived massive stars, and these do not have sufficient time to migrate significantly \citep[e.g.,][]{sanchez-blazquez:09, minchev:14}. 

Such a result where Mg stellar yields seem to be too low compared to observations (and, to a much lesser extent, Si) are not surprising, and they have been highlighted before in the literature. The artificial Mg and/or Si boosting is a well-known requirement of using e.g., the \cite{woosley:95} CCSN yields \citep[][]{gibson:97}. More recently, the same approach is implemented by \cite{spitoni:21} with \cite{woosley:95} yields, where Mg from CCSNe are boosted up to a factor of seven.

The distribution of the CNO elements is shown in the upper-right panel of Figure~\ref{fig: tk_plots/ratios_mup40}. Models oR18no and OR18dno show an [N/O] ratio increasing with the evolution time and with [Fe/H]. Yet, we find that they both reach the solar [N/O] ratio too early, more than 4 Gyr before the formation of the Sun. The other models instead reproduce the solar ratio to within 0.1 dex. For oL18no, the [N/O] ratio changes little with galactic time, remaining $40-80\%$ super-solar for the duration of the GCE. 
Several challenges need to be considered for the GCE of CNO elements and their stable isotopes \citep[e.g.,][]{kobayashi:20}. The contingent relevance of fast rotating stars (not included in our models) was highlighted by several previous works to reproduce the abundance patterns in the Milky Way \citep[e.g.,][]{chiappini:06, chiappini:08, prantzos:18, romano:19}. Additionally, \cite{pignatari:15} discussed the contribution of H-ingestion events in massive stars for the GCE of N (and in particular of the N isotopic ratio), using stellar models consistent with abundance measurements in presolar grains made by CCSNe just before the formation of the Sun. 

The predictions for N evolution from our GCE models have problems in reproducing the ratios in the Figure~\ref{fig: tk_plots/ratios_mup40}, lower-left panel. Contrary to most of the observations, all the GCE models produce super-solar [S/N] ratios, except for the oL18no model, which produces a [S/N] ratio significantly lower than solar, but still does not cover the full observational range down to [S/N]$\sim$-0.5 reached by a large number of stars. Most of the stars indeed exhibit a subsolar elemental ratio, with a scatter of the order of a factor of three. 
The observed [C/N] scatter is instead at least partially reproduced by most of GCE models, where this ratio decreases with evolution time. Note that this does not have to be the right physical reason for the observed [C/N] scatter. As we mentioned earlier, the N evolution is a well-known challenge for GCE, where standard CCSN models underproduce N compared to observations. 
Finally, model oK10no shows a smaller variation than the other models within sub-solar [C/N] values around $-$0.3 dex.

In the lower-right panel of Figure~\ref{fig: tk_plots/ratios_mup40}, stars show an observational scatter larger than a factor of two for both [Mg/O] and [S/Si] ratios. Such a variation is only marginally captured by the GCE models: for both ratios, variations at relevant timescale are in the order of 20$\%$ or less. The oK06no, oK10no and oL18no models reproduce the solar ratios within 0.1 dex, the other models are more consistent with the bulk of stars that are mildly S-rich compared to the Sun, up to [S/Si]$\sim$+0.3. If we consider all the GCE models, it may seem that the observed range of [S/Si] is reproduced. However, if we factor in single GCE models we notice that the [S/Si] variation within the acceptable evolution time frame is less than 0.1 dex. The only exception is oR18hno, where between 1Gyr 
and today the [S/Si] ratio increases by about 0.2 dex. Still, such an increase is much smaller than the scatter observed in the Milky Way disk. Such a dispersion may, in part, be due to observational uncertainties \citep[][]{bedell:18}. Chemo-dynamical simulations of the Milky Way disk would be needed to provide a realistic prediction for the expected [S/Si] dispersion \citep[e.g.,][]{thompson:18}, which is beyond the goal of this paper.   

Alternatively, as we discussed in Section \S~\ref{sec: stars}, this dispersion may instead be an indication that CCSN ejecta are not always dominated by an explosive O-burning signature, but that they vary between different supernova events. We have seen that some variations are obtained between different SNIa explosions (Figure~\ref{fig: snia_prodfac}), although the progenitor mass does not affect the S/Si ratio in the ejecta very much, and the same can be said for the initial metallicity \citep[e.g.,][]{keegans:23}. Note that according to the nuclear sensitivity study by \cite{parikh:13}, there should be no relevant impact of nuclear uncertainties on the SNIa yields of Si and S.

\begin{figure*}
\begin{multicols}{2}
    \includegraphics[width=\linewidth]{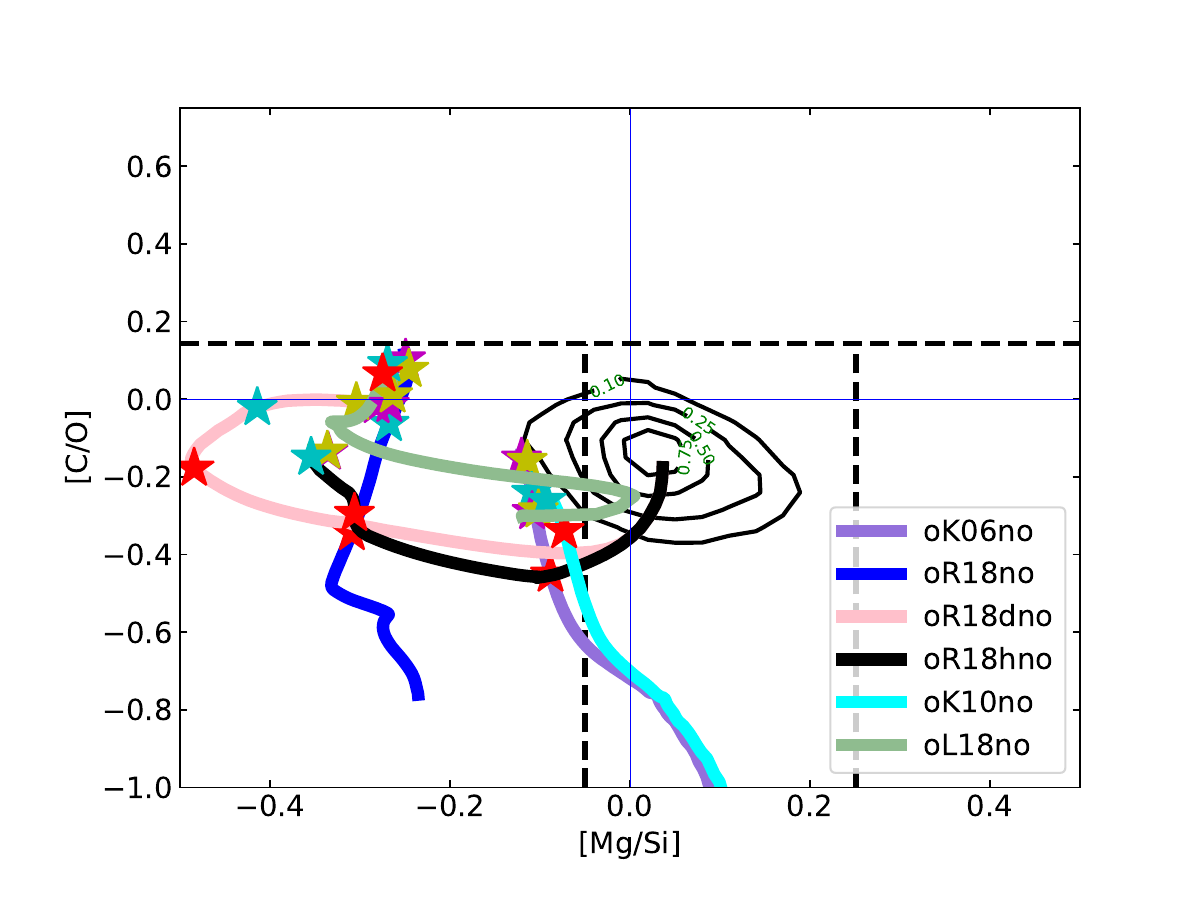}\par
    \includegraphics[width=\linewidth]{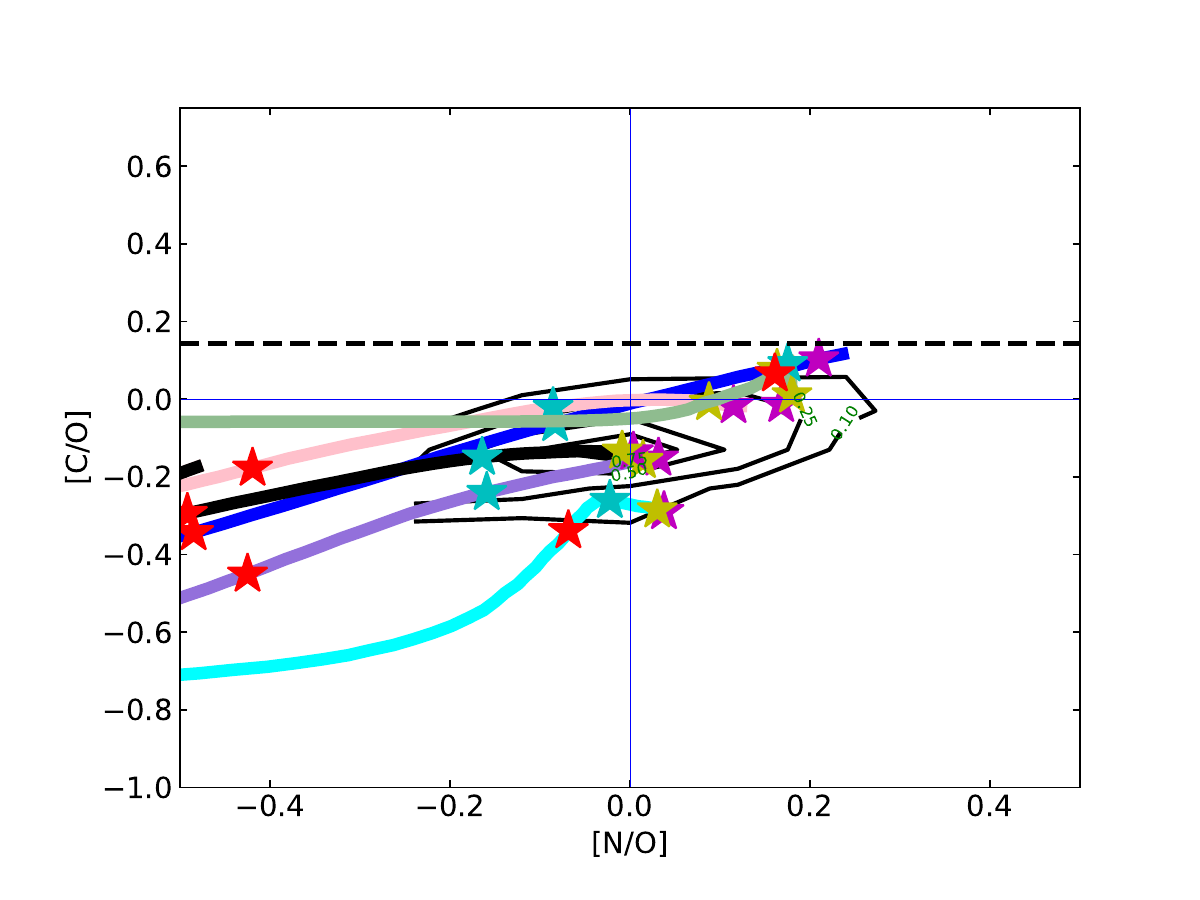}\par
    \end{multicols}
\begin{multicols}{2}
    \includegraphics[width=\linewidth]{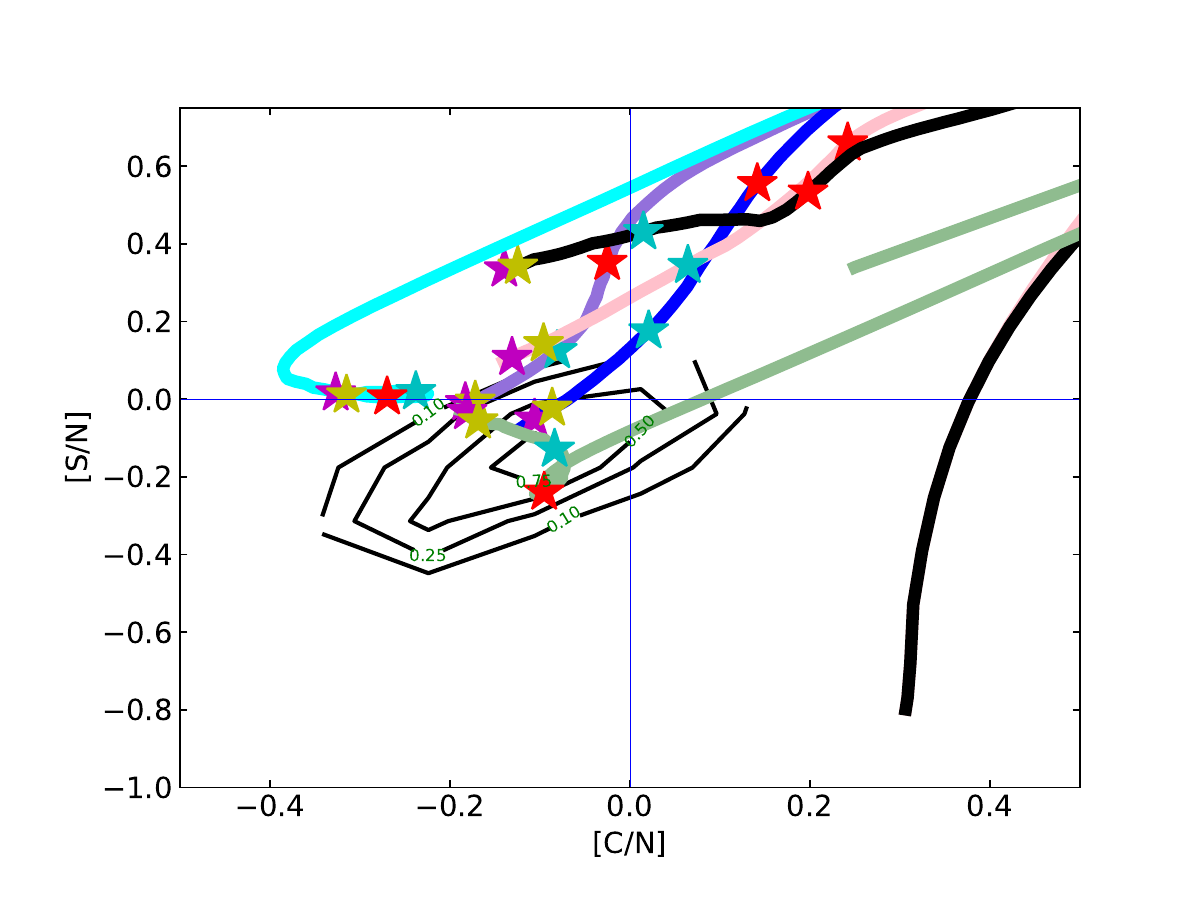}\par
    \includegraphics[width=\linewidth]{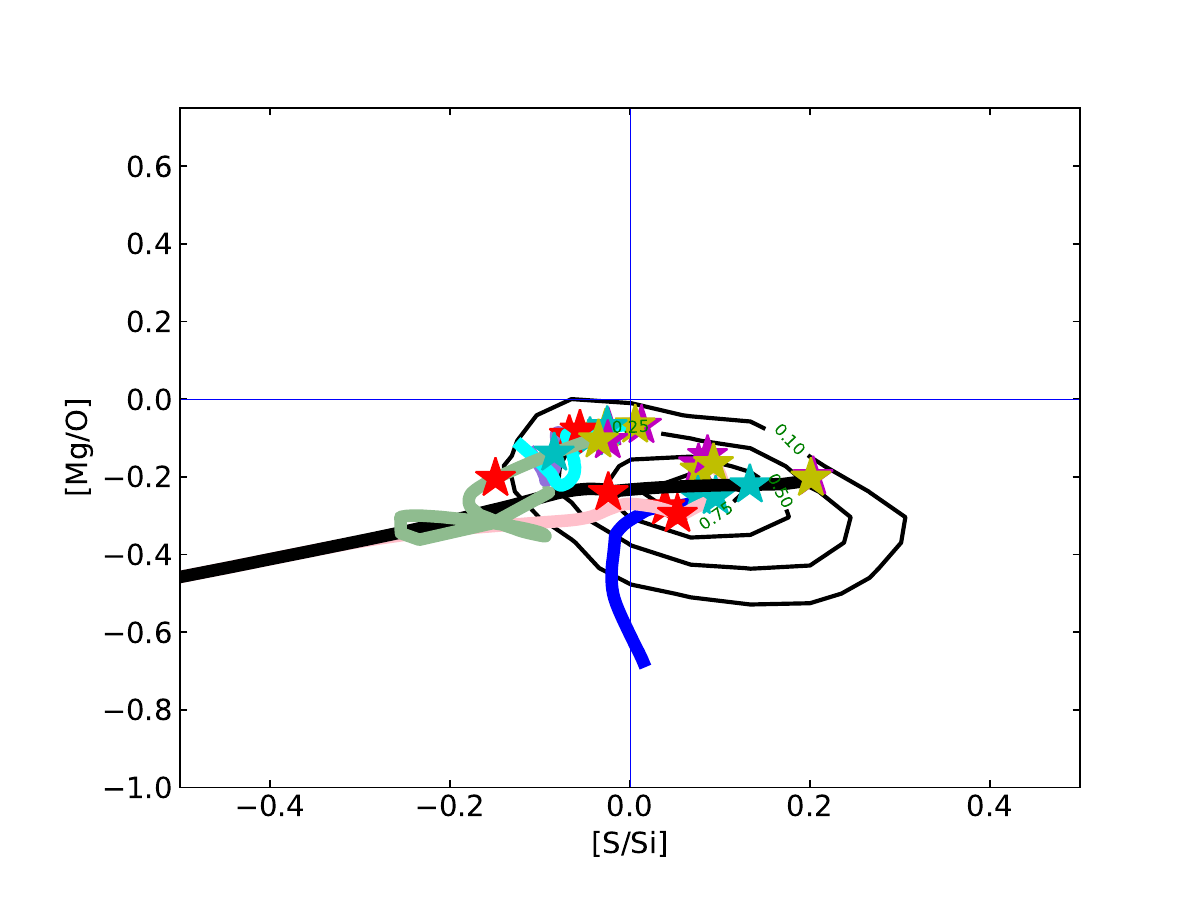}\par
\end{multicols}
    \caption{Selected elemental ratios normalized to the solar (L19$\_$3D by \protect\cite{lodders:19}, see Figure~\ref{fig: solar_ratios}) are plotted against each other for the GCE model sets oK06no, oR18no, oK10no and oL18no, in which CCSN contribution is provided up to M = 40M$_{\odot}$ (see Table \ref{tab: list_tk_plots/models}). Beyond this mass, we assume that no CCSN material is ejected. Time coordinates for each models are reported using star points of different colors, as in Figure~\ref{fig: tk_plots/ok06no_xh}. We compare the simulations to observations from solar neighbourhood stars by R03 and R06, 
    by means of contours of observational data mapping the distribution density of the stellar abundances. The green number on each contour line represents the normalised stellar counts represented. 
    Ratios discussed by \protect\cite{bond:10} as crucial for chemistry and mineralogy of rocky planets are also reported in the upper panels normalized to the solar values (L19$\_$3D, 
    black dashed lines). Note that the contour lines are not fully closed for both the two plots including N, due to the more limited number of stars with measured N abundances in the stellar sample considered and the consequent small statistics at the edges of the stellar density distribution}.
    \label{fig: tk_plots/ratios_mup40}
\end{figure*}

\begin{figure*}
\begin{multicols}{2}
    \includegraphics[width=\linewidth]{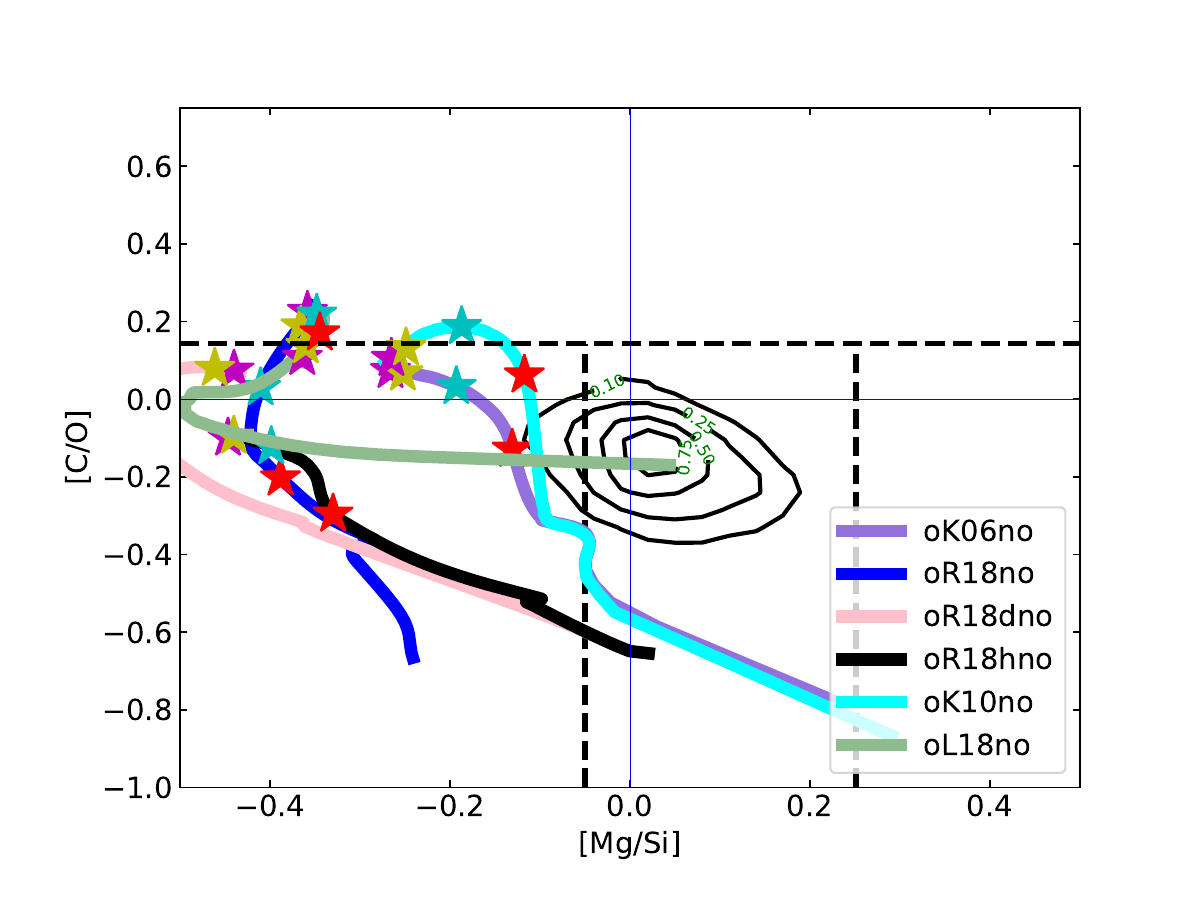}\par
    \includegraphics[width=\linewidth]{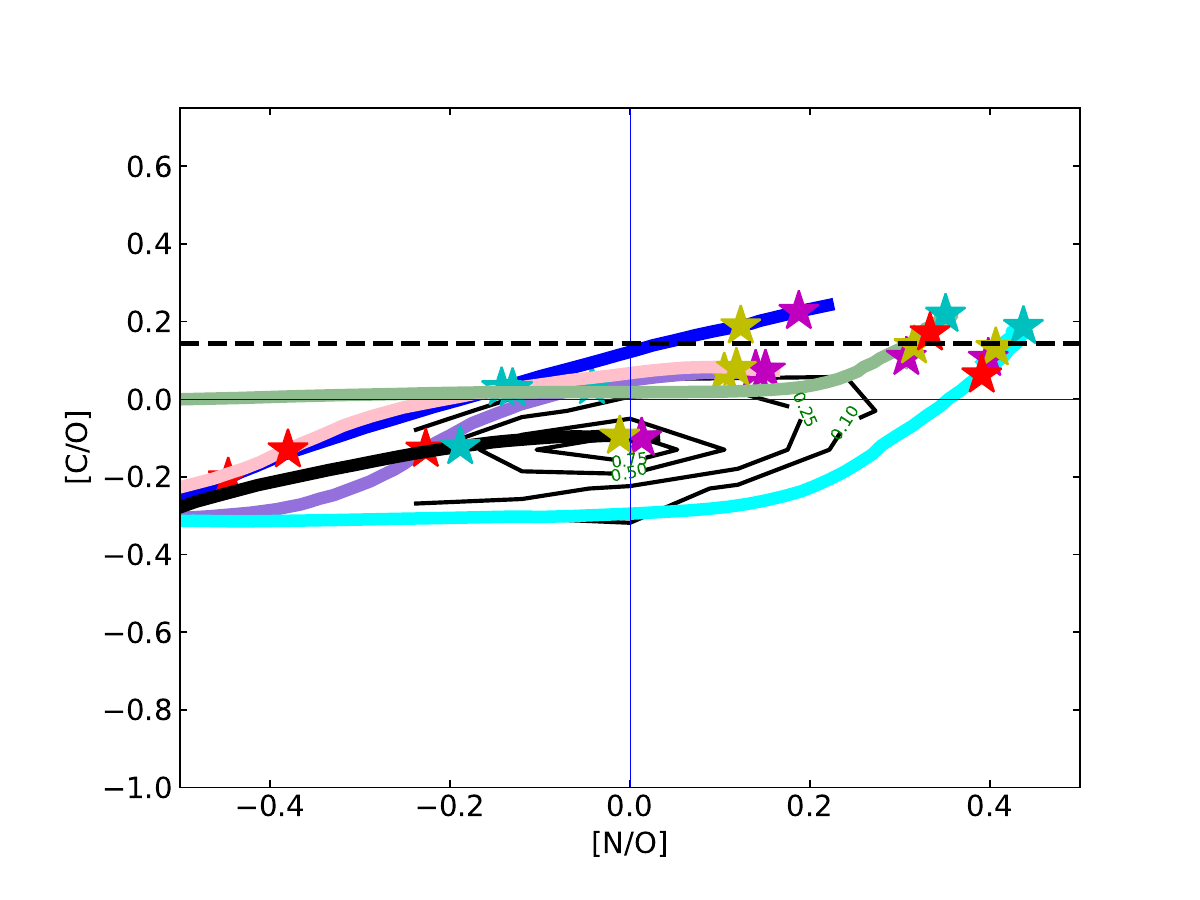}\par
    \end{multicols}
\begin{multicols}{2}
    \includegraphics[width=\linewidth]{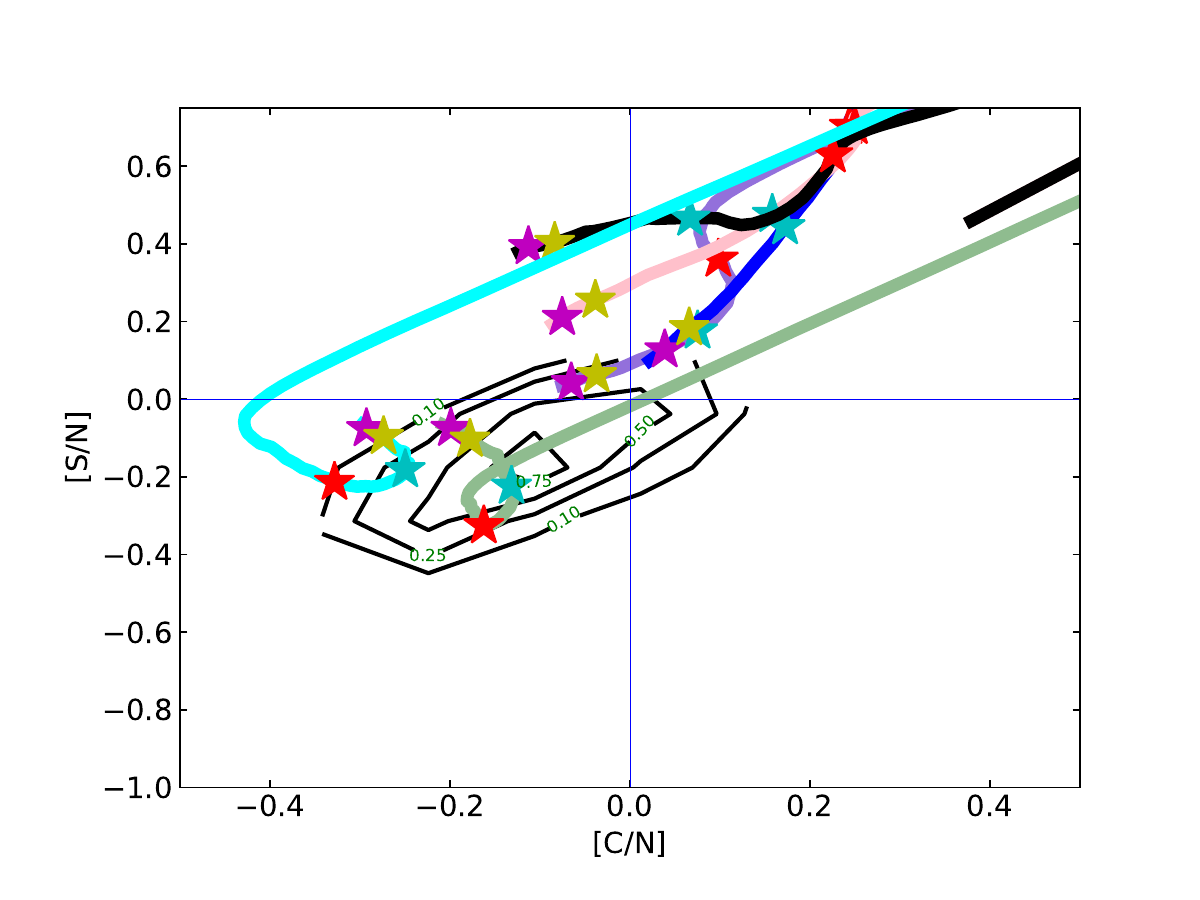}\par
    \includegraphics[width=\linewidth]{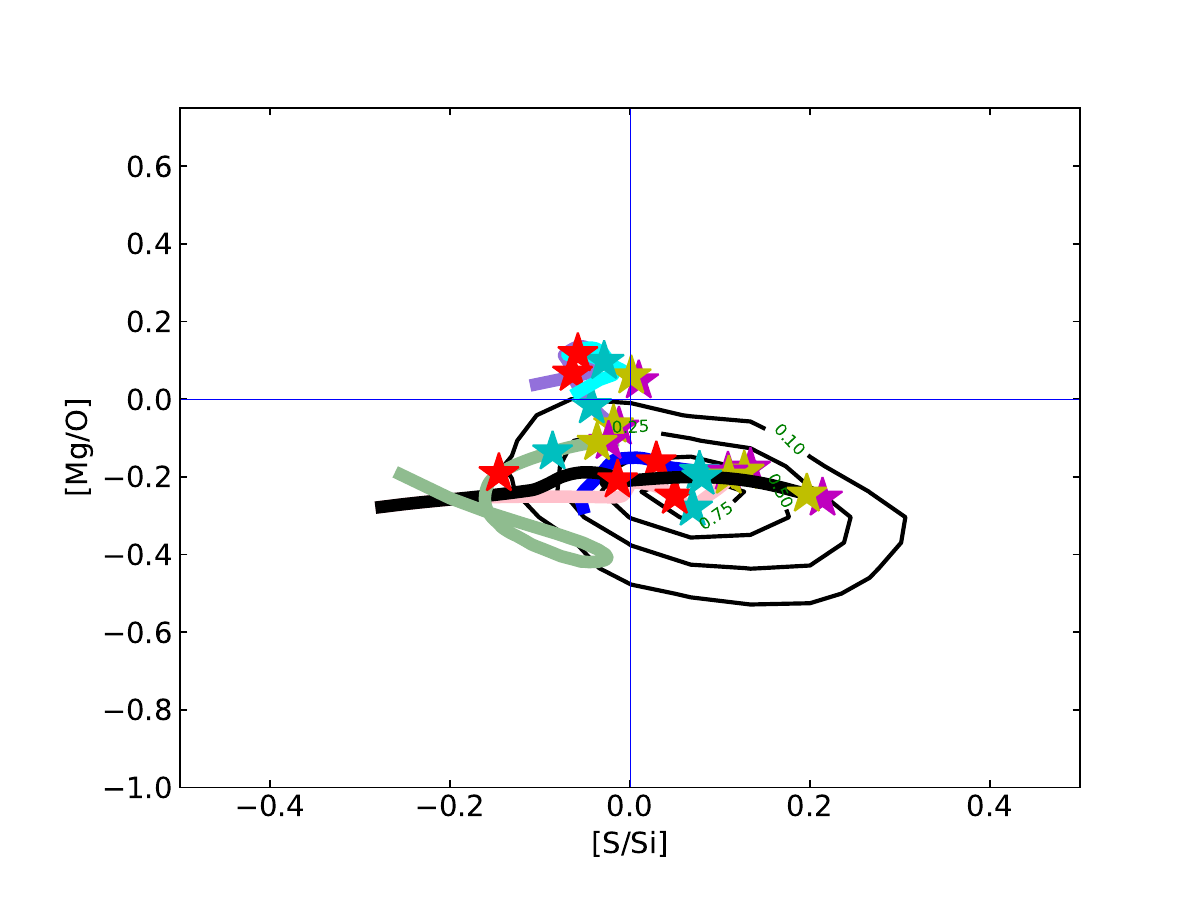}\par
\end{multicols}
    \caption{The same as in Figure~\ref{fig: tk_plots/ratios_mup40}, but with massive stars contributing to GCE up to M$_{\rm up}$ = 20M$_{\odot}$     
    }
    \label{fig: tk_plots/ratios_mup20}
\end{figure*}

\begin{figure*}
\begin{multicols}{2}
    \includegraphics[width=\linewidth]{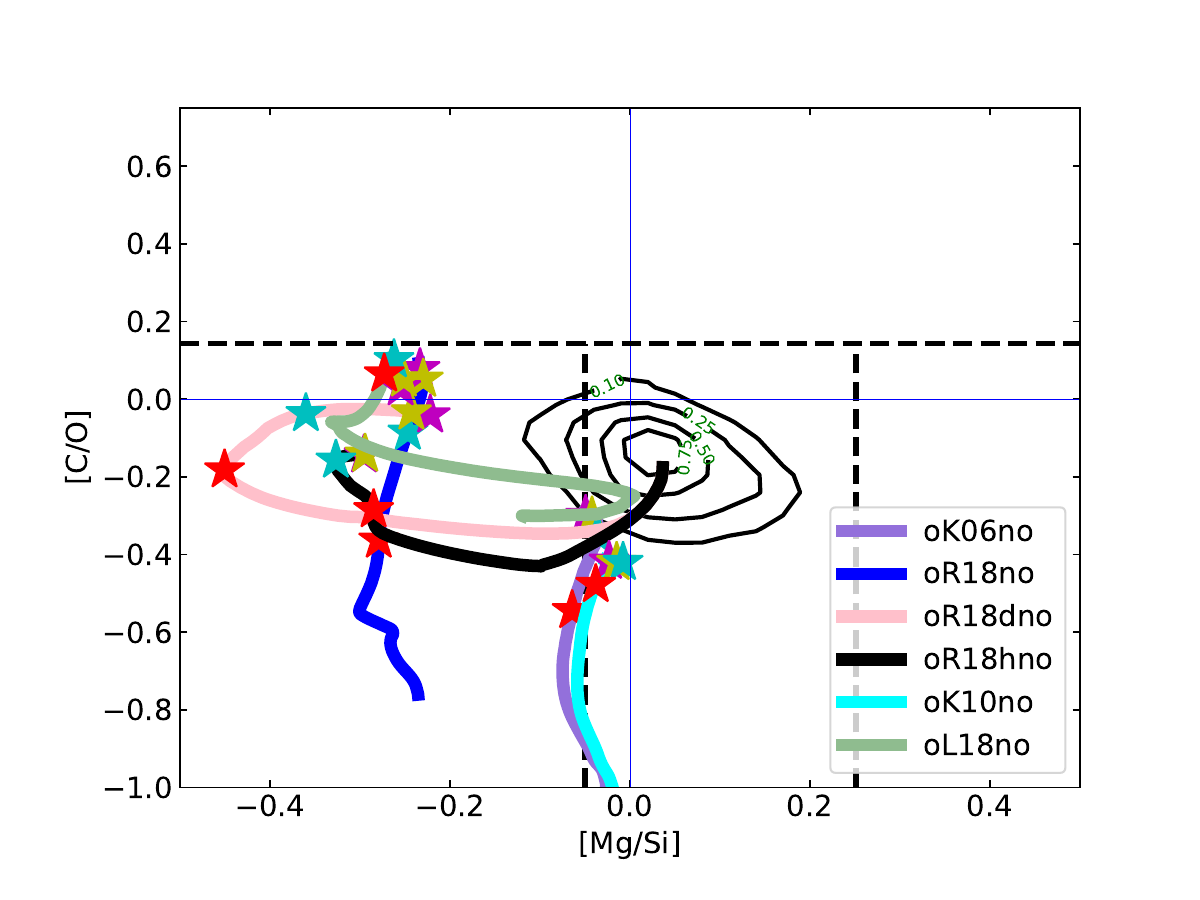}\par
    \includegraphics[width=\linewidth]{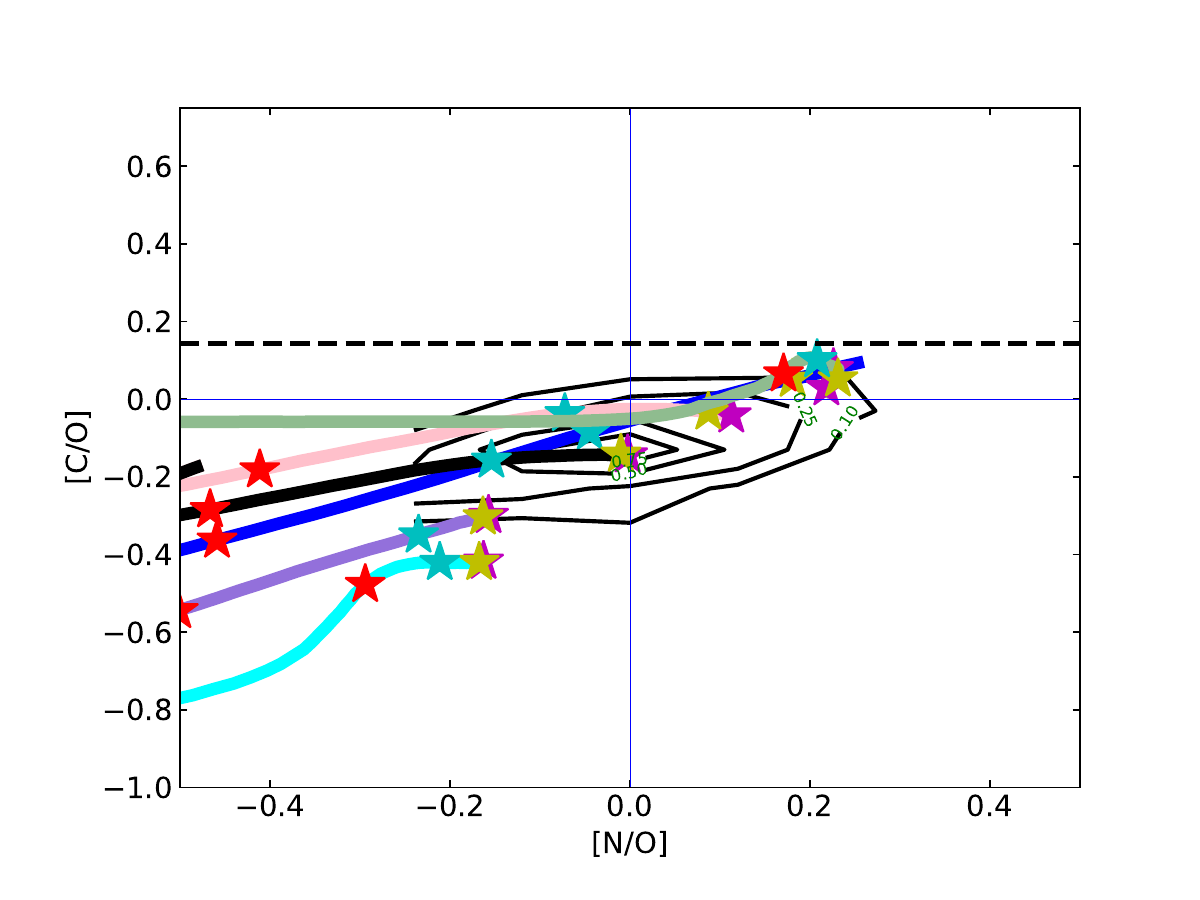}\par
    \end{multicols}
\begin{multicols}{2}
    \includegraphics[width=\linewidth]{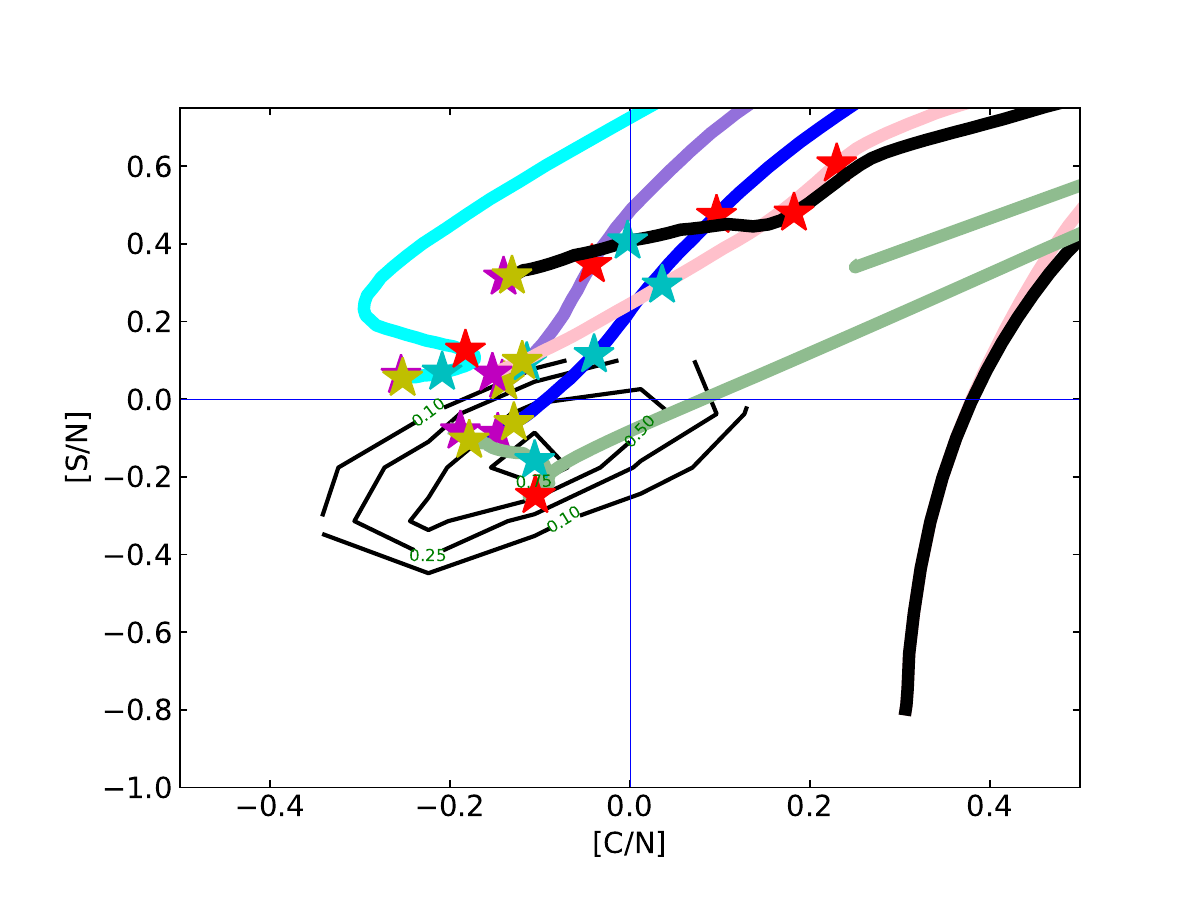}\par
    \includegraphics[width=\linewidth]{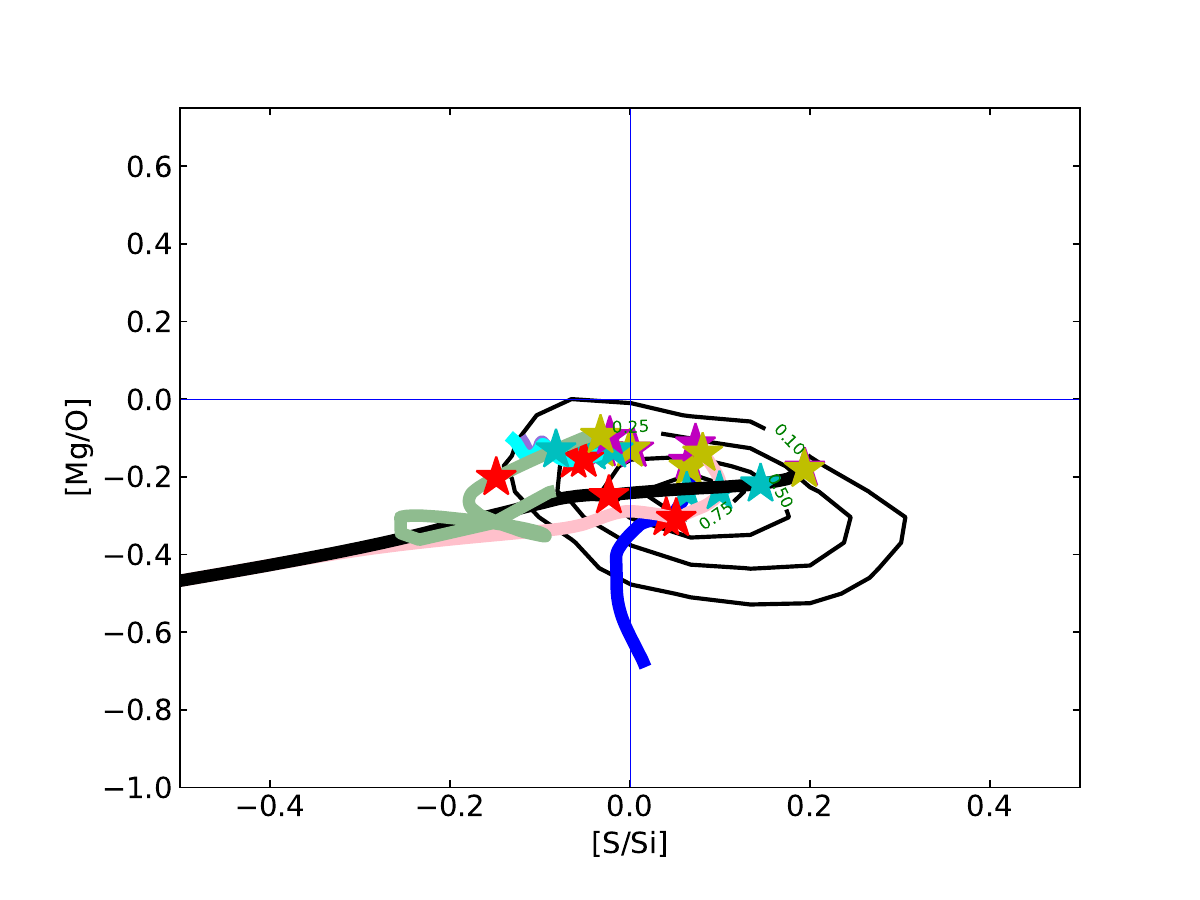}\par
\end{multicols}
    \caption{The same as in Figure~\ref{fig: tk_plots/ratios_mup40}, but with massive stars contributing to GCE up to M$_{\rm up}$ = 100M$_{\odot}$. 
    }
    \label{fig: tk_plots/ratios_mup100}
\end{figure*}

\subsection{The effect of changing M$_{\rm up}$ in GCE simulations}
\label{subsec: mup_change}

The models shown in Figure \ref{fig: tk_plots/ratios_mup40} used as mass upper limit M$_{\rm up}$ = 40M$_{\odot}$ (Table~\ref{tab: list_tk_plots/models}).
In Figure~\ref{fig: tk_plots/ratios_mup20} and Figure~\ref{fig: tk_plots/ratios_mup100} we have explored the impact of the M$_{\rm up}$ parameter space on the results, by considering M$_{\rm up}$ = 20 M$_{\odot}$ and M$_{\rm up}$ = 100 M$_{\odot}$, respectively.
The upper-left panels of the two figures show that by increasing M$_{\rm up}$, the predicted [Mg/Si] increases by up to 0.15 dex, but [C/O] decreases by up to 0.1 dex. For instance, with M$_{\rm up}$= 100 M$_{\odot}$, the models oK06no and oK10no approach the solar [Mg/Si] ratio (the super-solar ratios observed are still not reproduced). Rather, the predicted [C/O] is about 0.3-0.4 dex lower than solar. 
On the other hand, with the extreme case M$_{\rm up}$= 20 M$_{\odot}$, oK06no and oK10no predict a [Mg/Si]$\sim$-0.2 for the Sun, and for oL18no and all the oR18 models the same ratio is reduced down to about [Mg/Si]$\sim$-0.4. 
If we compare the upper-right panels, for models oK06no and oK10no the [N/O] typically decreases by 0.6 dex with increasing M$_{\rm up}$, while there is mostly no effect in all the R18 models. 
This is because in our GCE models the total mass ejected for the higher masses is extrapolated from the list of available models, and the yields abundance pattern is kept identical to that of the highest mass model, which is the M=25M$_{\odot}$ progenitor for \citeauthor{ritter:18} and the M=40M$_{\odot}$ progenitor for \citeauthor{kobayashi:06} yields. The 25M$_{\odot}$ models by \citeauthor{ritter:18} are all weak explosions, leaving large remnants. Thus, stars between 25 and 100 M$_{\odot}$ only eject limited amounts of elements such as O and Mg. Notice that since L18 yields have large remnants for stars above 30 M$_{\odot}$, the M$_{\rm up}$ impact will be limited in these cases too. On the other hand, the yields by \cite{kobayashi:06} are all made of successful CCSN explosions with low remnant masses, and increasing M$_{\rm up}$ of 20 to 100 M$_{\odot}$ makes a huge difference since more massive progenitors do contribute significantly to the chemical evolution. 

The lower-left panel of Figure~\ref{fig: tk_plots/ratios_mup20} shows that with M$_{\rm up}$ =20 M$_{\odot}$ the model oK10no can reach a sub-solar [S/N]$\sim$$-$0.2 dex, which would be consistent with the bulk of local stars, but with a [C/N]$\sim$-0.3 dex. The model oL18no can also reach a sub-solar [S/N]$\sim$$-$0.3 dex, with a [C/N]$\sim$-0.2 dex. All the other models exhibit a [S/N] range between solar and 2.5 times solar (oR18hno). With M$_{\rm up}$ = 100M$_{\odot}$ (Figure~\ref{fig: tk_plots/ratios_mup100}), the models closest to the observations are oR18no and oL18no with predicted [S/N] and [C/N] ratios in the range of $-$0.1 $-$ $-$0.2 dex, since the time of the formation of the Sun. Finally, if we compare the bottom right panels of Figure~\ref{fig: tk_plots/ratios_mup20} and Figure~\ref{fig: tk_plots/ratios_mup100}, the only significant variation is a decrease of the [Mg/O] ratio of about 0.1 dex or less for all models with increasing M$_{\rm up}$. Such a small variation is not surprising. Indeed, both O and Mg are mainly products of massive stars, they are made during the pre-SN stage and their pre-SN ratio is not significantly modified by the CCSN explosion \citep[e.g.,][]{thielemann:96, pignatari:16}. 

In summary, from exploring the impact of the M$_{\rm up}$ parameter space we do not see a clear effect of using a value different from the default M$_{\rm up}$ =40 M$_{\odot}$. While a higher M$_{\rm up}$ slightly increases the [Mg/Si] ratio, it would still not cover the Sun and most of the stars, 
with the [C/O] ratio too low as compared to observations. The impact on the [N/O] and [S/N] ratios is model dependent. Therefore, in the following part of the section we will discuss only the models with M$_{\rm up}$ = 40M$_{\odot}$. Results for the same models but with different M$_{\rm up}$ are available in Appendix~\S~\ref{app_1: plots extra}.

\subsection{The impact of Faint Supernovae}
\label{subsec: fainSN}

To study the impact of faint CCSNe we focus our analysis to the two set of models using the yields K10 and R18 (Table~\ref{tab: list_tk_plots/models}). Results of other models are consistent with the sample of simulations considered here and are available in Appendix~\S~\ref{app_1: plots extra}.

\begin{figure*}
\begin{multicols}{2}
    \includegraphics[width=0.8\linewidth]{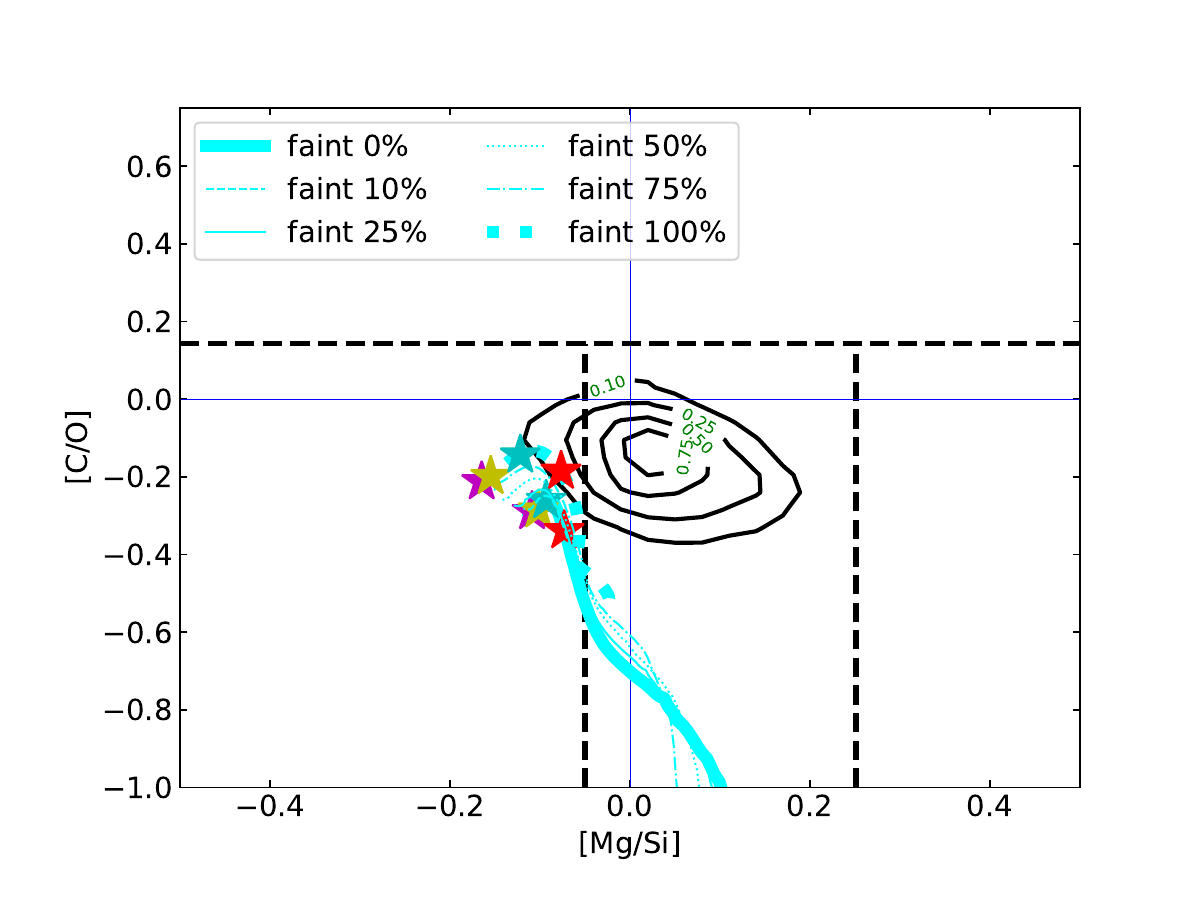}\par
    \includegraphics[width=0.8\linewidth]{tk_plots/gce_oK10_reddy_mgsi_co_m14.9_m40_type20.pdf}\par
    \end{multicols}
\begin{multicols}{2}
    \includegraphics[width=0.8\linewidth]{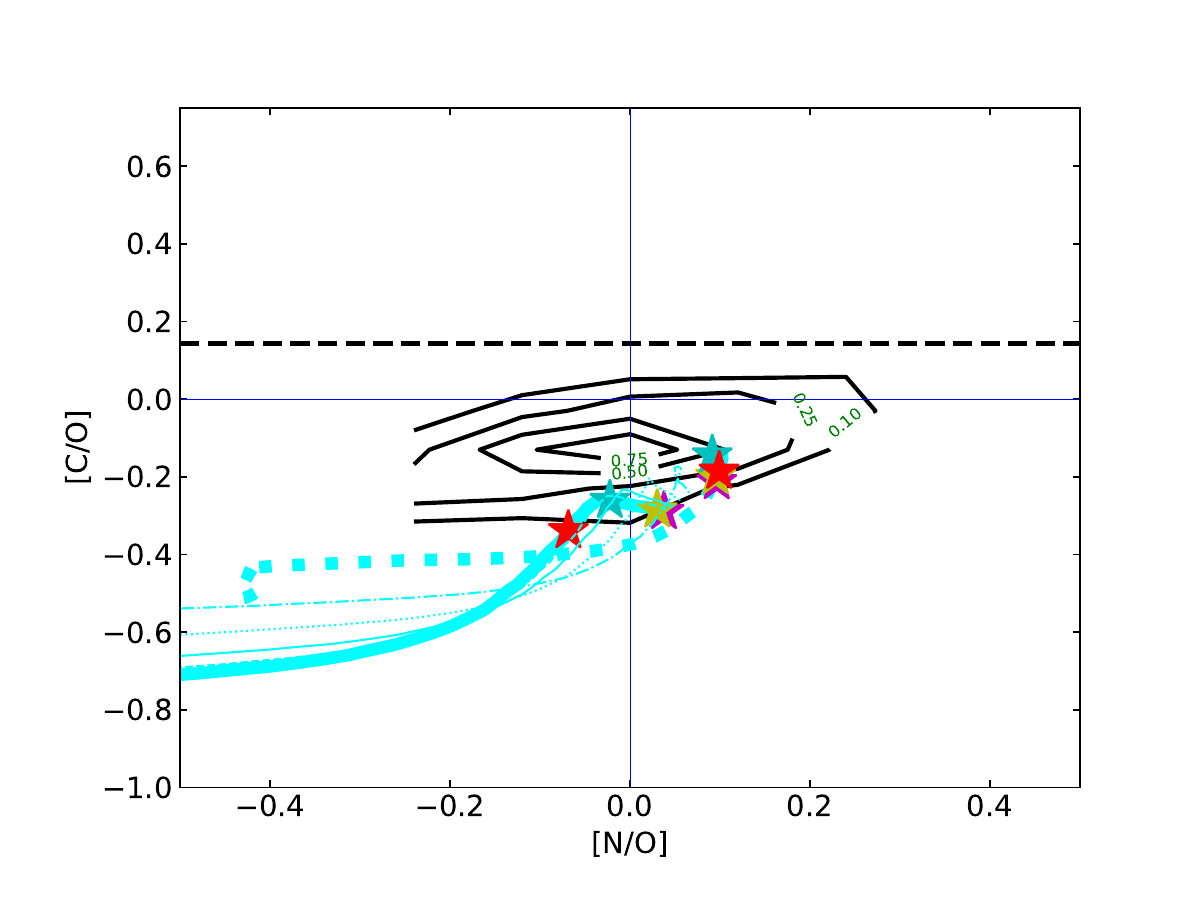}\par
    \includegraphics[width=0.8\linewidth]{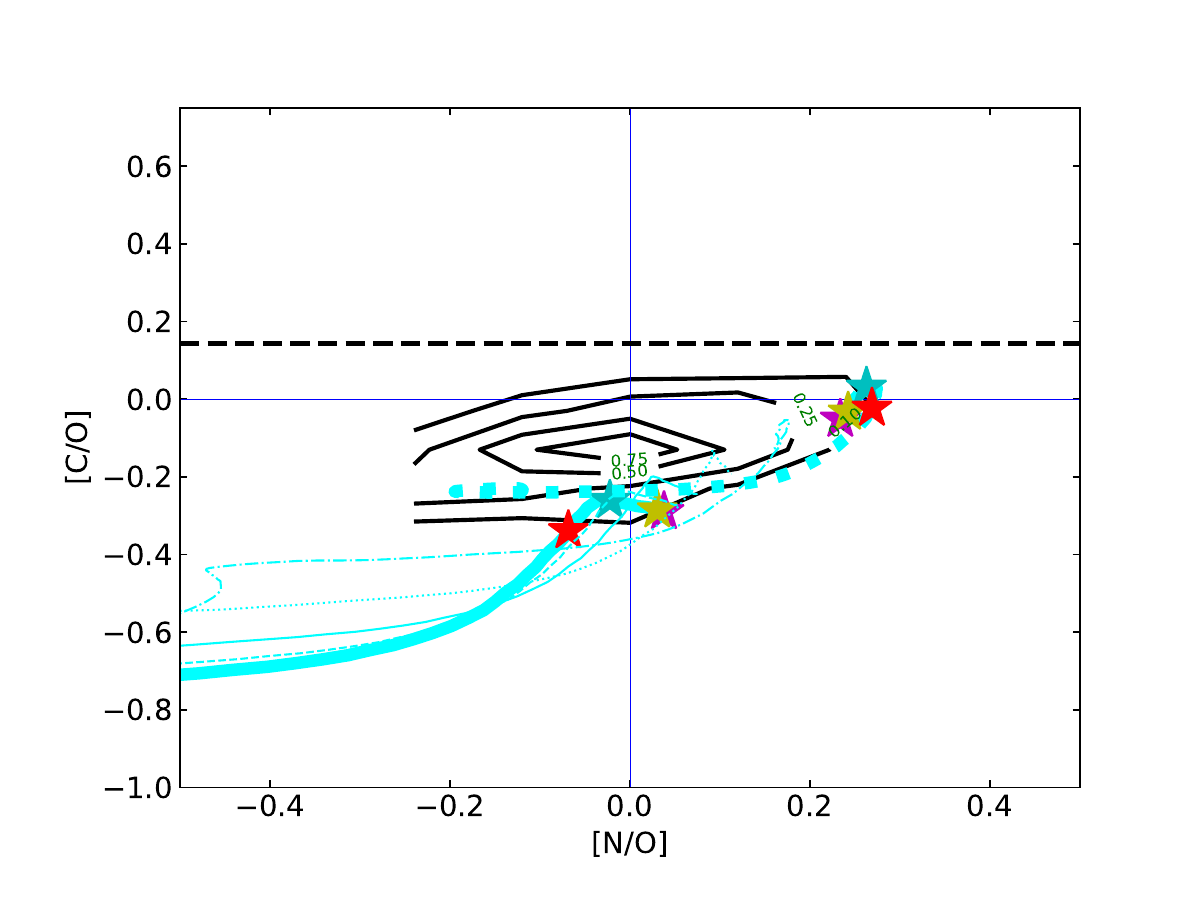}\par
\end{multicols}
\begin{multicols}{2}
    \includegraphics[width=0.8\linewidth]{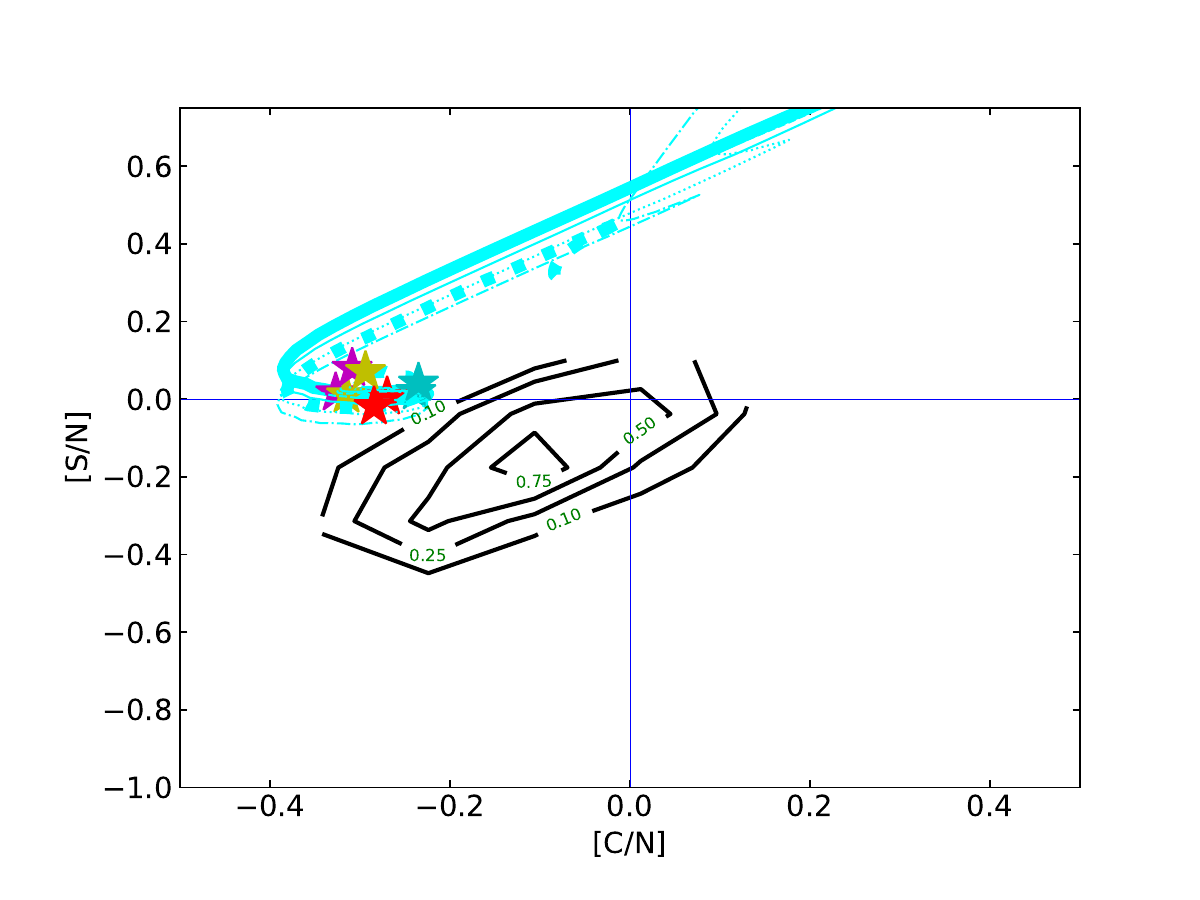}\par
    \includegraphics[width=0.8\linewidth]{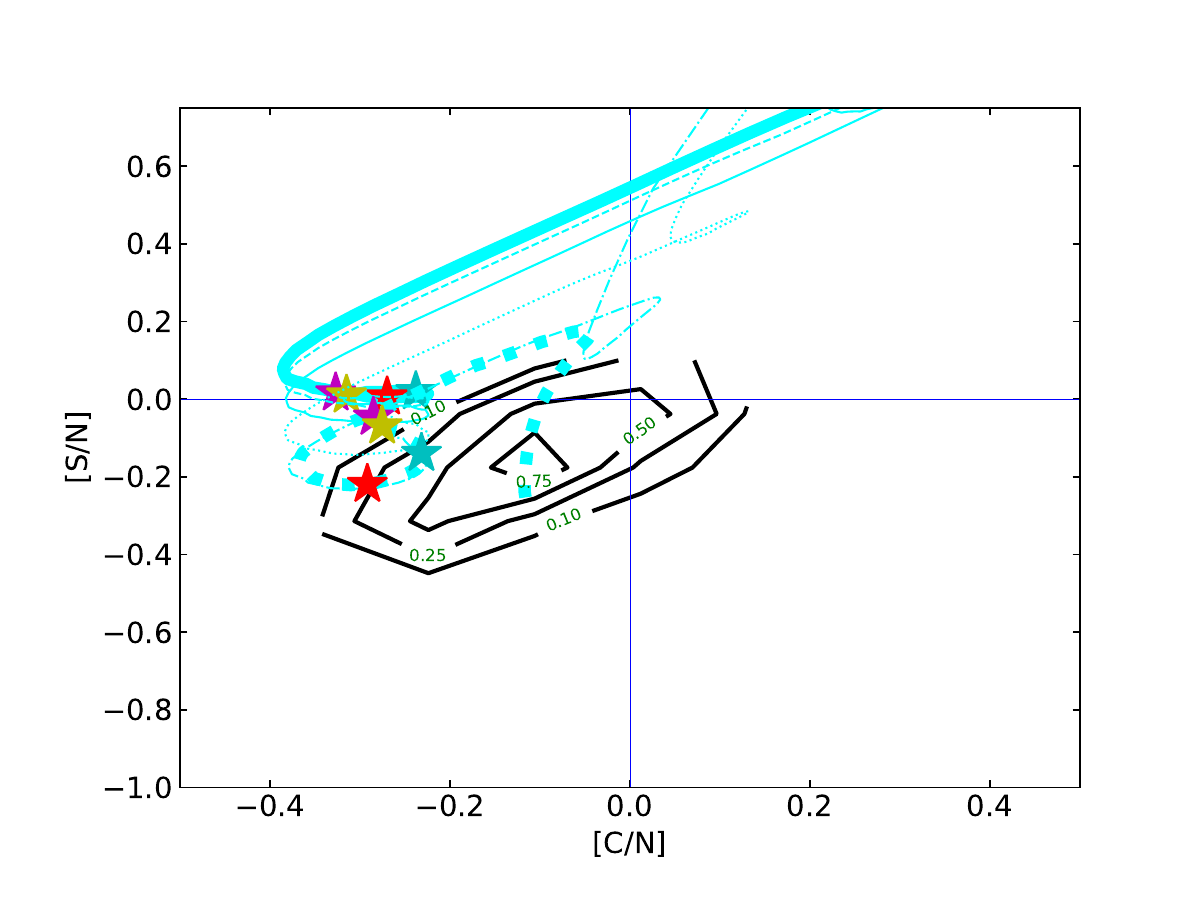}\par
\end{multicols}
\begin{multicols}{2}
    \includegraphics[width=0.8\linewidth]{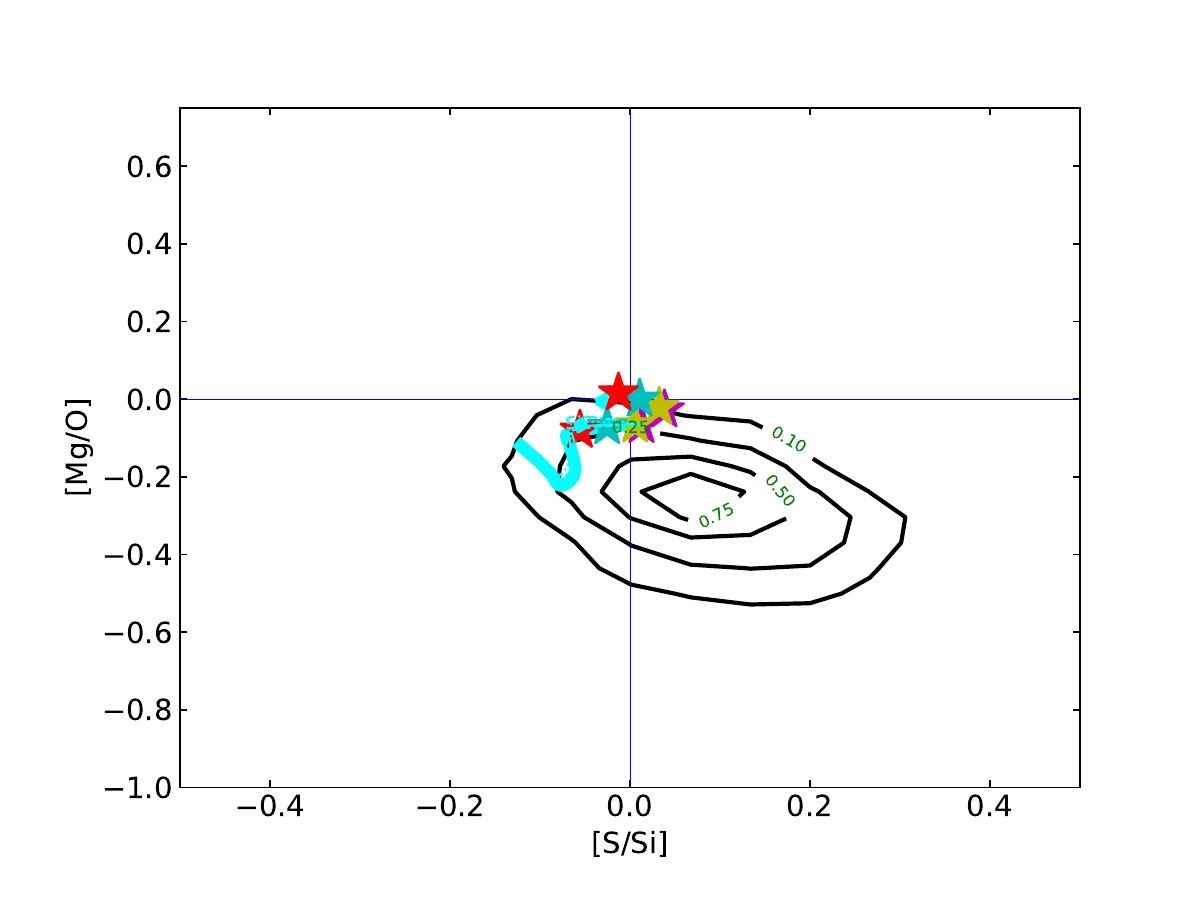}\par
    \includegraphics[width=0.8\linewidth]{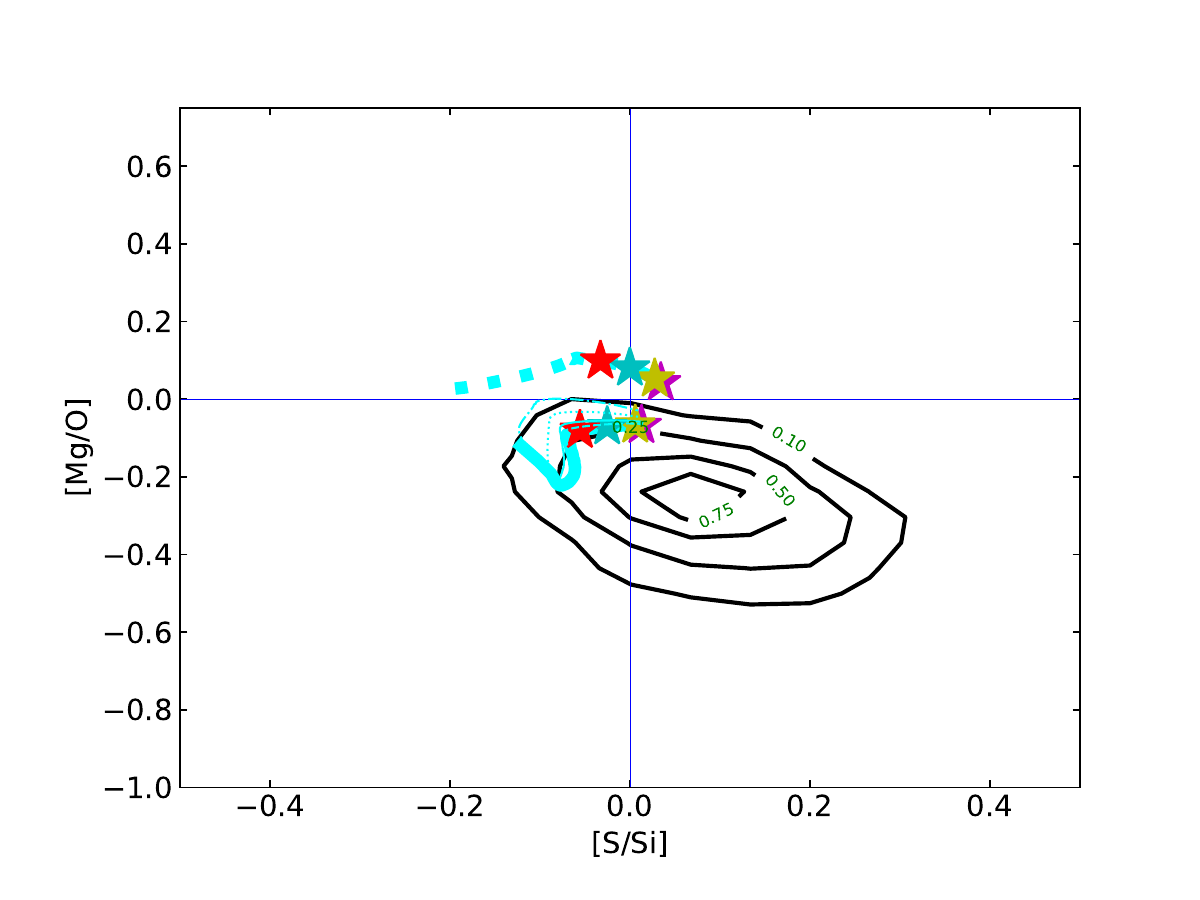}\par
\end{multicols}
    \caption{Selected elemental ratios normalized to the solar (L19$\_$3D by \protect\cite{lodders:19}, see Figure~\ref{fig: solar_ratios})} are plotted against each other for the GCE model set oK10, with CCSNe contribution up to M$_{\rm up}$ = 40 M$_{\odot}$ and using different fraction of faint CCSNe. As representatives of faint CCSNe, we use a 20M$_{\odot}$ model and a 25M$_{\odot}$ model (left and right panels, respectively, see Table \ref{tab: list_tk_plots/models}). For comparison, observations from the solar neighbourhood stars are reported as in the previous figures. 
    \label{fig: tk_plots/ratios_gce_oK10m40}
\end{figure*}

\begin{figure*}
\begin{multicols}{2}
    \includegraphics[width=0.8\linewidth]{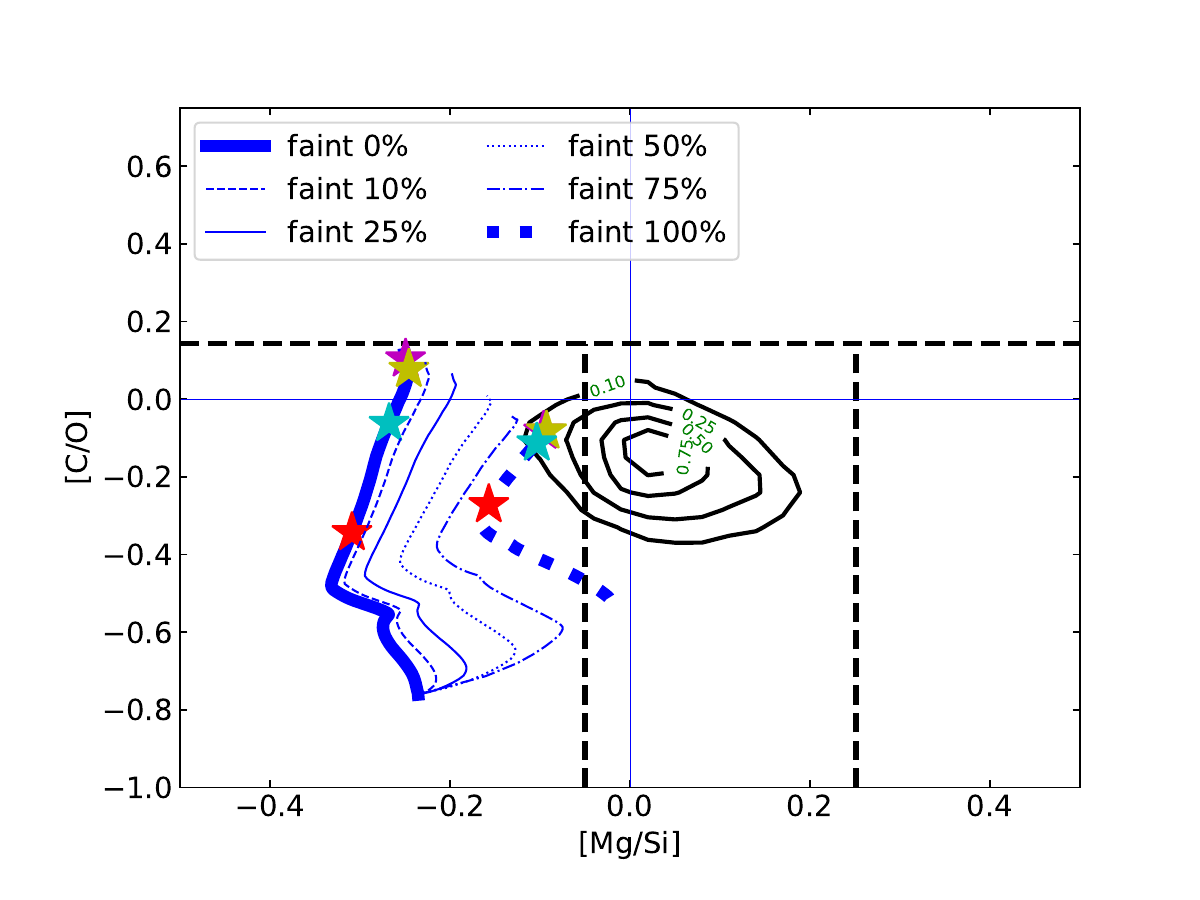}\par
    \includegraphics[width=0.8\linewidth]{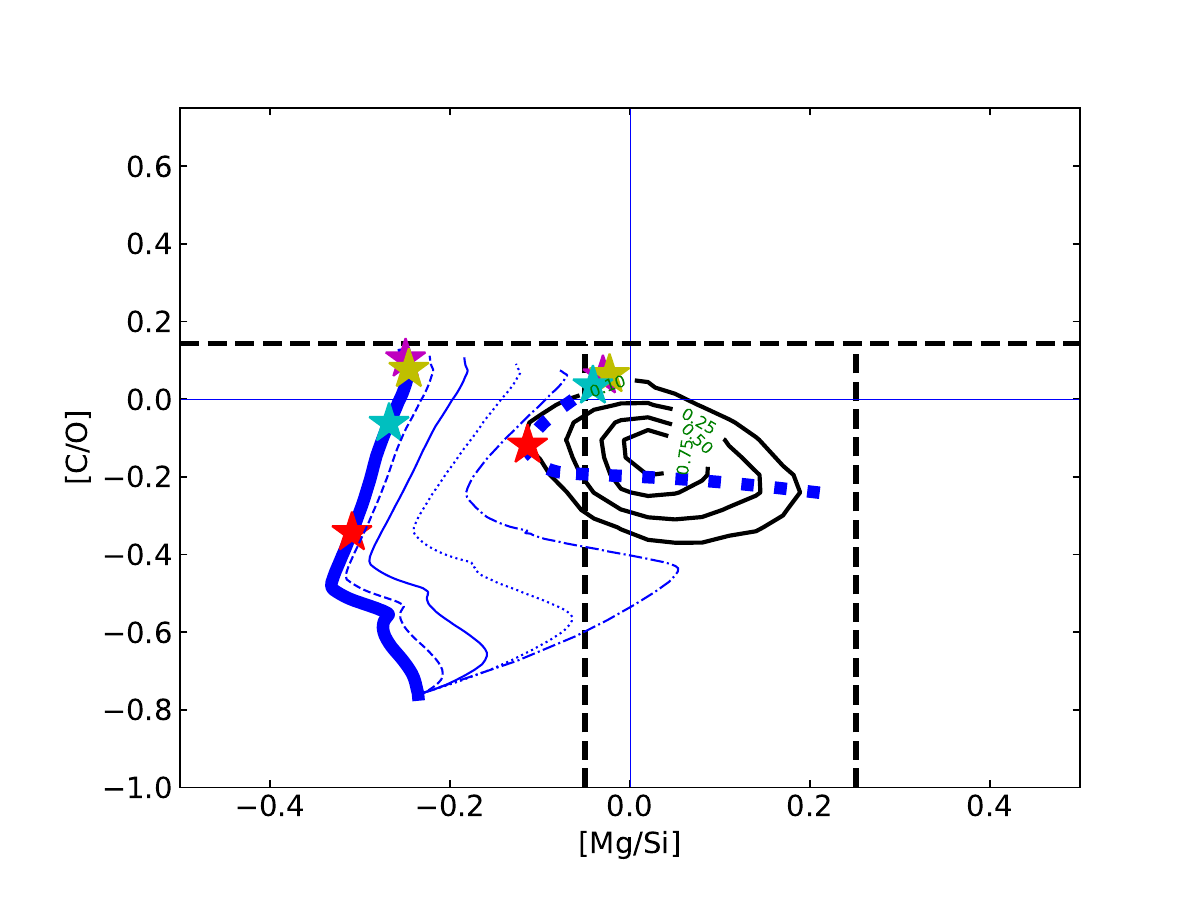}\par
    \end{multicols}
\begin{multicols}{2}
    \includegraphics[width=0.8\linewidth]{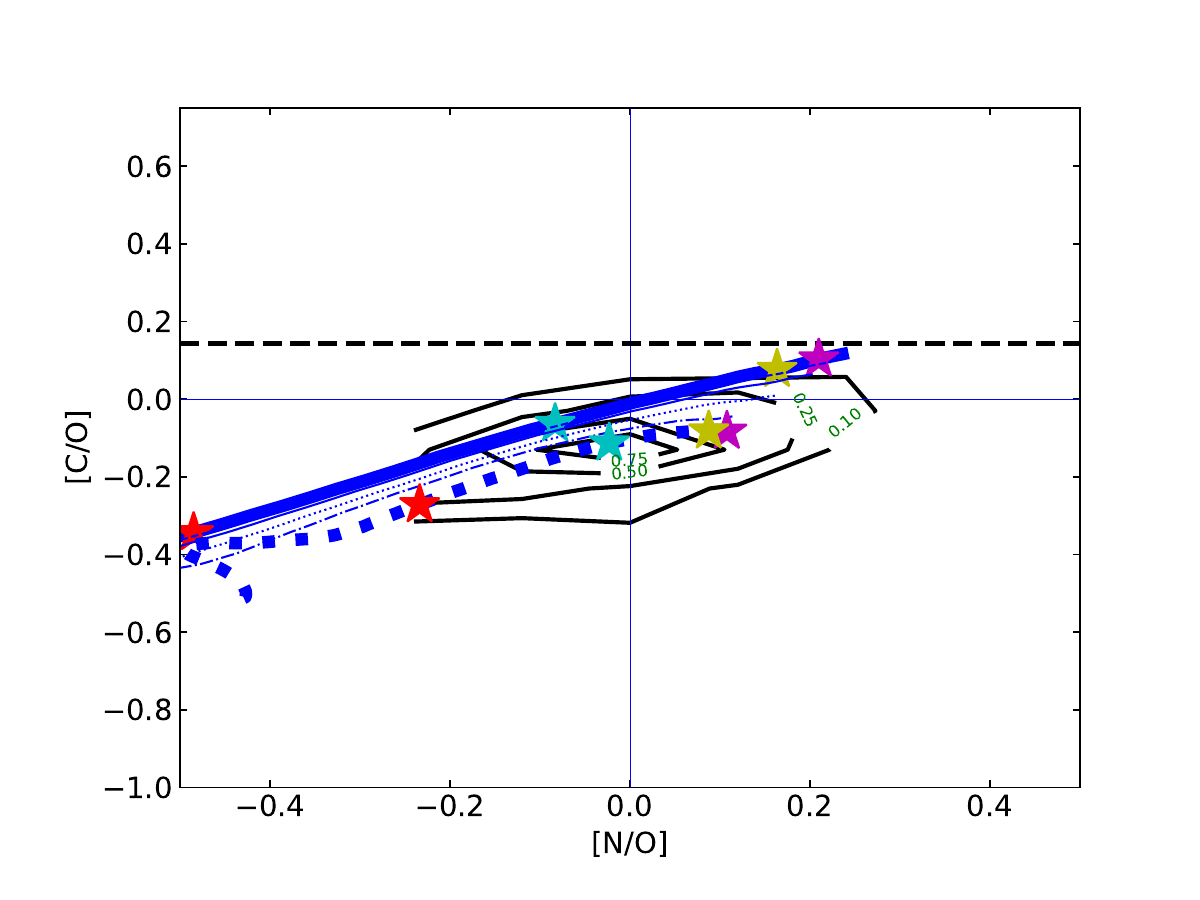}\par
    \includegraphics[width=0.8\linewidth]{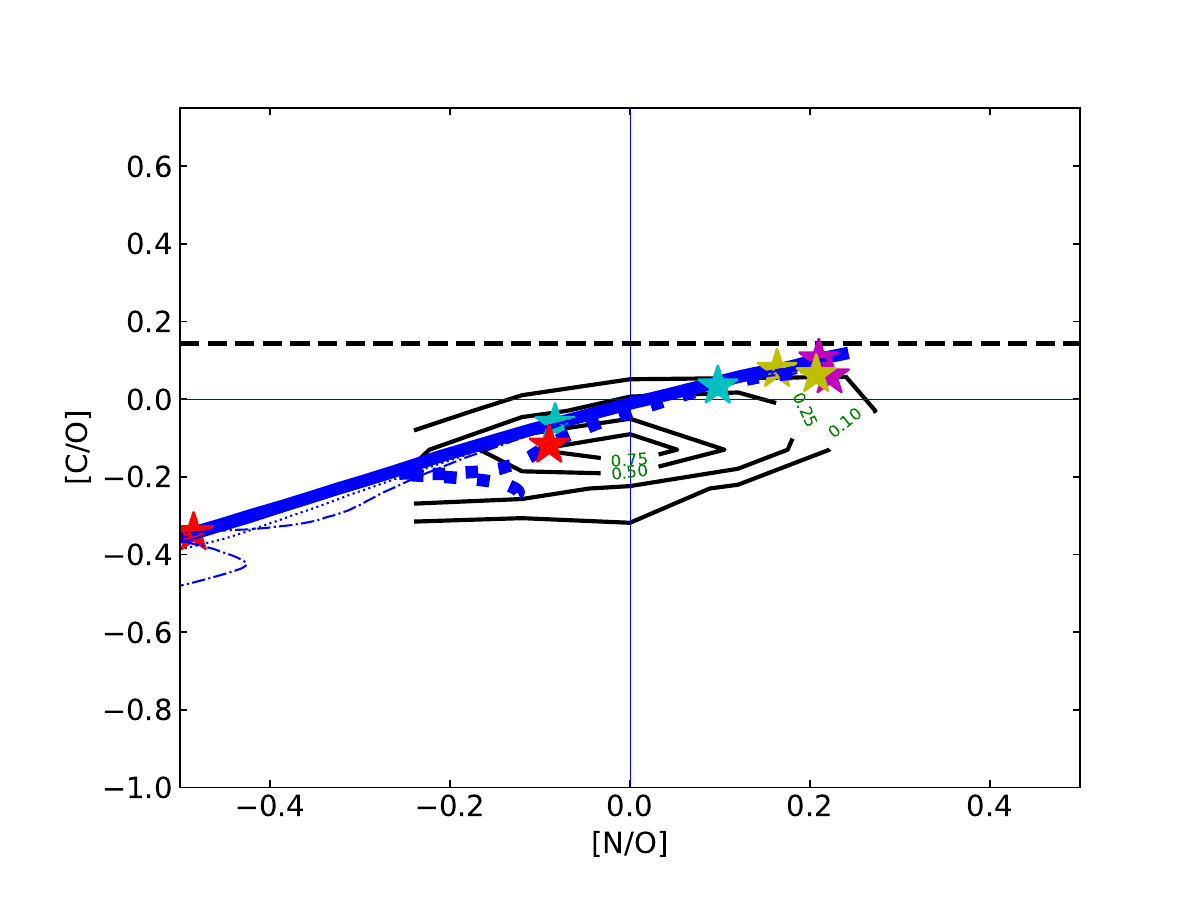}\par
\end{multicols}
\begin{multicols}{2}
    \includegraphics[width=0.8\linewidth]{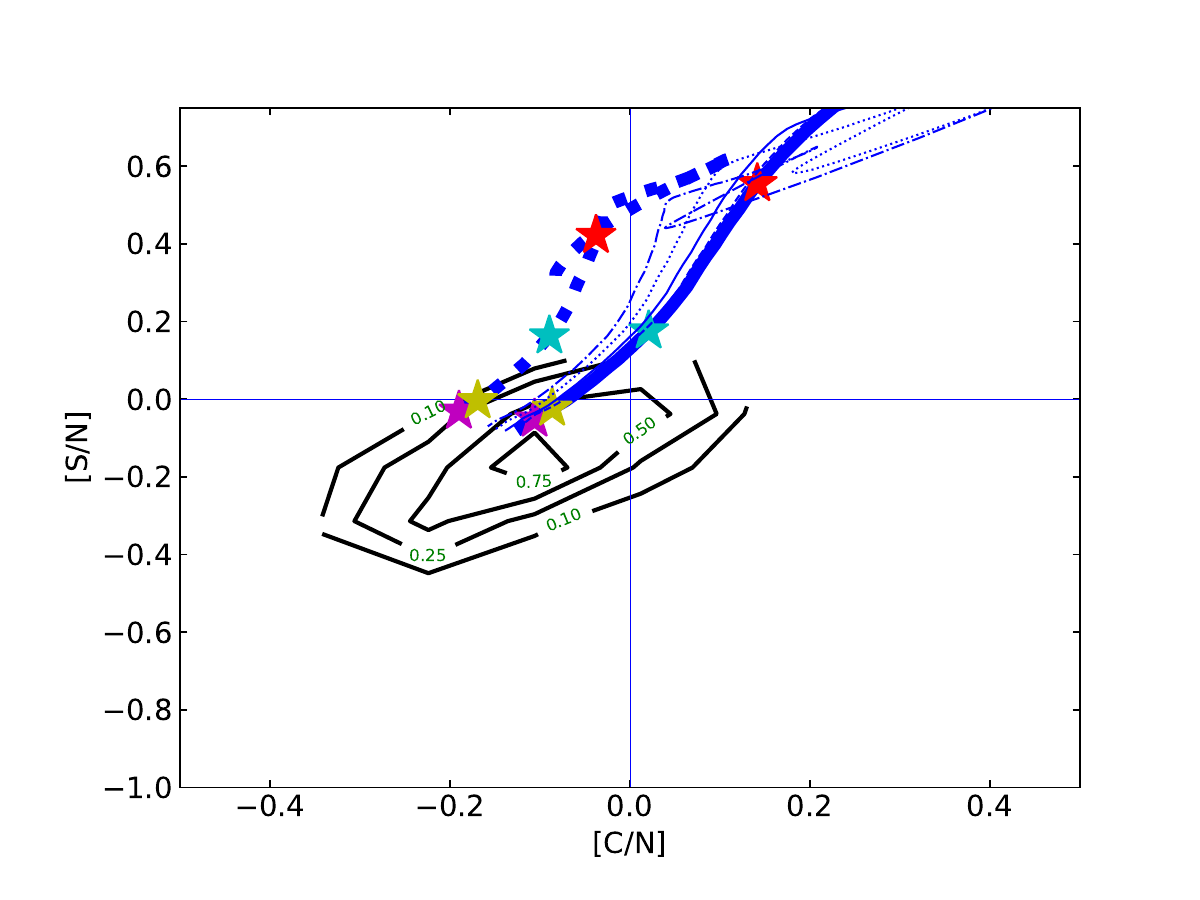}\par
    \includegraphics[width=0.8\linewidth]{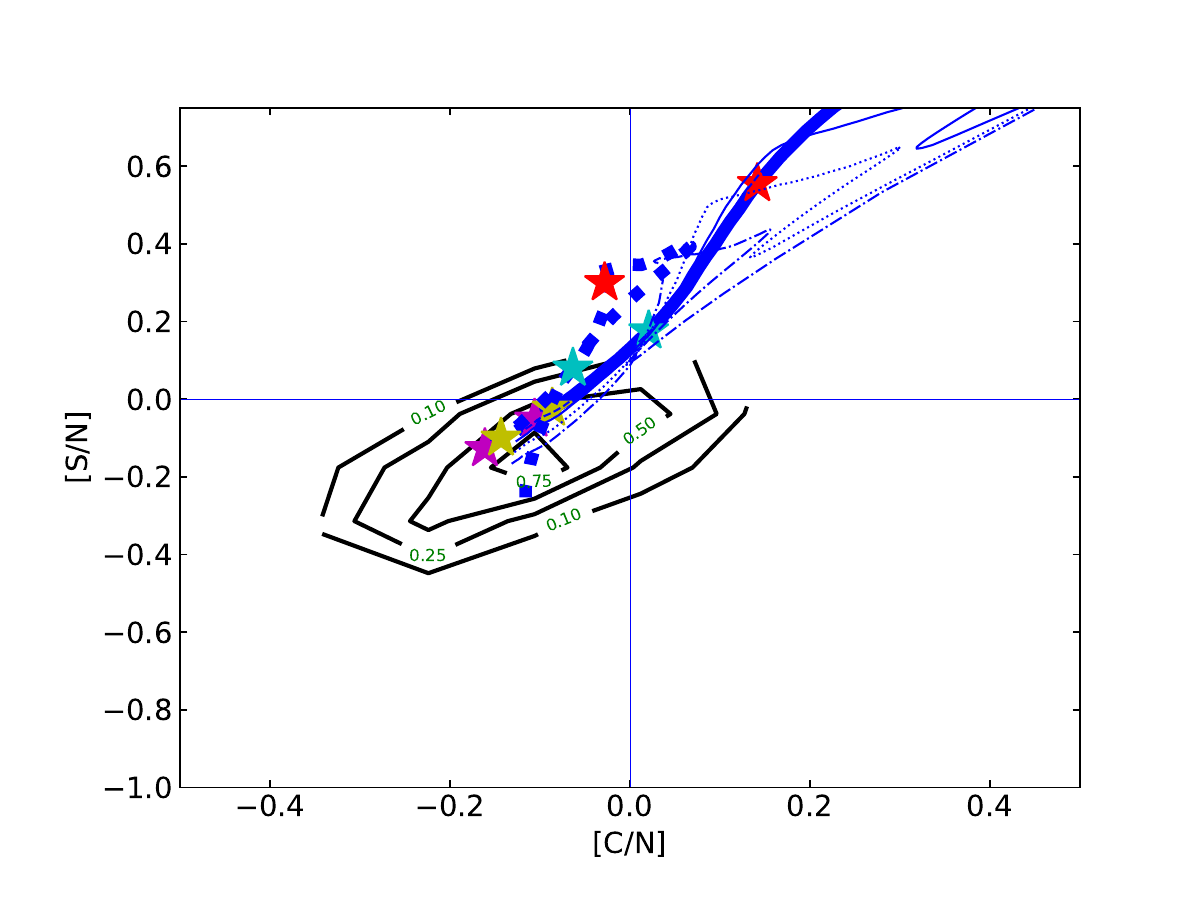}\par
\end{multicols}
\begin{multicols}{2}
    \includegraphics[width=0.8\linewidth]{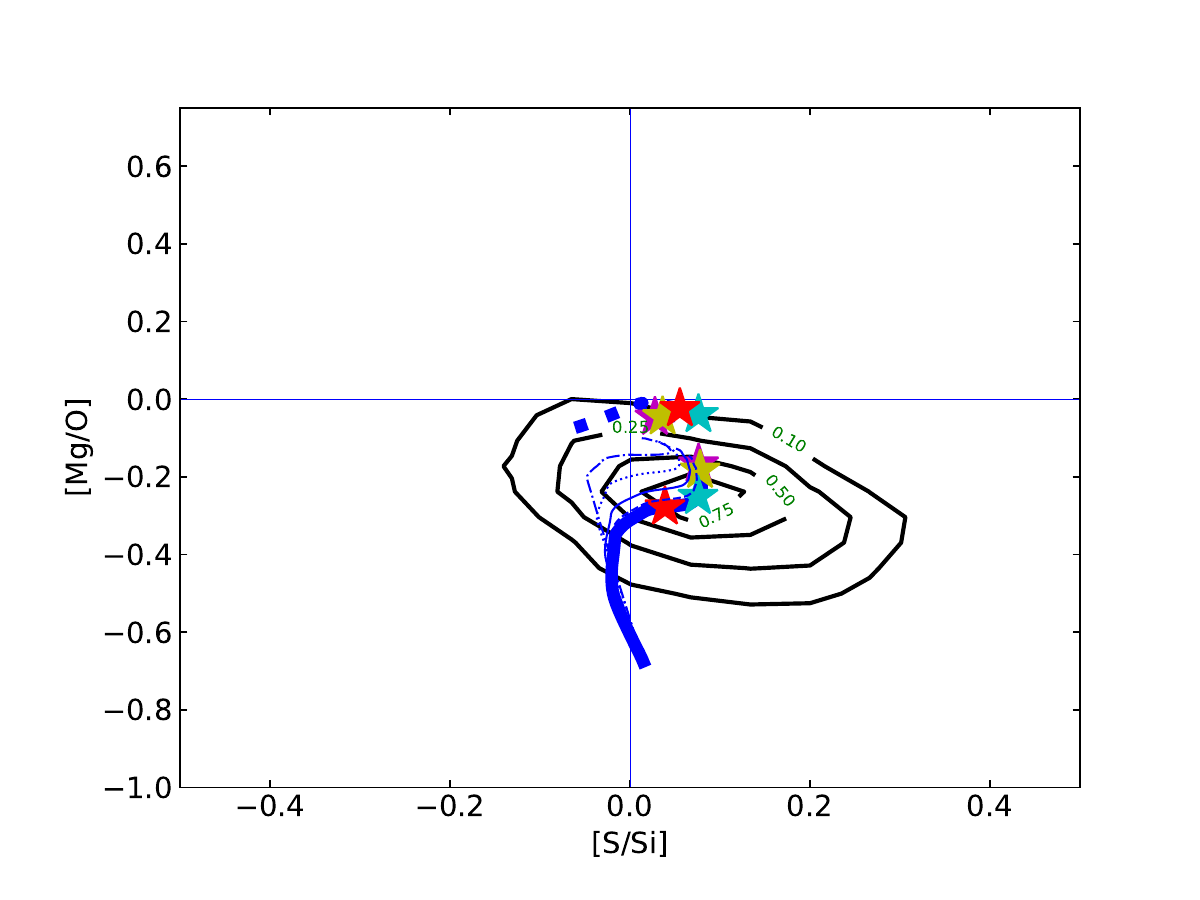}\par
    \includegraphics[width=0.8\linewidth]{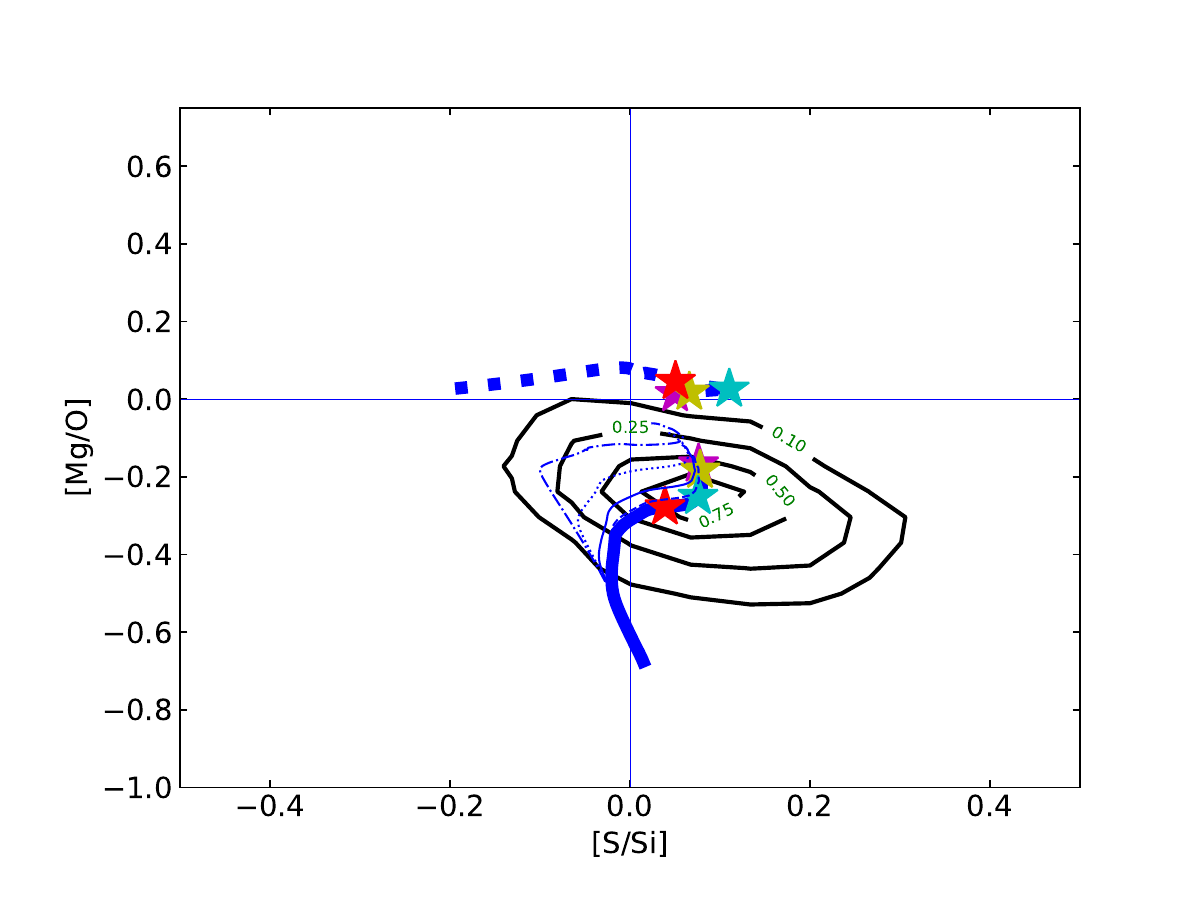}\par
\end{multicols}
    \caption{Same as in Figure~\ref{fig: tk_plots/ratios_gce_oK10m40}, but for the GCE model set oR18. 
    }
    \label{fig: tk_plots/ratios_oR18_m40}
\end{figure*}

Figure~\ref{fig: tk_plots/ratios_gce_oK10m40} reports the impact of faint CCSNe for the oK10 models. The full parameter space is considered, with the frequency of faint CCSNe from 0\% (which would correspond to the oK10no model shown in Figure~\ref{fig: tk_plots/ratios_mup40}) to 100\% (models oK10<faintSN$\_$model>f1p00). As representative of faint CCSNe models, we used the 20 M$_{\odot}$ (m20) and 25 M$_{\odot}$ (m25) CCSN models by \cite{ritter:18} shown in Figure~\ref{fig: ccsn_elements} (models oK10m20<faintSN$\_$weight> and oK10m25<faintSN$\_$weight>, respectively). As seen in Section~\S~\ref{sec: stars}, the m20 model still ejects some material carrying the signature of O-burning, while there is no Fe-rich Si-burning ejecta. The m25 model does not eject products of either Si-burning or O-burning. Note that considering present uncertainties in CCSN explosion and the wide zoo of CCSN remnants presently observed, we may expect the real fraction of faint CCSNe to be somewhere in between the two extreme cases, oK10no and oK10<faintSN$\_$model>f1p00. In the left panels of Figure~\ref{fig: tk_plots/ratios_gce_oK10m40}, no substantial effect is observed from considering faint CCSNe. The first reason is that CNO elements are not substantially affected by the CCSN explosion as they are mostly produced during stellar evolution before core collapse. Therefore, their relative abundances do not change significantly in faint CCSNe, as compared to standard CCSNe. The second reason is that although Si and S are O-burning products, the m20 faint CCSNe model used in these tests still eject some Si-rich and S-rich material, without affecting the [S/N] and the [S/Si] ratios much. 

The impact of faint CCSNe becomes more relevant in the GCE models where m25 is used. In the upper-right panel of Figure~\ref{fig: tk_plots/ratios_gce_oK10m40}, [C/O] and [Mg/Si] increase by about 0.2 dex and 0.1 dex respectively, when comparing oK10m25f0p50 to oK10no. Models with higher faint CCSNe frequency oK10m25f0p75, oK10m25f0p90 and oK10m25f1p00 have even larger increases, up to solar ratios. We would, however, consider these last models of the parametric study as less realistic. The reason for the impact on [C/O] is that a significant part of the former O-rich C shell is not ejected by m25 (Figure~\ref{fig: ccsn_elements}). Therefore, the overall galactic enrichment is driven to higher C/O-ratios. The impact on [Mg/Si] is instead smaller, since, as with O, Mg-rich material is not fully ejected by this model. A faint CCSN model with a lower masscut allows for more O and Mg ejection and still no O-burning products like Si. This is still realistic to consider within the uncertainties \citep[e.g.,][]{fryer:18} and would, in principle, achieve larger [Mg/Si]. The impact on the [C/O] would be small. 
In the second from top right panel of Figure~\ref{fig: tk_plots/ratios_gce_oK10m40}, [N/O] increases with increasing the faint CCSNe frequency, up to +0.35. The effect on the [N/O] ratio is the same as discussed for the [C/O] ratio, where N is mostly made in the most external layers of CCSNe. 
The model oK10m25f0p50 shows a reduction of the [S/N] ratio down to $-$0.15, providing a possible explanation of the observation range. As we mentioned in the previous section, however, the nucleosynthesis of N in CCSNe may be affected by physics not considered in this work, like stellar rotation or H-ingestion events, which would both increase the N yields as compared to S. Finally, also in this case there are only minor effects on the [Mg/O] and [S/Si] ratios.
 
Figure~\ref{fig: tk_plots/ratios_oR18_m40} presents the equivalent oR18 models. Qualitatively, there are similar effects using the m20 and m25 faint CCSNe as debated above for the oK10 models, with overall a more significant impact in the oR18m25<faintSN$\_$weight> models. The top panels confirm the increasing trend of [Mg/Si] with the faint SN frequency, with a ratio higher than about 0.1 dex in the oR18m20f0p50 and oR18m25f0p50 with respect to oR18no.
Instead, the [C/O] increase is more limited, as compared to the oK10 models, and more dependent on the evolution time of the model. The evolution of the [N/O] ratio in the oR18 models is also different compared to the effect seen in the oK10 models. While the final abundances vary by less than 0.1 dex between oR18m20no and oR18m20f1p00 and are mostly unaffected when using the m25 faint CCSN model, the ratio starts to increase at much earlier times, with the [N/O] ratio higher up to 0.4 dex. The impact on the final [S/N] in the oR18 models including faint CCSNe is less than 0.1 dex, while it was more significant in oK10 models using an m25 faint CCSN. Finally, like for the oK10 models, there is no effect on the [S/Si] ratio. There is instead an increase of the [Mg/O] ratio, by less than 0.1 dex, between the oR18m20f0p50 and oR18m25f0p50 with respect to oR18no, with an increase up to 0.2 dex reaching the solar ratio for oR18m20f1p00 and oR18m25f1p00. 

\section{Summary and Conclusions}
\label{sec: summary}

We presented 198 new GCE simulations of the solar neighbourhood, with the main goal to study the production and evolution of the important planet-building nuclides C, N, O, Mg, Si, and S, in comparison to stellar observations. We chose these elements because results from simulations of planet formation and evolution depend on their initial abundances \citep[e.g.,][]{frank:14}. One of the fundamental purposes of this work is also to provide an accessible (but comprehensive) picture about the challenges and the uncertainties in stellar simulations, observations and GCE, and at the same time we also want to make clear what are the needs of planet-formation and evolution studies from observations and from GCE, in the light of present and future opportunities to unfold thanks to the new generation of observation facilities. 
The new data coming from TESS \citep[][]{ricker:15} and from JWST \citep{beichman:14}, as well as from future facilities like ARIEL \citep[][]{tinetti:18,turrini:18,turrini:21a}, will complement those being provided by ongoing ground-based efforts \citep[e.g.][]{giacobbe:21,carleo:22, guilluy:22} and we can anticipate a greatly enhanced window with which to study these processes in more detail.
In this context, GCE models provide the initial composition of 
stars and of their proto-planetary disks where planets are formed at different times and locations in the Galaxy for all the elements. Based on theoretical simulations and observations, we also expect that planet-formation processes will drastically affect some of the planet abundances measured today with respect to the pristine abundances of the proto-planetary disk, while others will only be marginally affected. The results of GCE models provide therefore a crucial additional benchmark for simulations of planet formation, in particular when elemental abundances from the stellar host are uncertain or not available.

The GCE simulation framework presented here is made of five sets of models, where the impacts of stellar yields, of the stellar mass upper limit contributing to the chemical evolution (M$_{\rm up}$) and of faint CCSNe were explored. 
With our models, classical stellar sources used to reproduce the evolution of the elements C, N, O, Mg, Si and S are not able to fully reproduce the solar abundances, and/or the observed range in the solar neighbourhood, in particular for the [C/O] and [Mg/Si] diagram.
In our analysis, we did not apply any corrections to force the results from GCE simulations to match the solar abundances. 
The chemical enrichment history of these elements in the Milky Way is complicated, since all the contributions from CCSNe, SNIa and AGB stars must be taken into account, along with their different timescales \citep[e.g.,][]{molla:15,mishenina:17,prantzos:18,kobayashi:20}. 

We did not find a specific set of yields that is able to solve all the ratios considered in our analysis within the correct GCE evolution timescale. We also show that the impact of M$_{\rm up}$ is in general limited for these elements considered, and it is model dependent. 

By considering realistic frequencies of faint CCSNe, we obtain instead variations of elemental ratios in the order of 0.1-0.2 dex. In particular, we find that the increase of [C/O] and [Mg/Si] with increasing faint CCSN frequency may help to better reproduce the abundances 
observed in stars in the solar neighbourhood. The potential reduction of [S/N] in the order of 0.2 dex can also help match the range of observations, with its impact depending on the set of CCSN yields adopted.

The reduction of the observational uncertainties for the elements considered will be a crucial step towards solving present discrepancies between GCE simulations and observations. The more limited abundance dispersion in the stellar sample by \cite{bedell:18} compared to other analogous works requires independent verification. 
\cite{ramirez:14} discussed the star-to-star scatter for different elements, showing that while several elements (including O and Si discussed in this paper) present a variation compatible with the measurement errors, other elements not discussed here (e.g., Na, Al, V, Y, and Ba) may have larger discrepancies. From a similar analysis \cite{adibekyan:15a} instead found that the star-to-star scatter may simply increase with the decrease of the number of spectral lines used in the derivation of the abundances. This would indicate that a good fraction of the observed scatter is not astrophysical.
In more general terms, the comparison between data from different observational surveys obtained using different spectral lines and stellar parameters should be undertaken with caution. New comprehensive atomic physics investigations of transition probabilities for relevant spectral lines are further needed in order to improve the present results. 

We have highlighted how a different definition of solar references provide a major additional source of uncertainty. We have shown that spectroscopic observations vary significantly between different works once absolute abundances are compared instead of those normalised to solar. Planet-formation simulations, however, use absolute pristine stellar abundances as a starting point, and therefore they are directly affected \citep[][]{spaargaren:23}. Alongside C, N, O, Mg, Si, and S discussed in this work, elements of interest for planet-formation studies include lithophile elements such as Cl, Cr, K, Na, V, P, Ti, Al and Ca 
the abundances of which
can potentially be better constrained by future facilities such as ARIEL \citep{tinetti:18,turrini:18}. As discussed by \cite{turrini:21a} and \cite{turrini:22} each of these elements, being more refractory than O, can be used in place of S to study the planet-formation history together with C, N and O. As the accuracy of atmospheric retrieval methods for exoplanetary observations is currently capped to about 10-20\% \citep[see][for discussion]{barstow:20,turrini:22}, the characterisation of stellar abundances with the precision of 0.1 dex would provide a solid base for the next generation of planet-formation studies to compare with atmospheric data. 
Note that Fe is another element essential for mineralogy and planet formation, but it is not included in the analysis presented here. Undeniably, CCSNe yields for Fe are quite uncertain \citep[e.g.,][]{pignatari:16,sukhbold:16,curtis:19}, and stellar-yields uncertainties are then propagated to GCE, where additional uncertainties include for instance assigning the correct populations of SNIa contributing to GCE \citep[e.g.,][]{lach:20, grunov:21}. We therefore report the study of the GCE of Fe and of the Fe-group elements (for a consistent analysis they cannot be treated separately) in the Milky Way disk as a separate work (Trueman et al., submitted).

Once the uncertainties in spectroscopic observations and in the solar composition are sufficiently reduced, GCE simulations hold the potential to generate a 
robust 
fit to the compositional catalogue of stars in the solar neighbourhood. Notwithstanding, more powerful constraints need to be derived on the role of faint CCSNe required to cover the full range of observations. 




\section*{Acknowledgements}
We acknowledge support from STFC (through the University of Hull's Consolidated Grant ST/R000840/1), and access to {\sc viper}, the University of Hull HPC Facility. MP acknowledges the support to NuGrid from the National Science Foundation (NSF, USA) under grant No. PHY-1430152 (JINA Center for the Evolution of the Elements), the "Lendulet-2014" Program of the Hungarian Academy of Sciences (Hungary), the ERC Consolidator Grant funding scheme (Project RADIOSTAR, G.A. n. 724560, Hungary), the ChETEC COST Action (CA16117), supported by the European Cooperation in Science and Technology, and the IReNA network supported by NSF AccelNet. We acknowledges support from the ChETEC-INFRA project funded by the European Union’s Horizon 2020 Research and Innovation programme (Grant Agreement No 101008324). TCB also acknowledges partial support from PHY 14-30152; Physics Frontier Center/JINA Center for the Evolution of the Elements (JINA-CEE), and OISE-1927130: The International Research Network for Nuclear Astrophysics (IReNA), awarded by the US National Science Foundation. TCB and BKG thank the Leverhulme Trust for the award of a Visiting Professorship for TCB to Hull University. DT acknowledges the support of the Italian National Institute of Astrophysics (INAF) through the INAF Main Stream project ``Ariel and the astrochemical link between circumstellar discs and planets'' (CUP: C54I19000700005), and of the Italian Space Agency (ASI) through the ASI-INAF contract no. 2021-5-HH.0. SJM thanks the Alexander von Humboldt Foundation for support during significant phases of this writing. SJM is supported by the Research Centre for Astronomy and Earth Sciences, a Centre for Excellence of the Hungarian Academy of Sciences.

\section*{Data Availability}

The data generated for this article will be shared on reasonable request to the corresponding author.




\bibliographystyle{mnras}
\bibliography{main.bbl} 



\appendix
\section{Complete list of figures for GCE simulations}
\label{app_1: plots extra}

In this section the full list of figures exploring the impact of both M$_{\rm up}$ and faint supernovae parameter spaces in GCE simulations are provided. 

\begin{figure*}
\begin{multicols}{2}
    \includegraphics[width=0.8\linewidth]{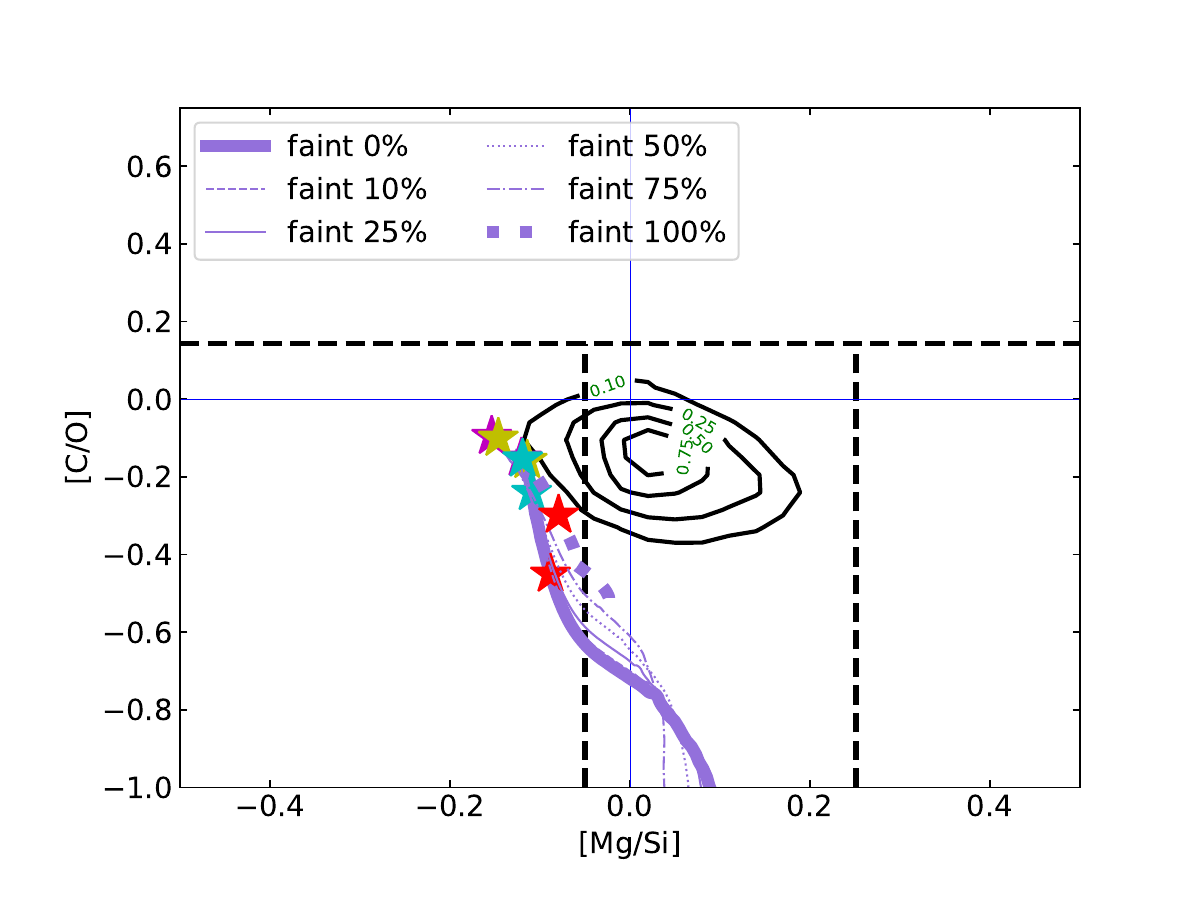}\par
    \includegraphics[width=0.8\linewidth]{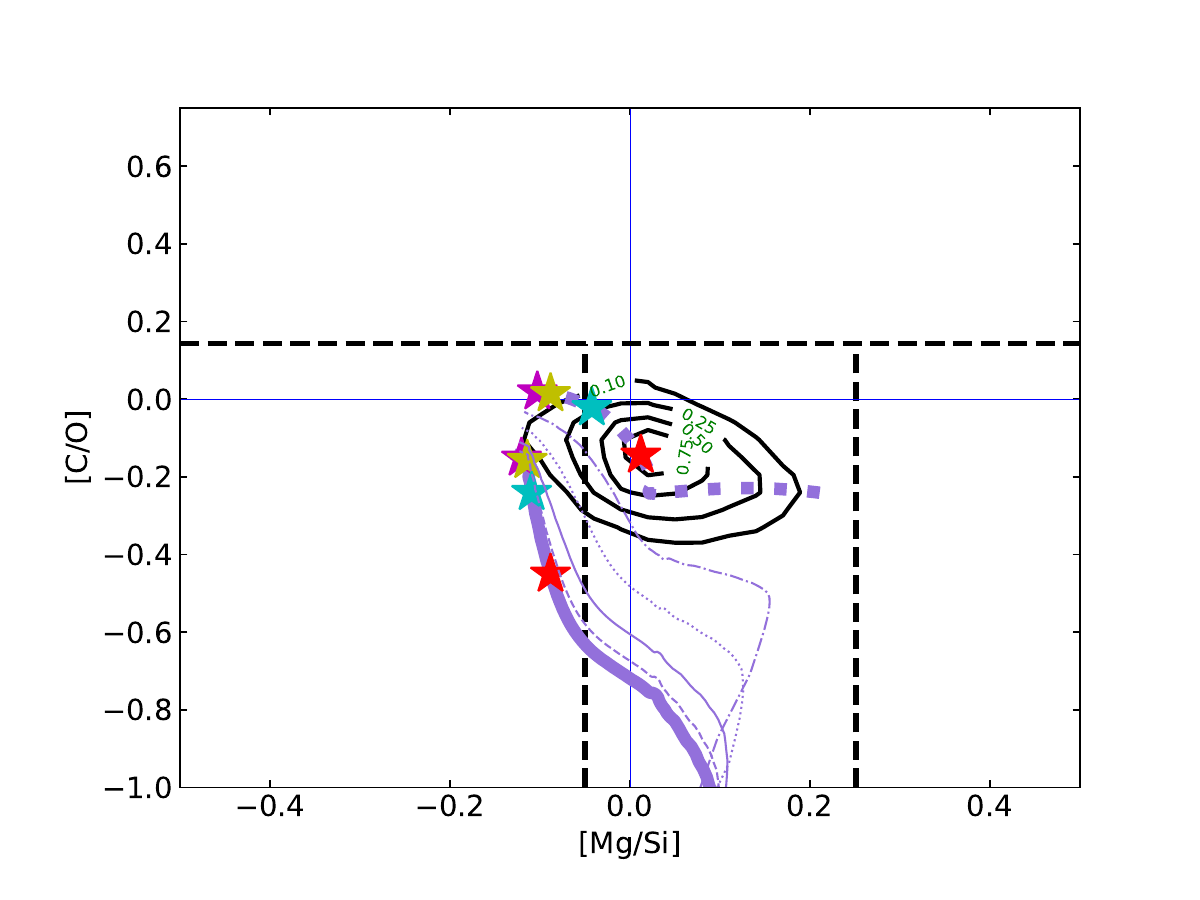}\par
    \end{multicols}
\begin{multicols}{2}
    \includegraphics[width=0.8\linewidth]{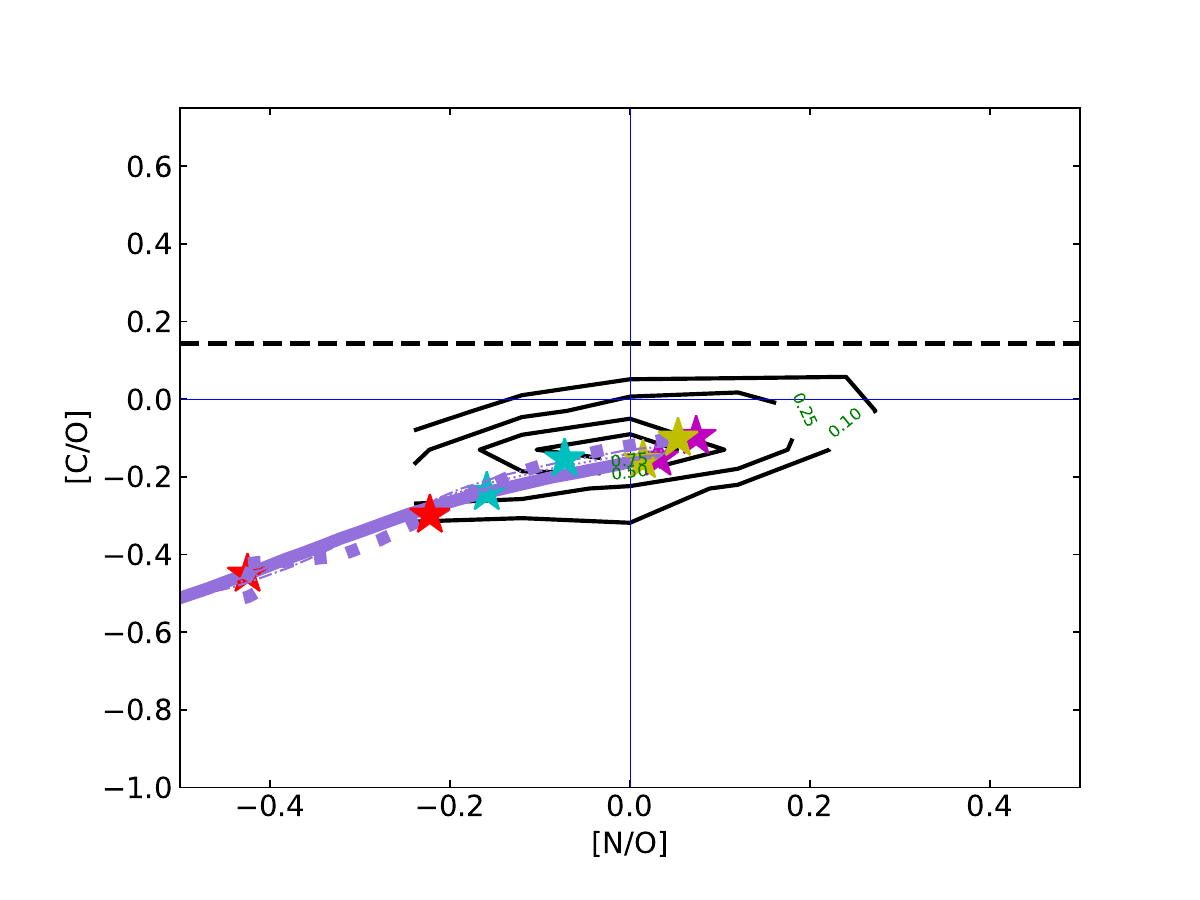}\par
    \includegraphics[width=0.8\linewidth]{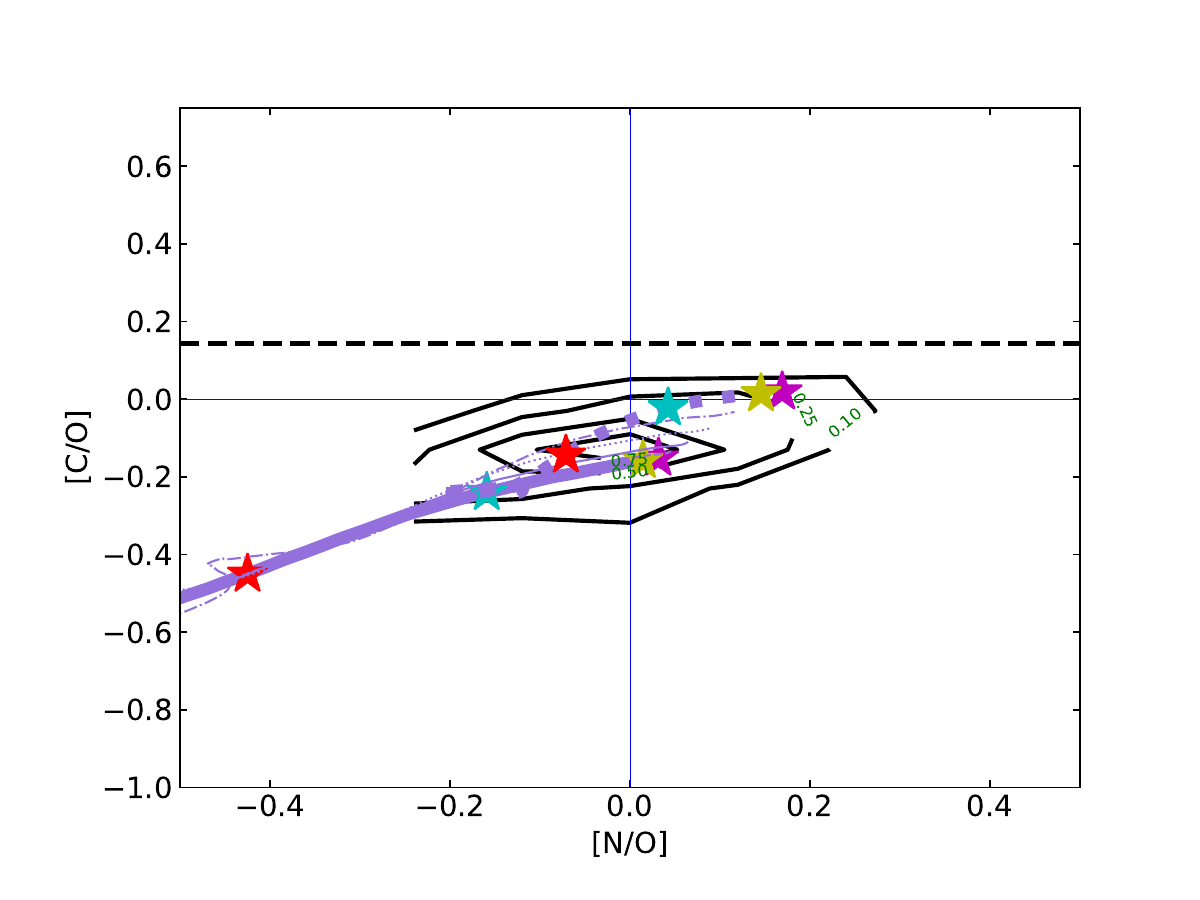}\par
\end{multicols}
\begin{multicols}{2}
    \includegraphics[width=0.8\linewidth]{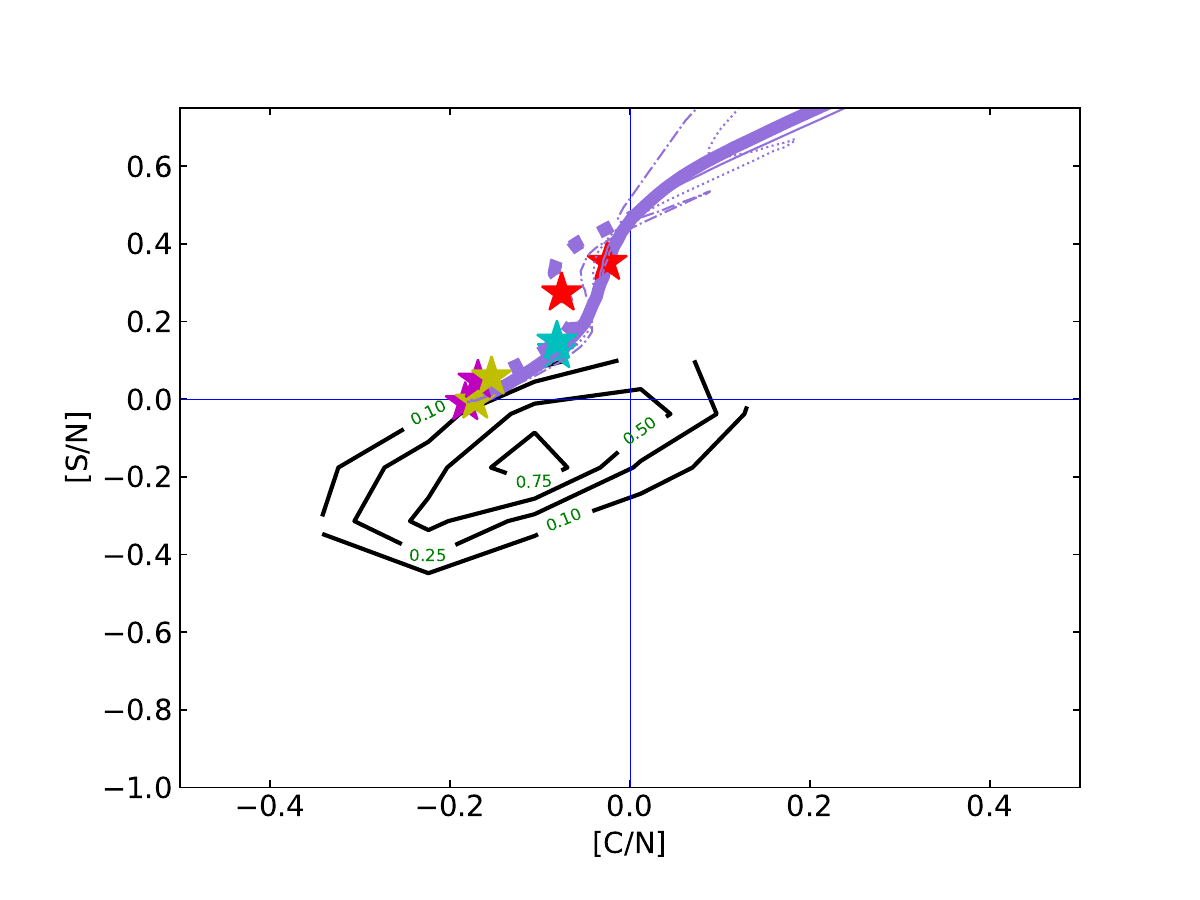}\par
    \includegraphics[width=0.8\linewidth]{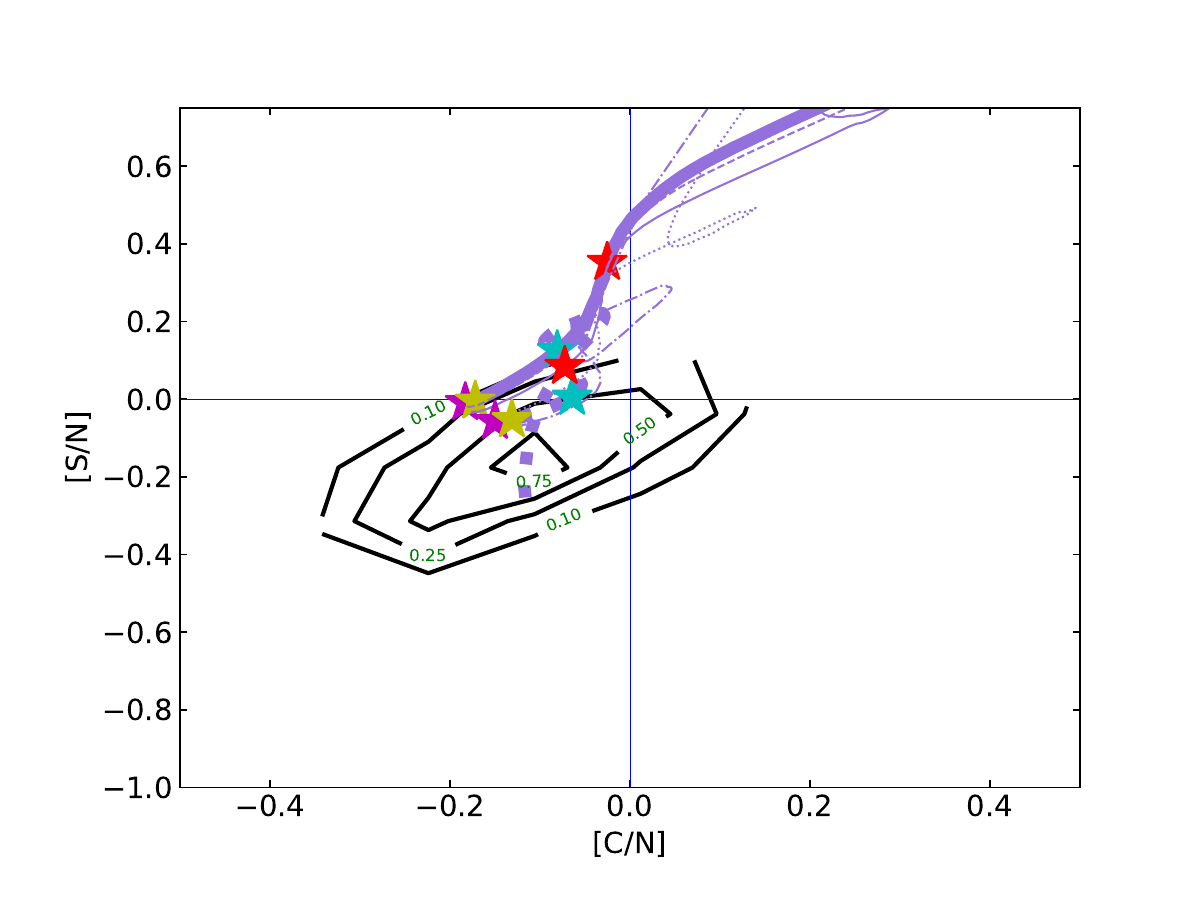}\par
\end{multicols}
\begin{multicols}{2}
    \includegraphics[width=0.8\linewidth]{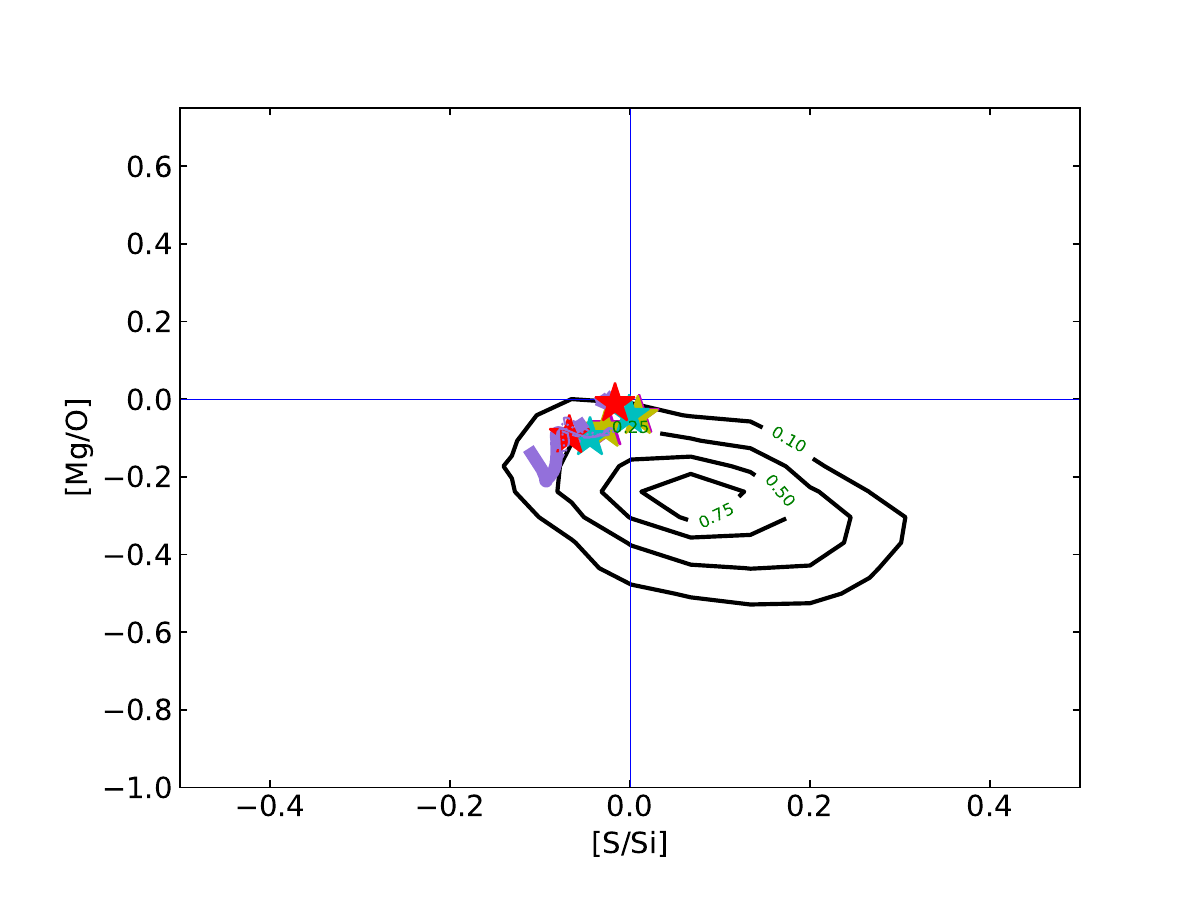}\par
    \includegraphics[width=0.8\linewidth]{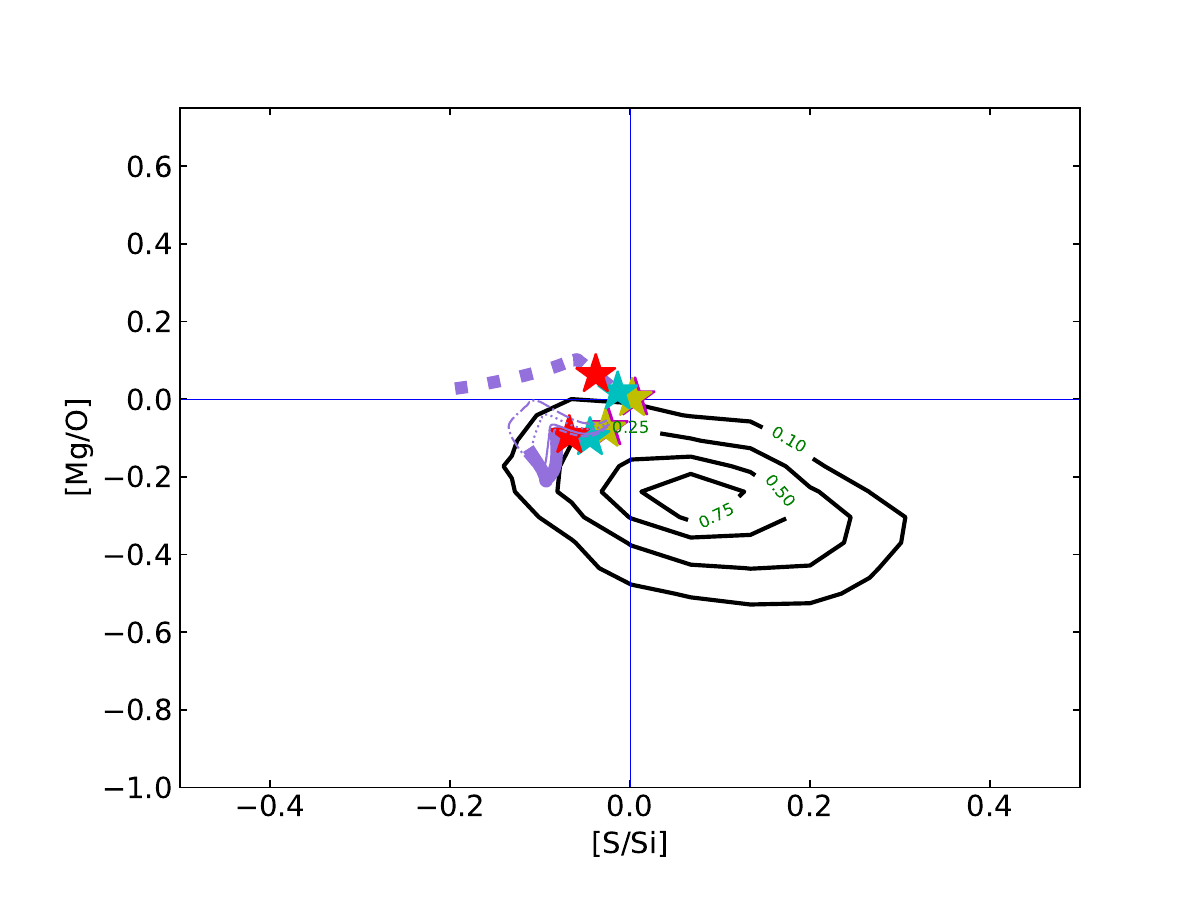}\par
\end{multicols}
    \caption{Same as in Figure~\ref{fig: tk_plots/ratios_gce_oK10m40}, but for the GCE model set oK06. 
    }
    \label{fig: tk_plots/ratios_oK06_m40}
\end{figure*}

\begin{figure*}
\begin{multicols}{2}
    \includegraphics[width=0.8\linewidth]{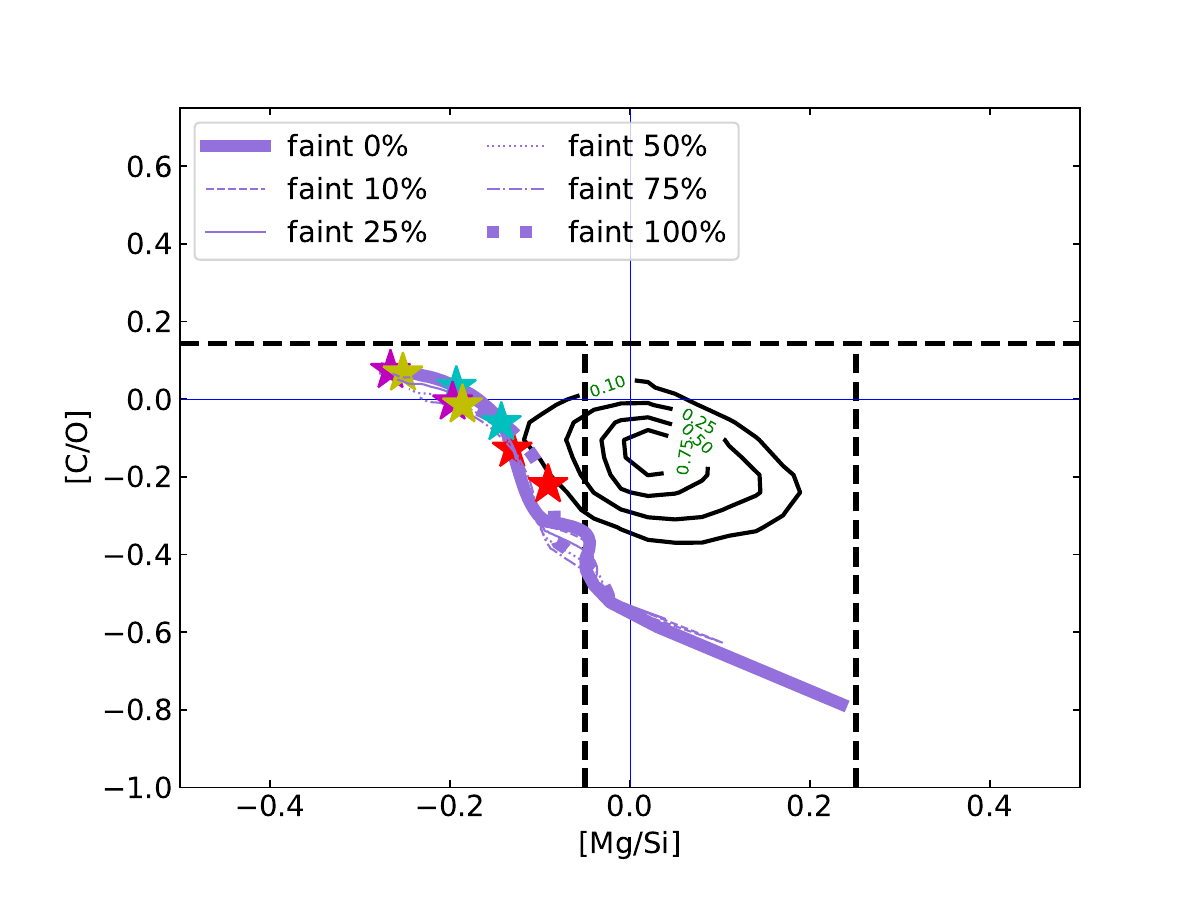}\par
    \includegraphics[width=0.8\linewidth]{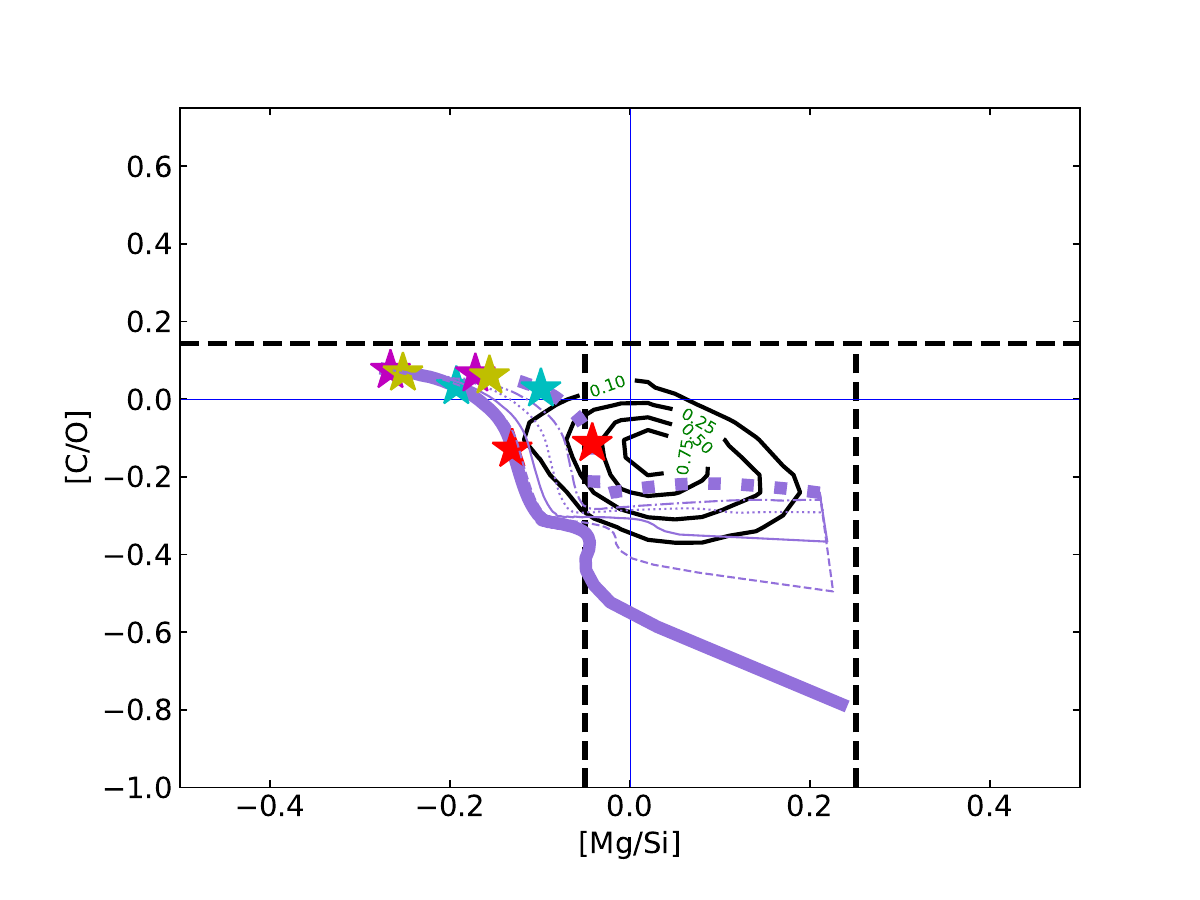}\par
    \end{multicols}
\begin{multicols}{2}
    \includegraphics[width=0.8\linewidth]{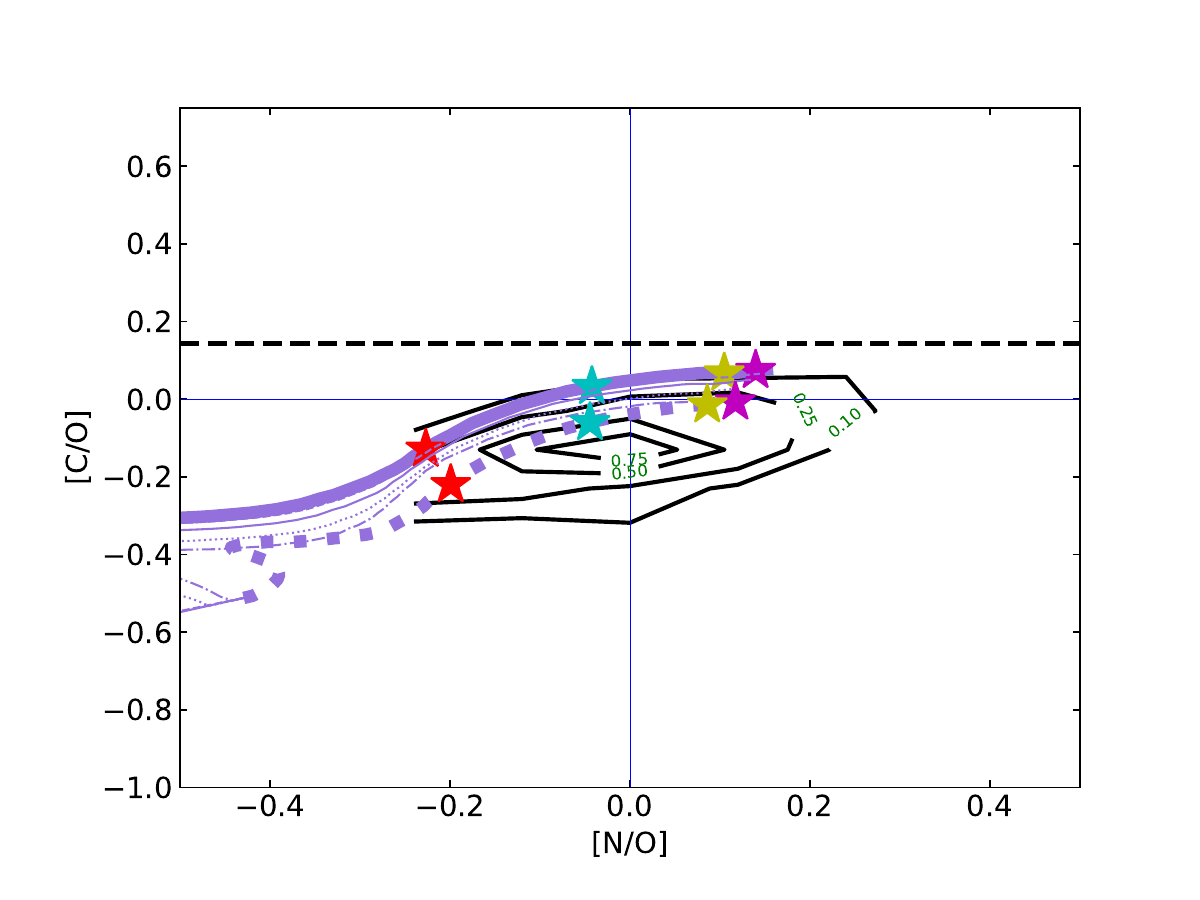}\par
    \includegraphics[width=0.8\linewidth]{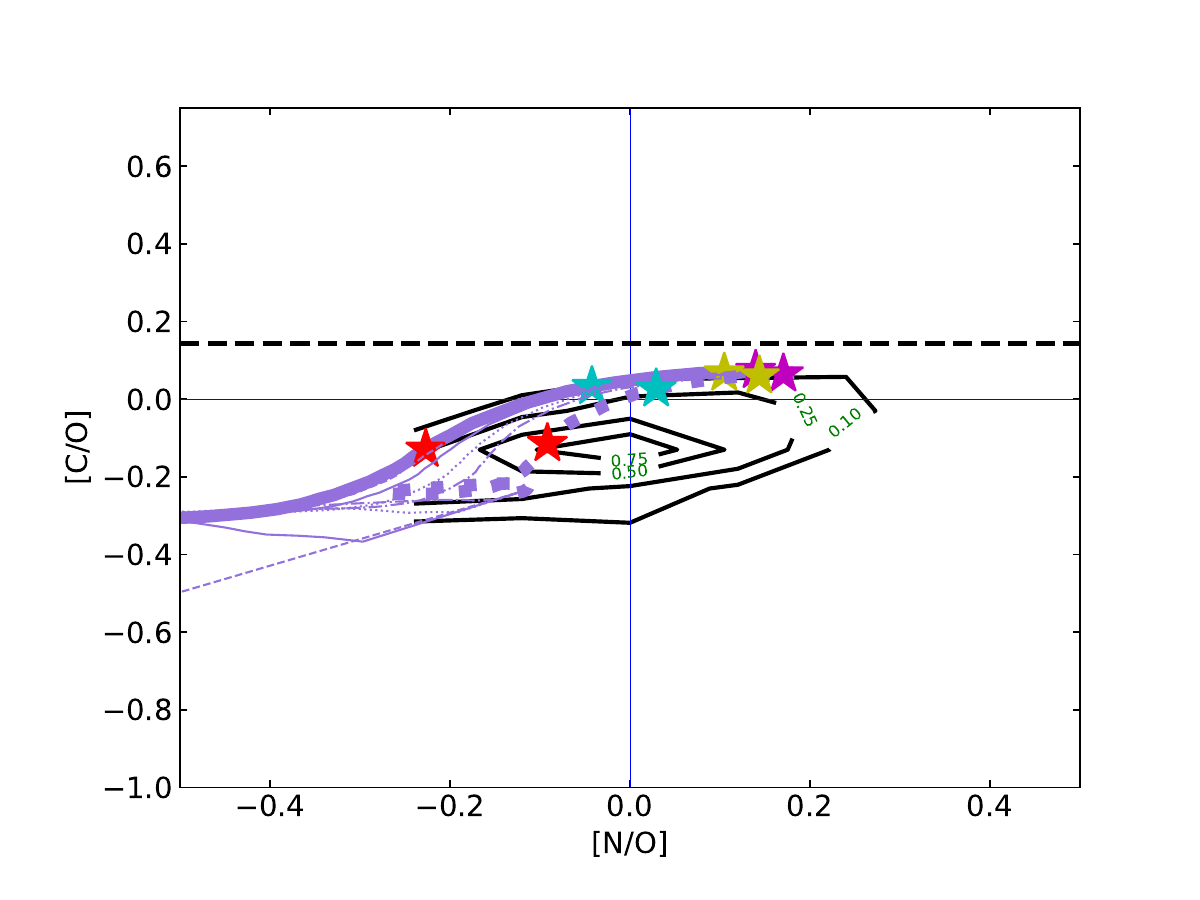}\par
\end{multicols}
\begin{multicols}{2}
    \includegraphics[width=0.8\linewidth]{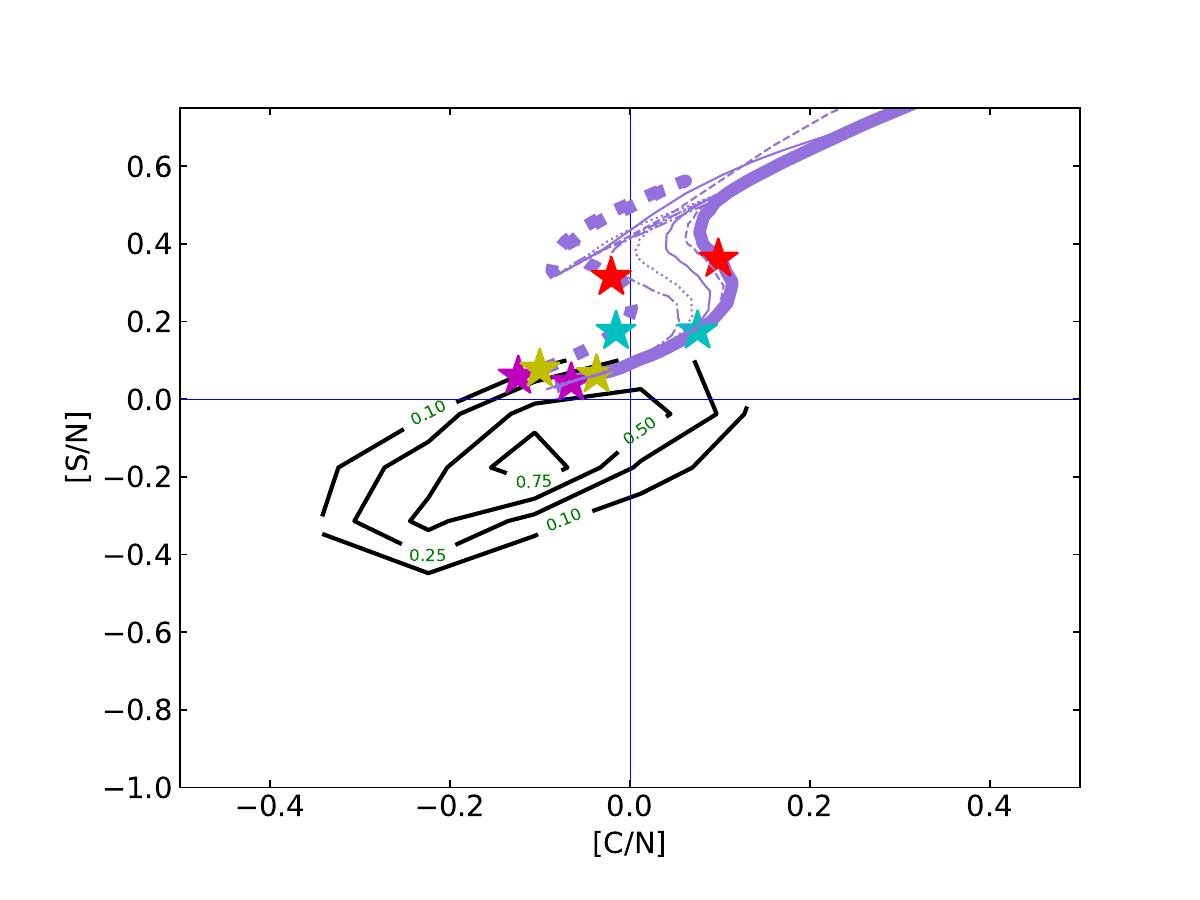}\par
    \includegraphics[width=0.8\linewidth]{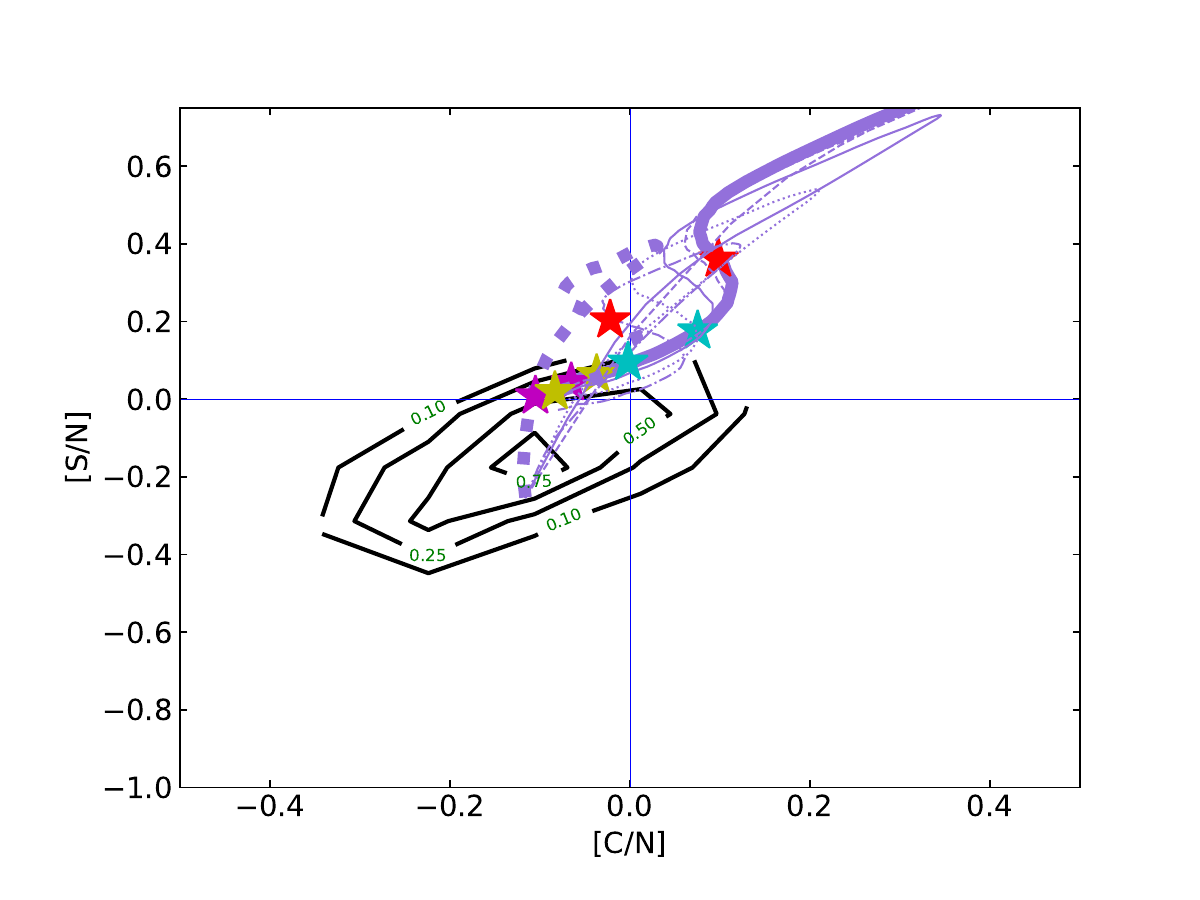}\par
\end{multicols}
\begin{multicols}{2}
    \includegraphics[width=0.8\linewidth]{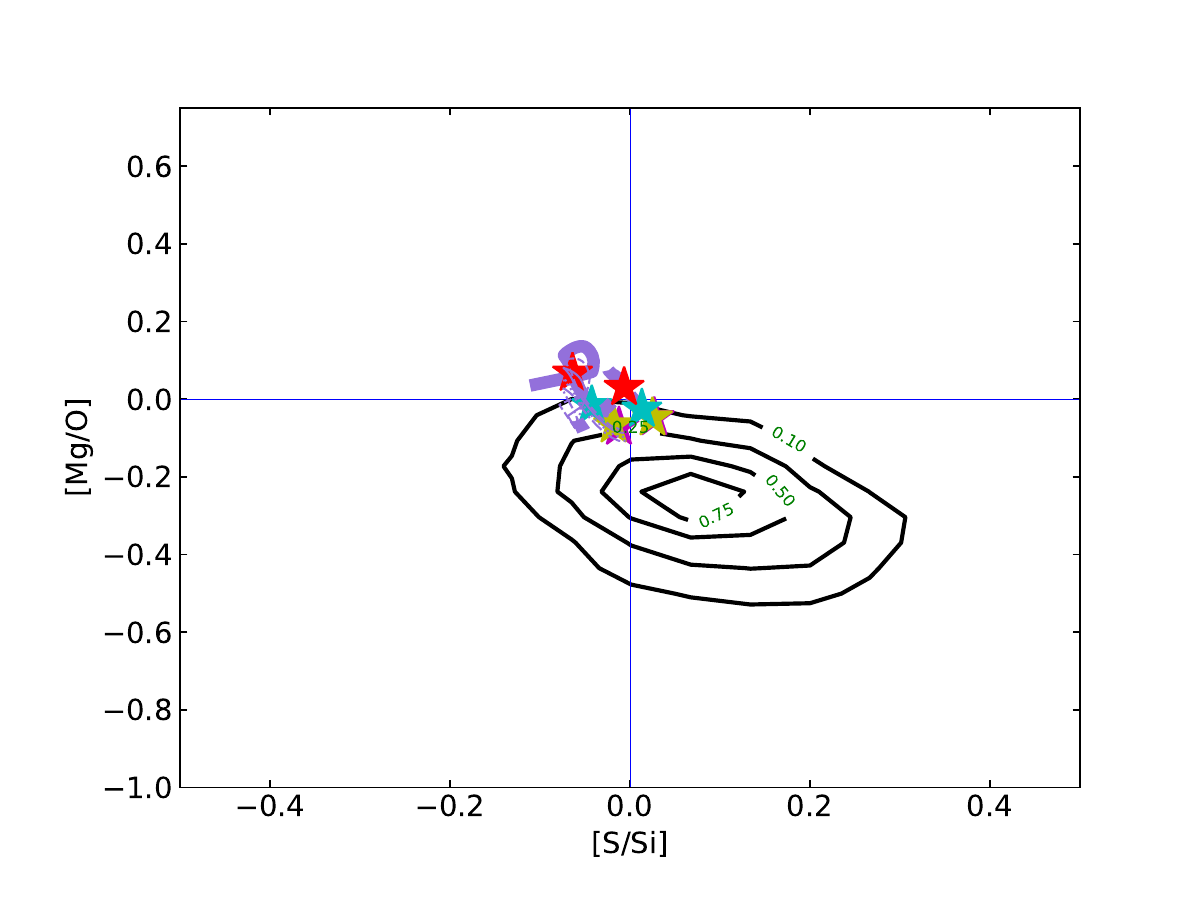}\par
    \includegraphics[width=0.8\linewidth]{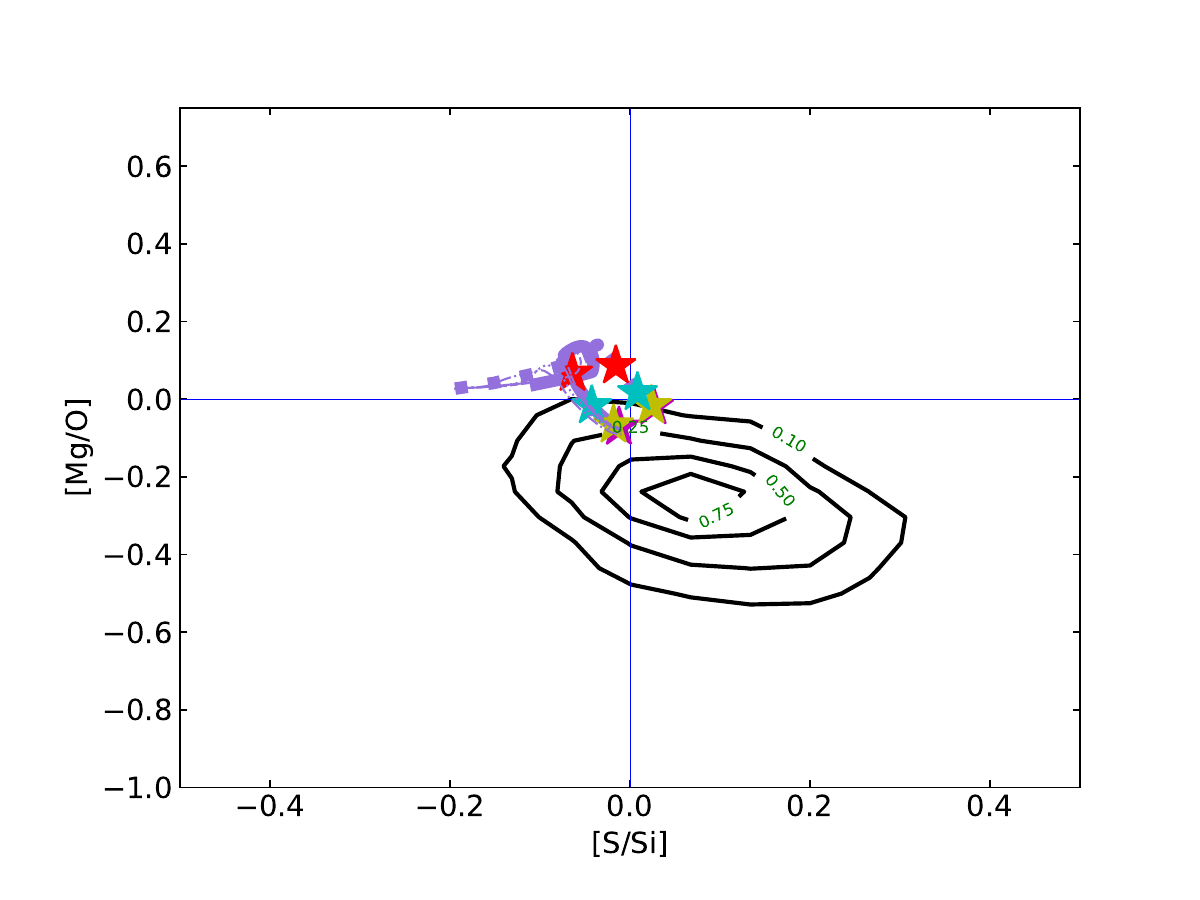}\par
\end{multicols}
    \caption{As in figure~\ref{fig: tk_plots/ratios_oK06_m40}, but models are shown with CCSN supernovae contribution up to M$_{\rm up}$ = 20 M$_{\odot}$.
    }
    \label{fig: tk_plots/ratios_oK06_m20}
\end{figure*}

\begin{figure*}
\begin{multicols}{2}
    \includegraphics[width=0.8\linewidth]{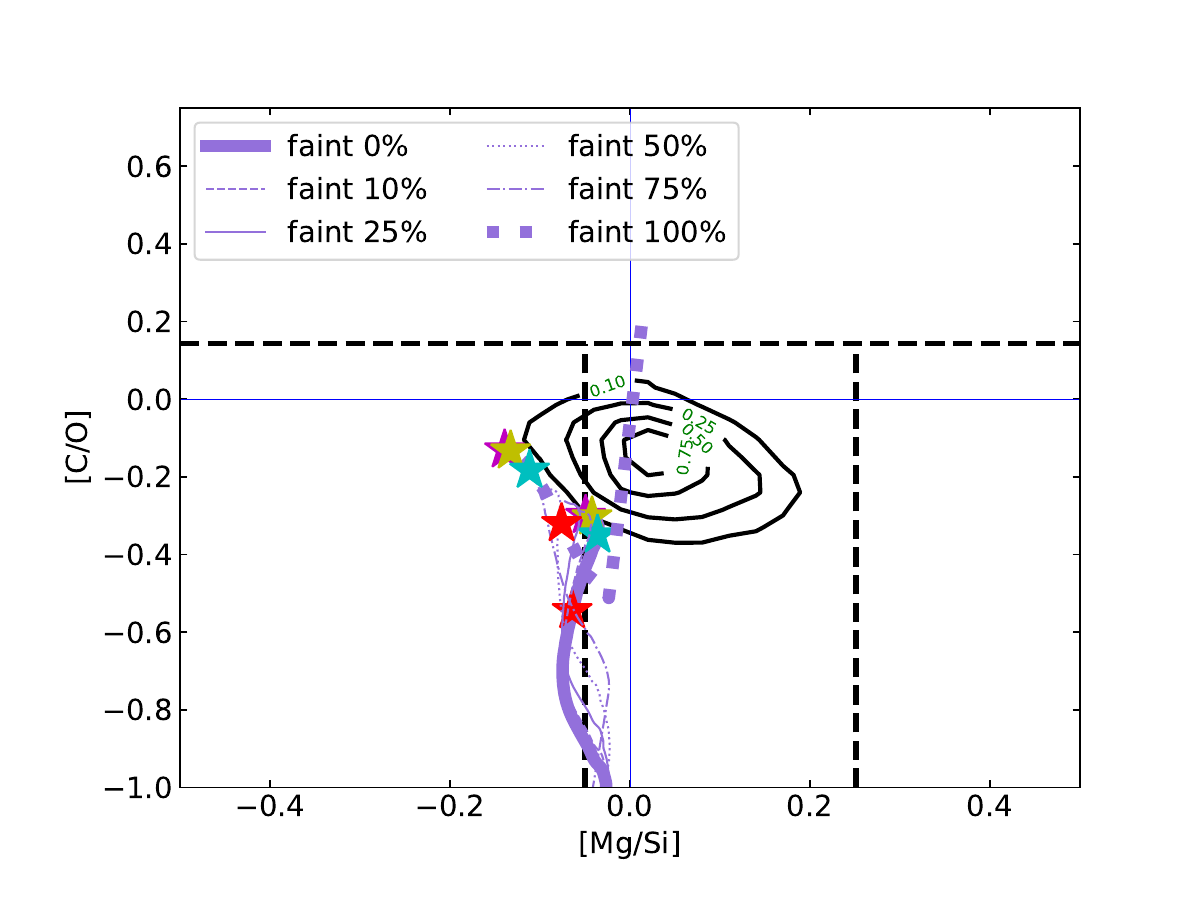}\par
    \includegraphics[width=0.8\linewidth]{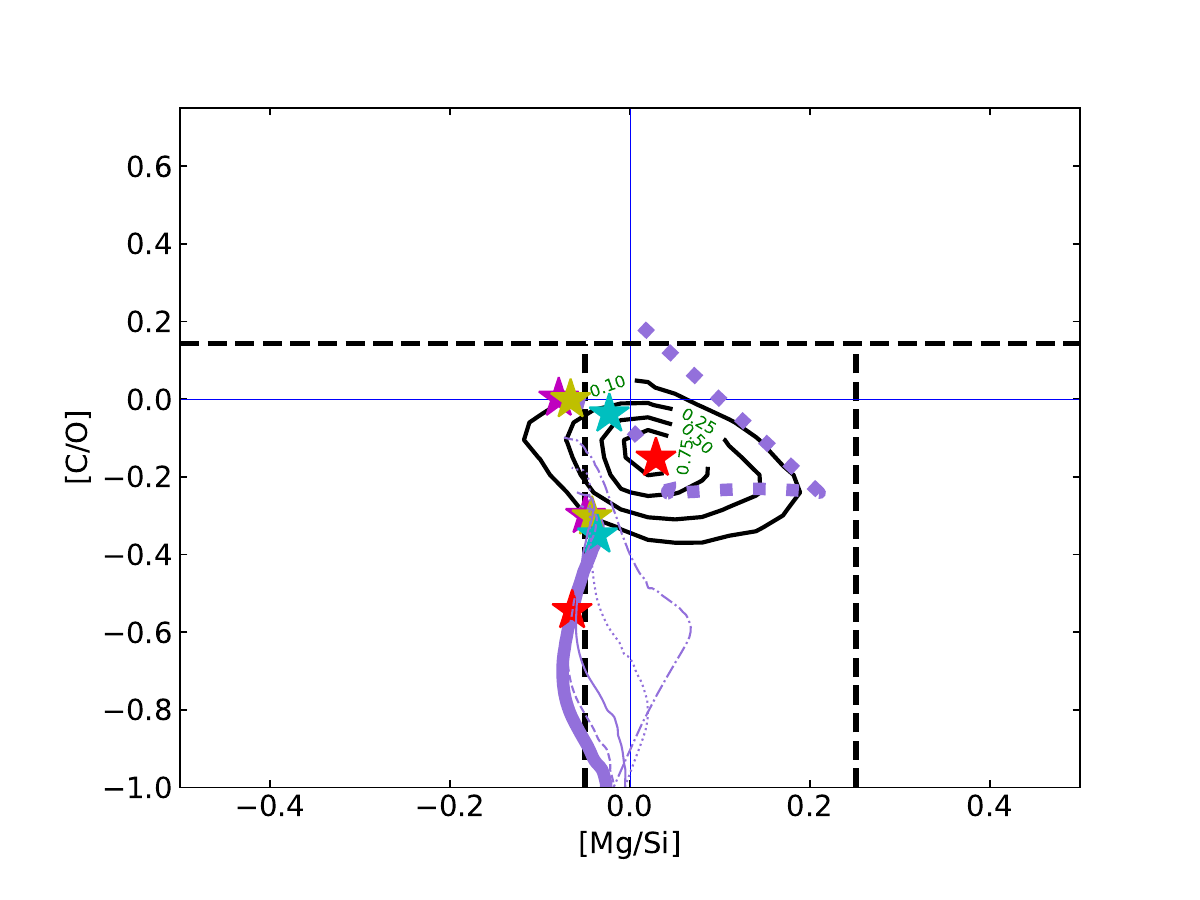}\par
    \end{multicols}
\begin{multicols}{2}
    \includegraphics[width=0.8\linewidth]{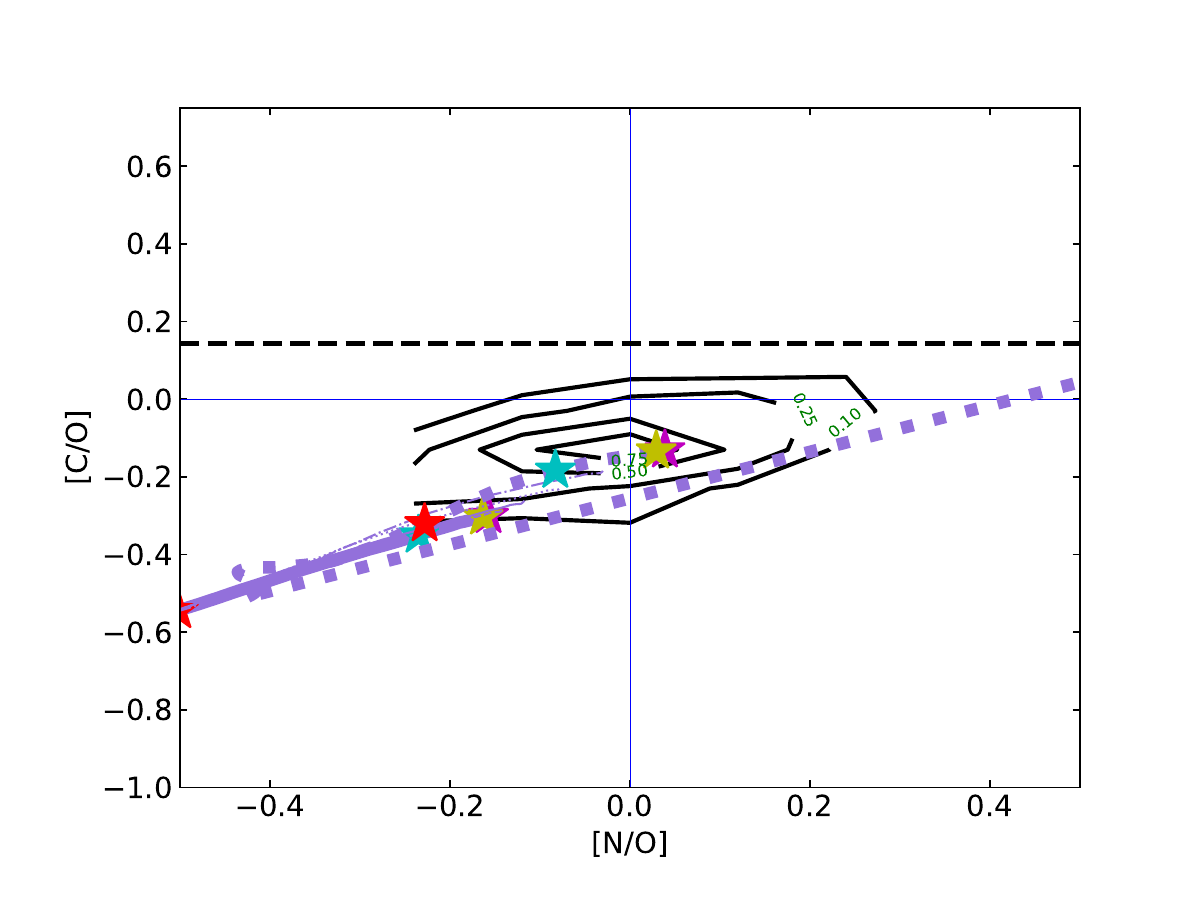}\par
    \includegraphics[width=0.8\linewidth]{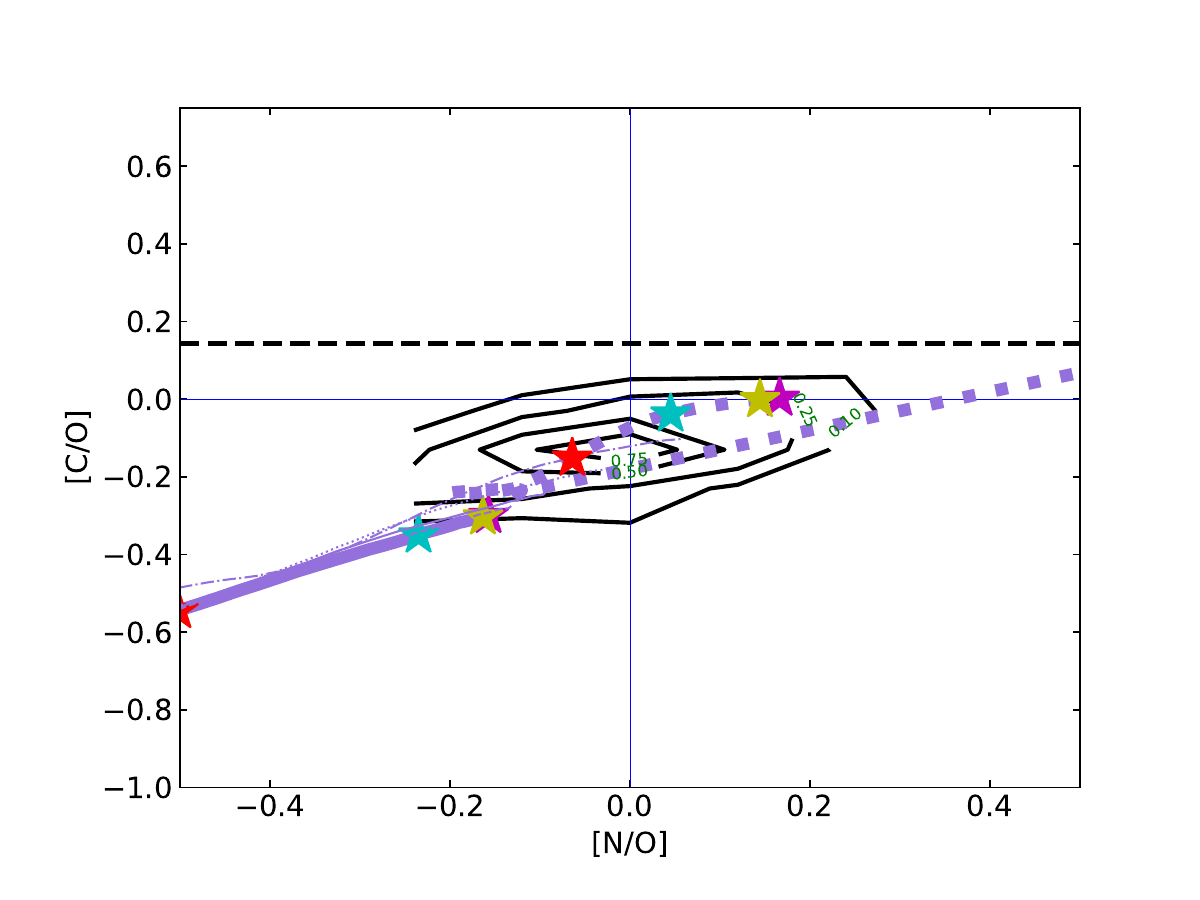}\par
\end{multicols}
\begin{multicols}{2}
    \includegraphics[width=0.8\linewidth]{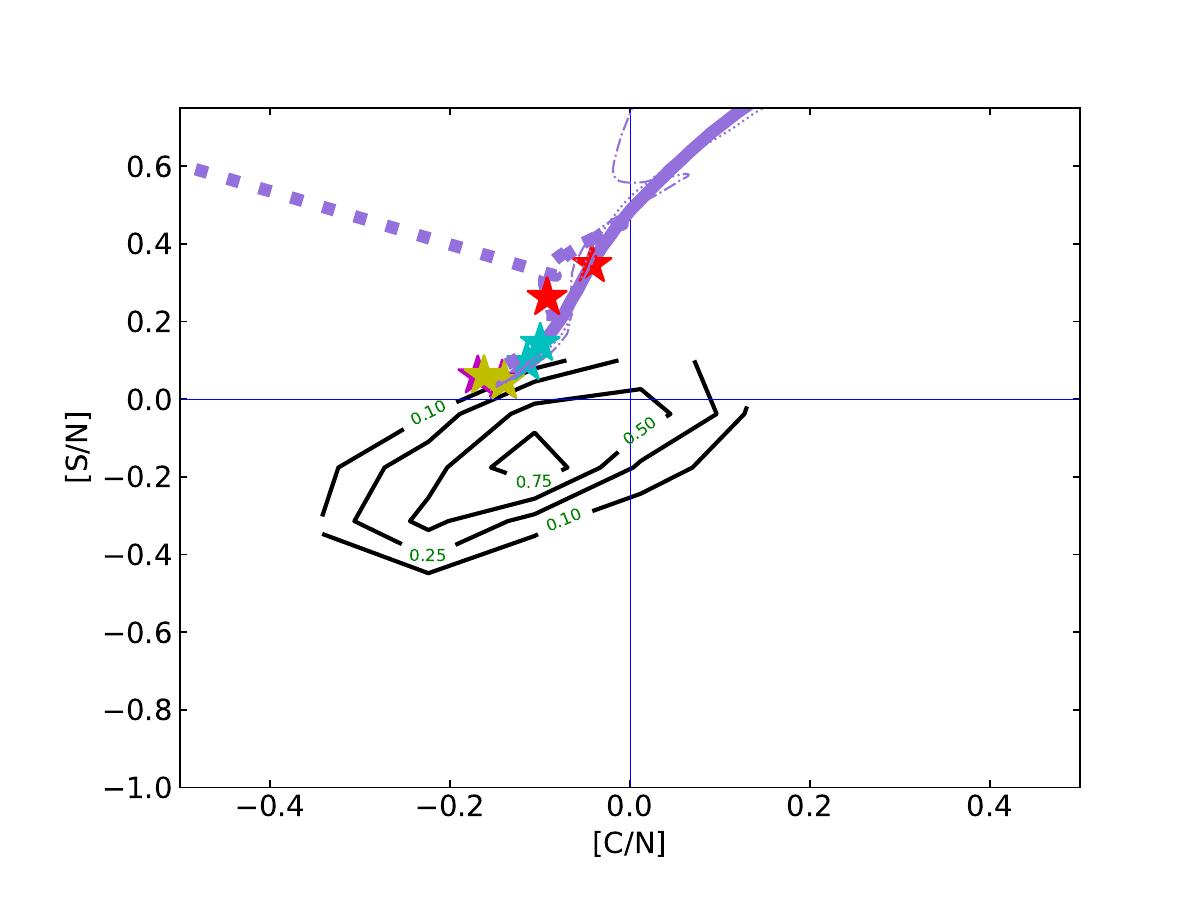}\par
    \includegraphics[width=0.8\linewidth]{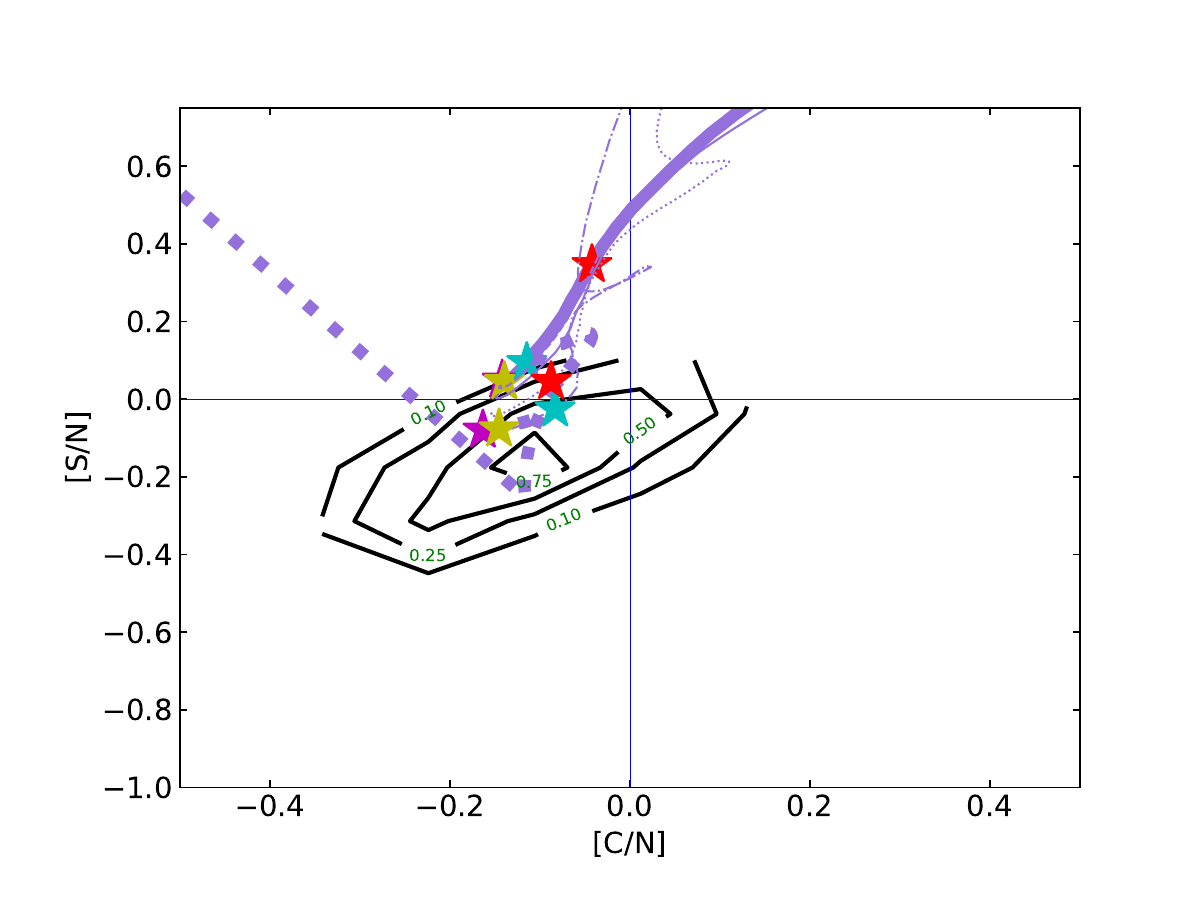}\par
\end{multicols}
\begin{multicols}{2}
    \includegraphics[width=0.8\linewidth]{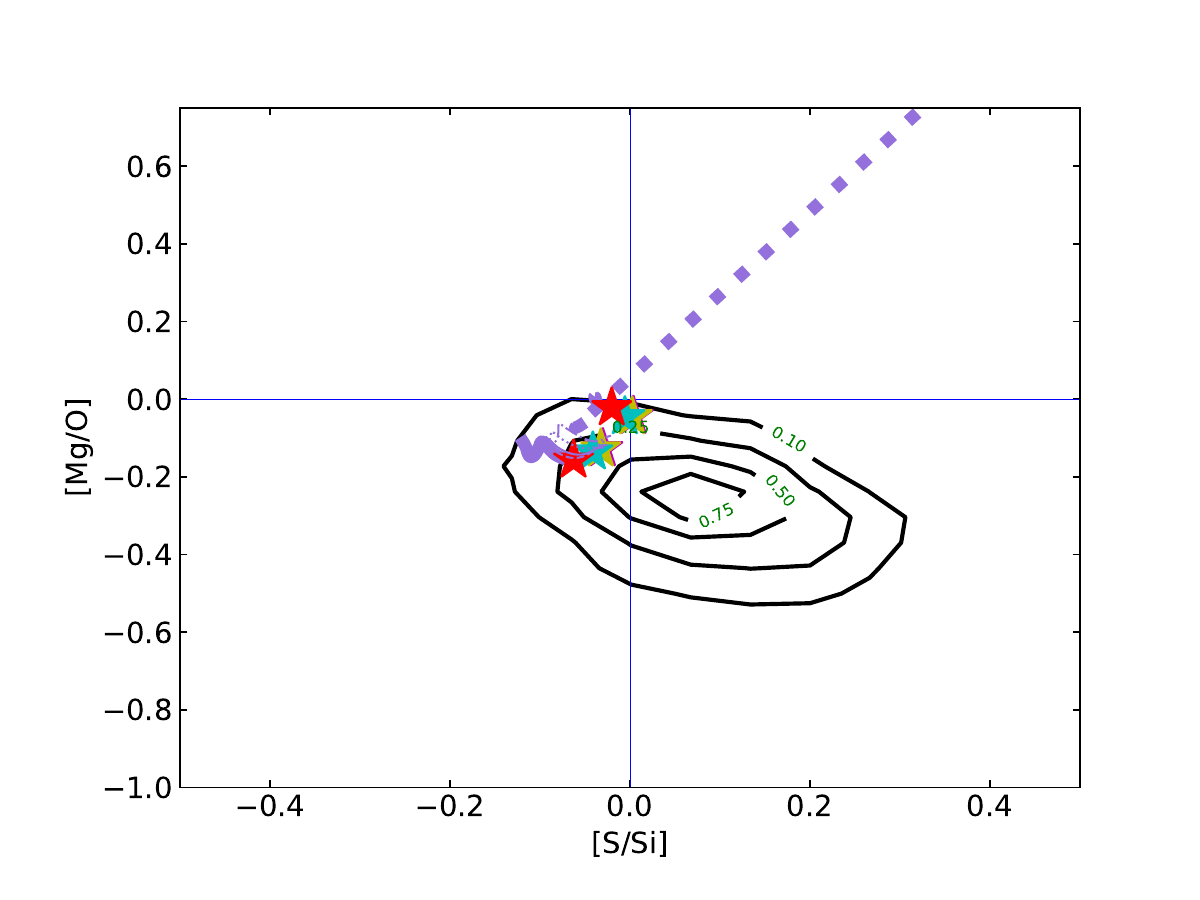}\par
    \includegraphics[width=0.8\linewidth]{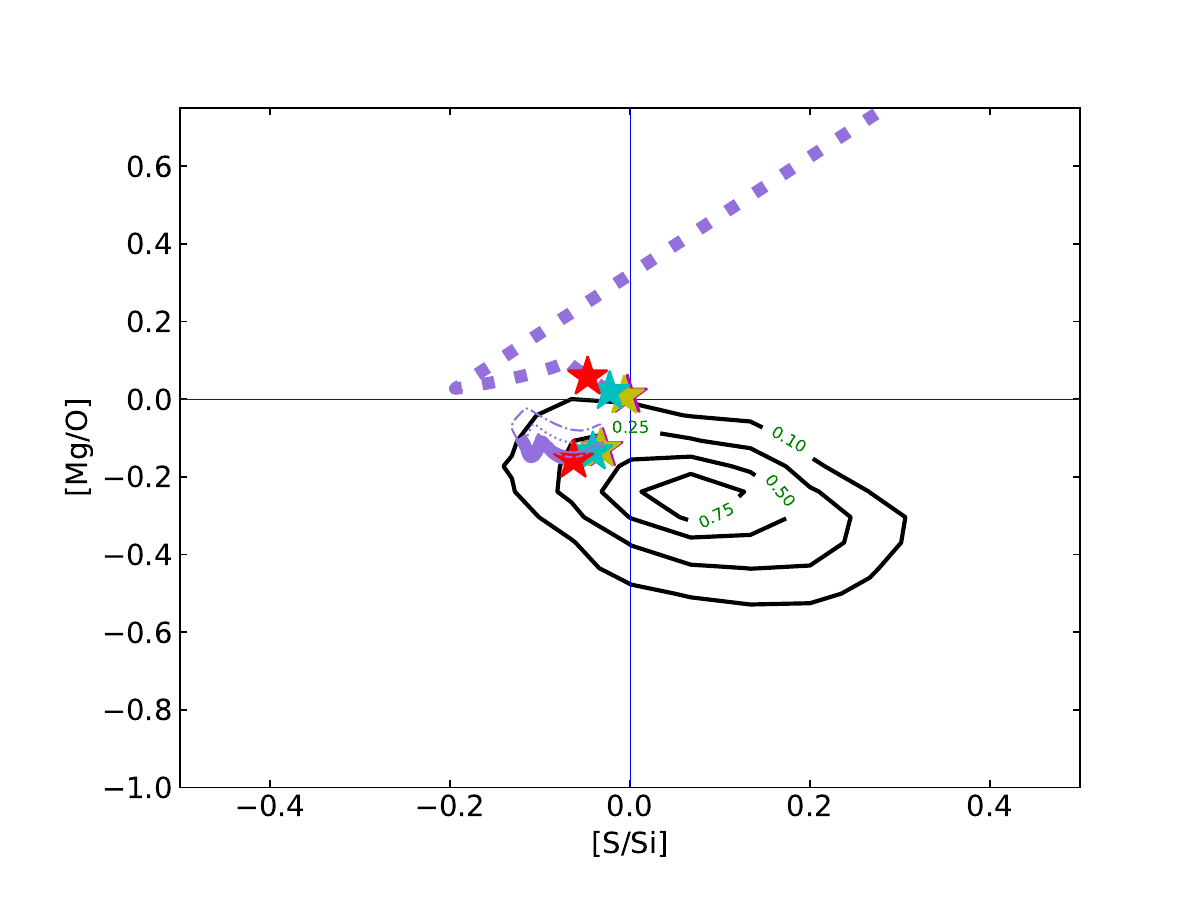}\par
\end{multicols}
    \caption{As in figure~\ref{fig: tk_plots/ratios_oK06_m40}, but models are shown with CCSN supernovae contribution up to M$_{\rm up}$ = 100 M$_{\odot}$.
    }
    \label{fig: tk_plots/ratios_oK06_m100}
\end{figure*}

As in Figure \ref{fig: tk_plots/ratios_gce_oK10m40} for the oK10 set, the results from oK06 models with M$_{\rm up}$=40M$_{\odot}$ are shown using faint supernova models m20 and m25 in Figure \ref{fig: tk_plots/ratios_oK06_m40}. The same is done for M$_{\rm up}$=20M$_{\odot}$ and M$_{\rm up}$=100M$_{\odot}$ in Figures \ref{fig: tk_plots/ratios_oK06_m20} and \ref{fig: tk_plots/ratios_oK06_m100}, respectively.

\begin{figure*}
\begin{multicols}{2}
    \includegraphics[width=0.8\linewidth]{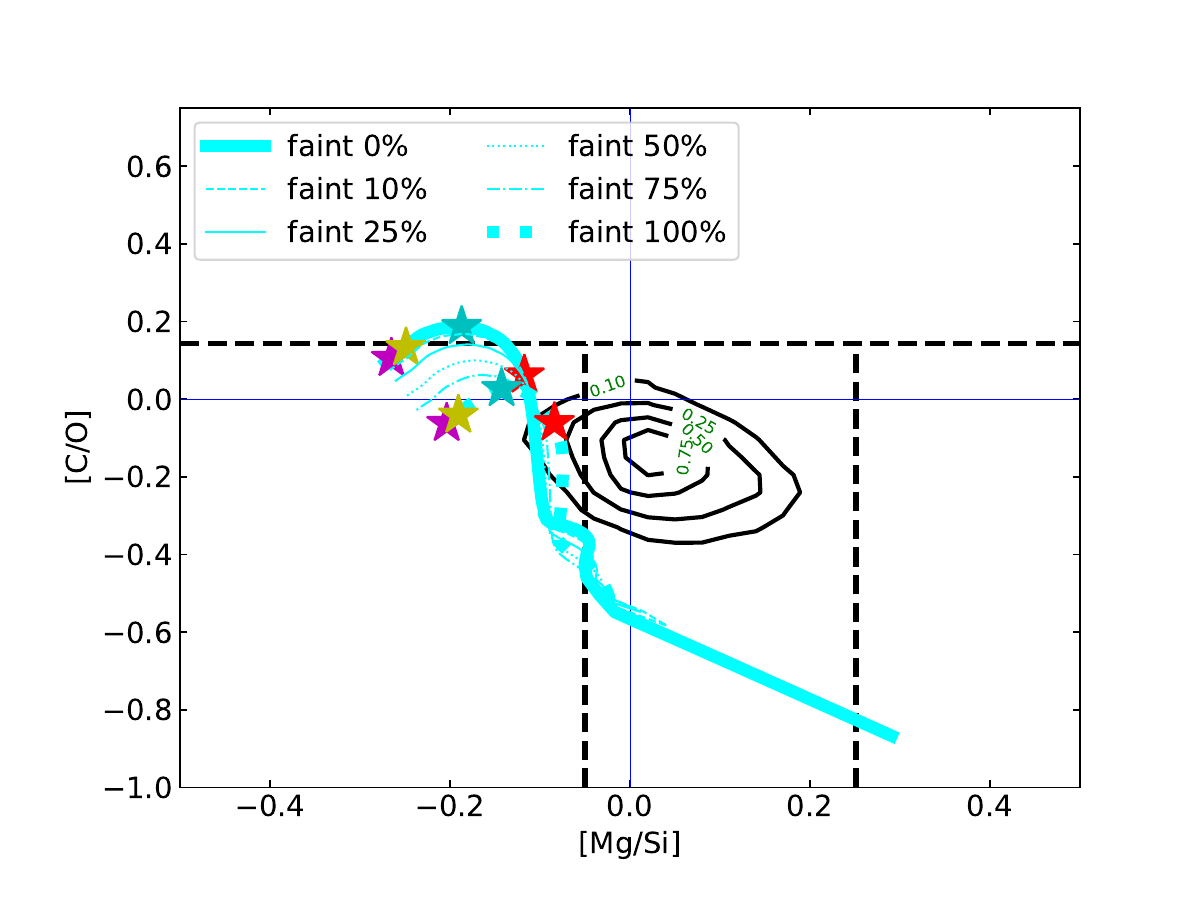}\par
    \includegraphics[width=0.8\linewidth]{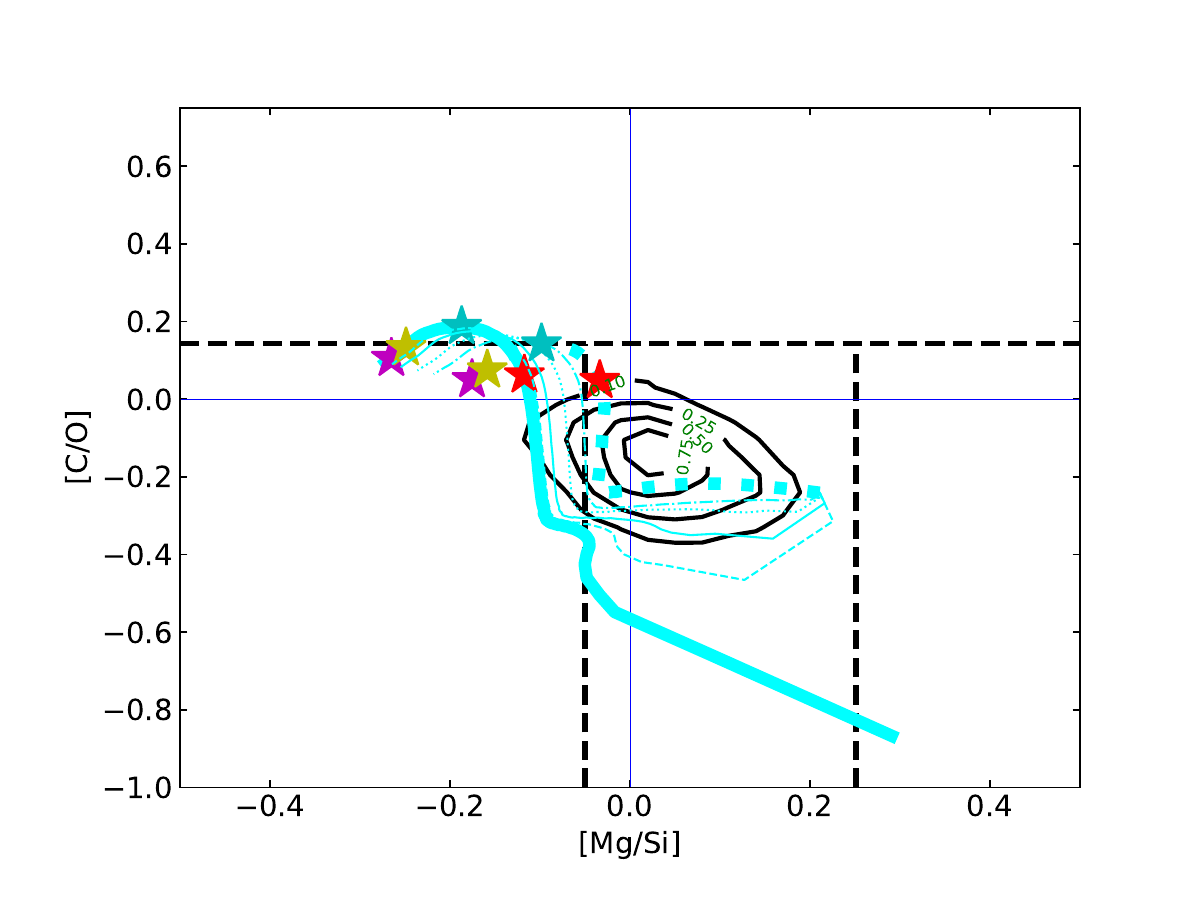}\par
    \end{multicols}
\begin{multicols}{2}
    \includegraphics[width=0.8\linewidth]{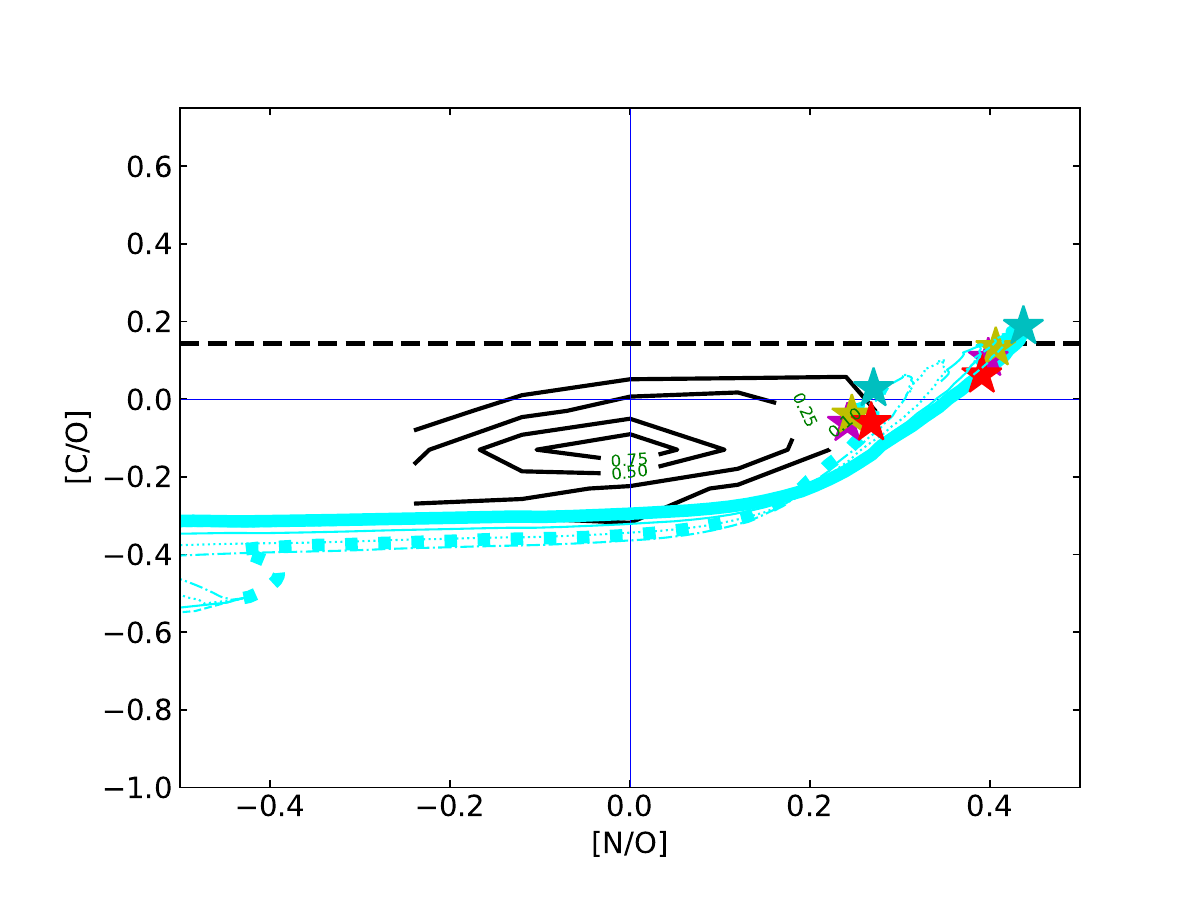}\par
    \includegraphics[width=0.8\linewidth]{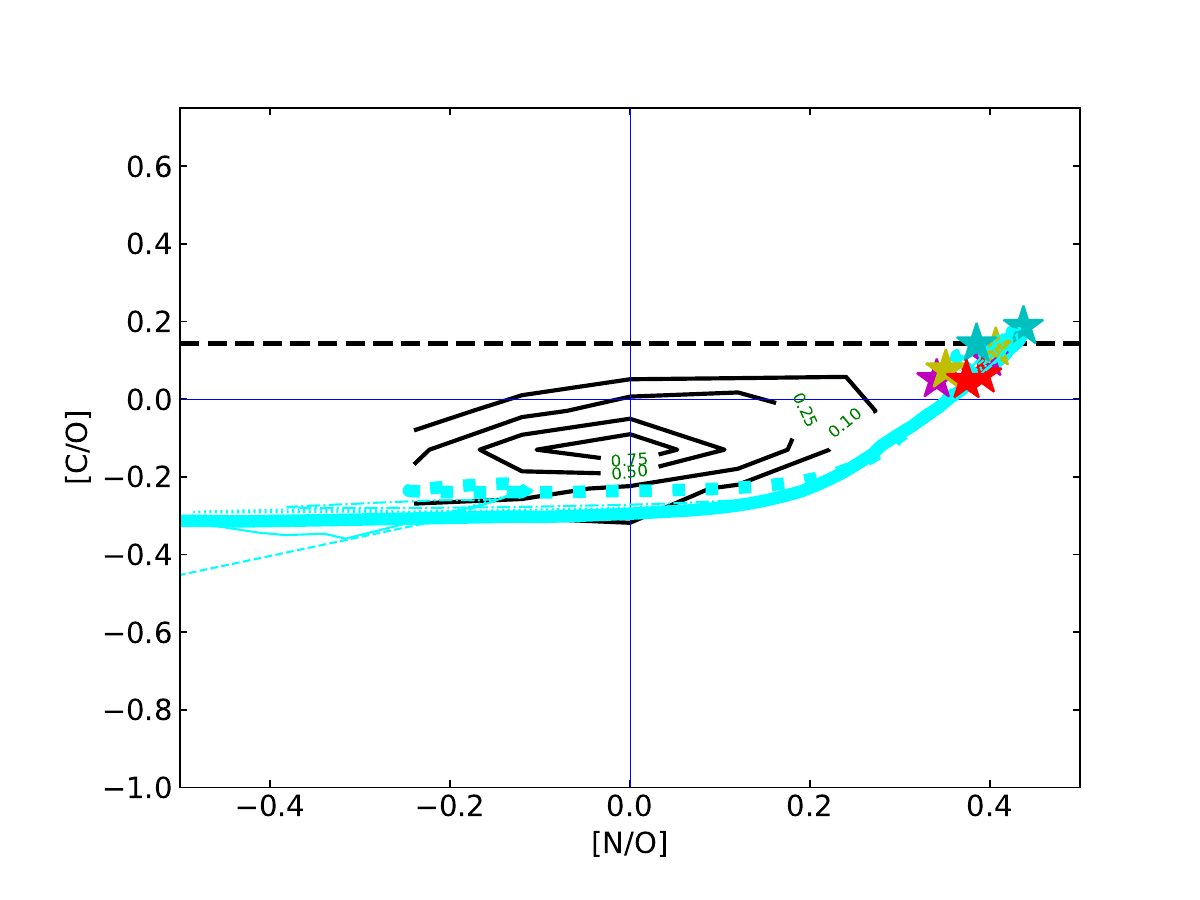}\par
\end{multicols}
\begin{multicols}{2}
    \includegraphics[width=0.8\linewidth]{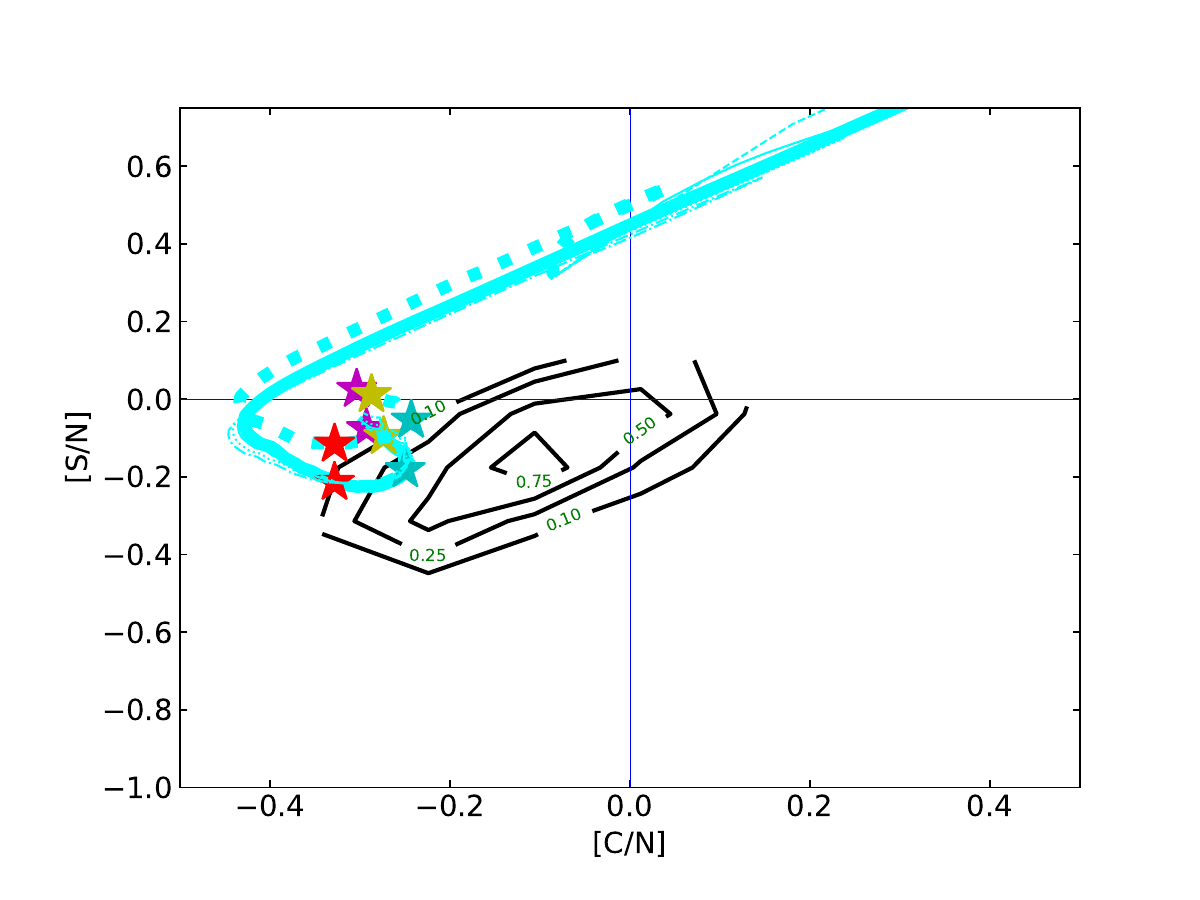}\par
    \includegraphics[width=0.8\linewidth]{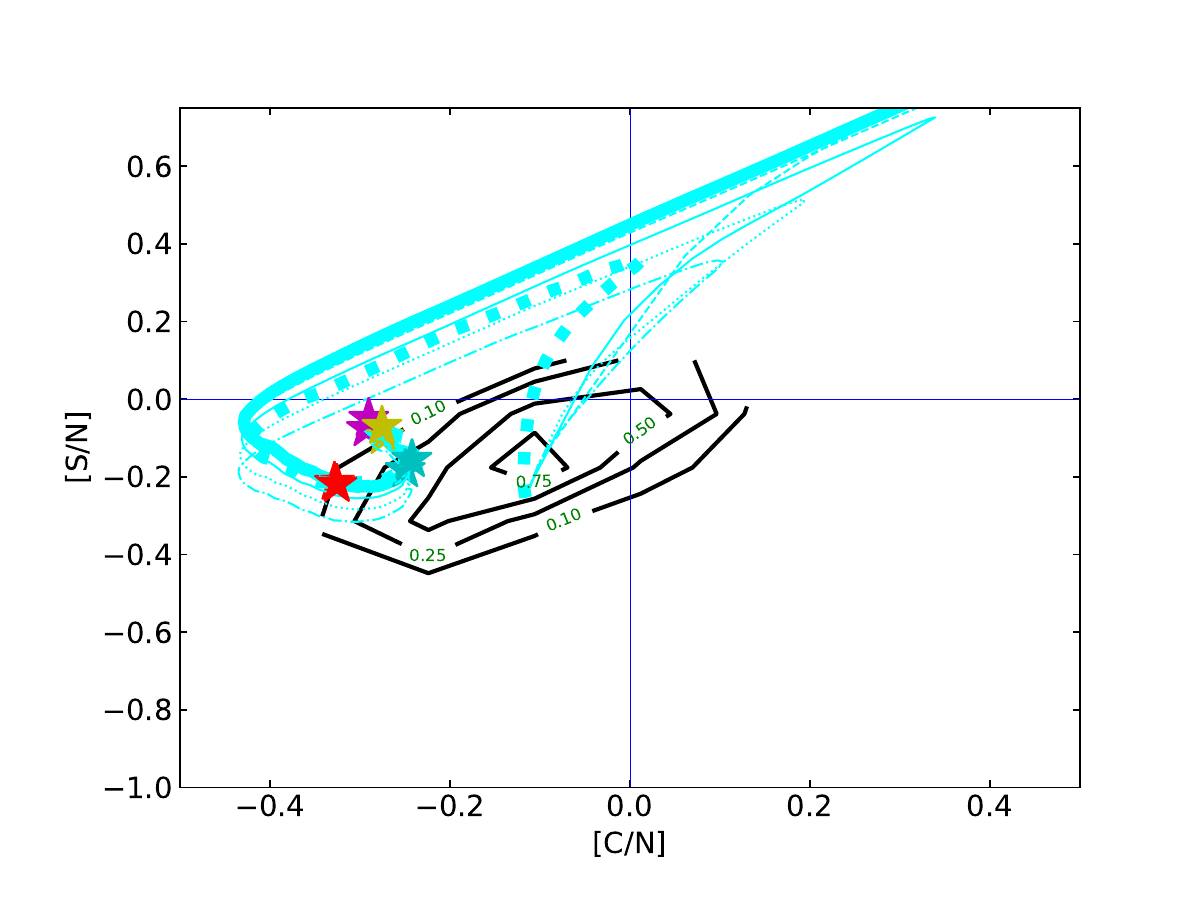}\par
\end{multicols}
\begin{multicols}{2}
    \includegraphics[width=0.8\linewidth]{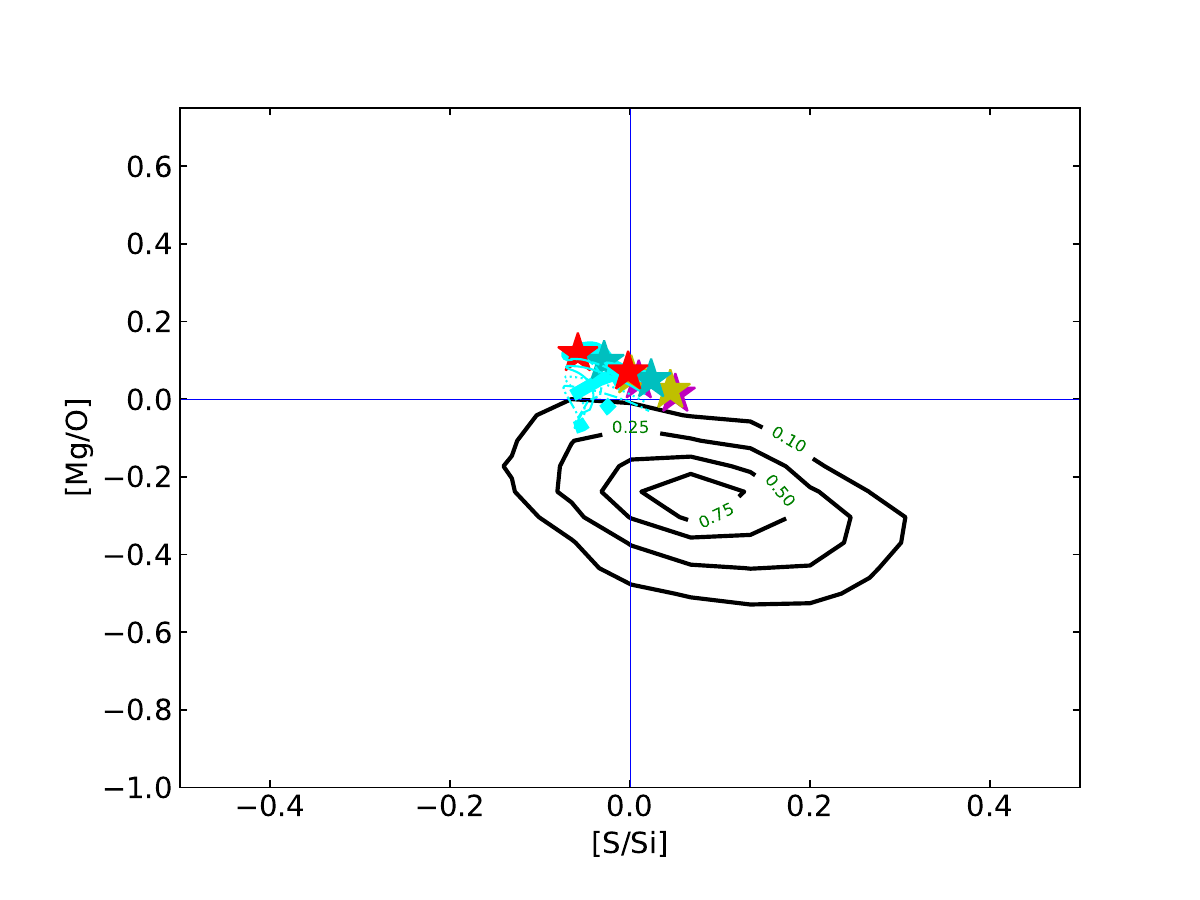}\par
    \includegraphics[width=0.8\linewidth]{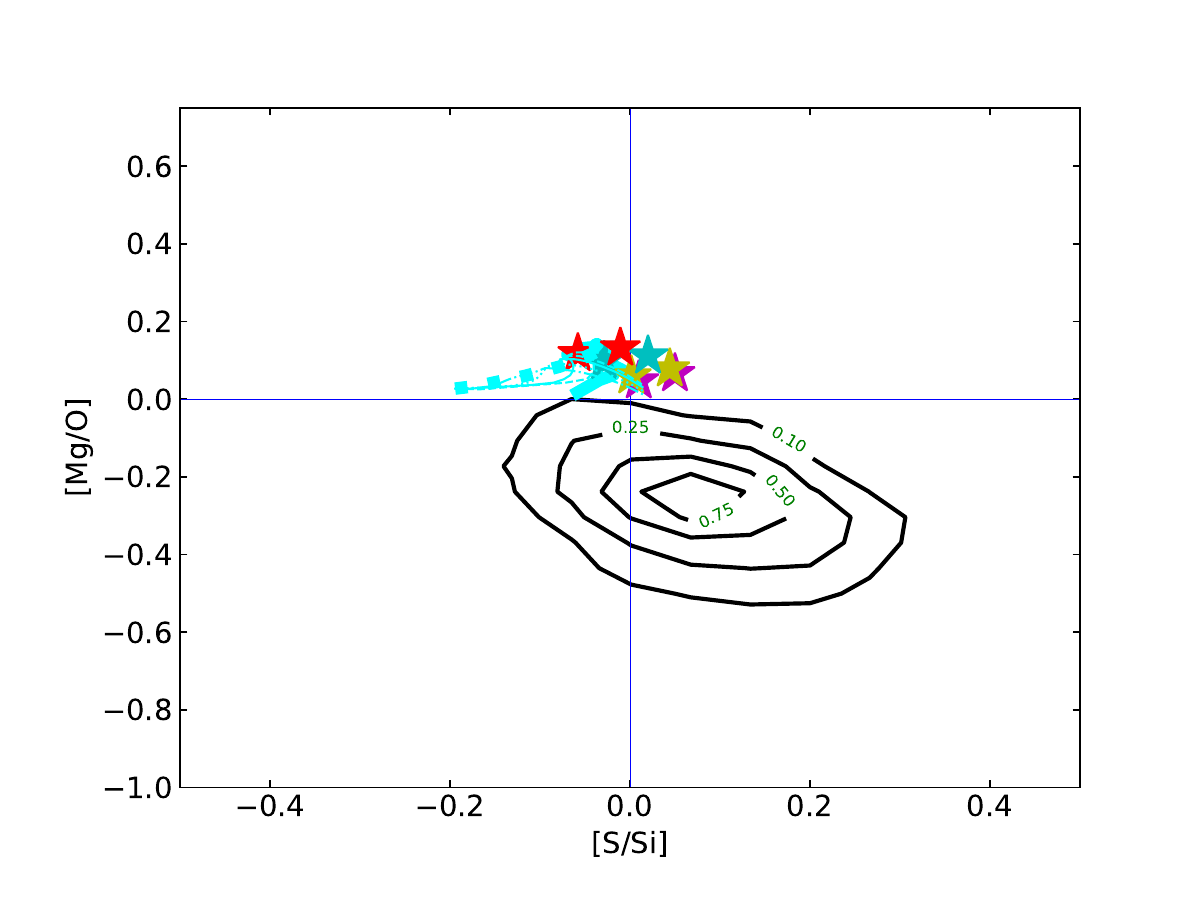}\par
\end{multicols}
    \caption{As in figure~\ref{fig: tk_plots/ratios_gce_oK10m40} for the oK10 model set, but models are shown with CCSN supernovae contribution up to M$_{\rm up}$ = 20 M$_{\odot}$.
    }
    \label{fig: tk_plots/ratios_gce_oK10m20}
\end{figure*}

\begin{figure*}
\begin{multicols}{2}
    \includegraphics[width=0.8\linewidth]{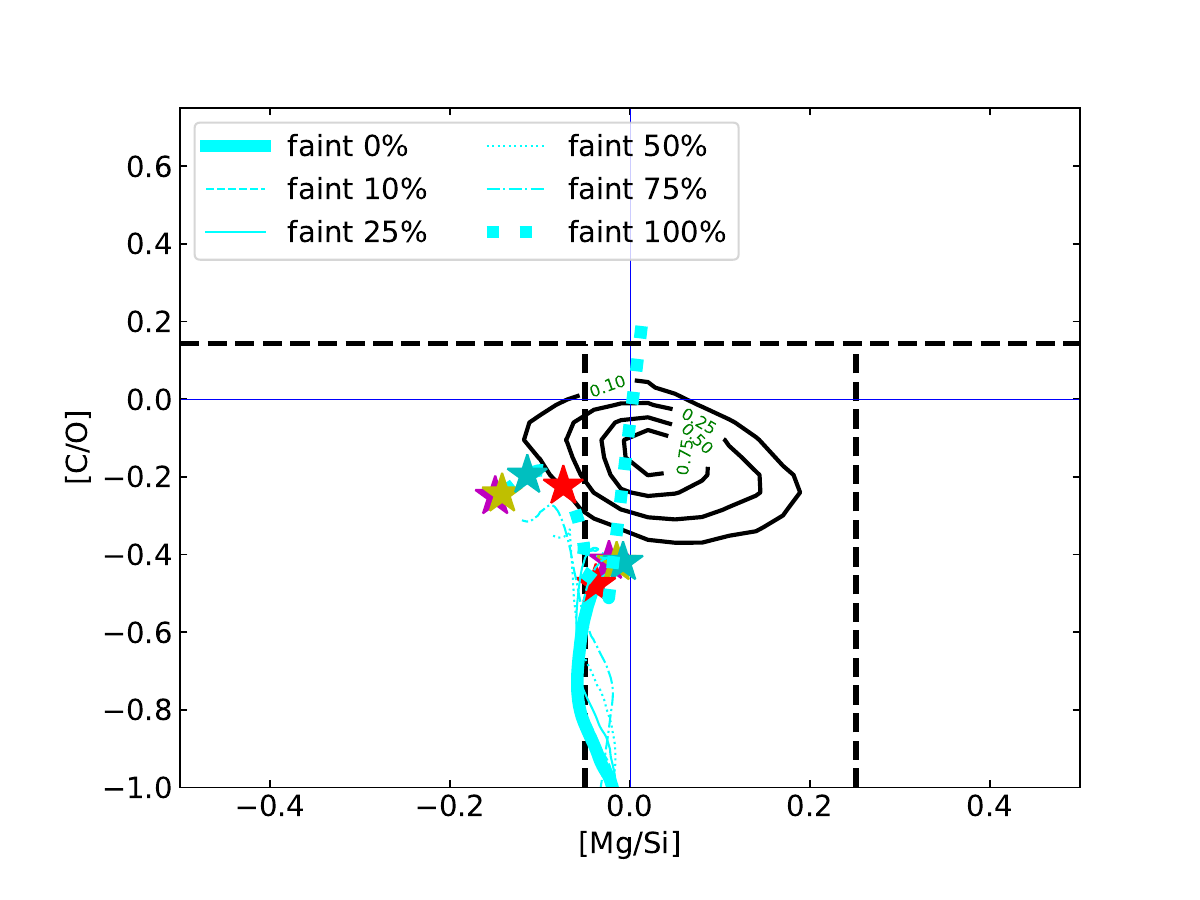}\par
    \includegraphics[width=0.8\linewidth]{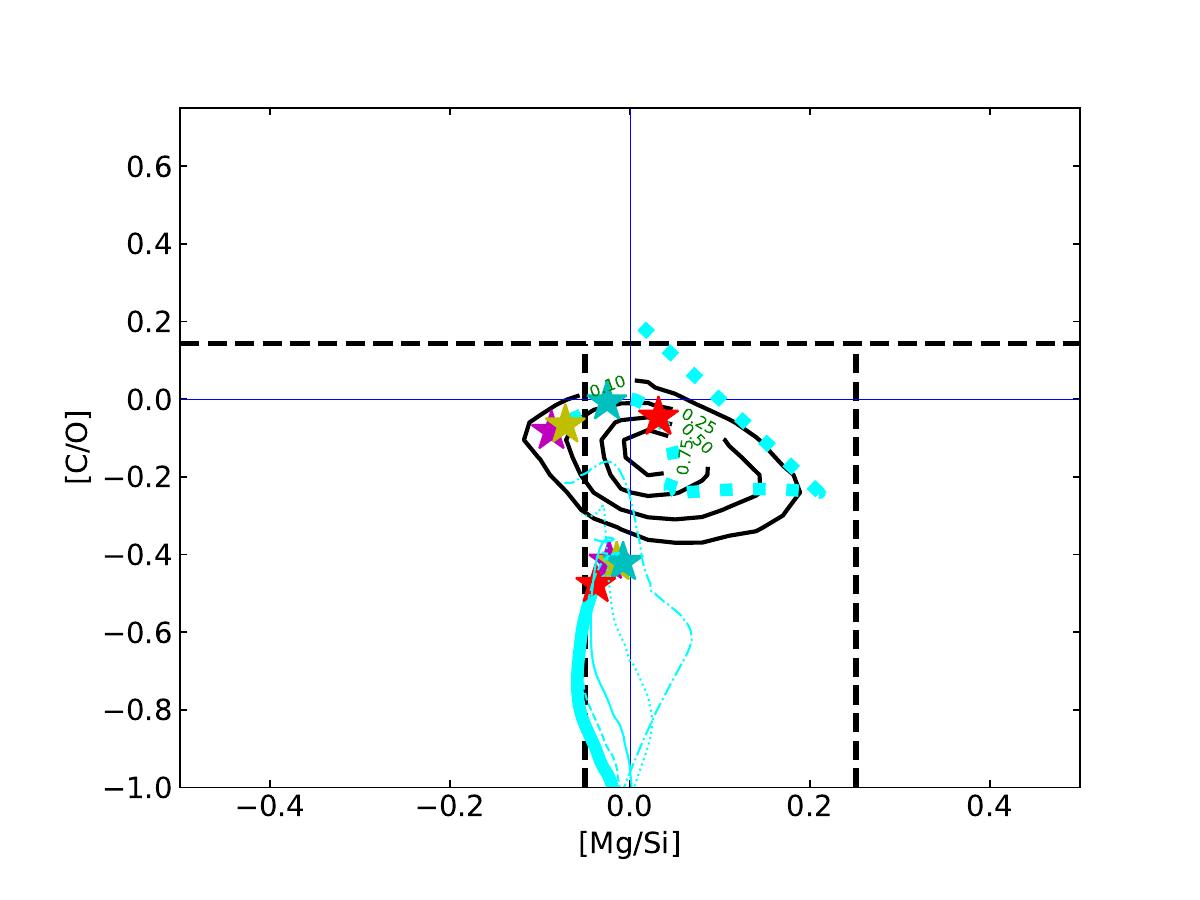}\par
    \end{multicols}
\begin{multicols}{2}
    \includegraphics[width=0.8\linewidth]{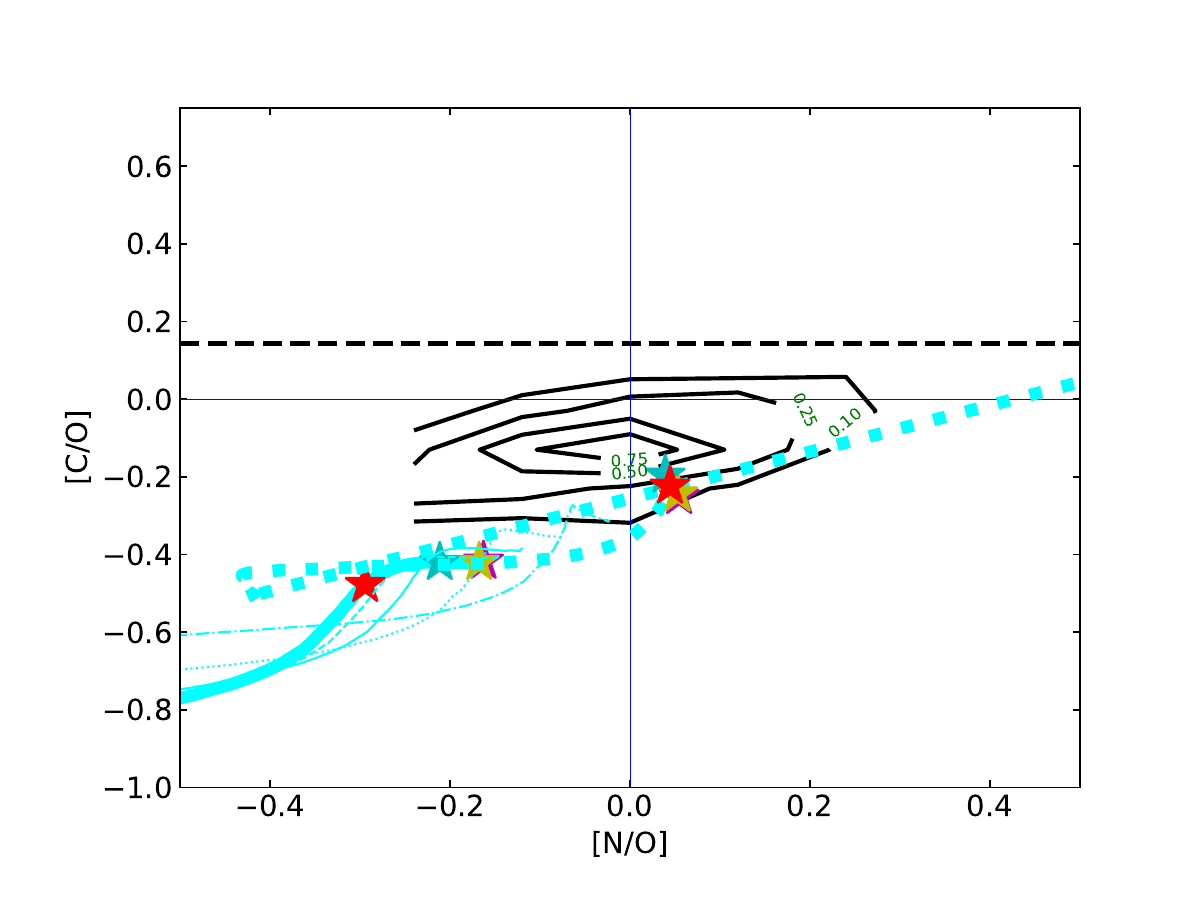}\par
    \includegraphics[width=0.8\linewidth]{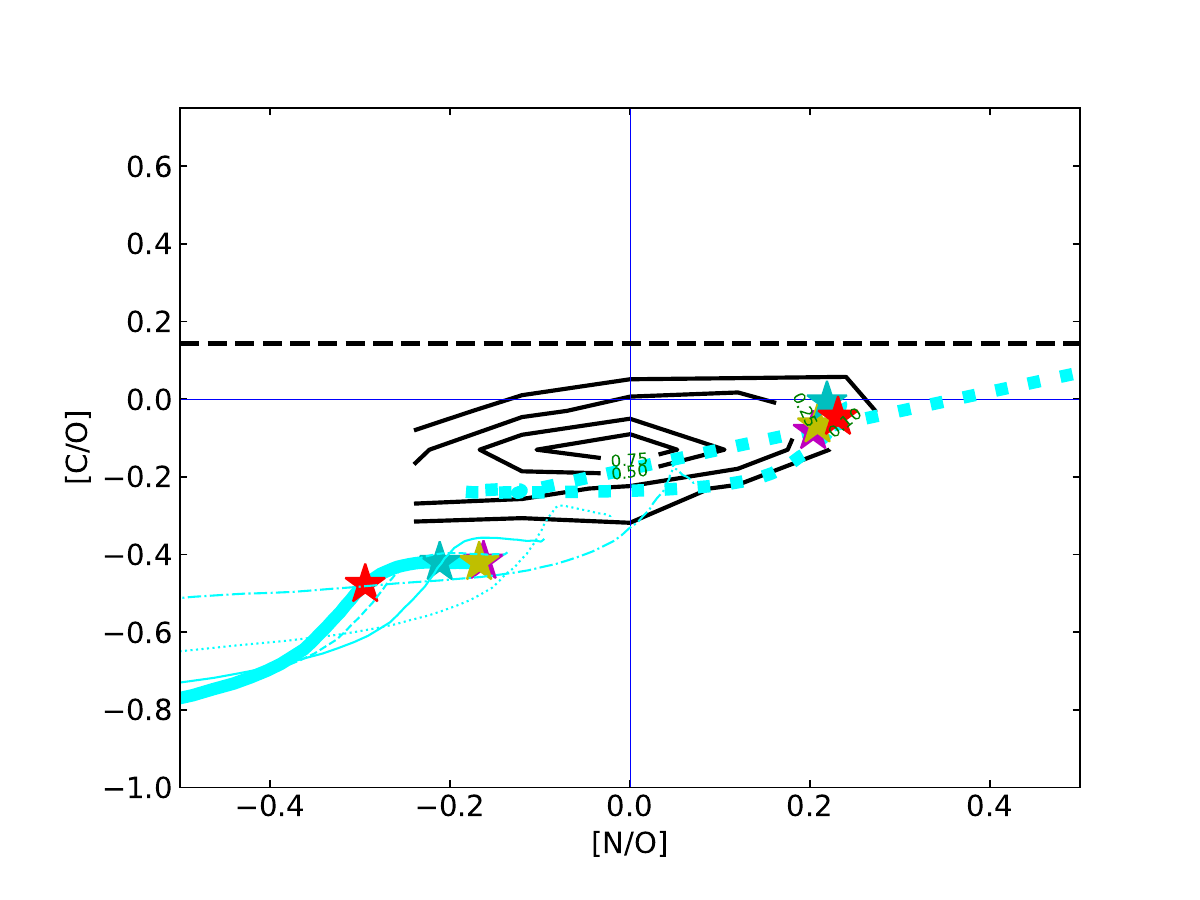}\par
\end{multicols}
\begin{multicols}{2}
    \includegraphics[width=0.8\linewidth]{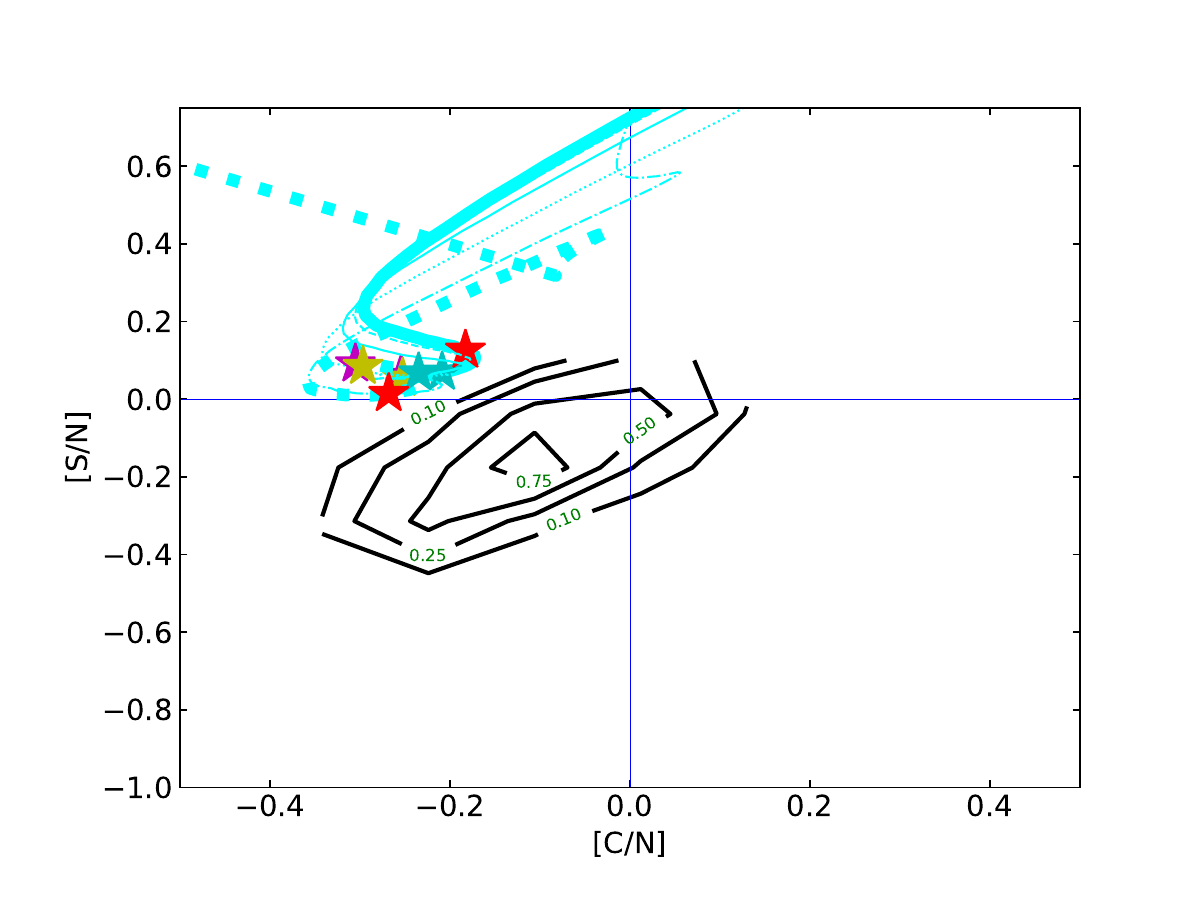}\par
    \includegraphics[width=0.8\linewidth]{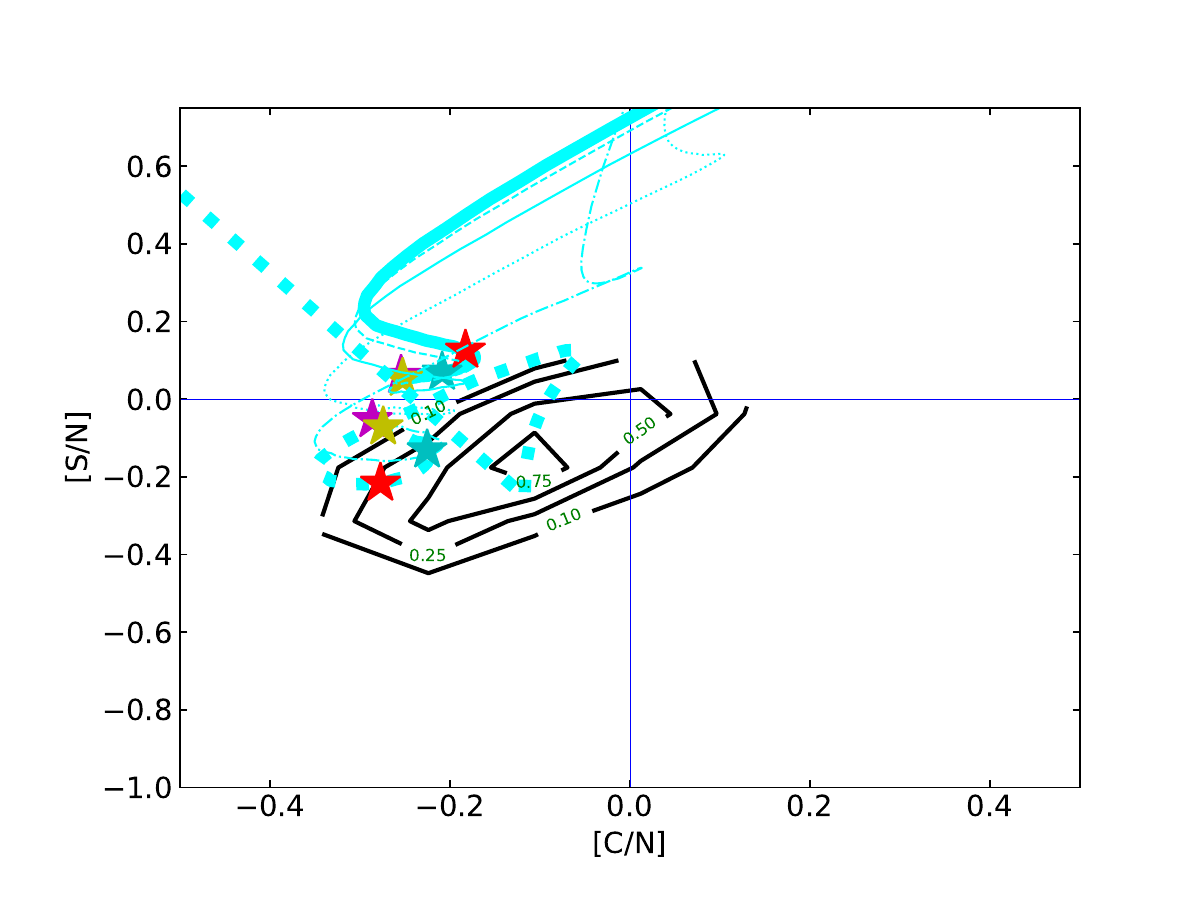}\par
\end{multicols}
\begin{multicols}{2}
    \includegraphics[width=0.8\linewidth]{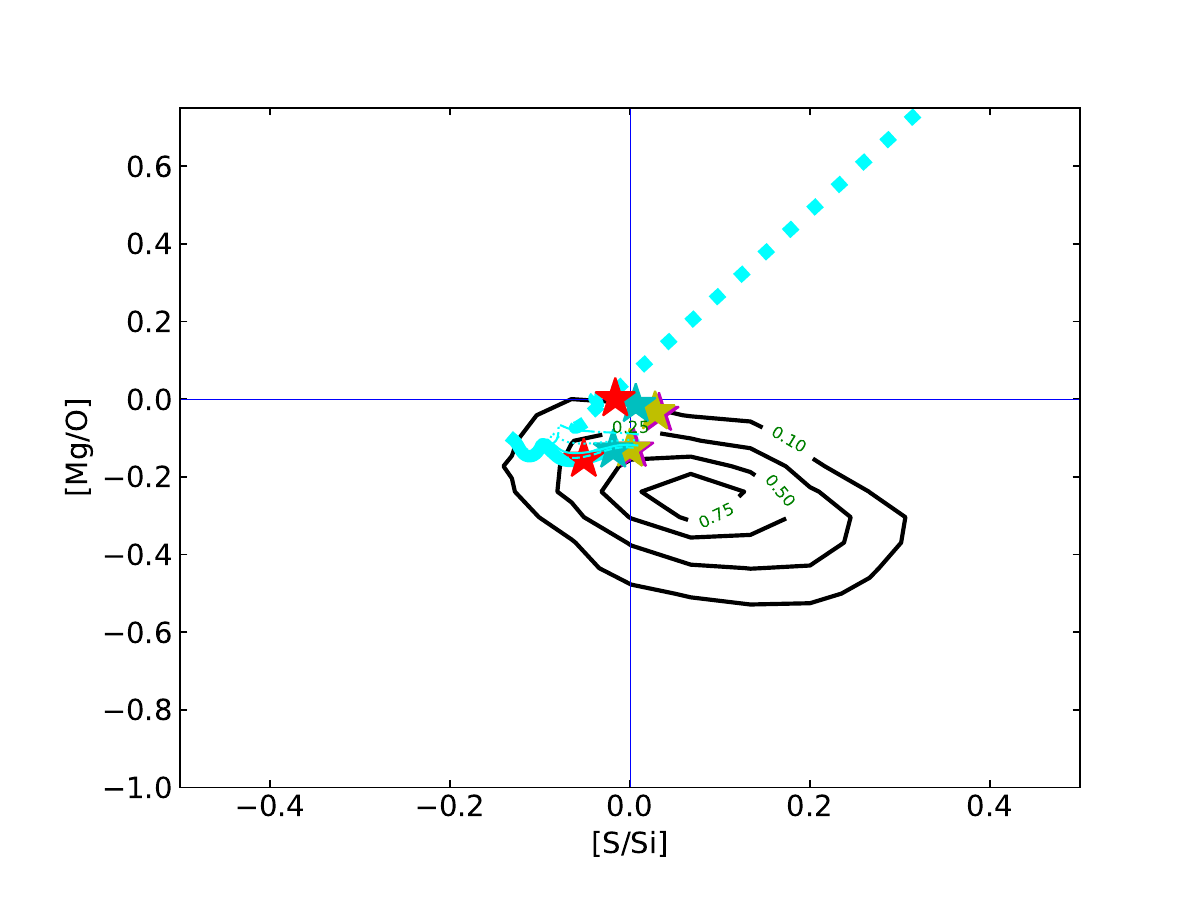}\par
    \includegraphics[width=0.8\linewidth]{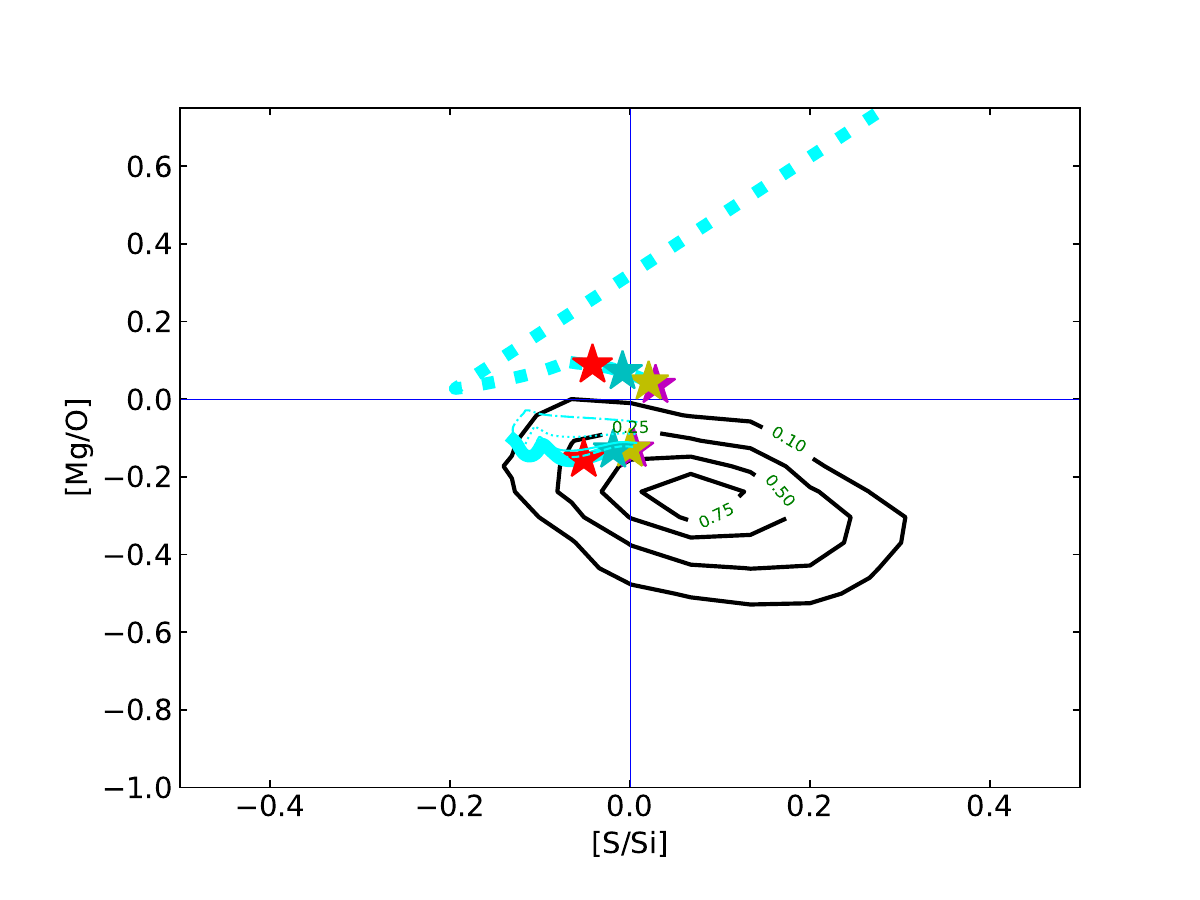}\par
\end{multicols}
    \caption{As in figure~\ref{fig: tk_plots/ratios_gce_oK10m40} for the oK10 model set, but models are shown with CCSN supernovae contribution up to M$_{\rm up}$ = 100 M$_{\odot}$.
    }
    \label{fig: tk_plots/ratios_gce_oK10m100}
\end{figure*}

For oK10, the results using M$_{\rm up}$=20M$_{\odot}$ and 100M$_{\odot}$ for different faint supernovae are given in Figures \ref{fig: tk_plots/ratios_gce_oK10m20} and \ref{fig: tk_plots/ratios_gce_oK10m100}. The same results for M$_{\rm up}$=40M$_{\odot}$ are discussed in section \S~\ref{sec: results} (Figure \ref{fig: tk_plots/ratios_gce_oK10m40}).

\begin{figure*}
\begin{multicols}{2}
    \includegraphics[width=0.8\linewidth]{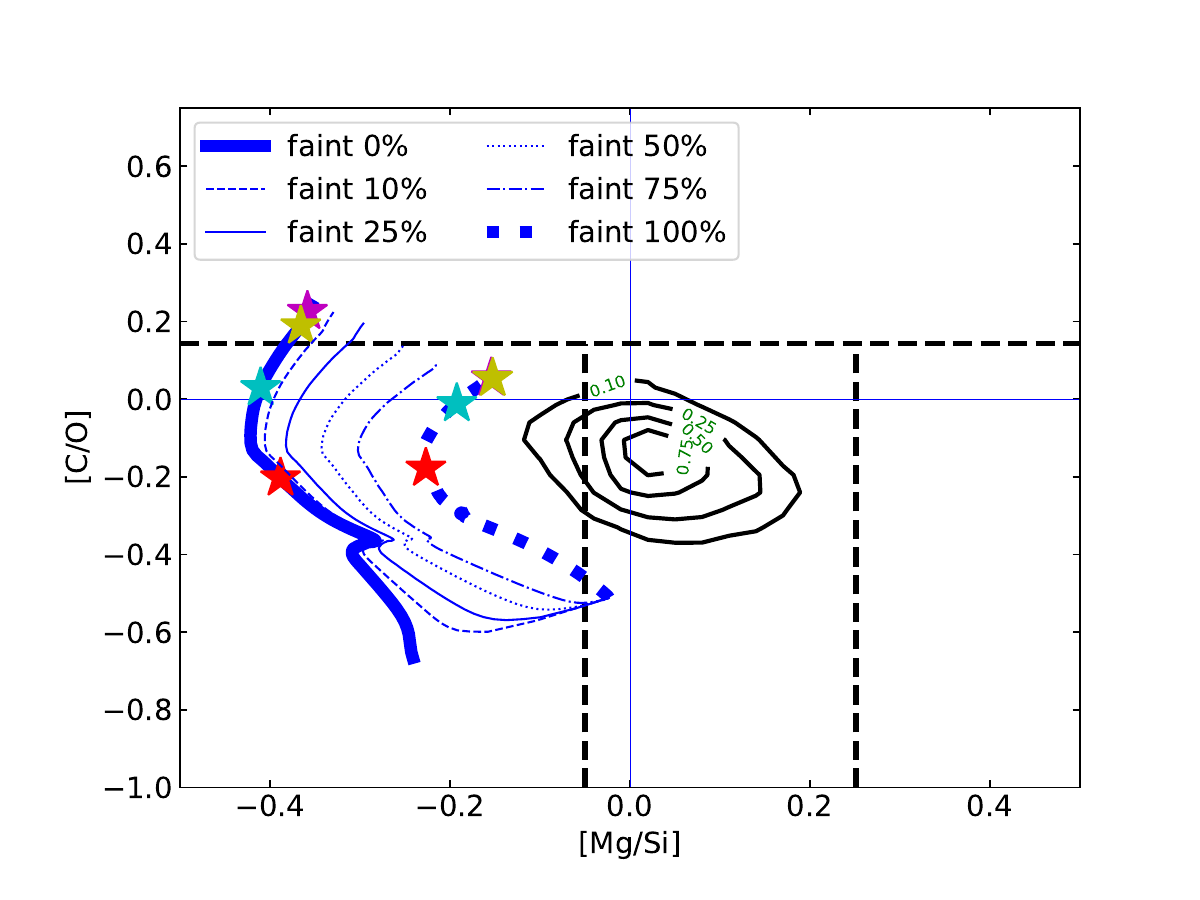}\par
    \includegraphics[width=0.8\linewidth]{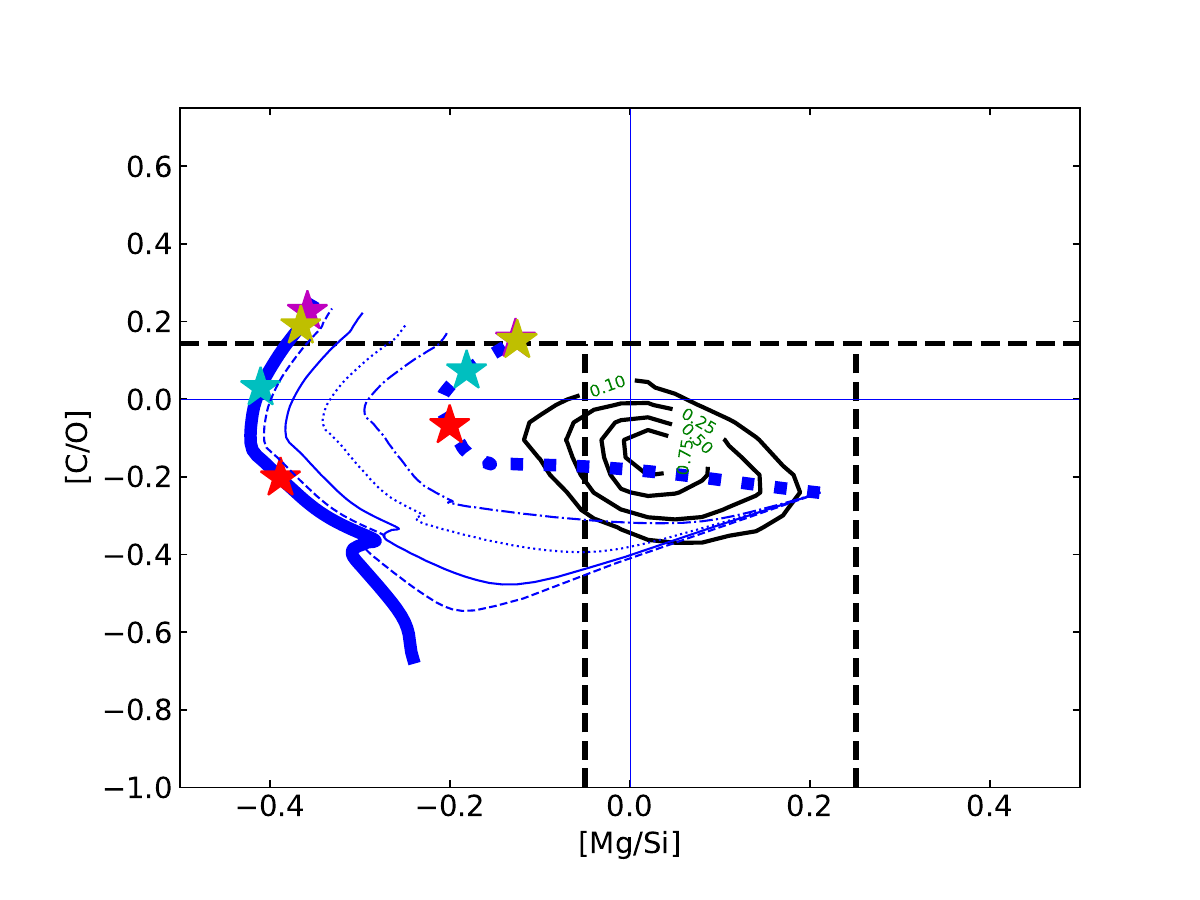}\par
    \end{multicols}
\begin{multicols}{2}
    \includegraphics[width=0.8\linewidth]{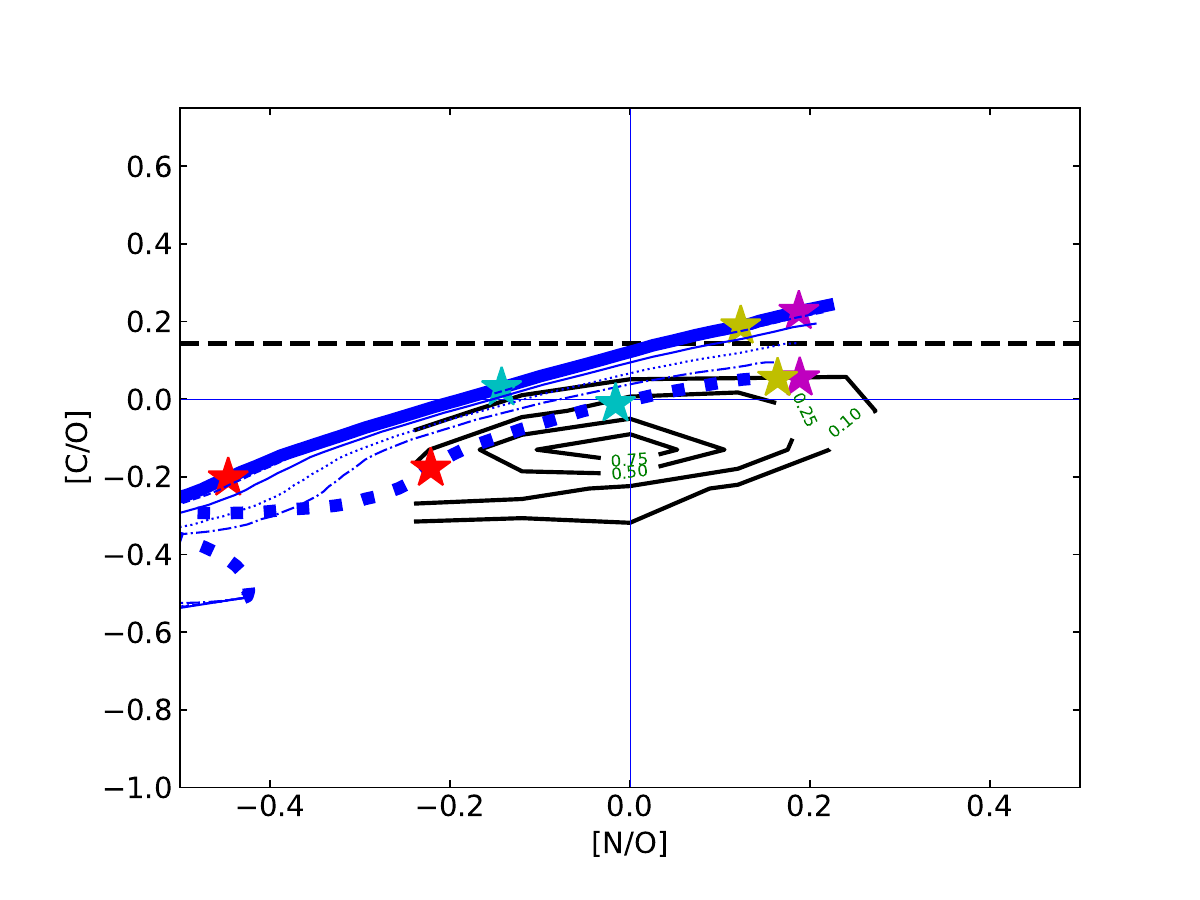}\par
    \includegraphics[width=0.8\linewidth]{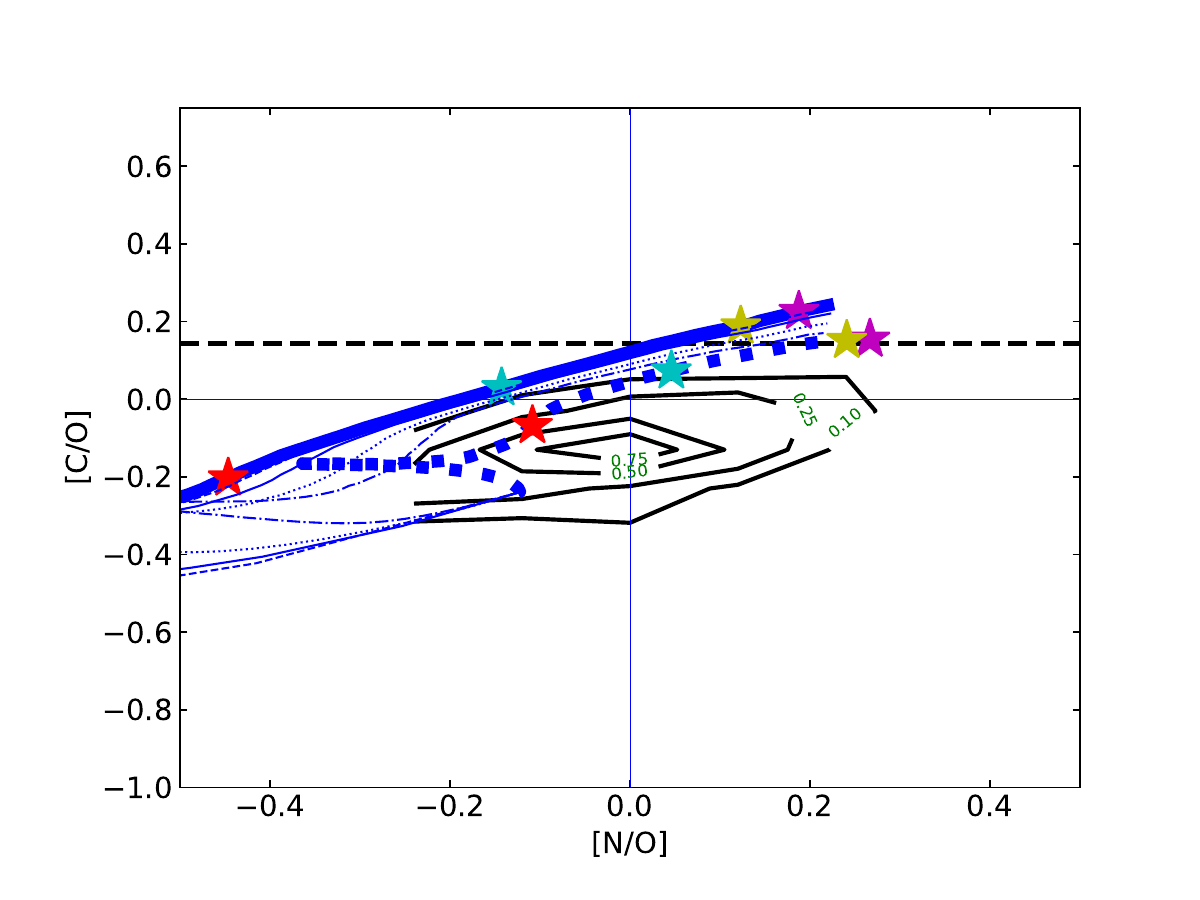}\par
\end{multicols}
\begin{multicols}{2}
    \includegraphics[width=0.8\linewidth]{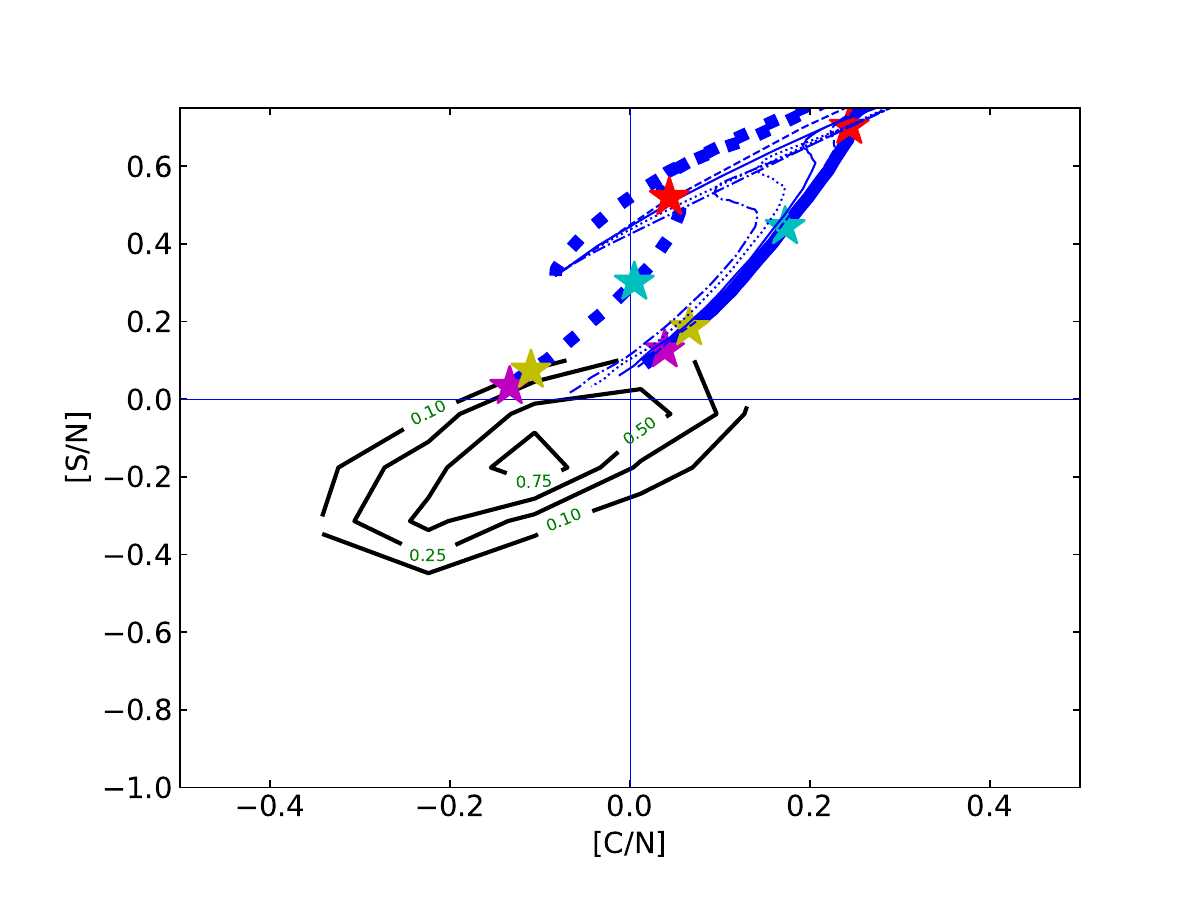}\par
    \includegraphics[width=0.8\linewidth]{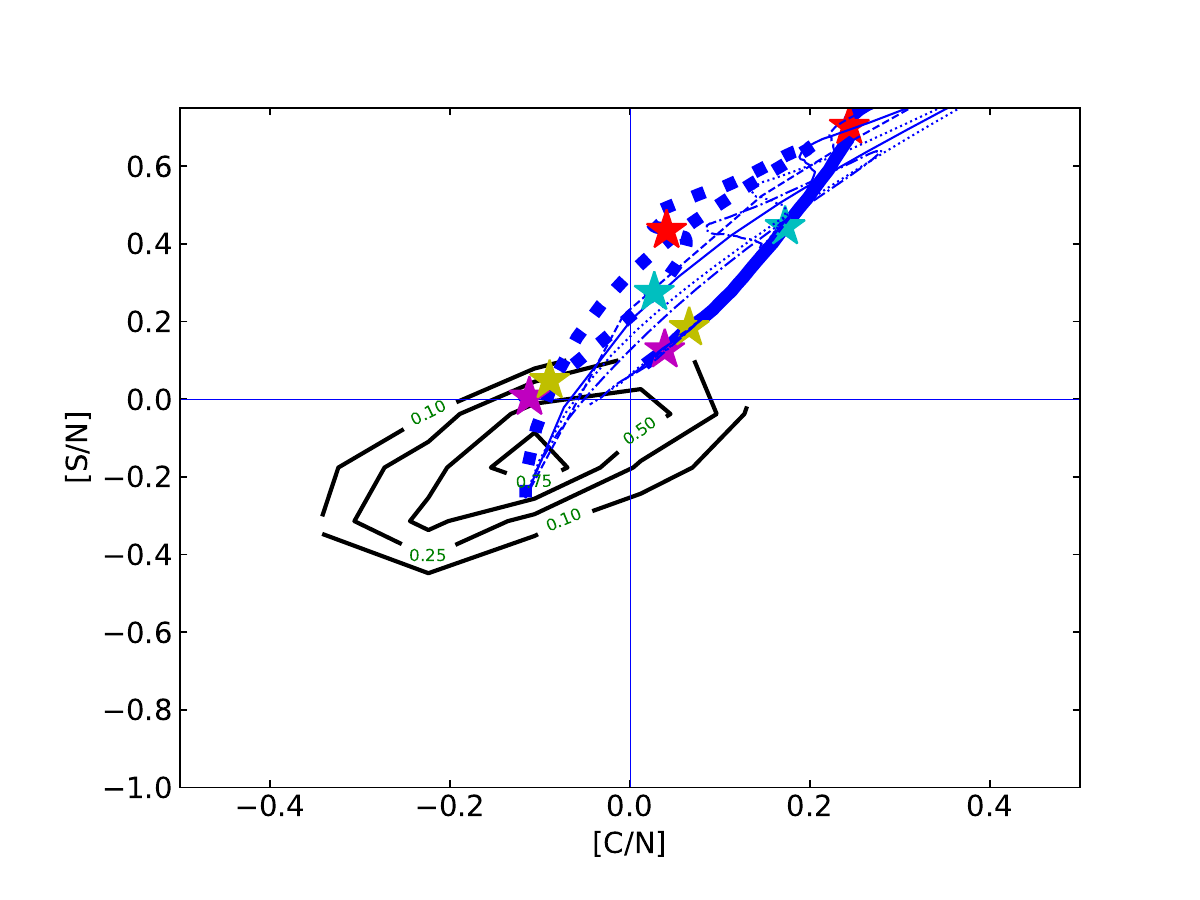}\par
\end{multicols}
\begin{multicols}{2}
    \includegraphics[width=0.8\linewidth]{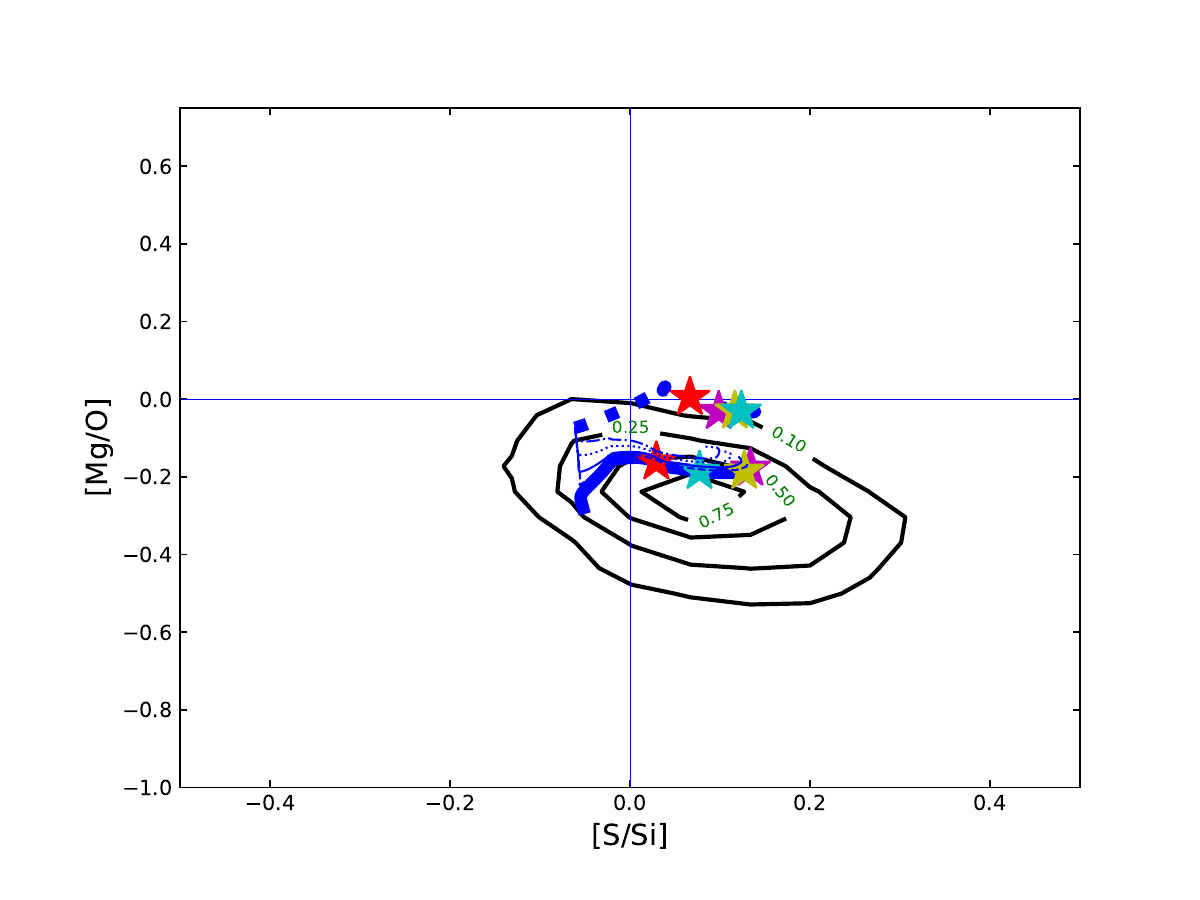}\par
    \includegraphics[width=0.8\linewidth]{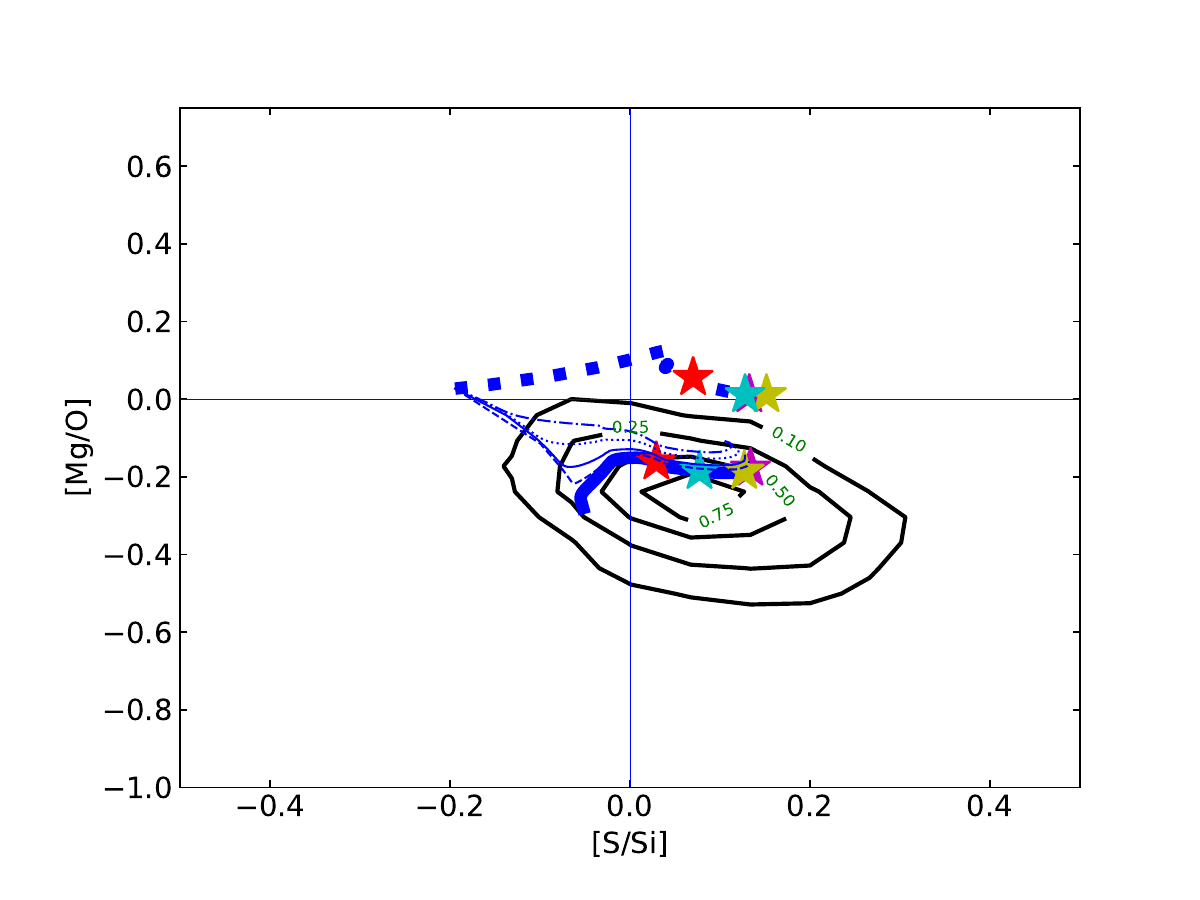}\par
\end{multicols}
    \caption{As in figure~\ref{fig: tk_plots/ratios_oR18_m40} for the oR18 set, but models are shown with CCSN supernovae contribution up to M$_{\rm up}$ = 20 M$_{\odot}$.
    }
    \label{fig: tk_plots/ratios_oR18_m20}
\end{figure*}

\begin{figure*}
\begin{multicols}{2}
    \includegraphics[width=0.8\linewidth]{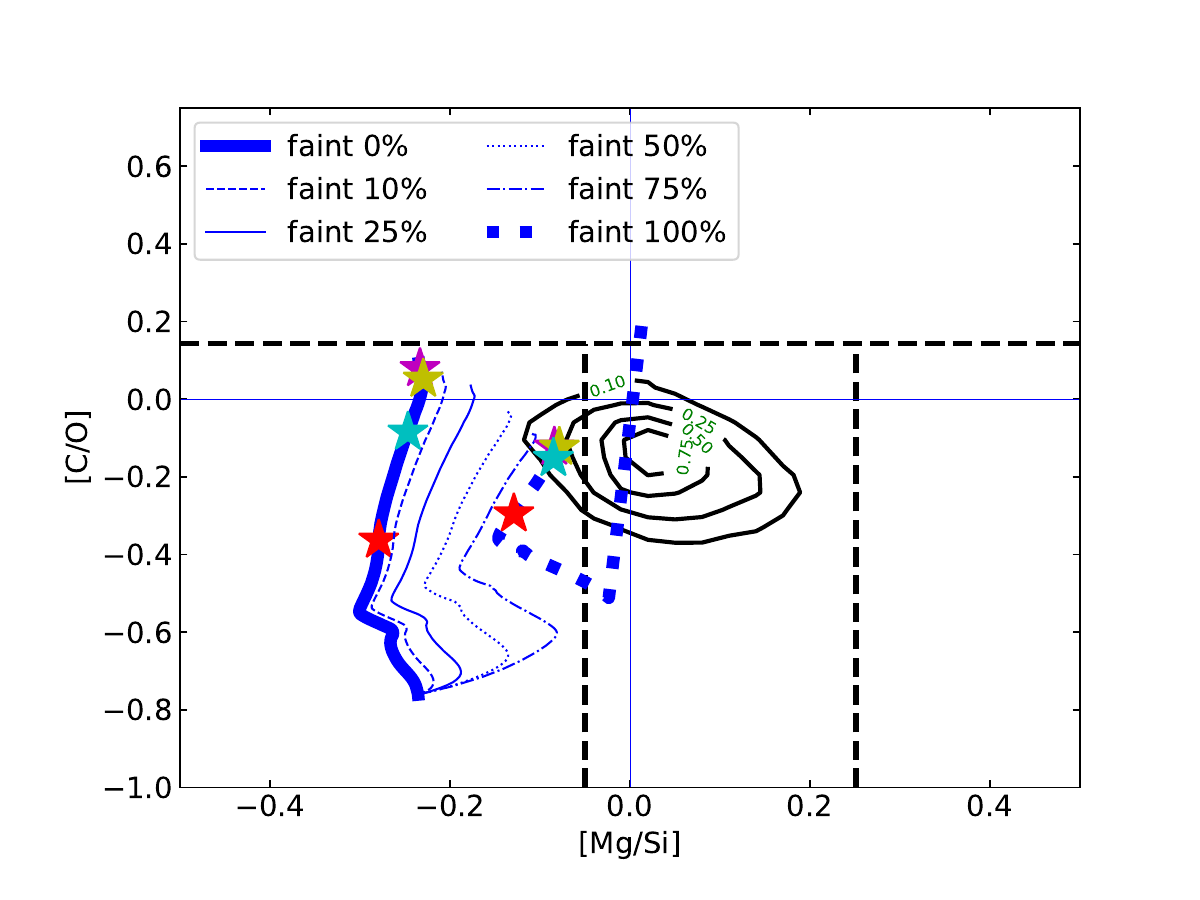}\par
    \includegraphics[width=0.8\linewidth]{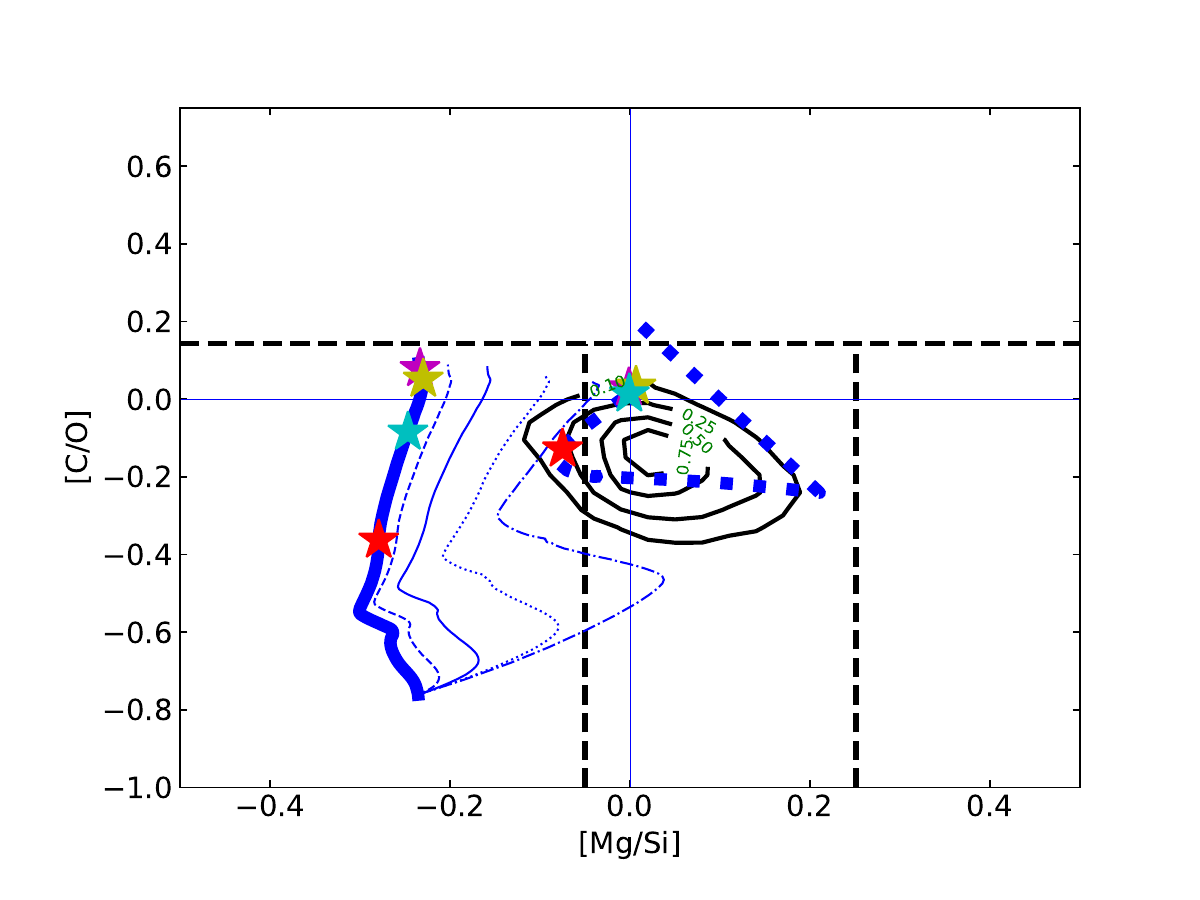}\par
    \end{multicols}
\begin{multicols}{2}
    \includegraphics[width=0.8\linewidth]{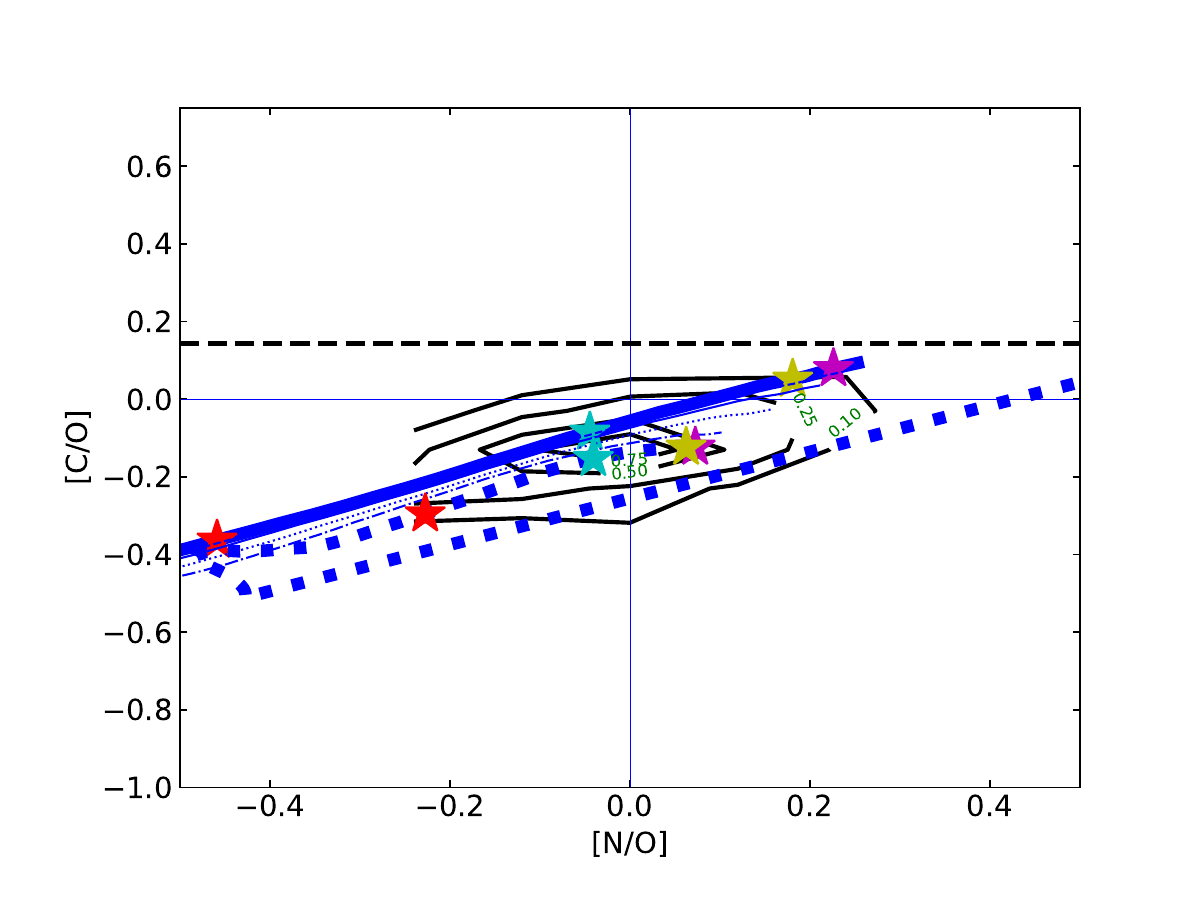}\par
    \includegraphics[width=0.8\linewidth]{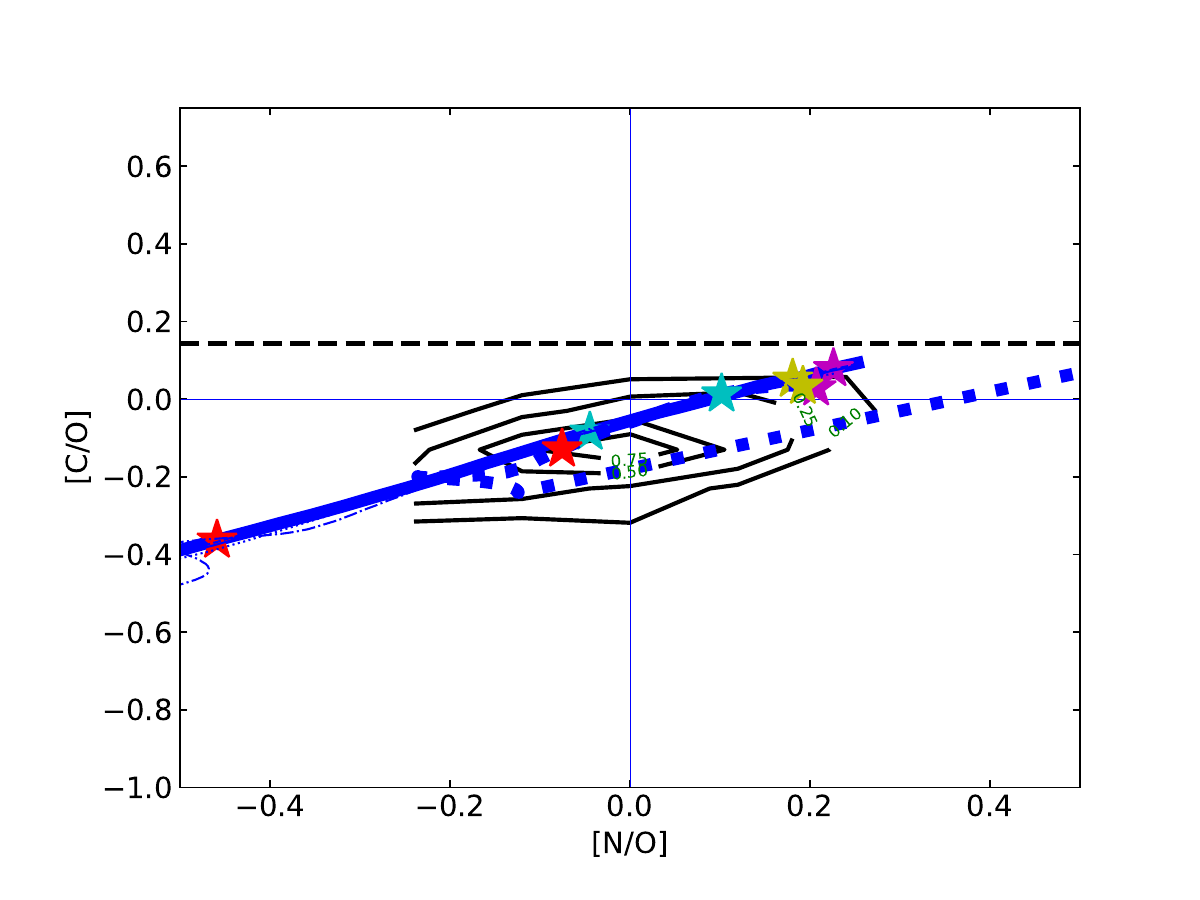}\par
\end{multicols}
\begin{multicols}{2}
    \includegraphics[width=0.8\linewidth]{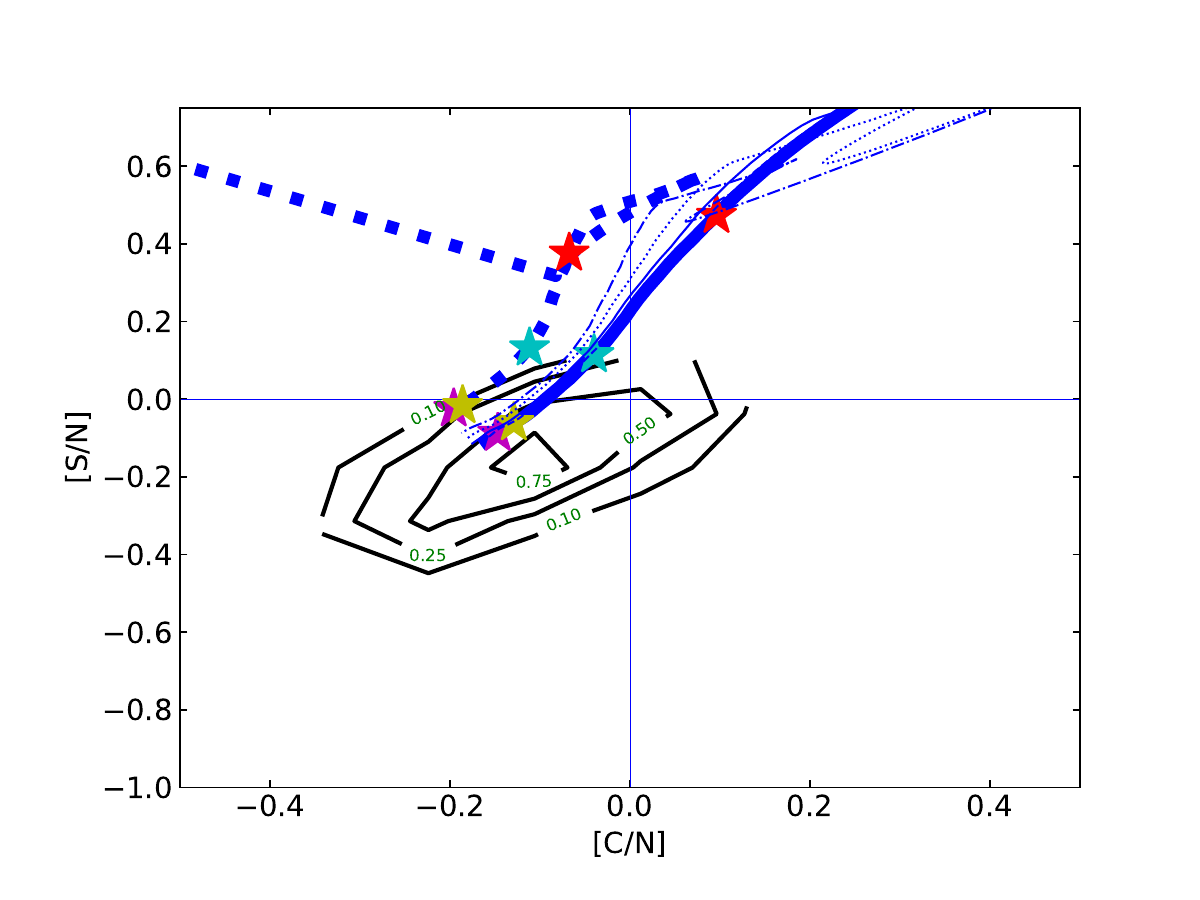}\par
    \includegraphics[width=0.8\linewidth]{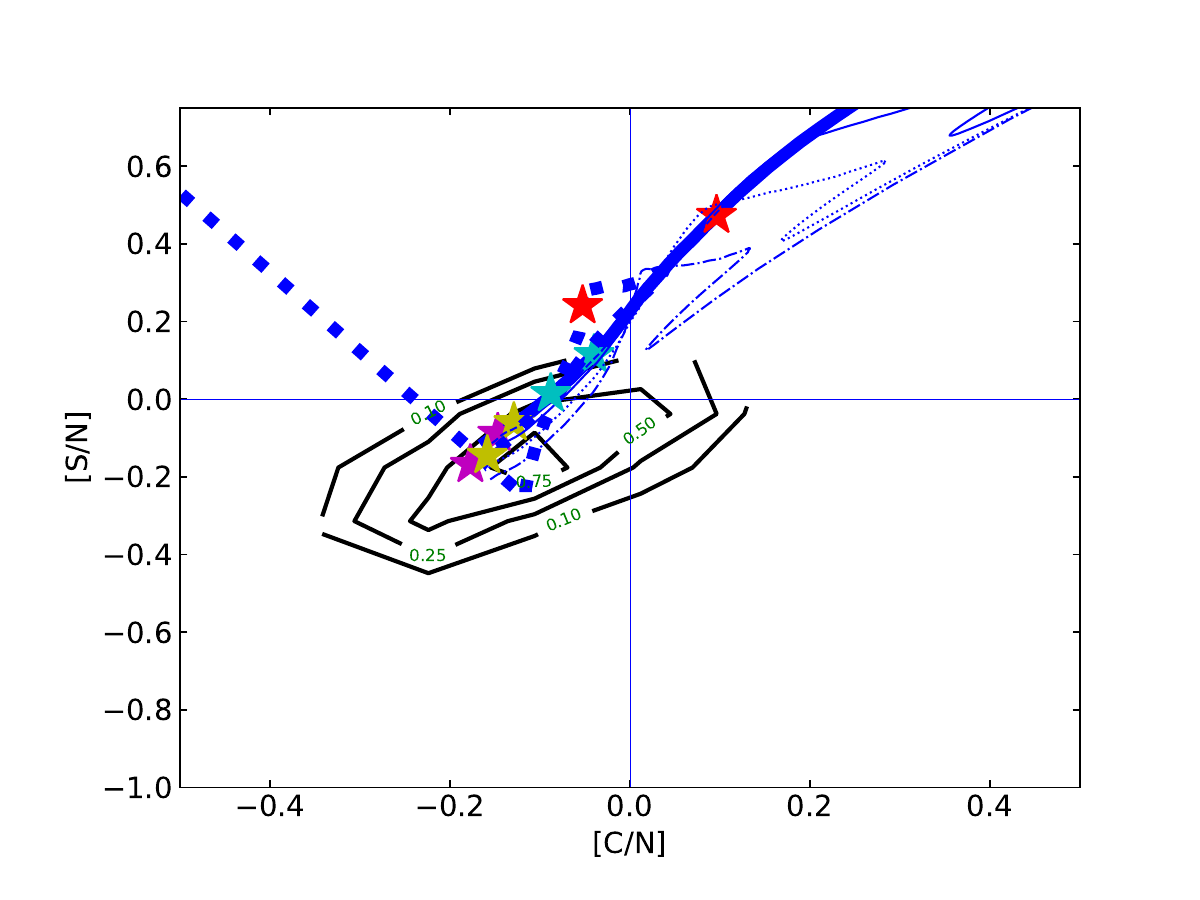}\par
\end{multicols}
\begin{multicols}{2}
    \includegraphics[width=0.8\linewidth]{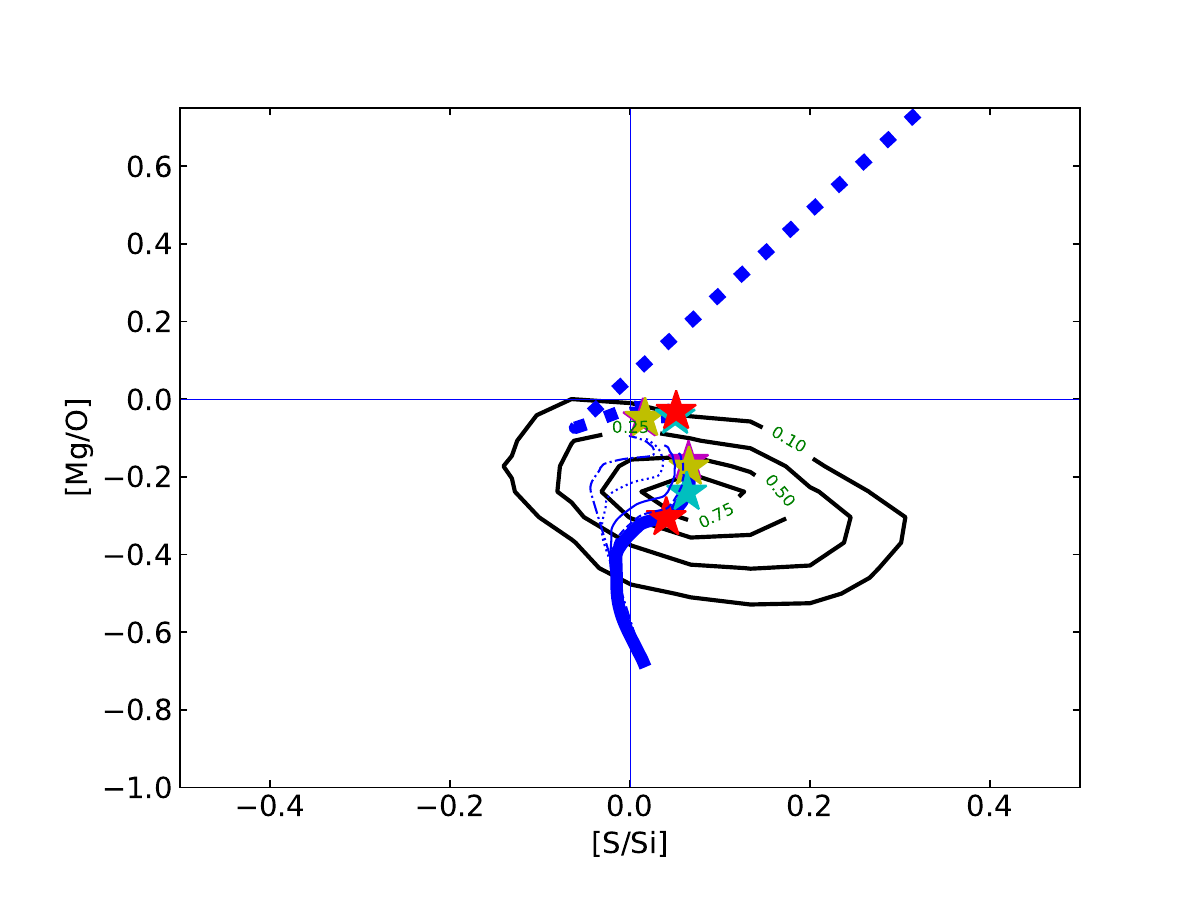}\par
    \includegraphics[width=0.8\linewidth]{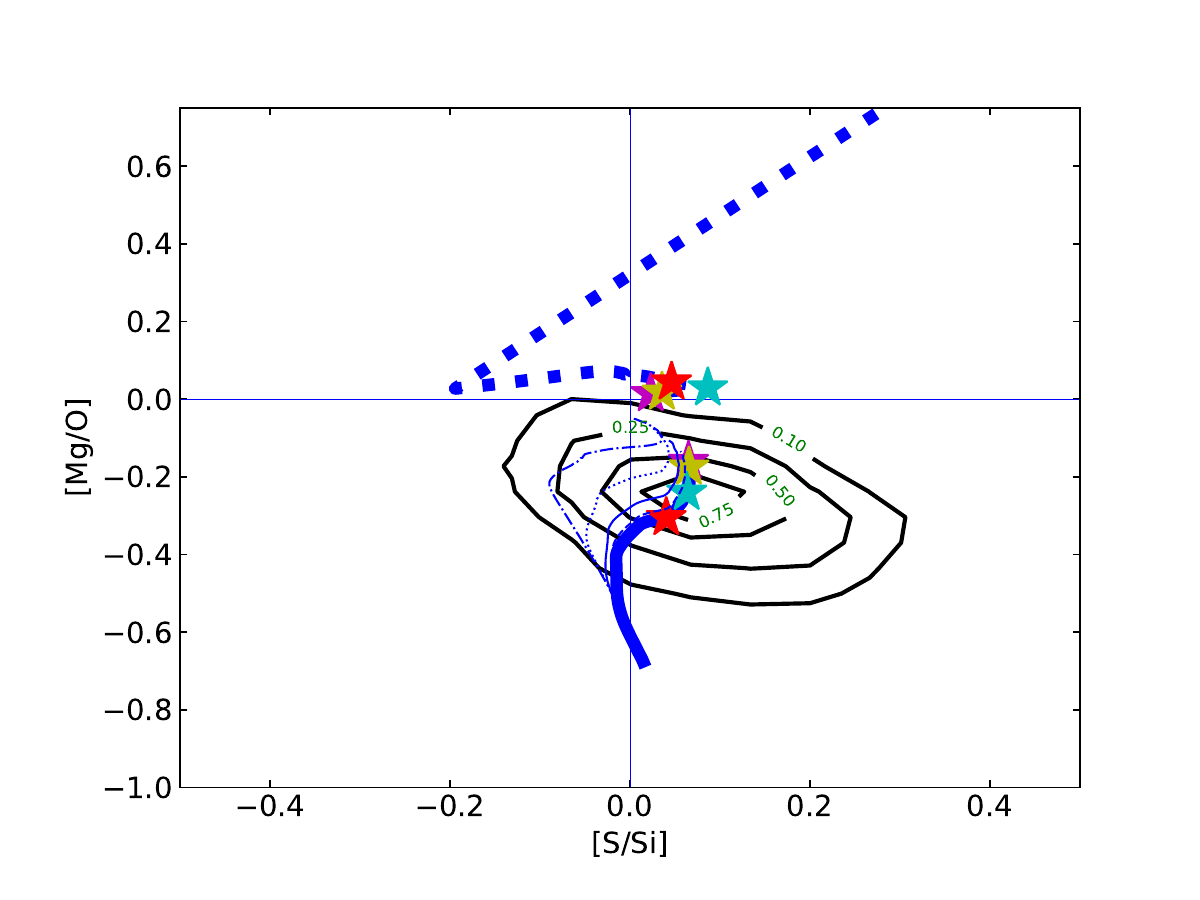}\par
\end{multicols}
    \caption{As in figure~\ref{fig: tk_plots/ratios_oR18_m40} for the oR18 set, but models are shown with CCSN supernovae contribution up to M$_{\rm up}$ = 100 M$_{\odot}$.
    }
    \label{fig: tk_plots/ratios_oR18_m100}
\end{figure*}

For the oR18 set, the results using M$_{\rm up}$=20M$_{\odot}$ and 100M$_{\odot}$ for different faint supernovae are given in Figures \ref{fig: tk_plots/ratios_oR18_m20} and \ref{fig: tk_plots/ratios_oR18_m100}.
The same results for M$_{\rm up}$=40M$_{\odot}$ are discussed in section \S~\ref{sec: results} (Figure \ref{fig: tk_plots/ratios_oR18_m40}).

\begin{figure*}
\begin{multicols}{2}
    \includegraphics[width=0.8\linewidth]{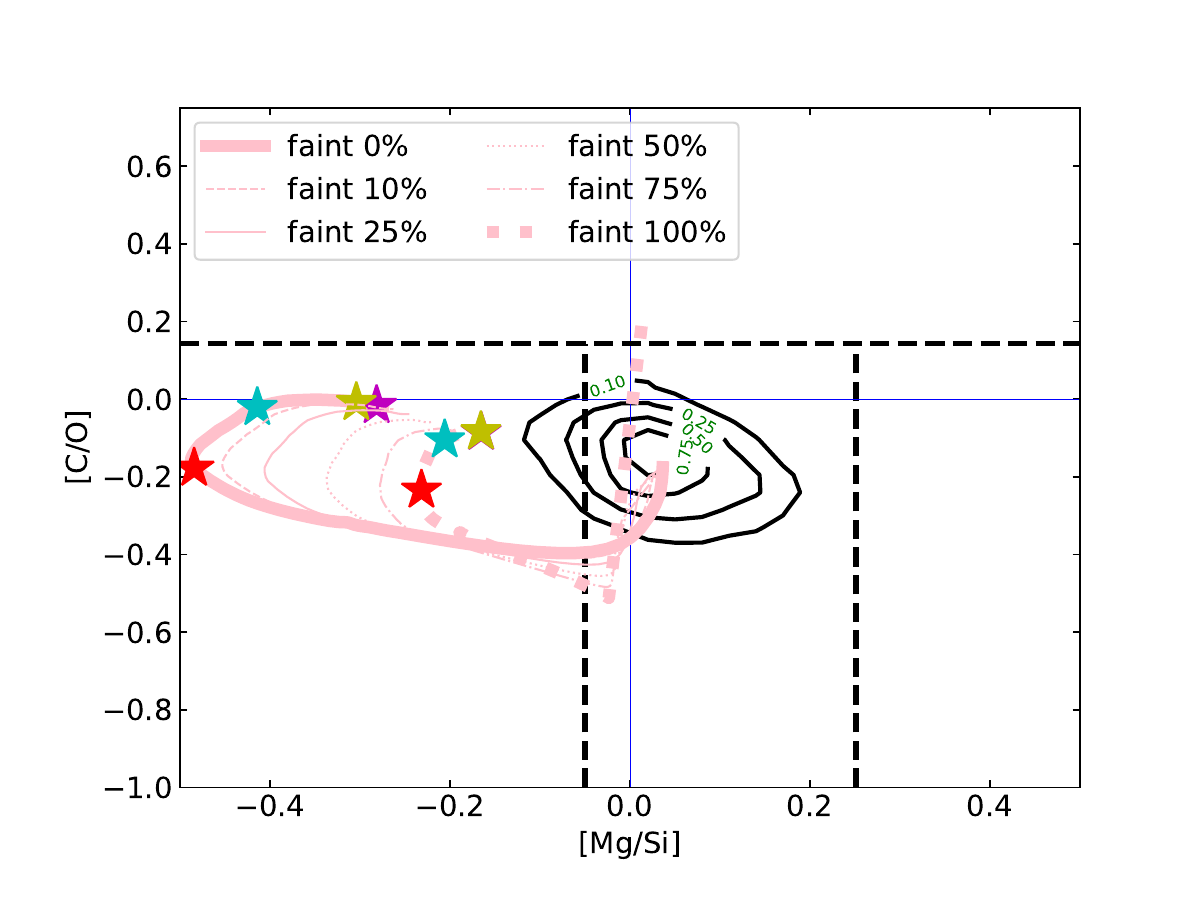}\par
    \includegraphics[width=0.8\linewidth]{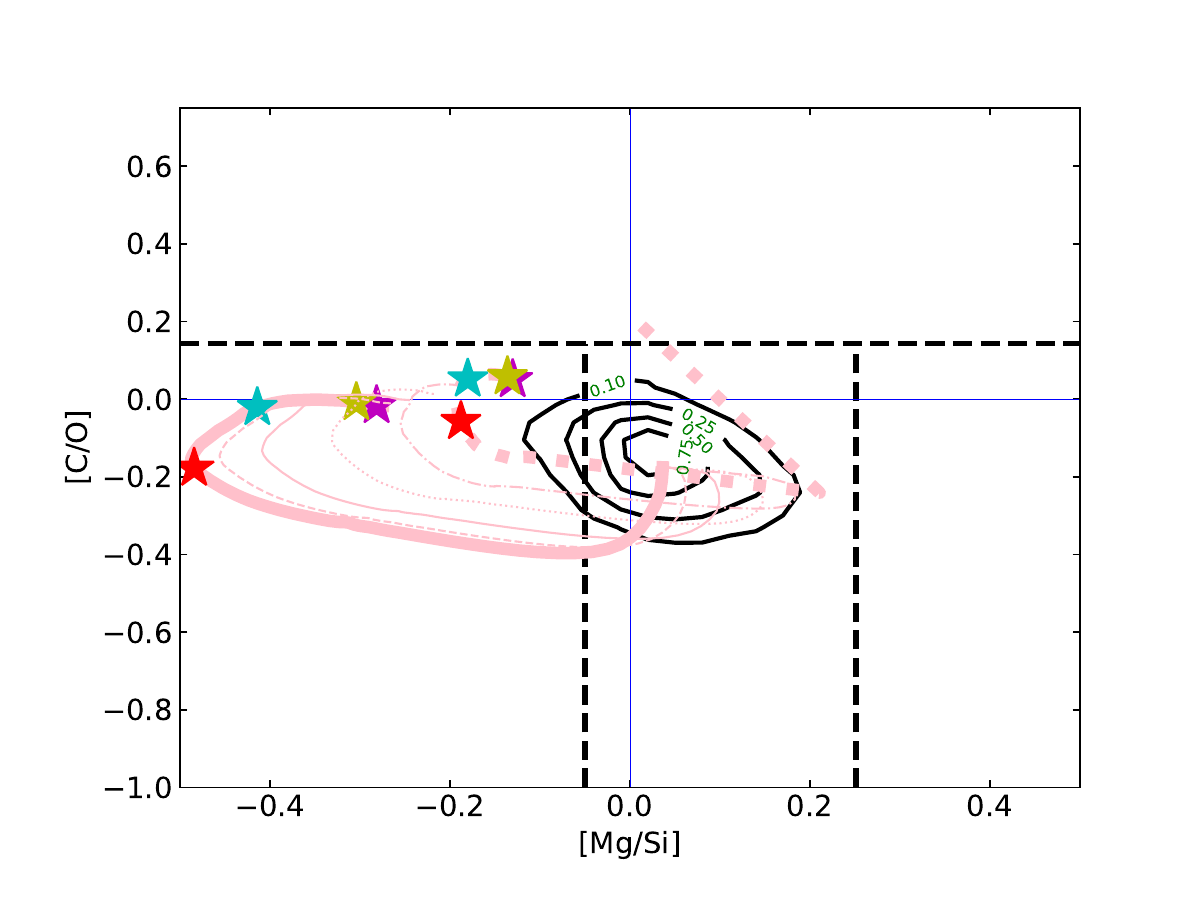}\par
    \end{multicols}
\begin{multicols}{2}
    \includegraphics[width=0.8\linewidth]{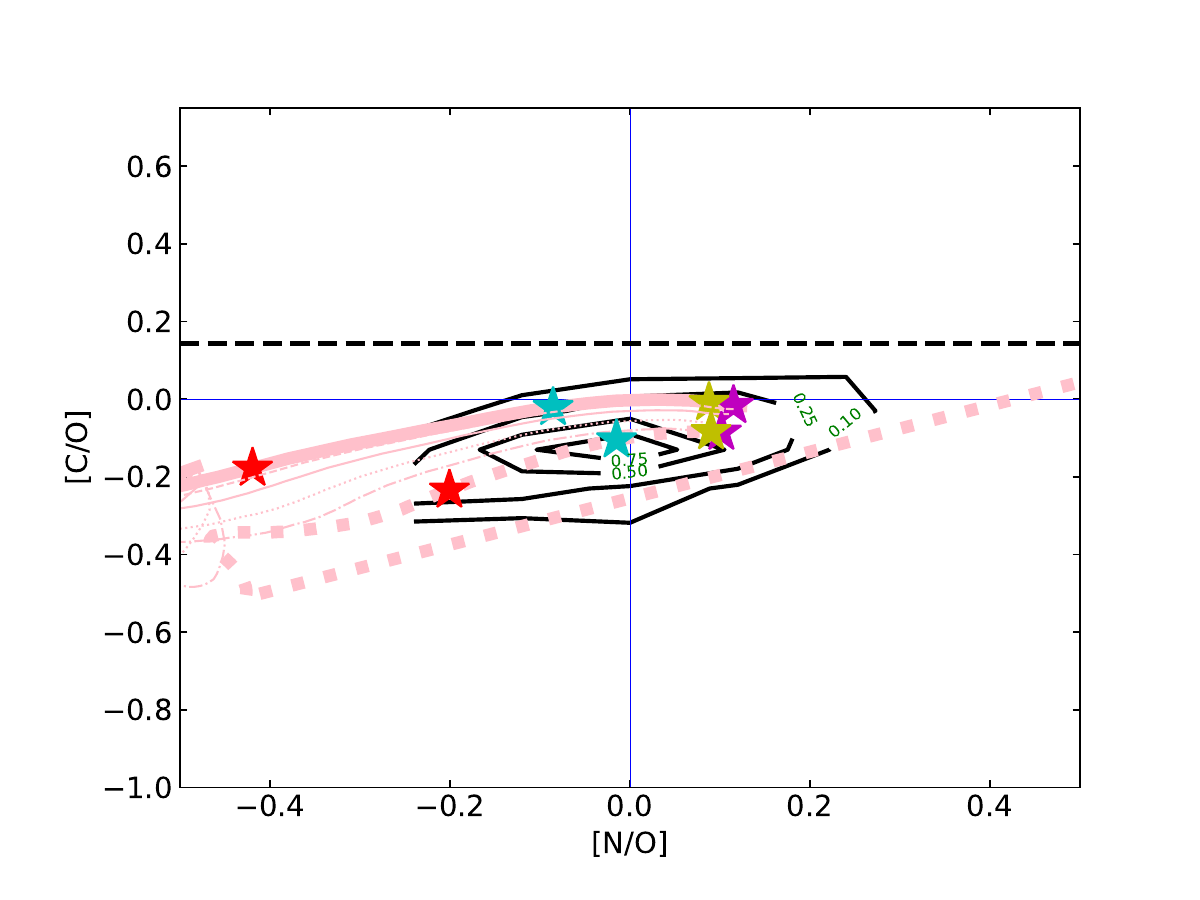}\par
    \includegraphics[width=0.8\linewidth]{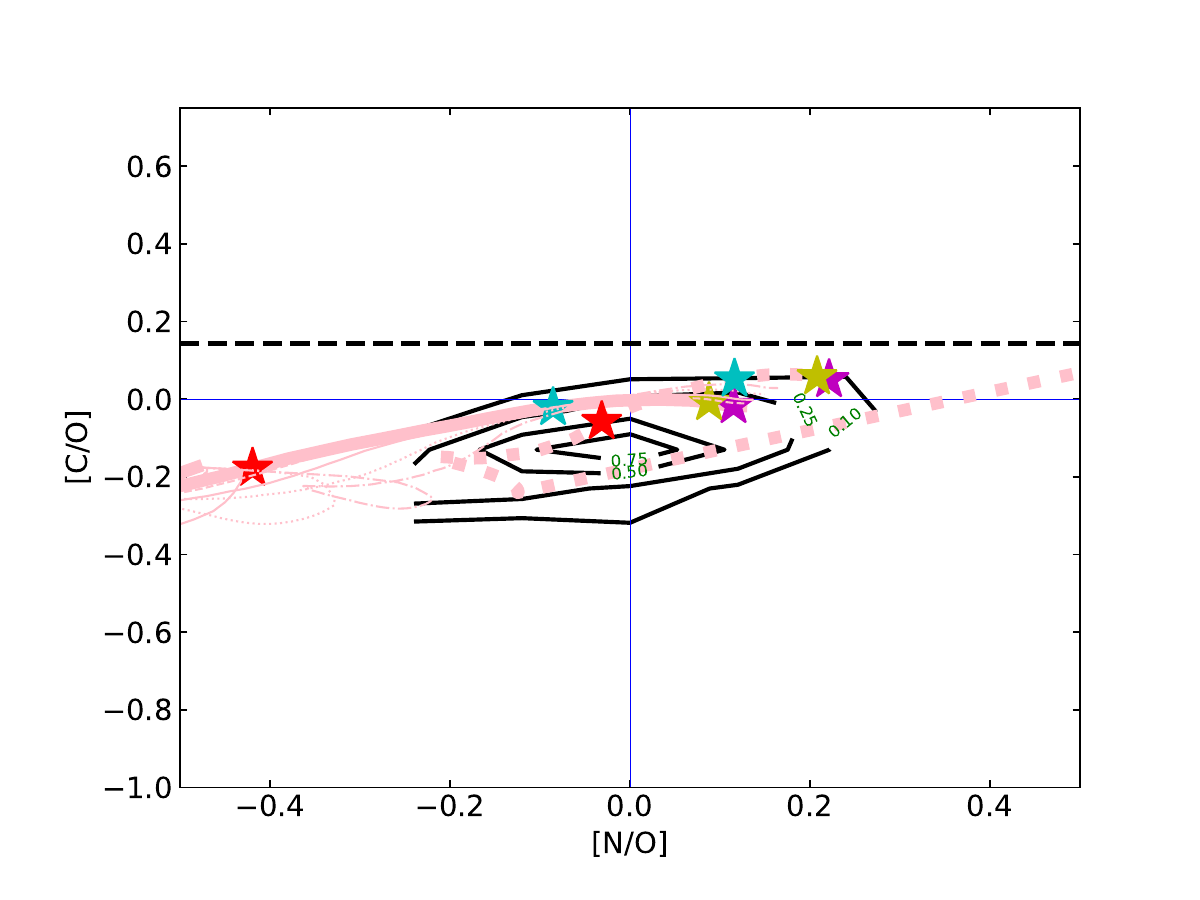}\par
\end{multicols}
\begin{multicols}{2}
    \includegraphics[width=0.8\linewidth]{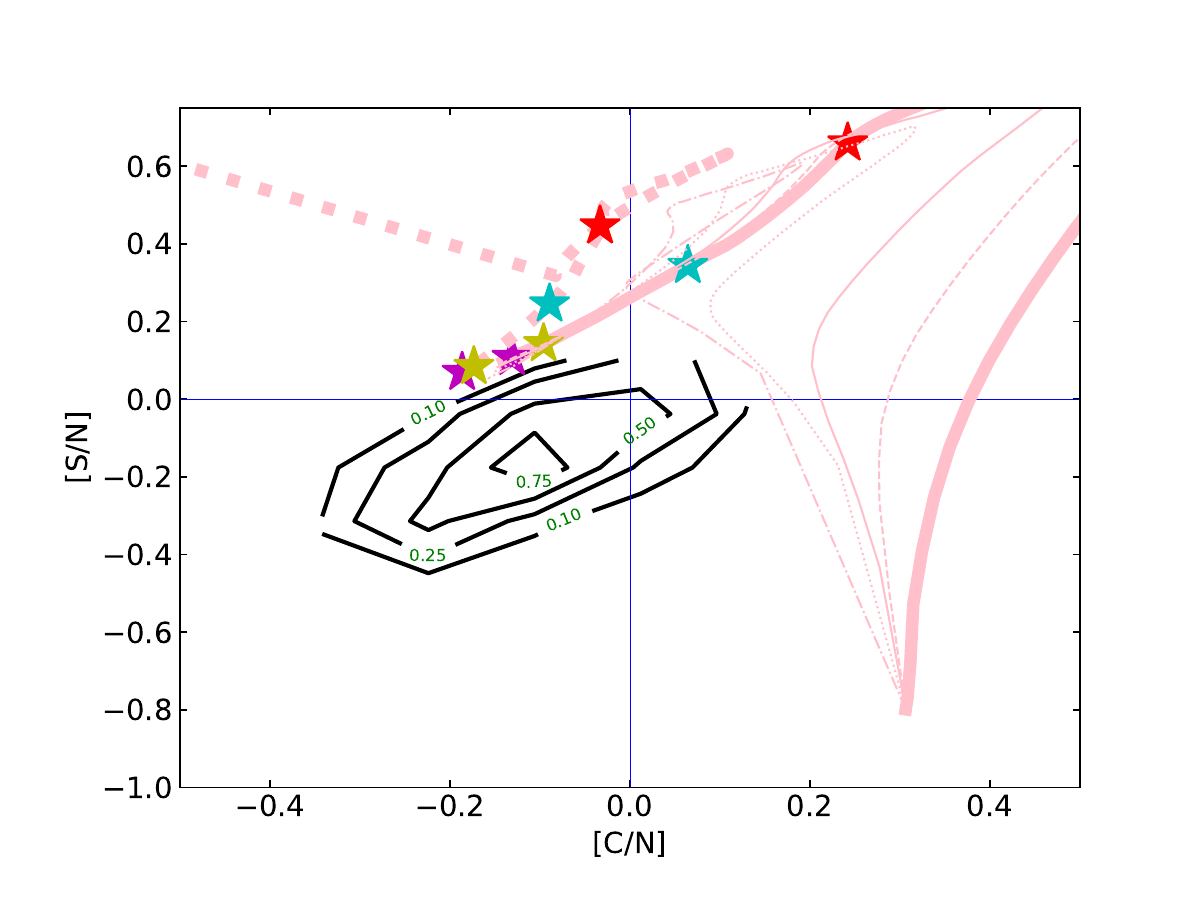}\par
    \includegraphics[width=0.8\linewidth]{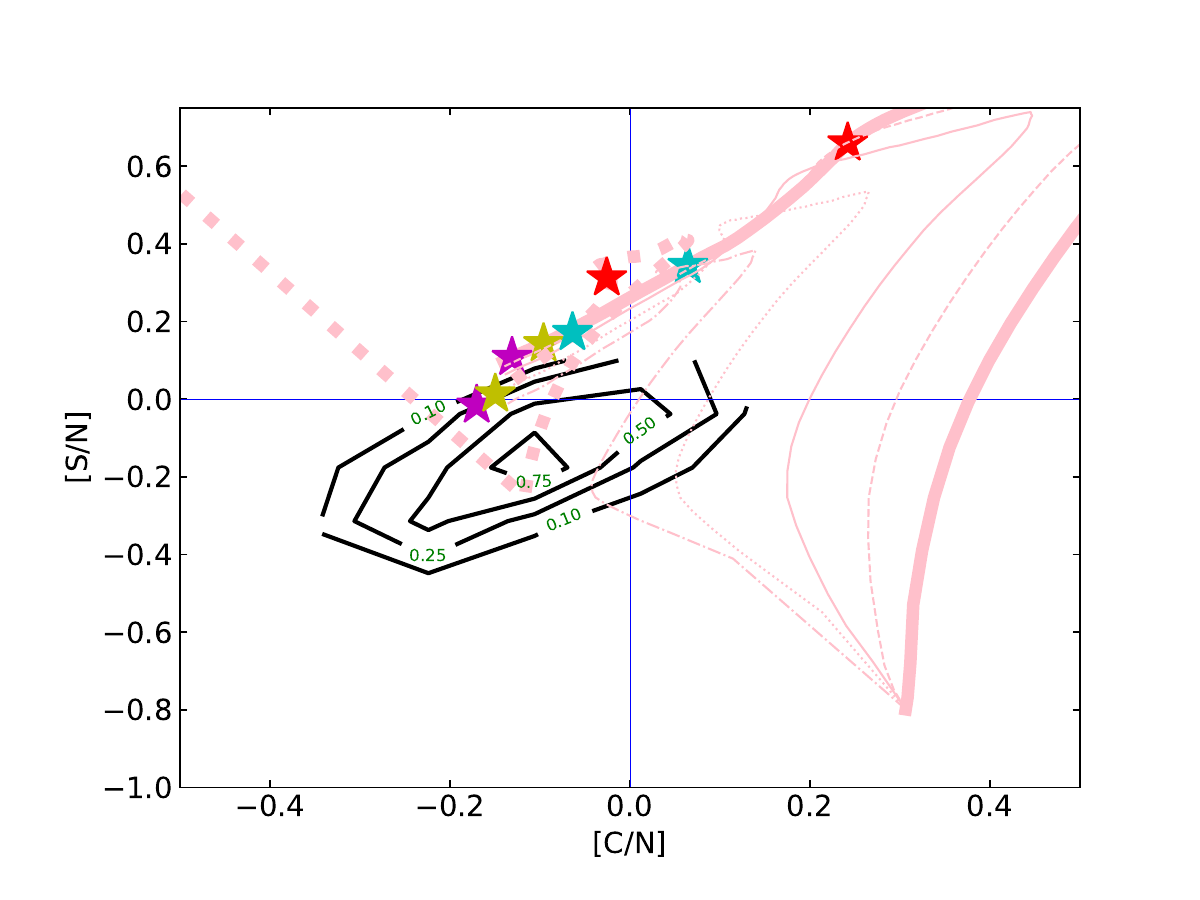}\par
\end{multicols}
\begin{multicols}{2}
    \includegraphics[width=0.8\linewidth]{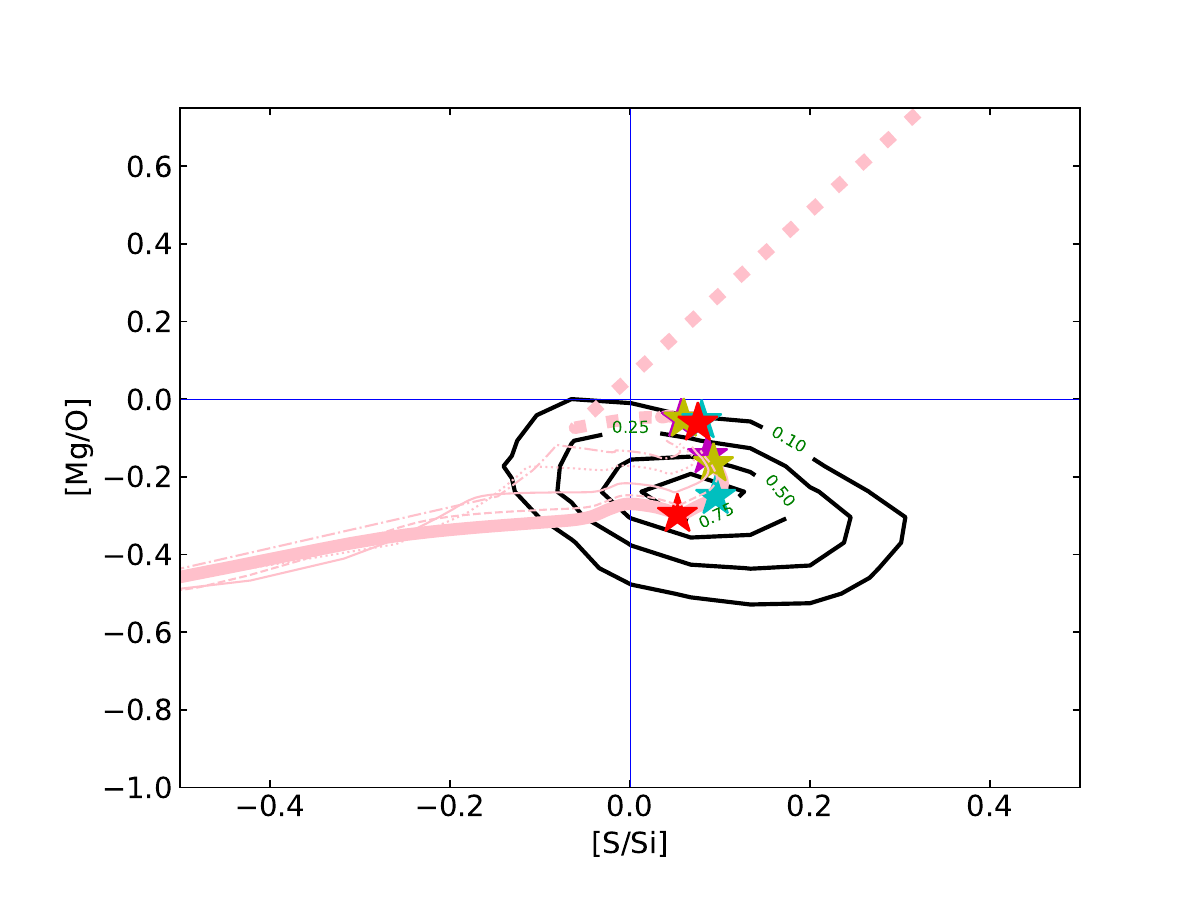}\par
    \includegraphics[width=0.8\linewidth]{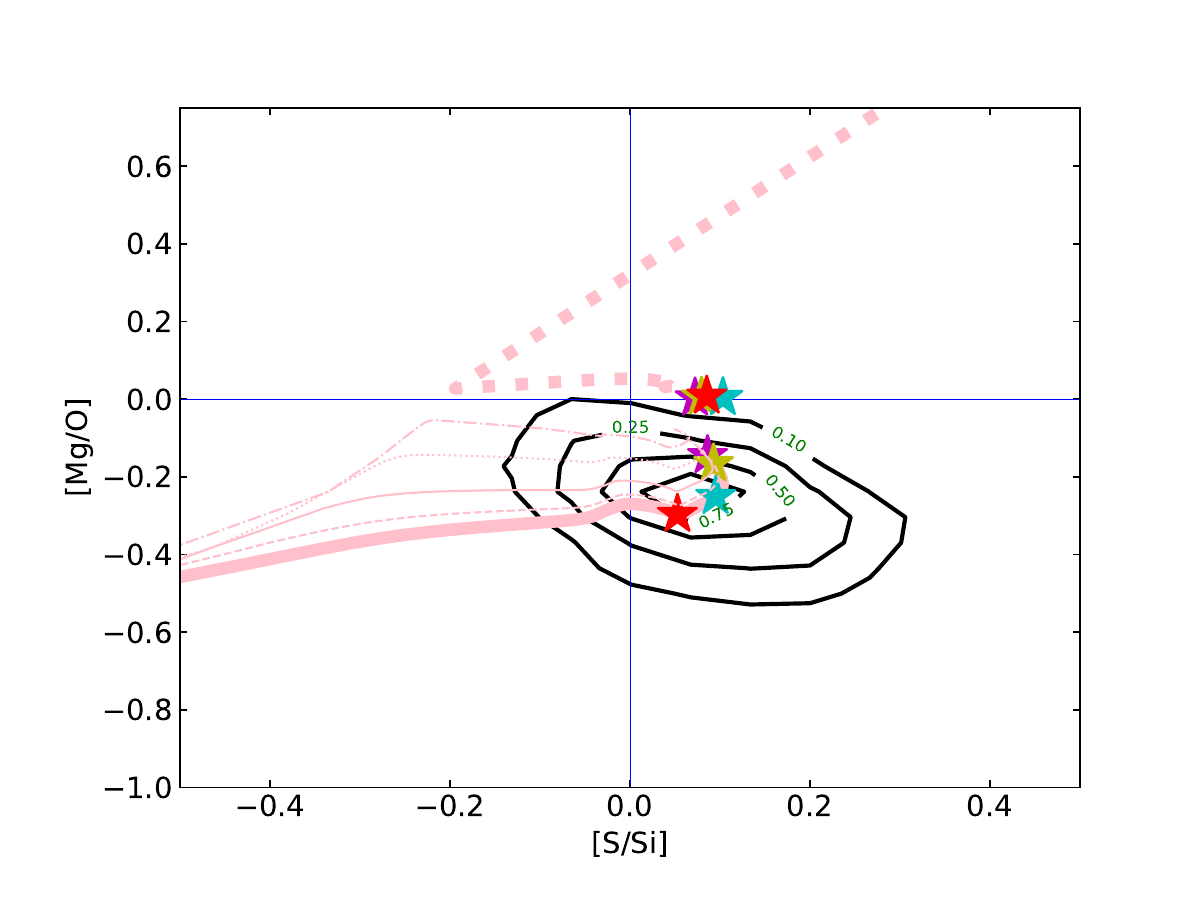}\par
\end{multicols}
    \caption{Same as in Figure~\ref{fig: tk_plots/ratios_gce_oK10m40}, but for the GCE model set oR18d.
    }
    \label{fig: tk_plots/ratios_oR18d_m40}
\end{figure*}

\begin{figure*}
\begin{multicols}{2}
    \includegraphics[width=0.8\linewidth]{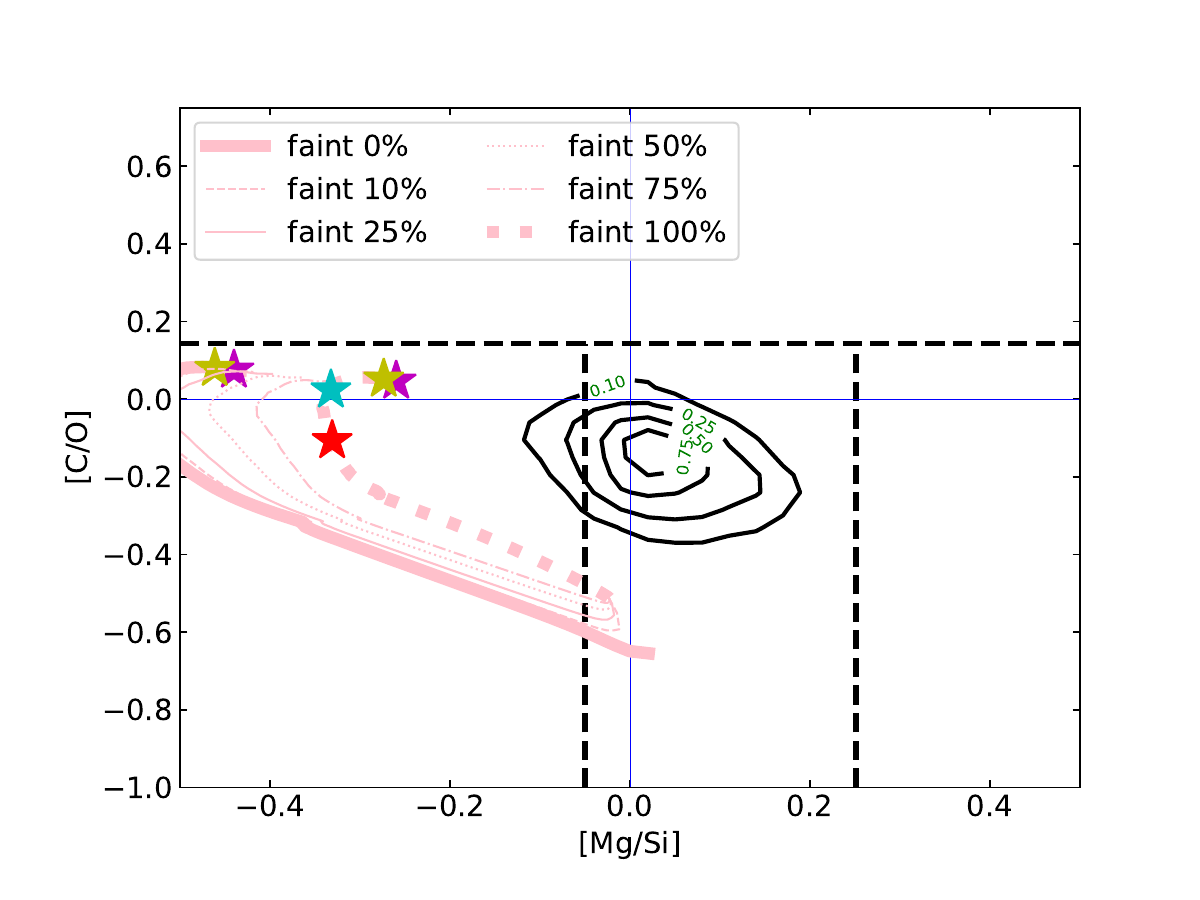}\par
    \includegraphics[width=0.8\linewidth]{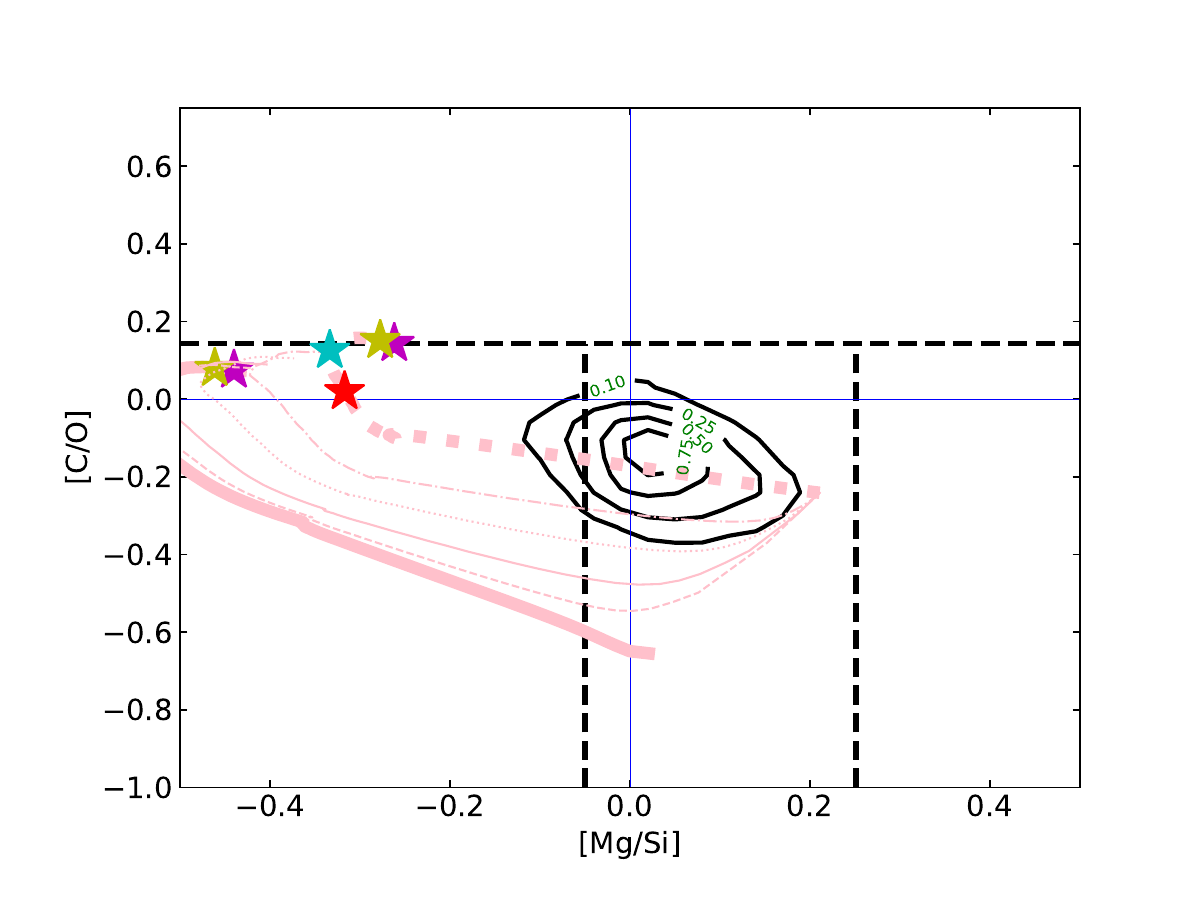}\par
    \end{multicols}
\begin{multicols}{2}
    \includegraphics[width=0.8\linewidth]{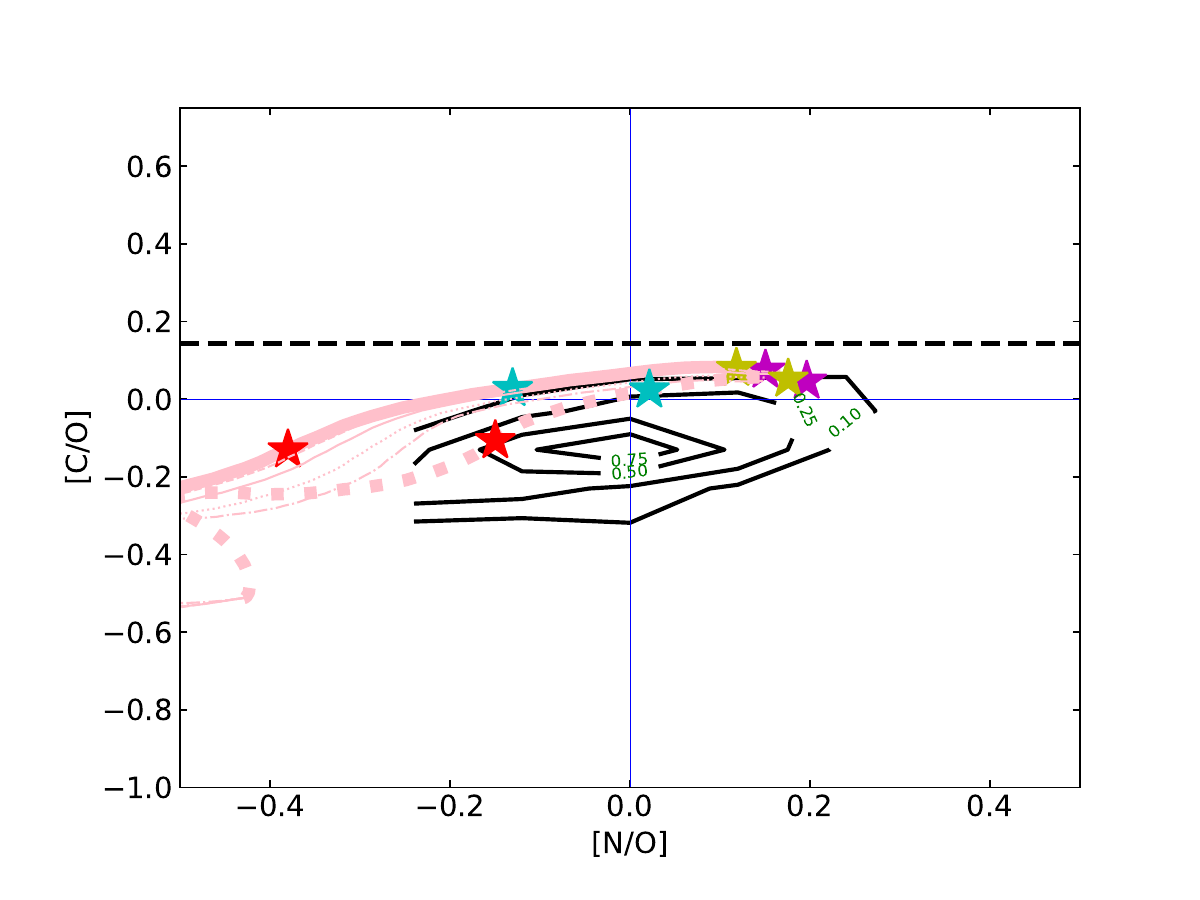}\par
    \includegraphics[width=0.8\linewidth]{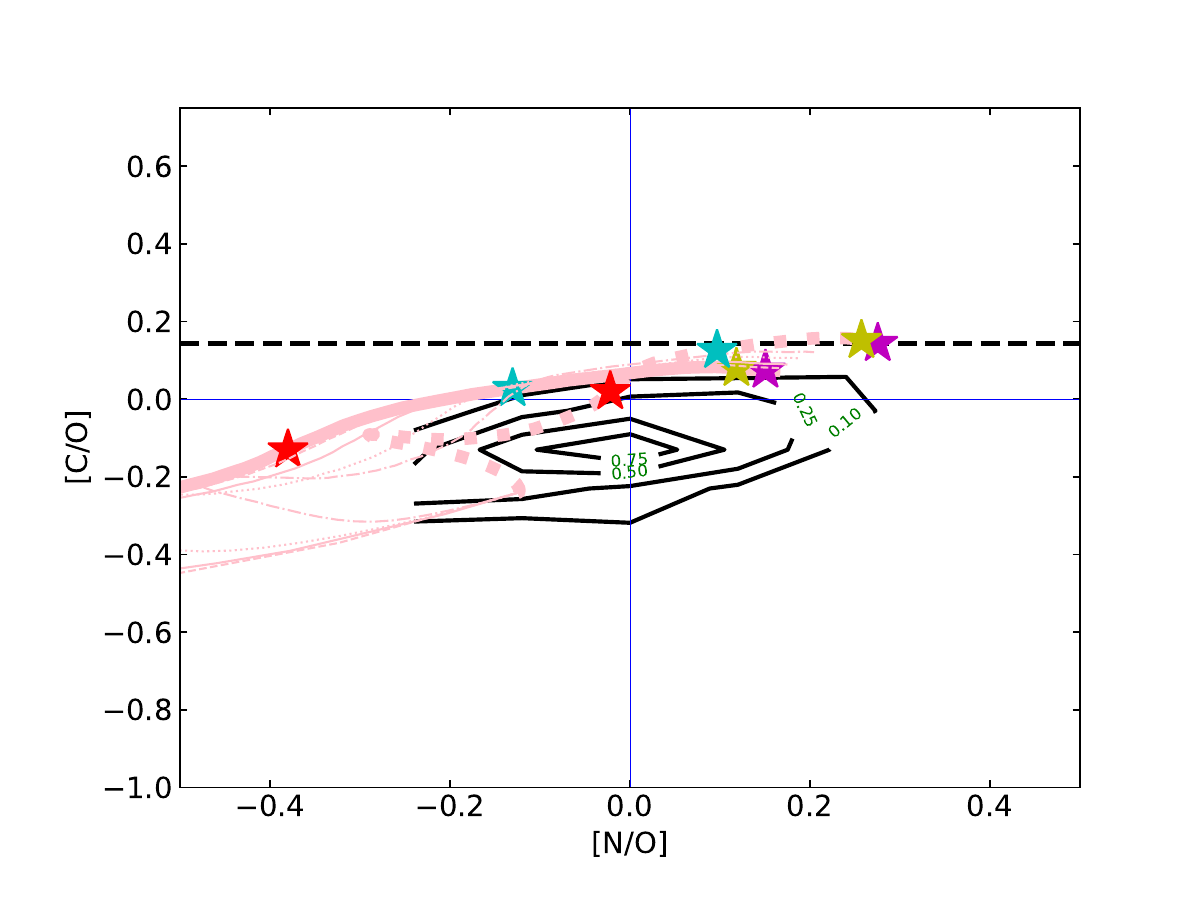}\par
\end{multicols}
\begin{multicols}{2}
    \includegraphics[width=0.8\linewidth]{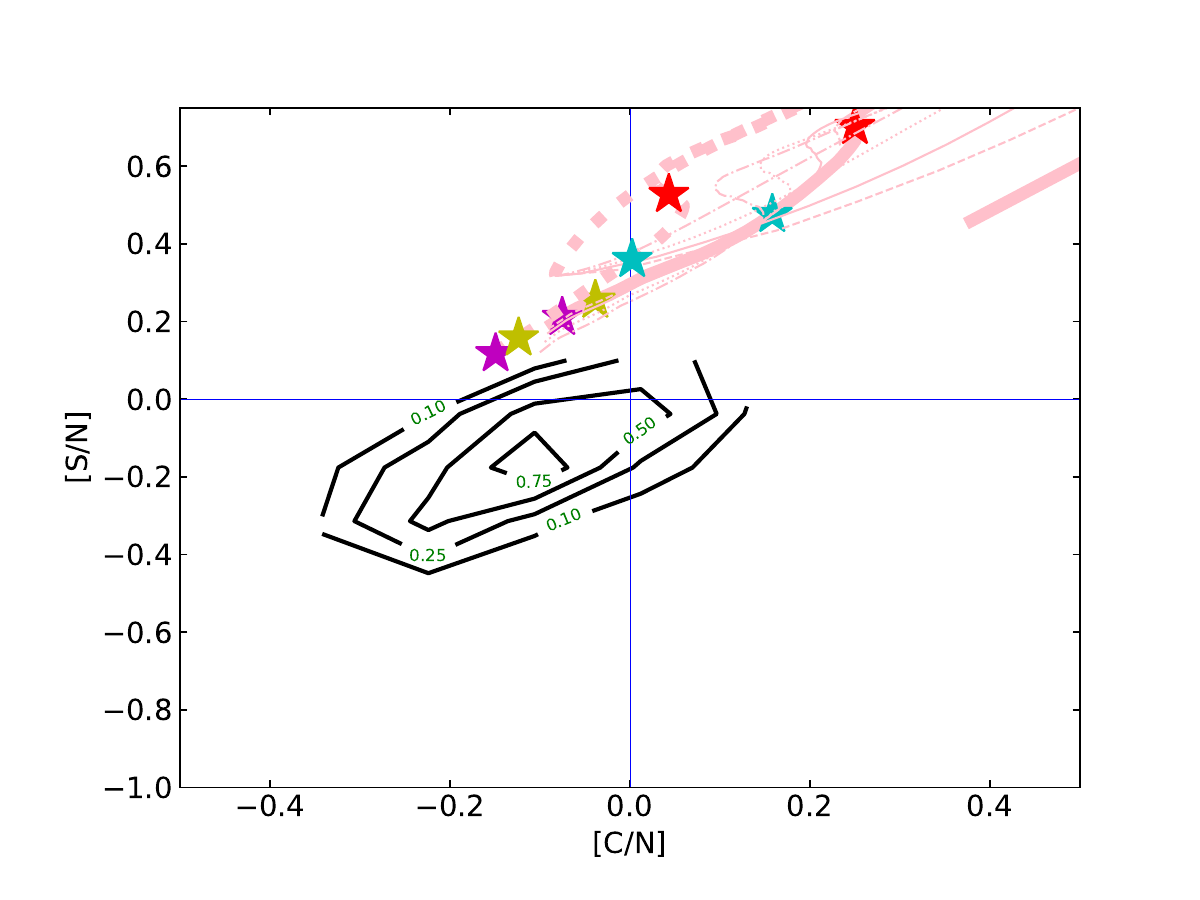}\par
    \includegraphics[width=0.8\linewidth]{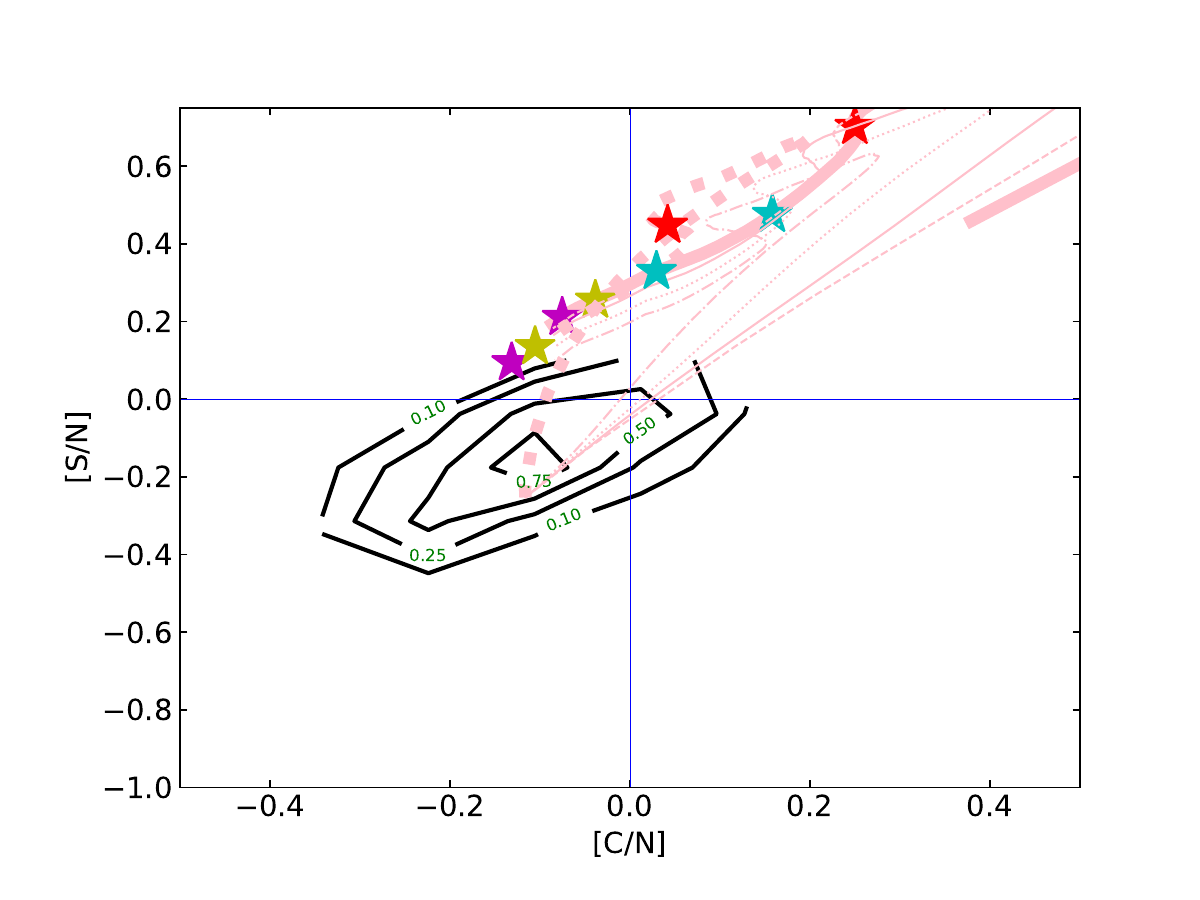}\par
\end{multicols}
\begin{multicols}{2}
    \includegraphics[width=0.8\linewidth]{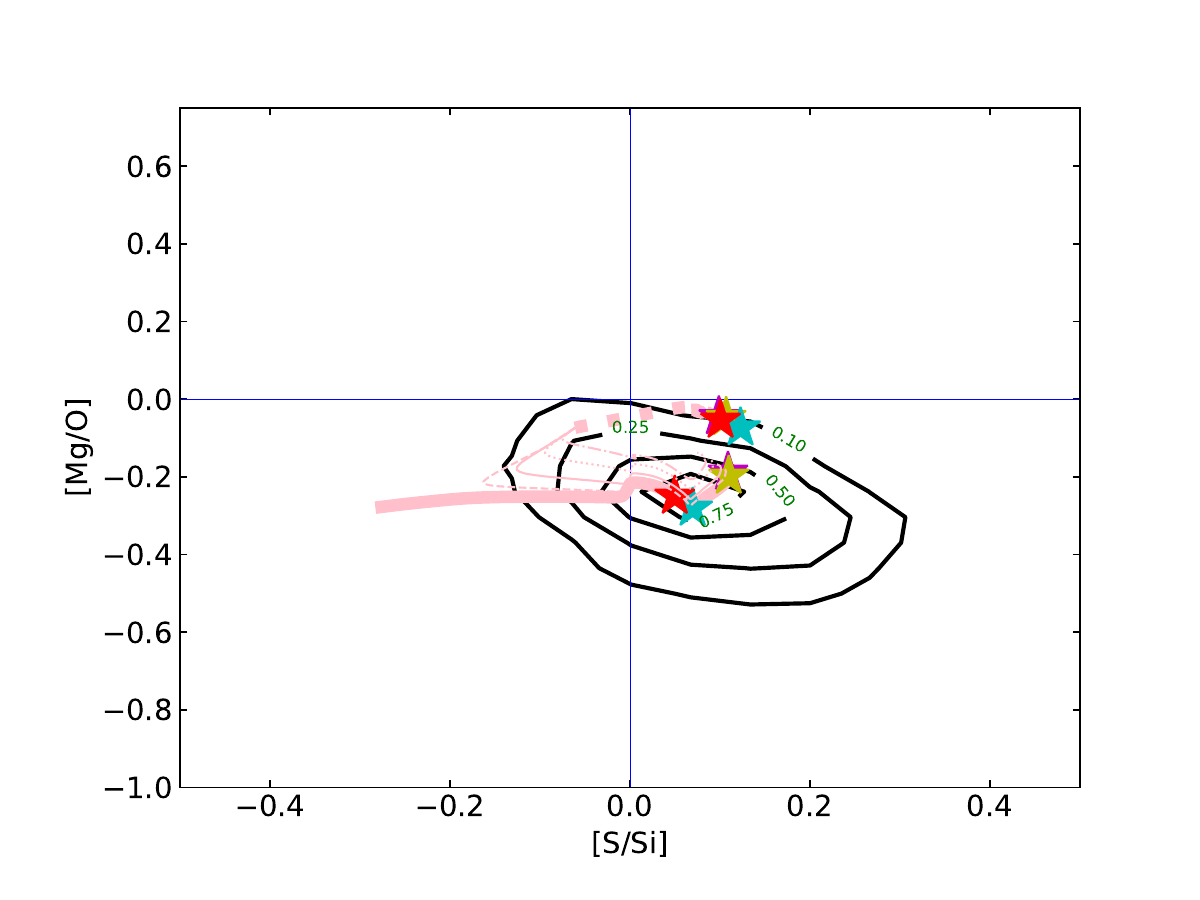}\par
    \includegraphics[width=0.8\linewidth]{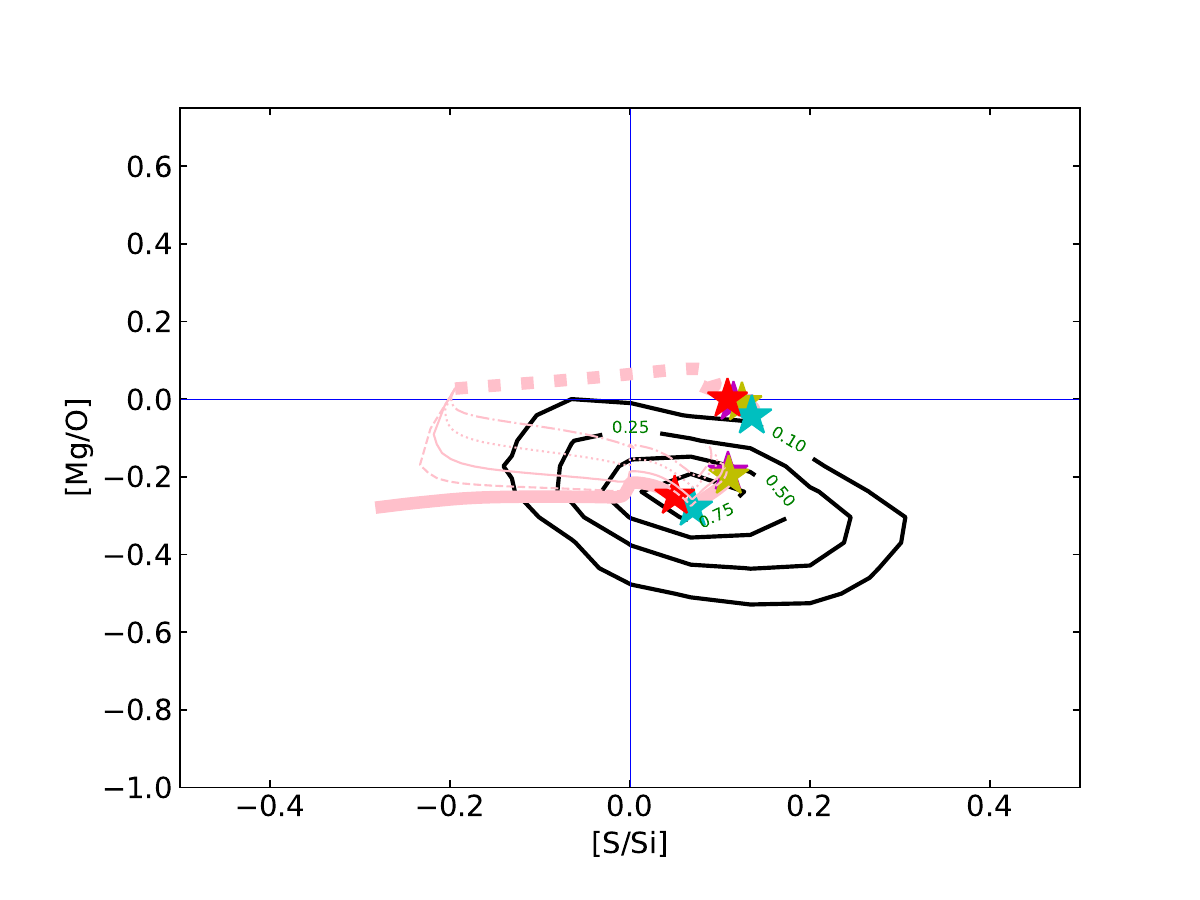}\par
\end{multicols}
    \caption{As in figure~\ref{fig: tk_plots/ratios_oR18d_m40}, but models are shown with CCSN supernovae contribution up to M$_{\rm up}$ = 20 M$_{\odot}$.
    }
    \label{fig: tk_plots/ratios_oR18d_m20}
\end{figure*}

\begin{figure*}
\begin{multicols}{2}
    \includegraphics[width=0.8\linewidth]{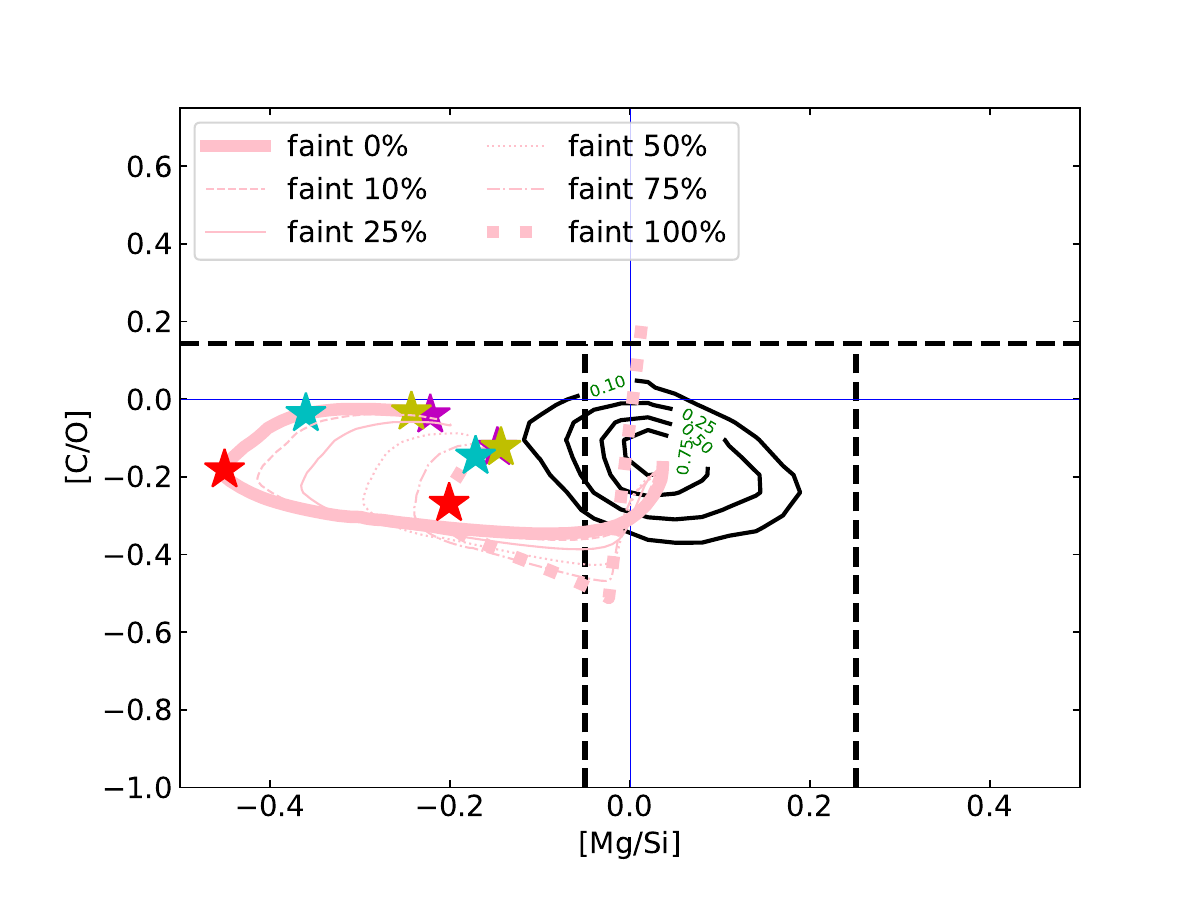}\par
    \includegraphics[width=0.8\linewidth]{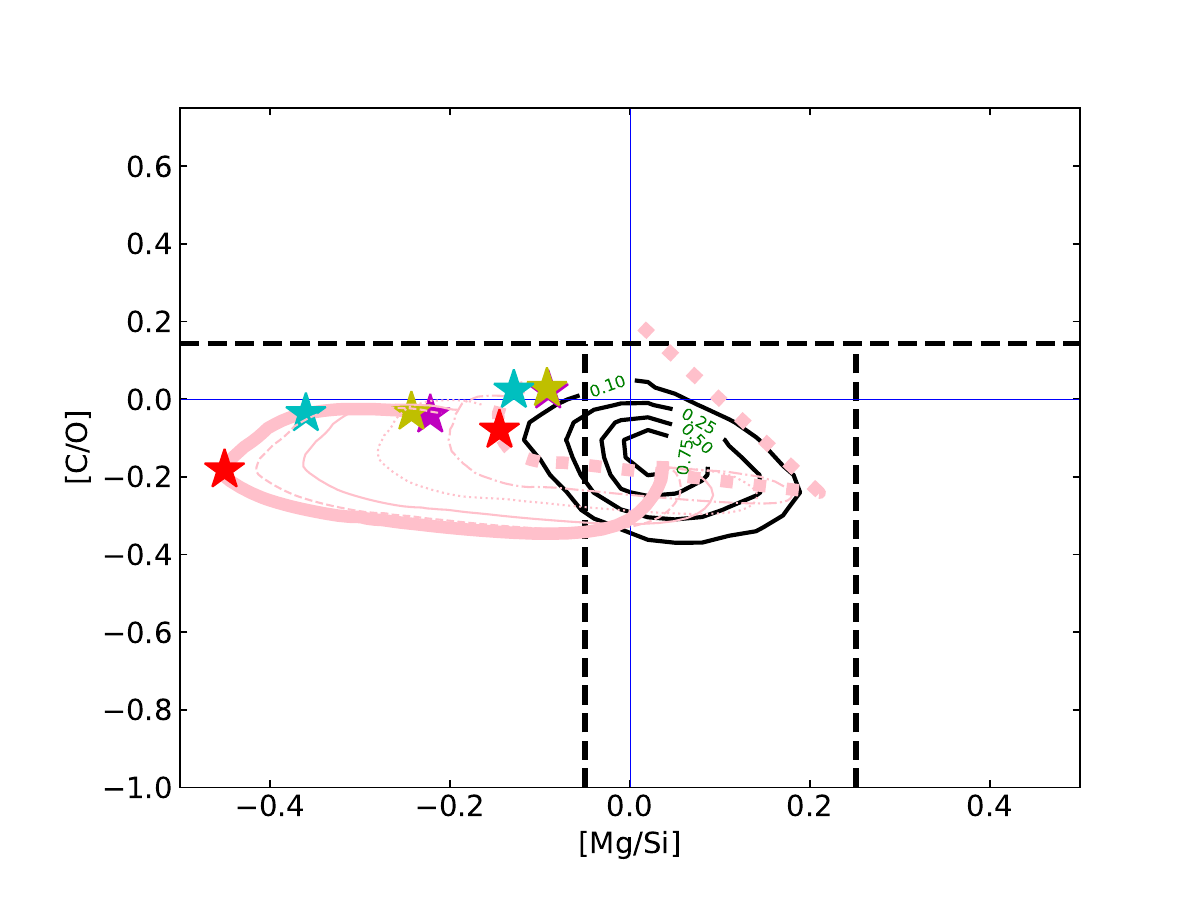}\par
    \end{multicols}
\begin{multicols}{2}
    \includegraphics[width=0.8\linewidth]{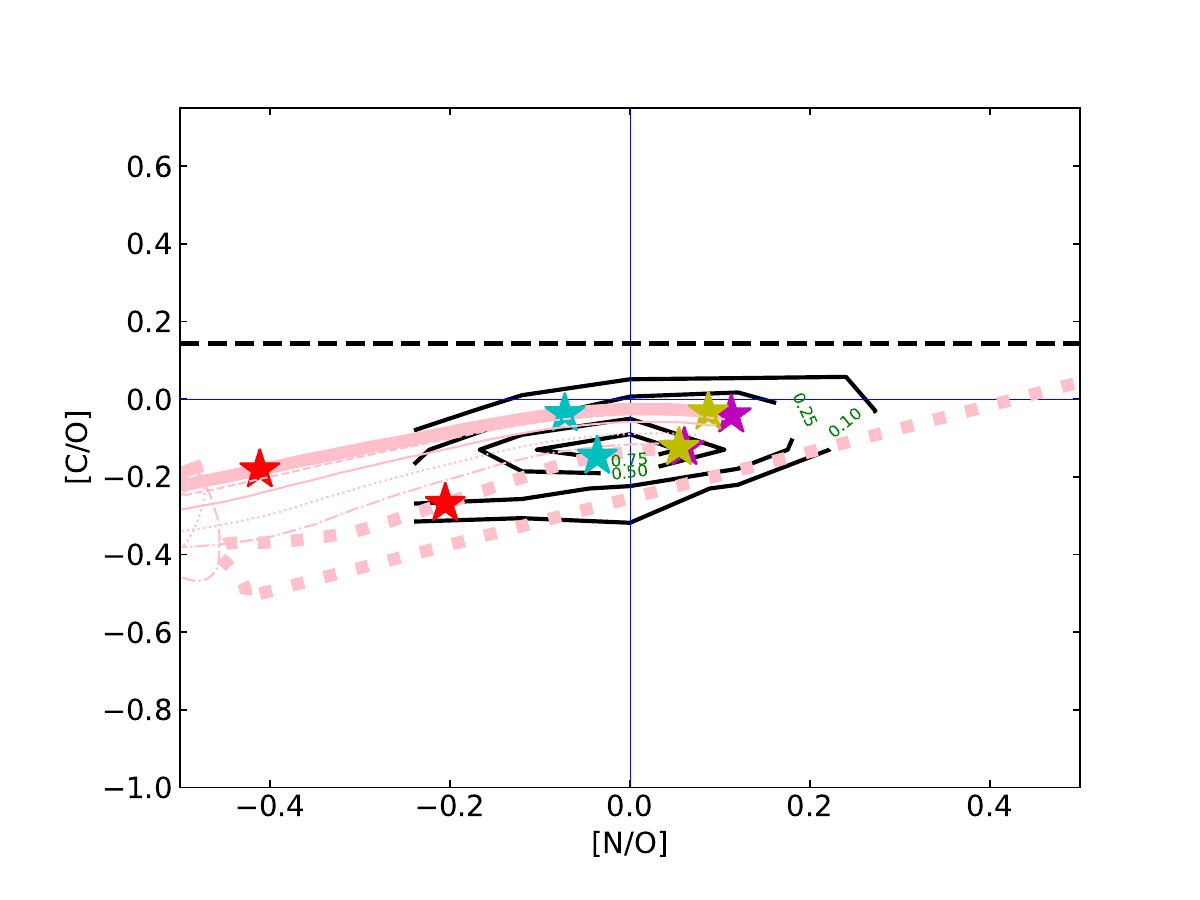}\par
    \includegraphics[width=0.8\linewidth]{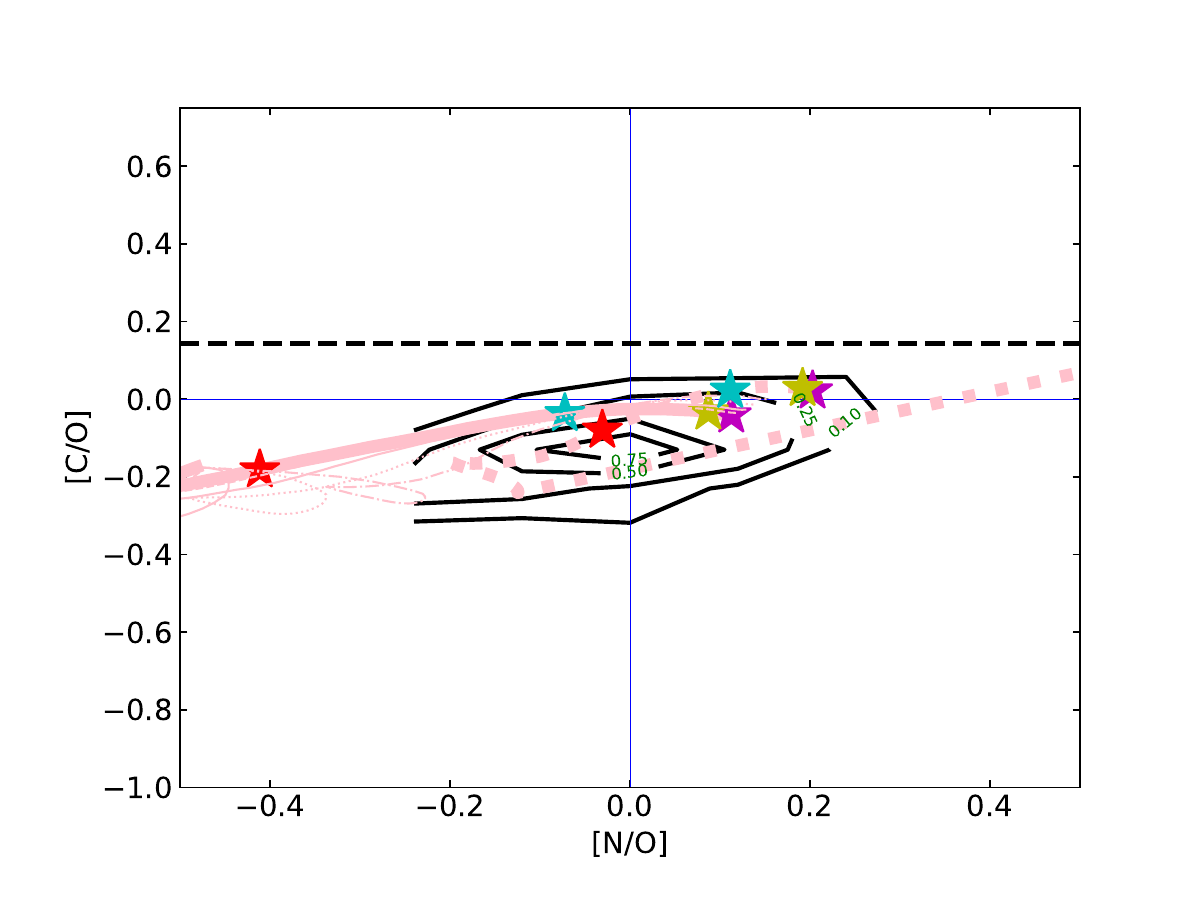}\par
\end{multicols}
\begin{multicols}{2}
    \includegraphics[width=0.8\linewidth]{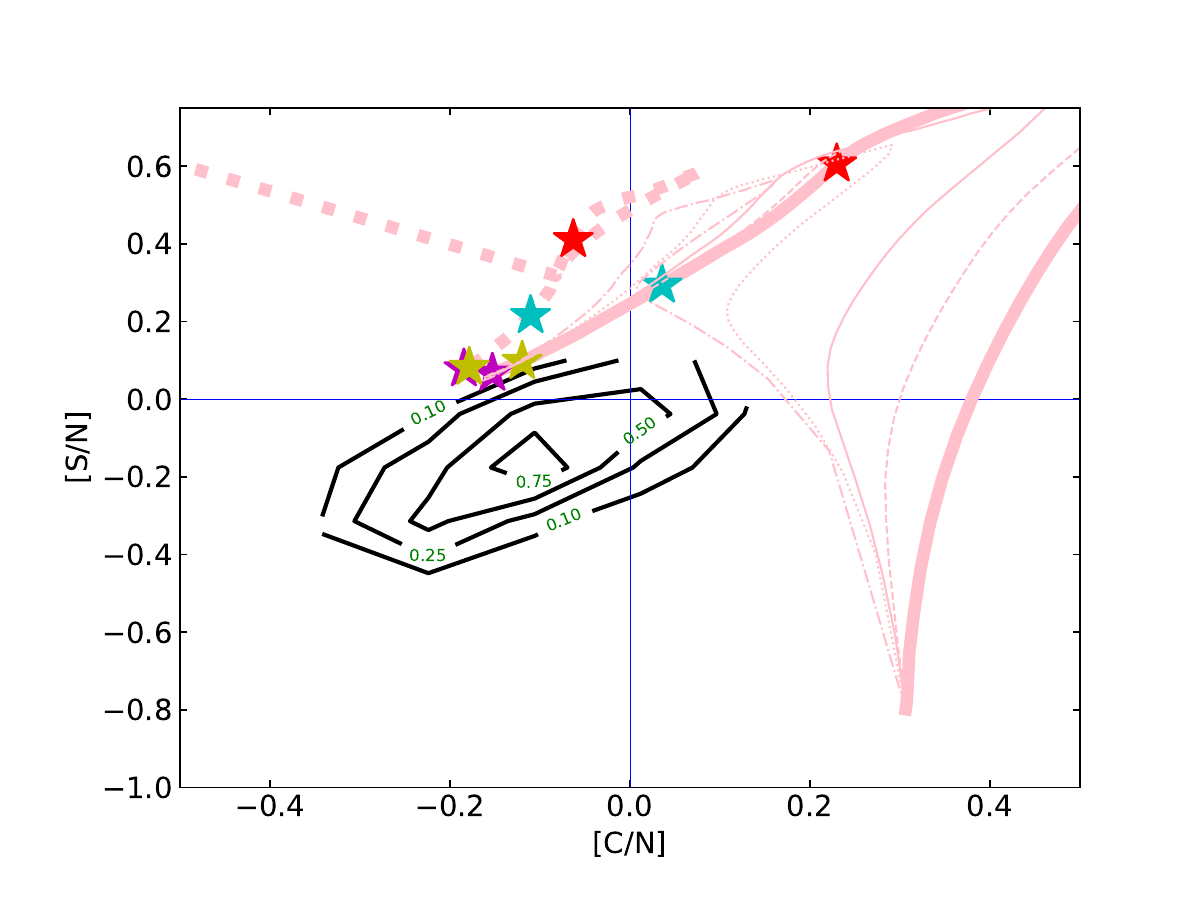}\par
    \includegraphics[width=0.8\linewidth]{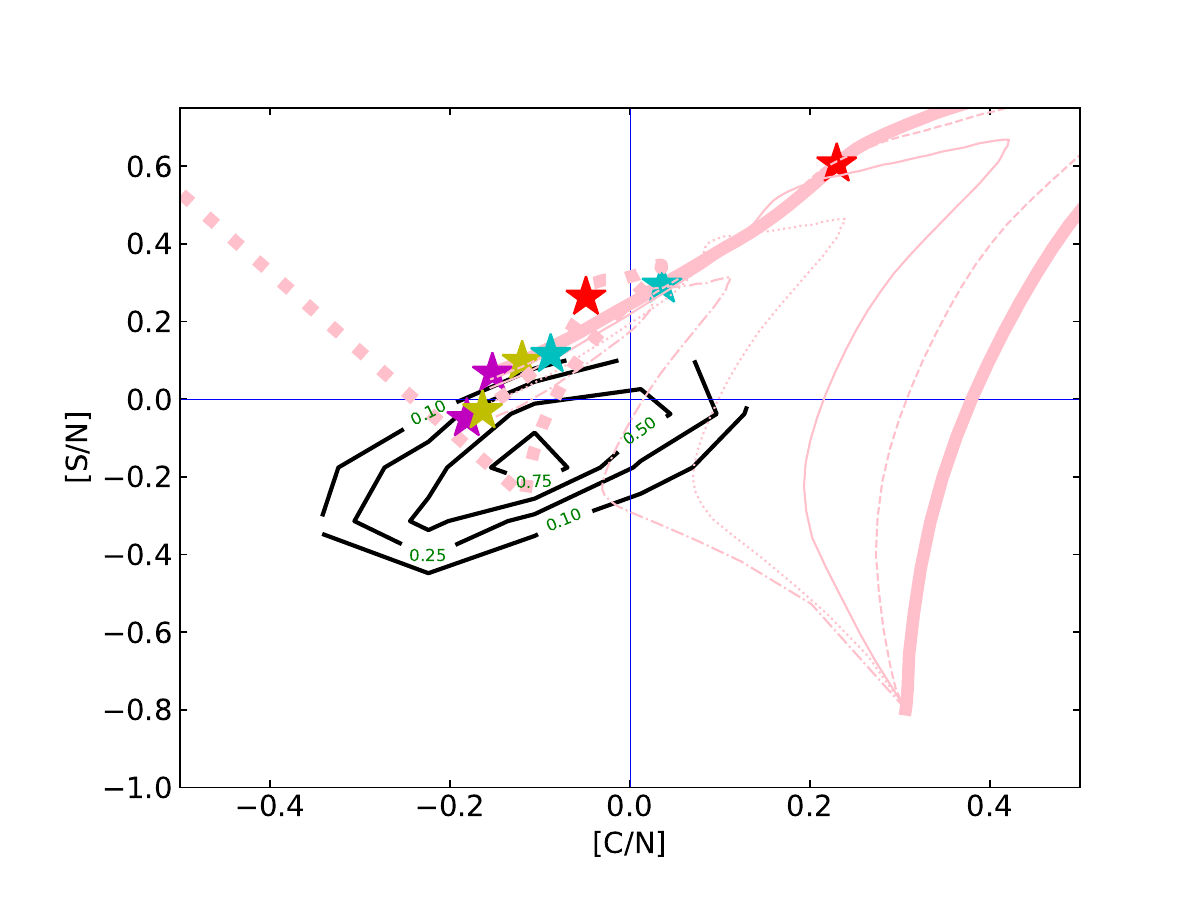}\par
\end{multicols}
\begin{multicols}{2}
    \includegraphics[width=0.8\linewidth]{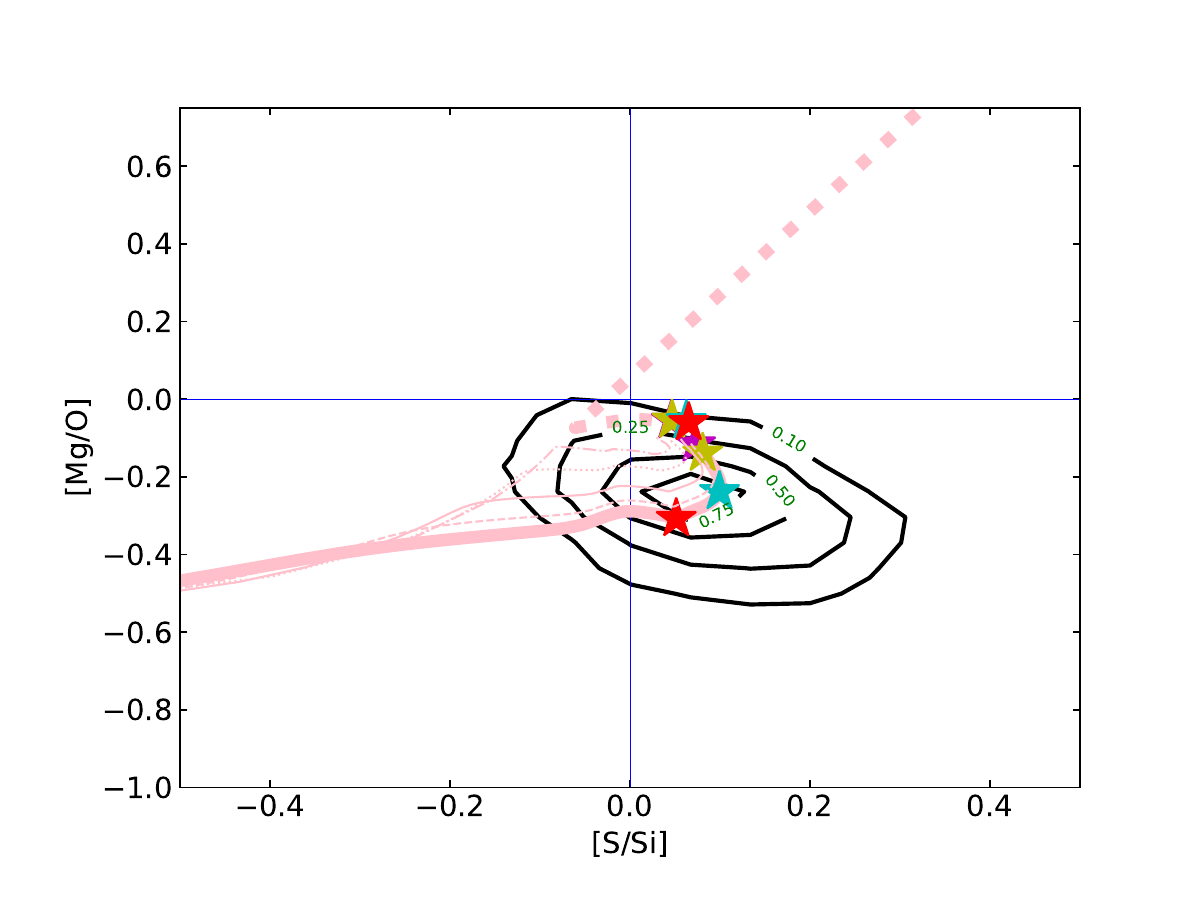}\par
    \includegraphics[width=0.8\linewidth]{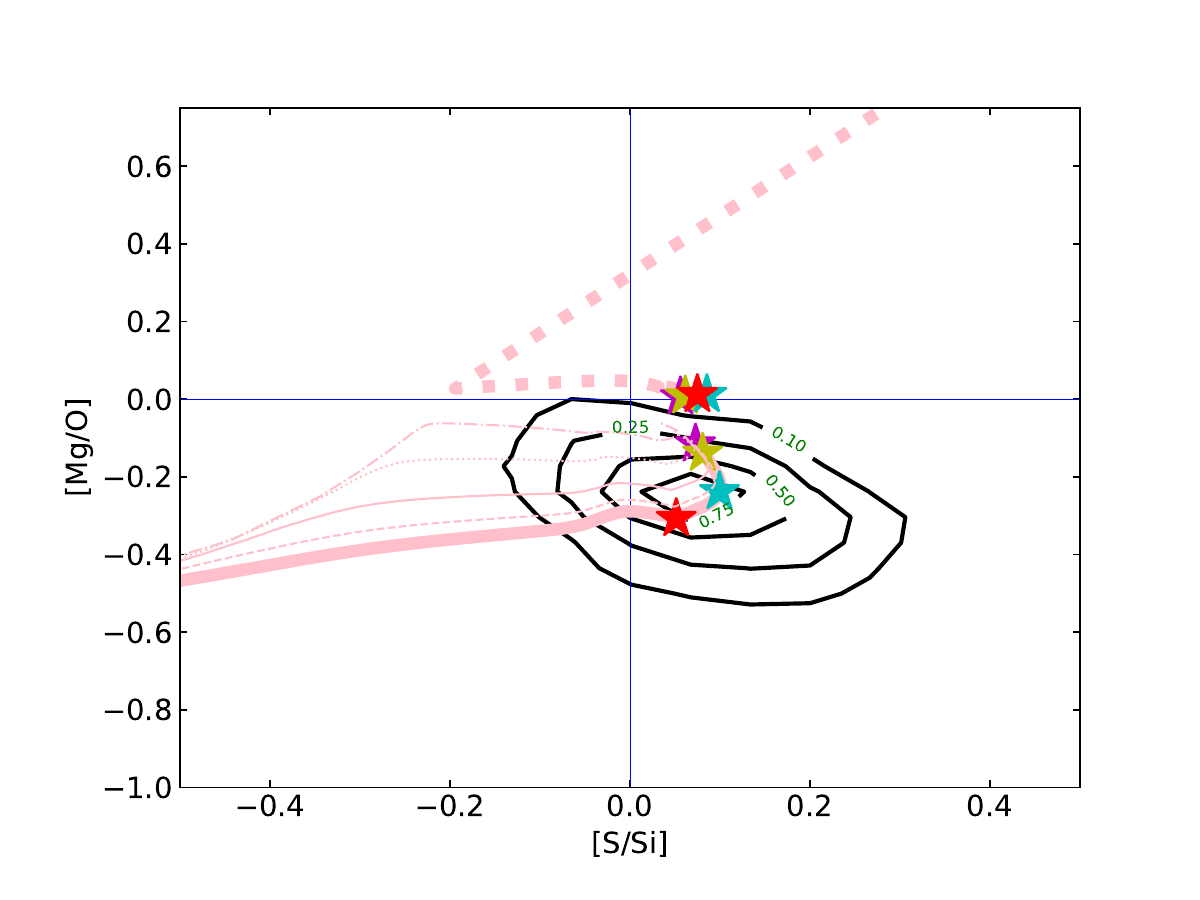}\par
\end{multicols}
    \caption{As in figure~\ref{fig: tk_plots/ratios_oR18d_m40}, but models are shown with CCSN supernovae contribution up to M$_{\rm up}$ = 100 M$_{\odot}$.
    }
    \label{fig: tk_plots/ratios_oR18d_m100}
\end{figure*}

\begin{figure*}
\begin{multicols}{2}
    \includegraphics[width=0.8\linewidth]{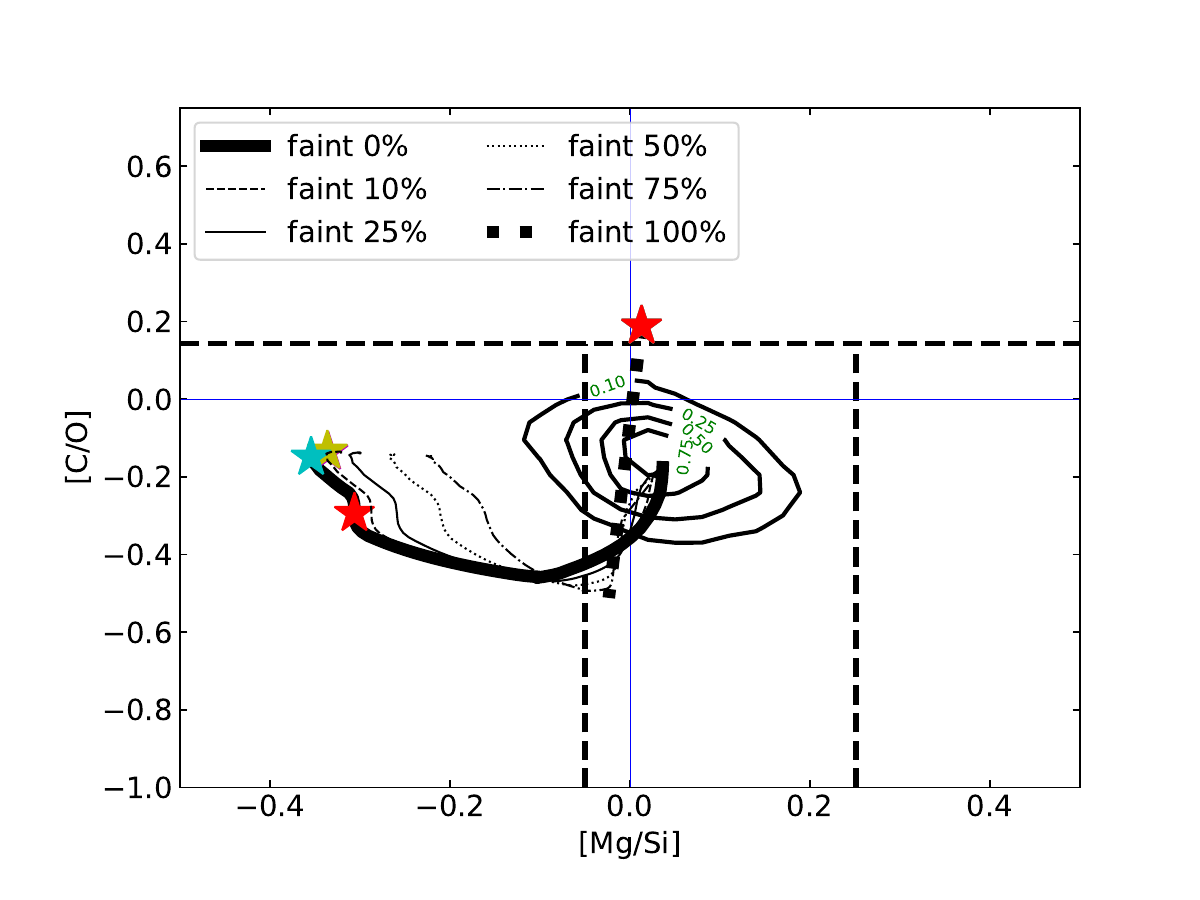}\par
    \includegraphics[width=0.8\linewidth]{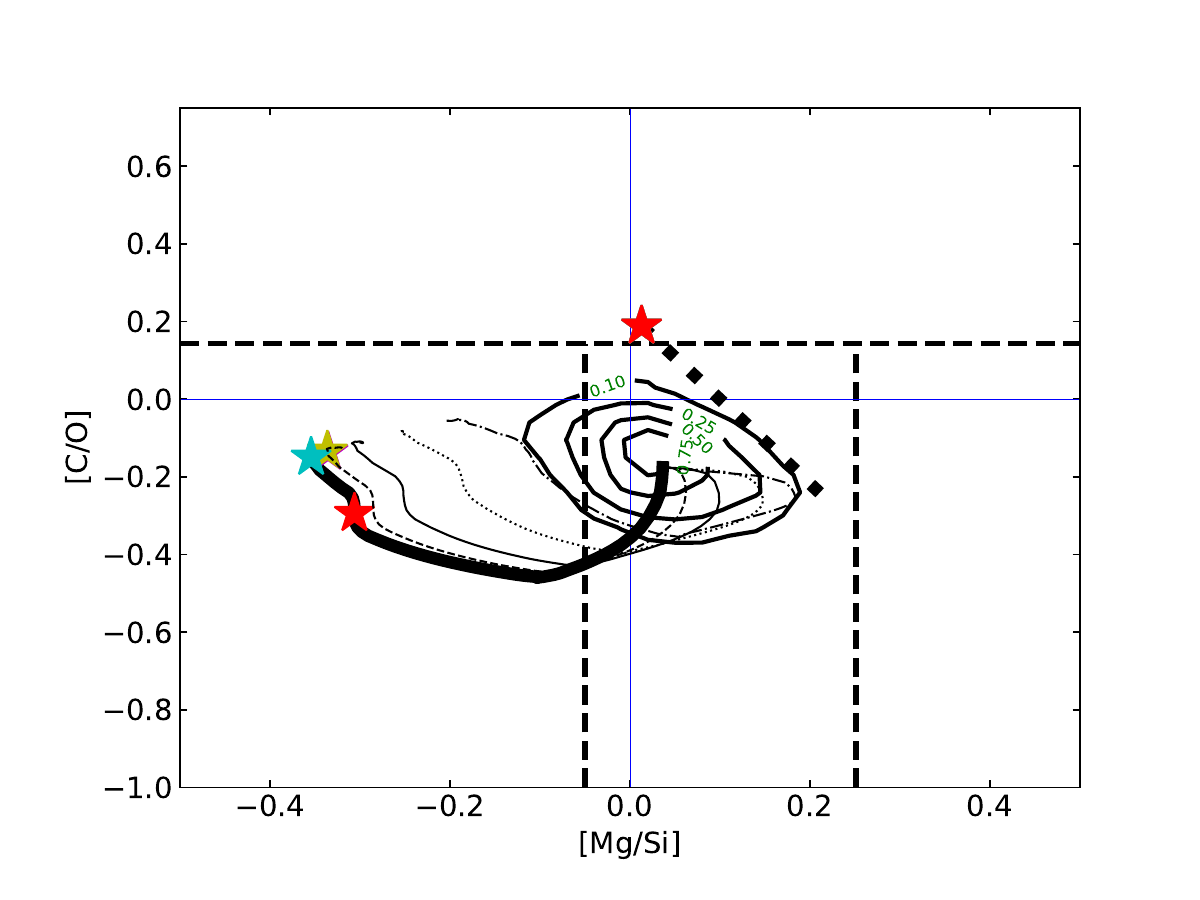}\par
    \end{multicols}
\begin{multicols}{2}
    \includegraphics[width=0.8\linewidth]{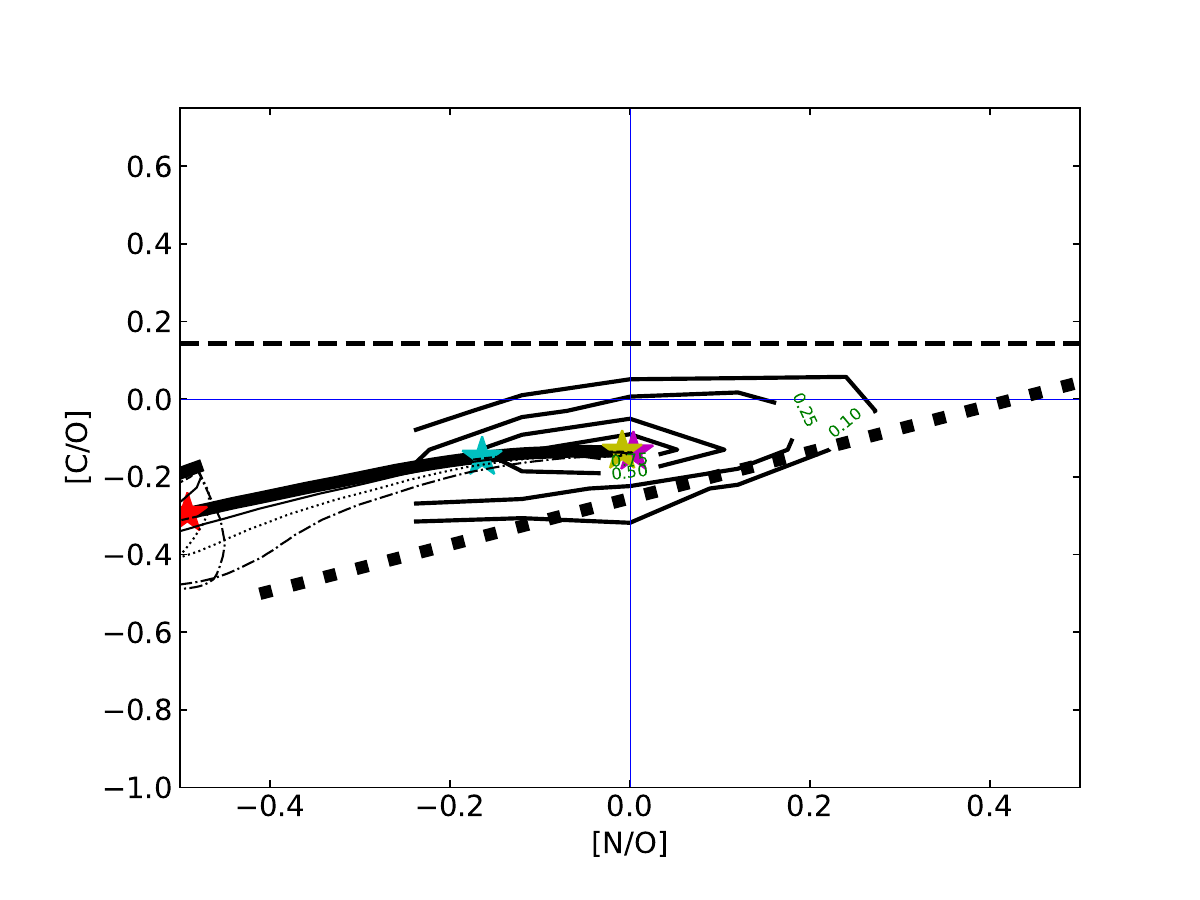}\par
    \includegraphics[width=0.8\linewidth]{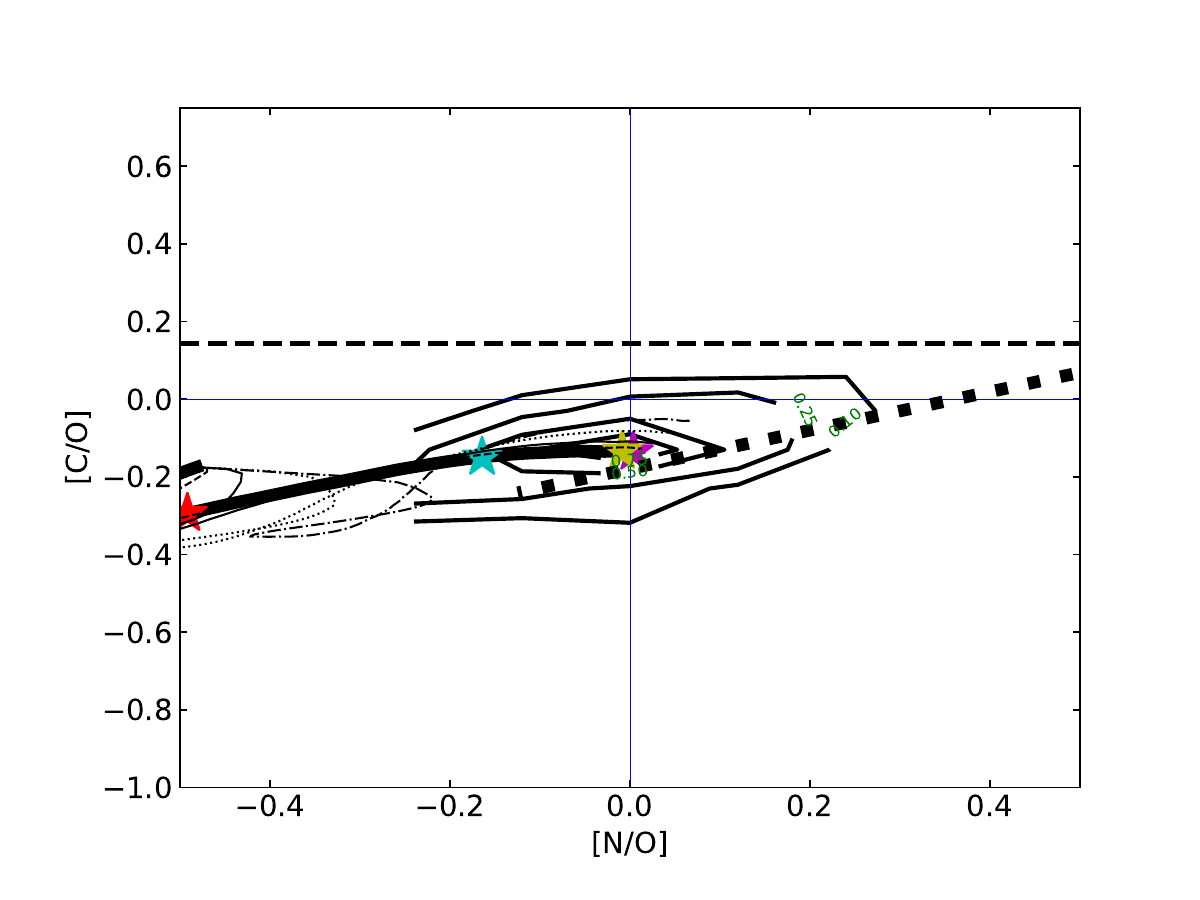}\par
\end{multicols}
\begin{multicols}{2}
    \includegraphics[width=0.8\linewidth]{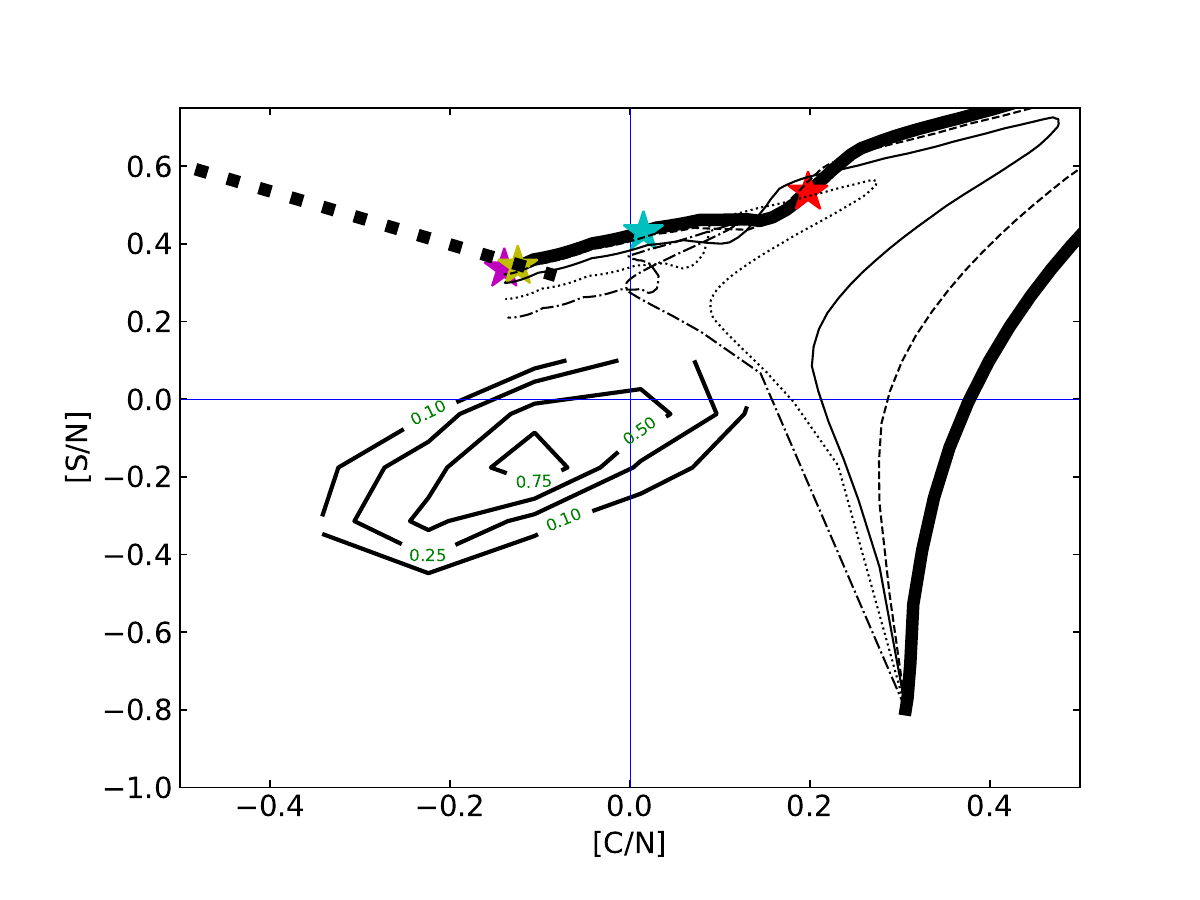}\par
    \includegraphics[width=0.8\linewidth]{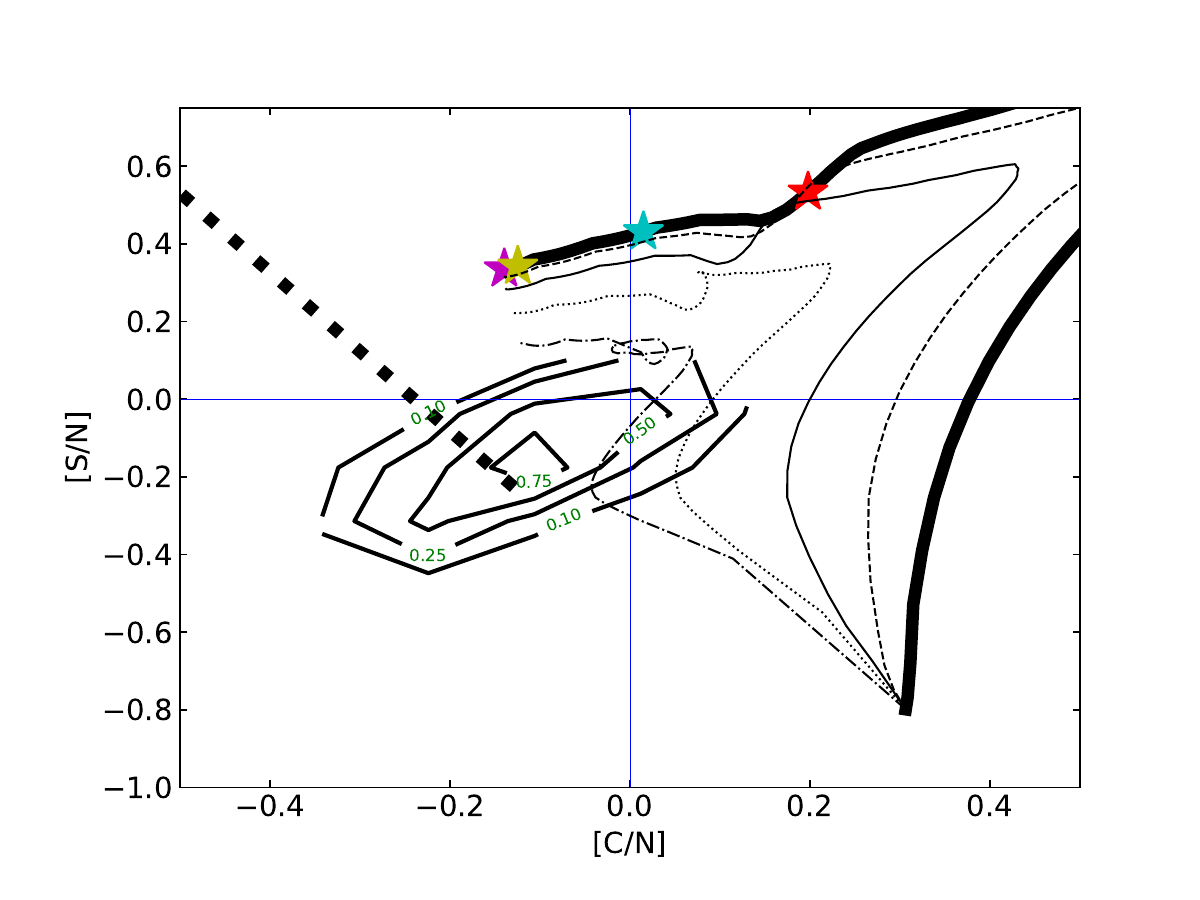}\par
\end{multicols}
\begin{multicols}{2}
    \includegraphics[width=0.8\linewidth]{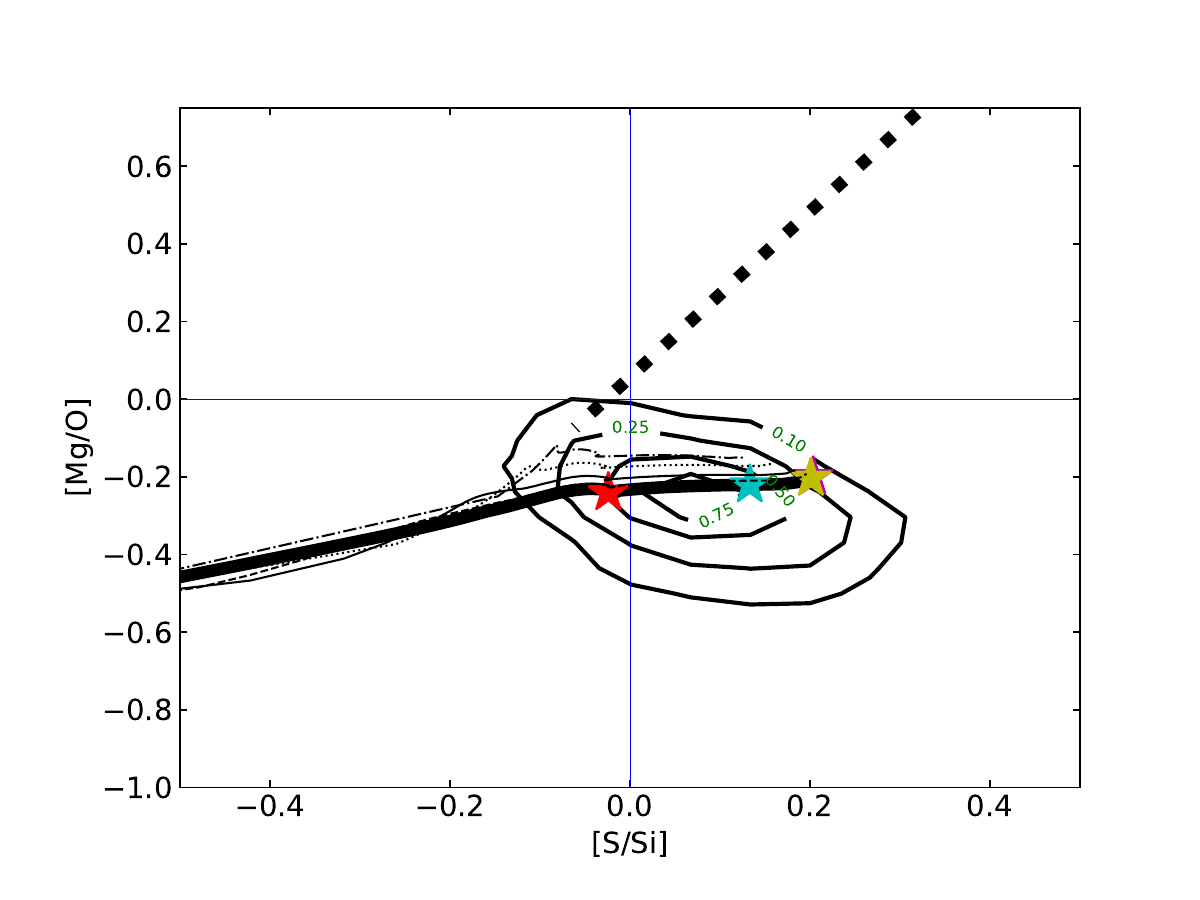}\par
    \includegraphics[width=0.8\linewidth]{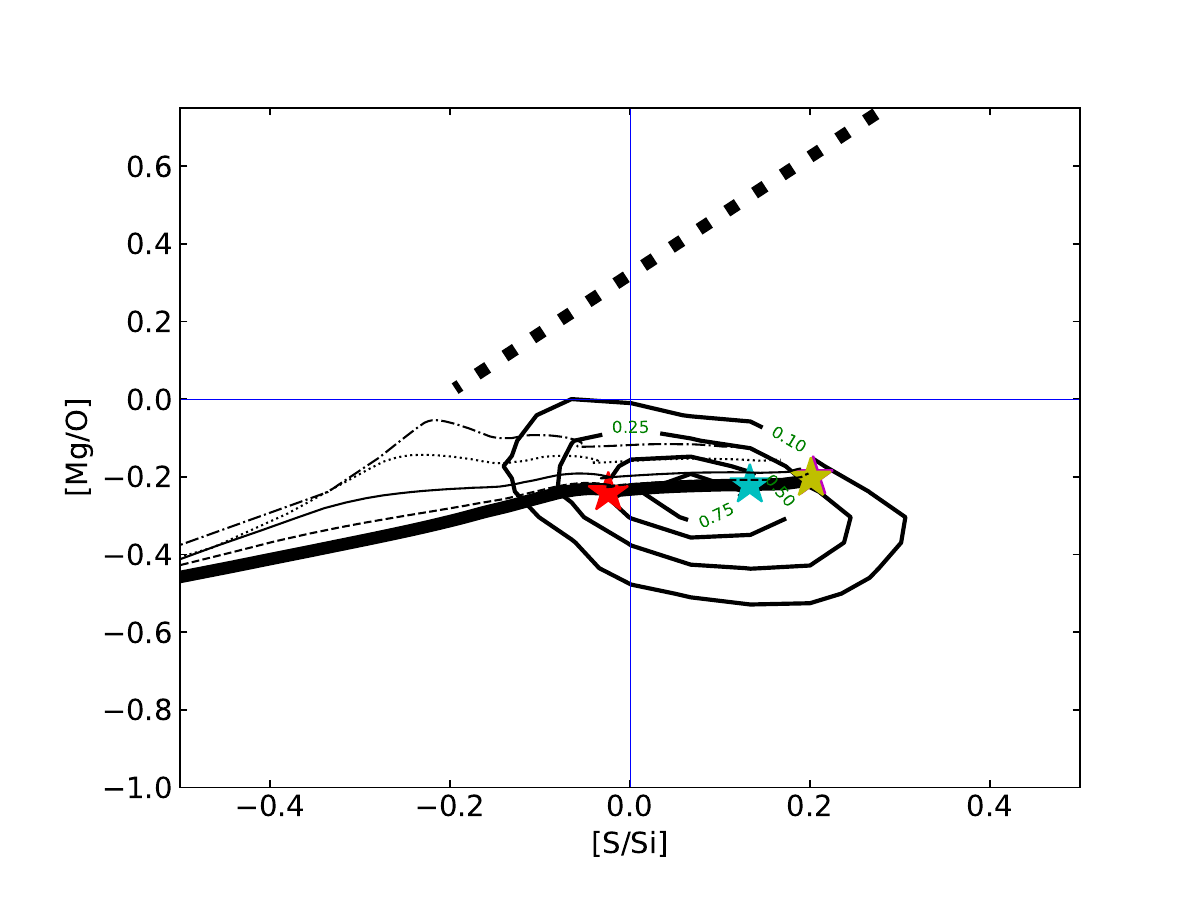}\par
\end{multicols}
    \caption{Same as in Figure~\ref{fig: tk_plots/ratios_gce_oK10m40}, but for the GCE model set oR18h.
    }
    \label{fig: tk_plots/ratios_oR18h_m40}
\end{figure*}

\begin{figure*}
\begin{multicols}{2}
    \includegraphics[width=0.8\linewidth]{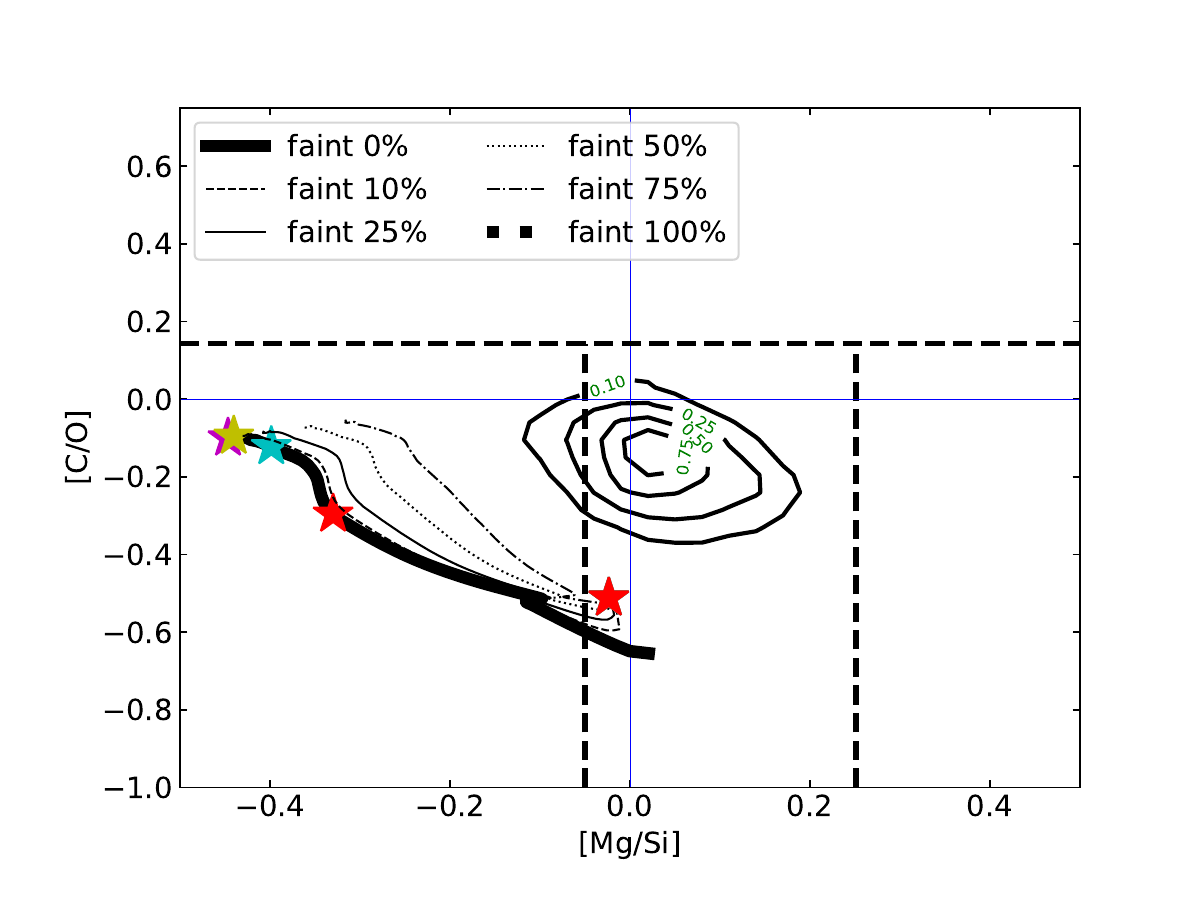}\par
    \includegraphics[width=0.8\linewidth]{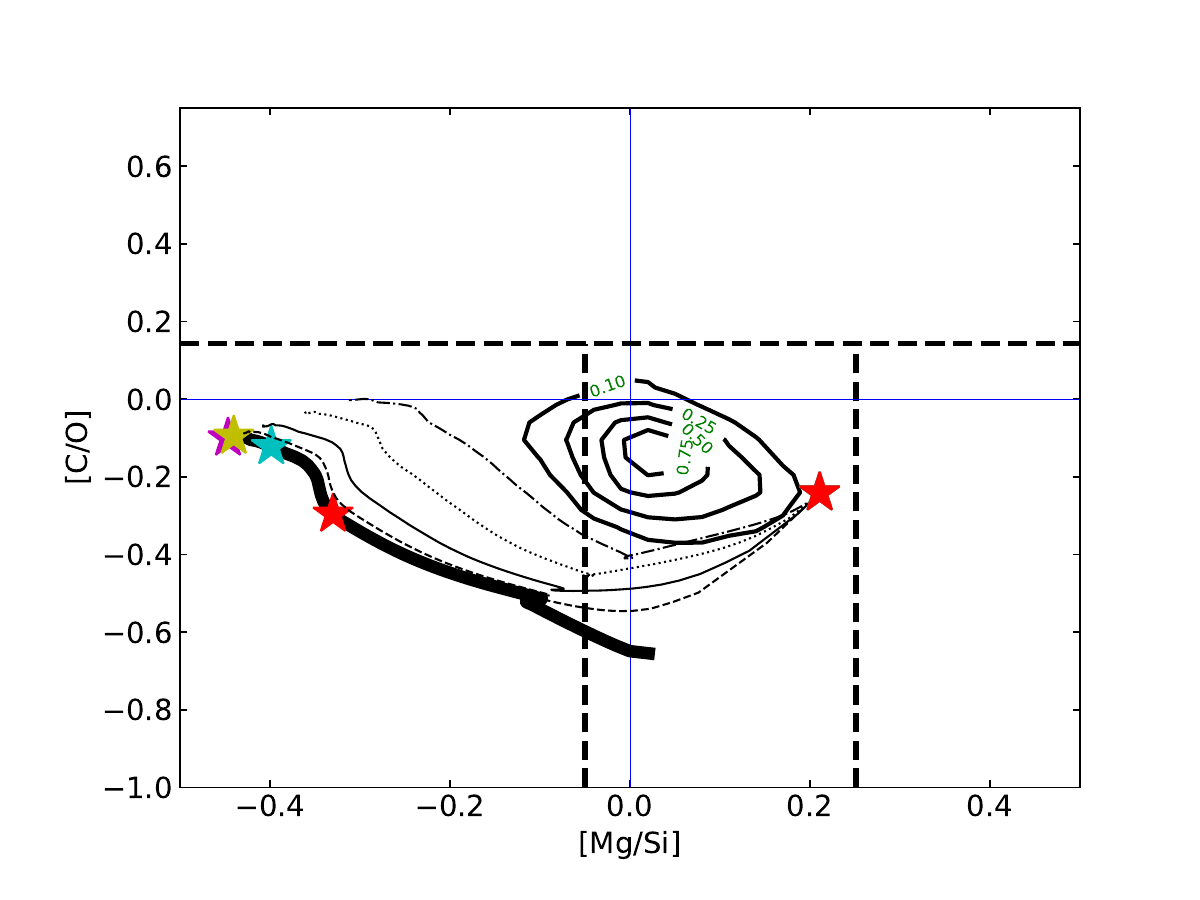}\par
    \end{multicols}
\begin{multicols}{2}
    \includegraphics[width=0.8\linewidth]{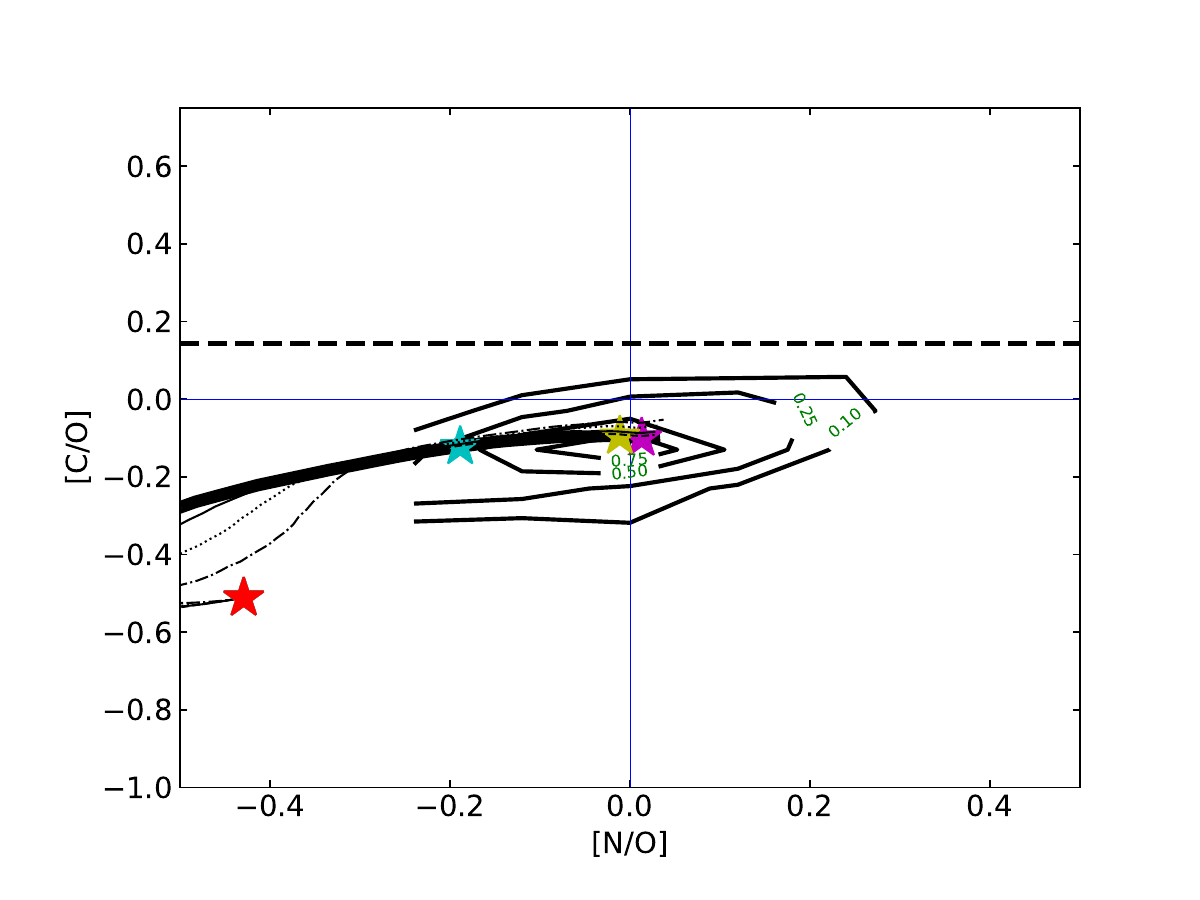}\par
    \includegraphics[width=0.8\linewidth]{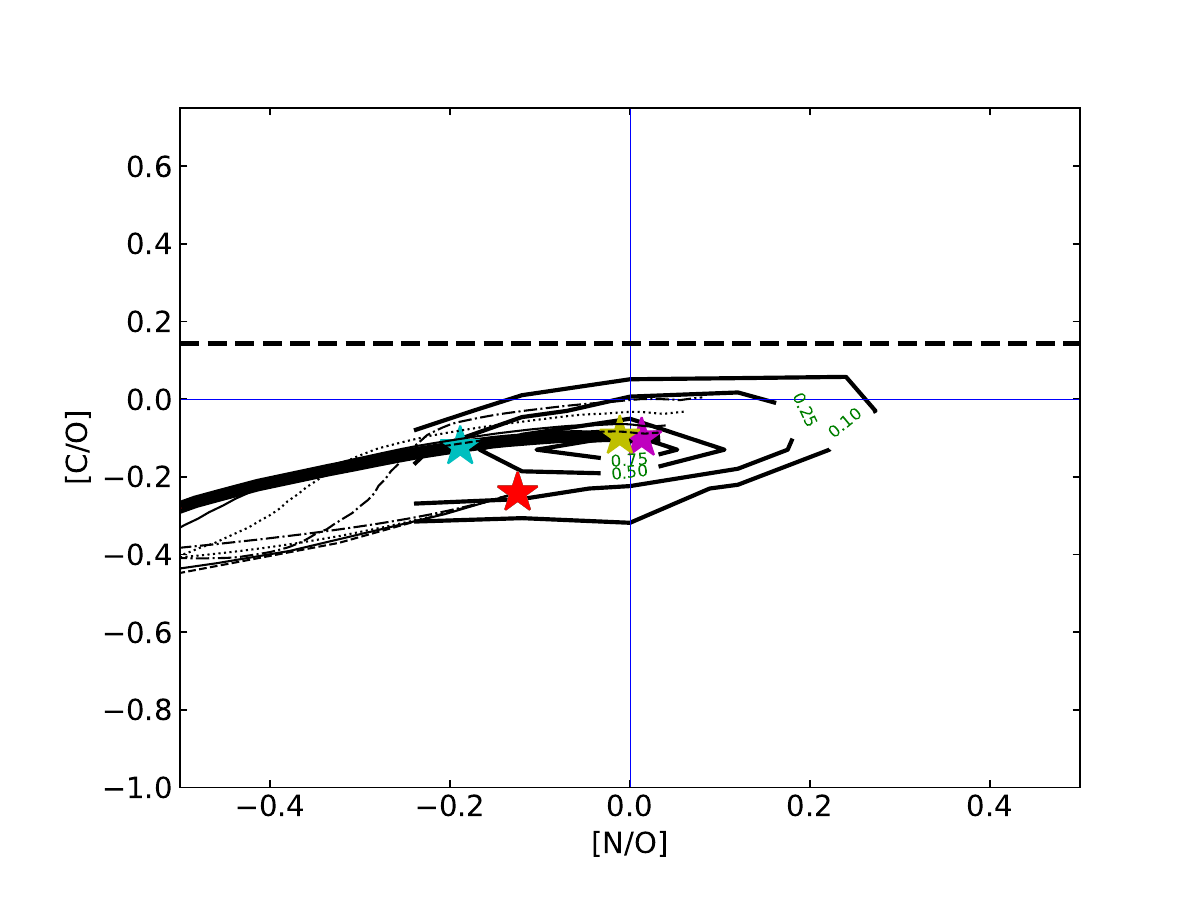}\par
\end{multicols}
\begin{multicols}{2}
    \includegraphics[width=0.8\linewidth]{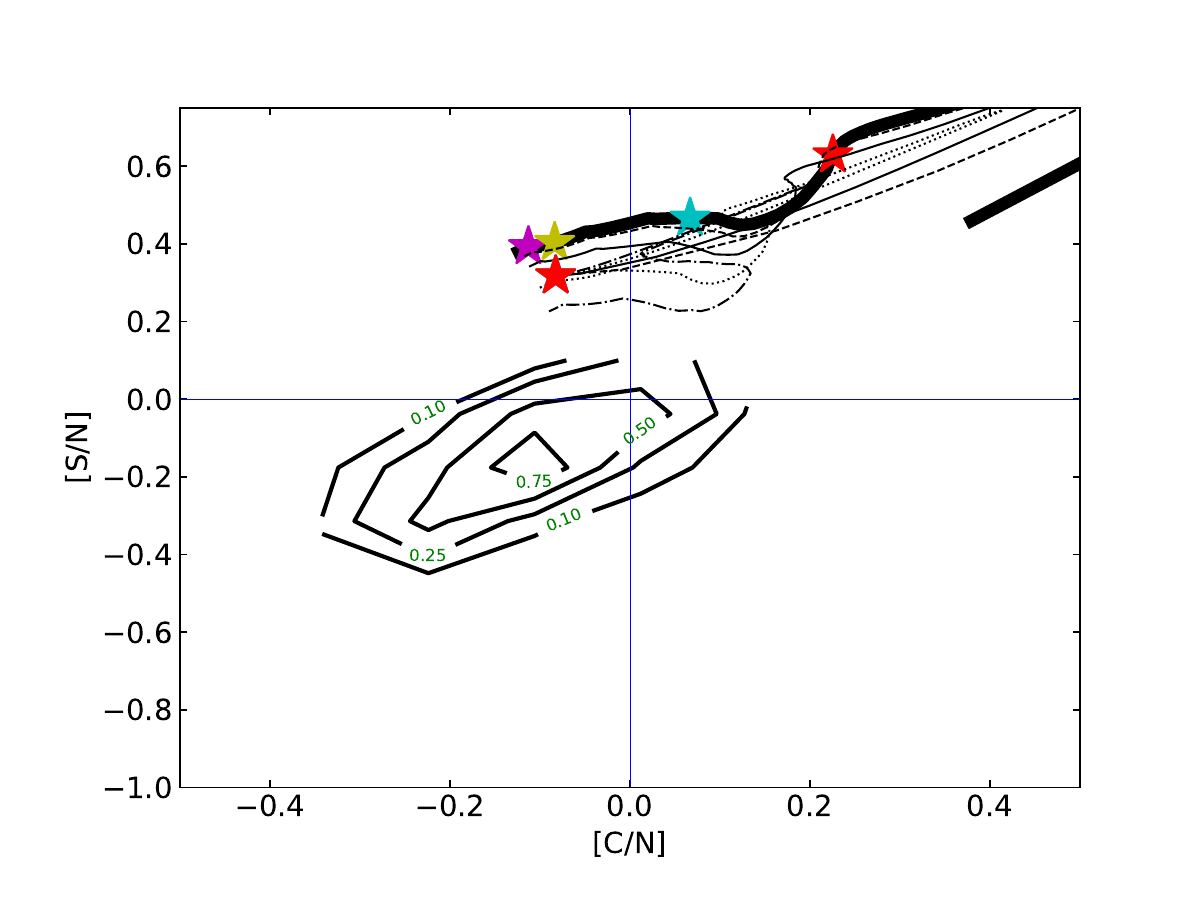}\par
    \includegraphics[width=0.8\linewidth]{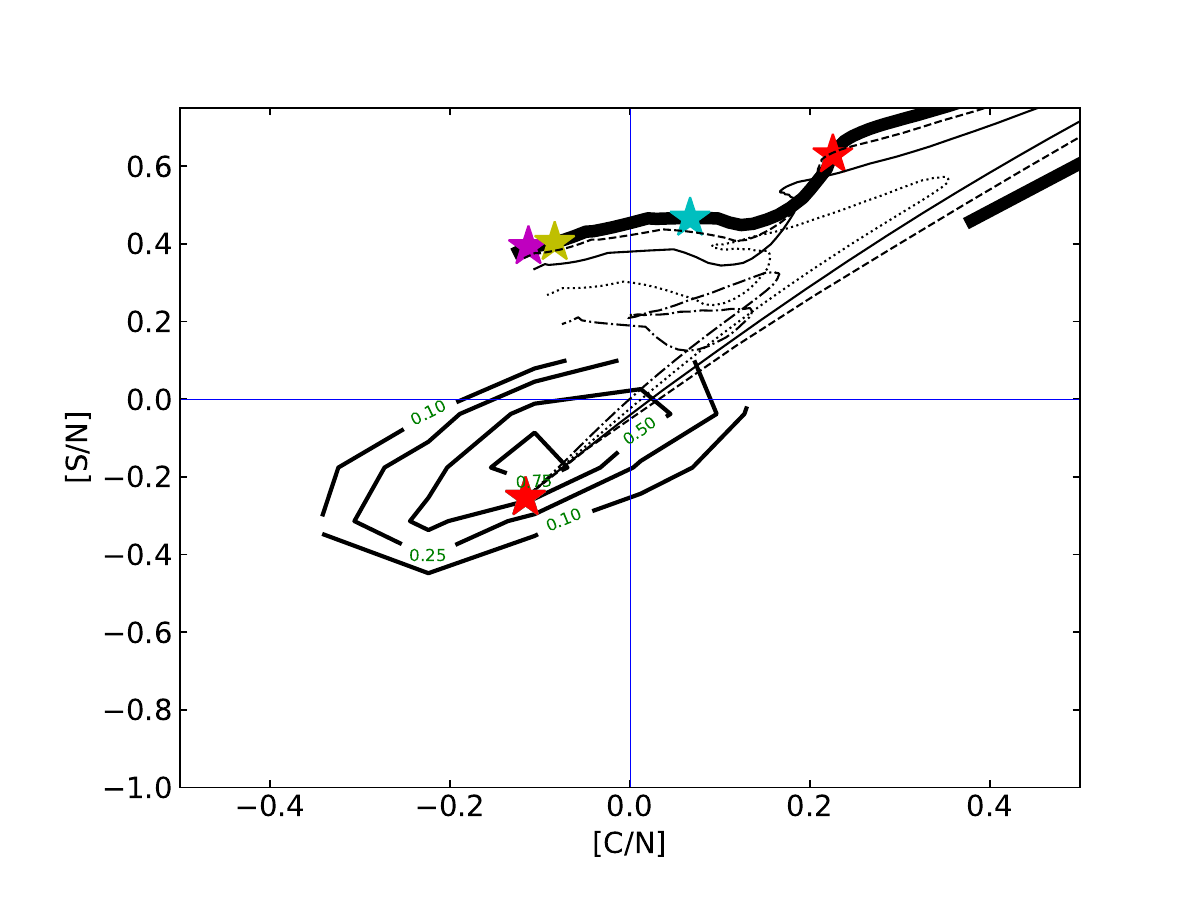}\par
\end{multicols}
\begin{multicols}{2}
    \includegraphics[width=0.8\linewidth]{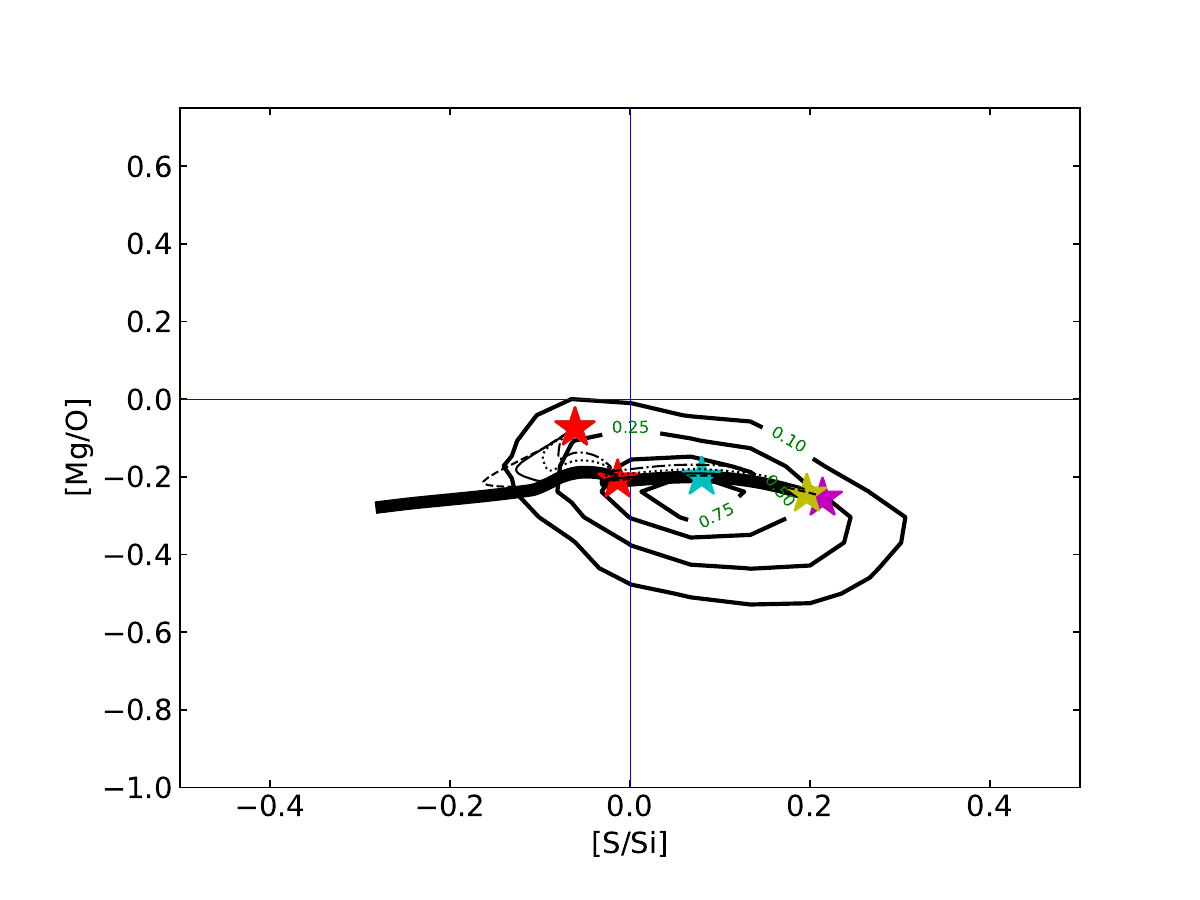}\par
    \includegraphics[width=0.8\linewidth]{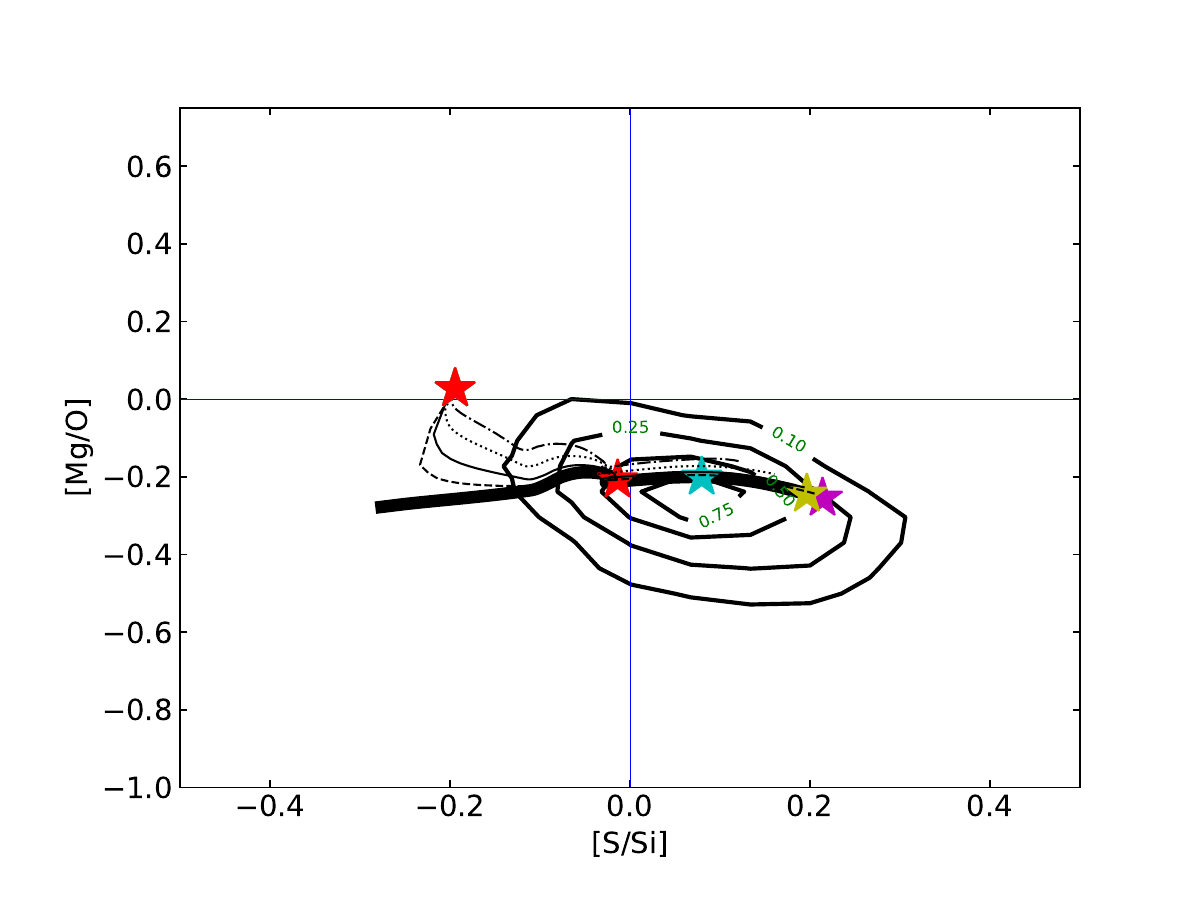}\par
\end{multicols}
    \caption{As in figure~\ref{fig: tk_plots/ratios_oR18h_m40}, but models are shown with CCSN supernovae contribution up to M=20M$_{\odot}$.
    }
    \label{fig: tk_plots/ratios_oR18h_m20}
\end{figure*}

\begin{figure*}
\begin{multicols}{2}
    \includegraphics[width=0.8\linewidth]{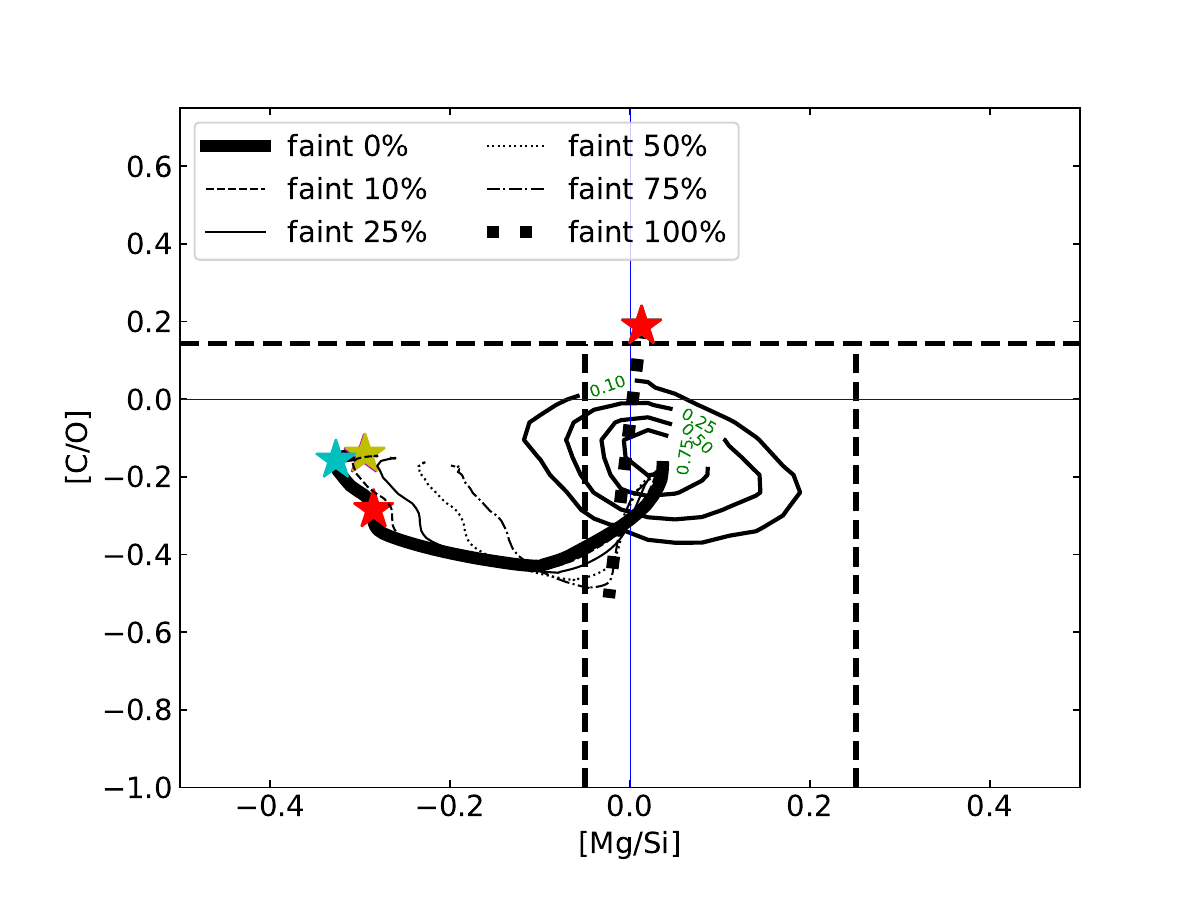}\par
    \includegraphics[width=0.8\linewidth]{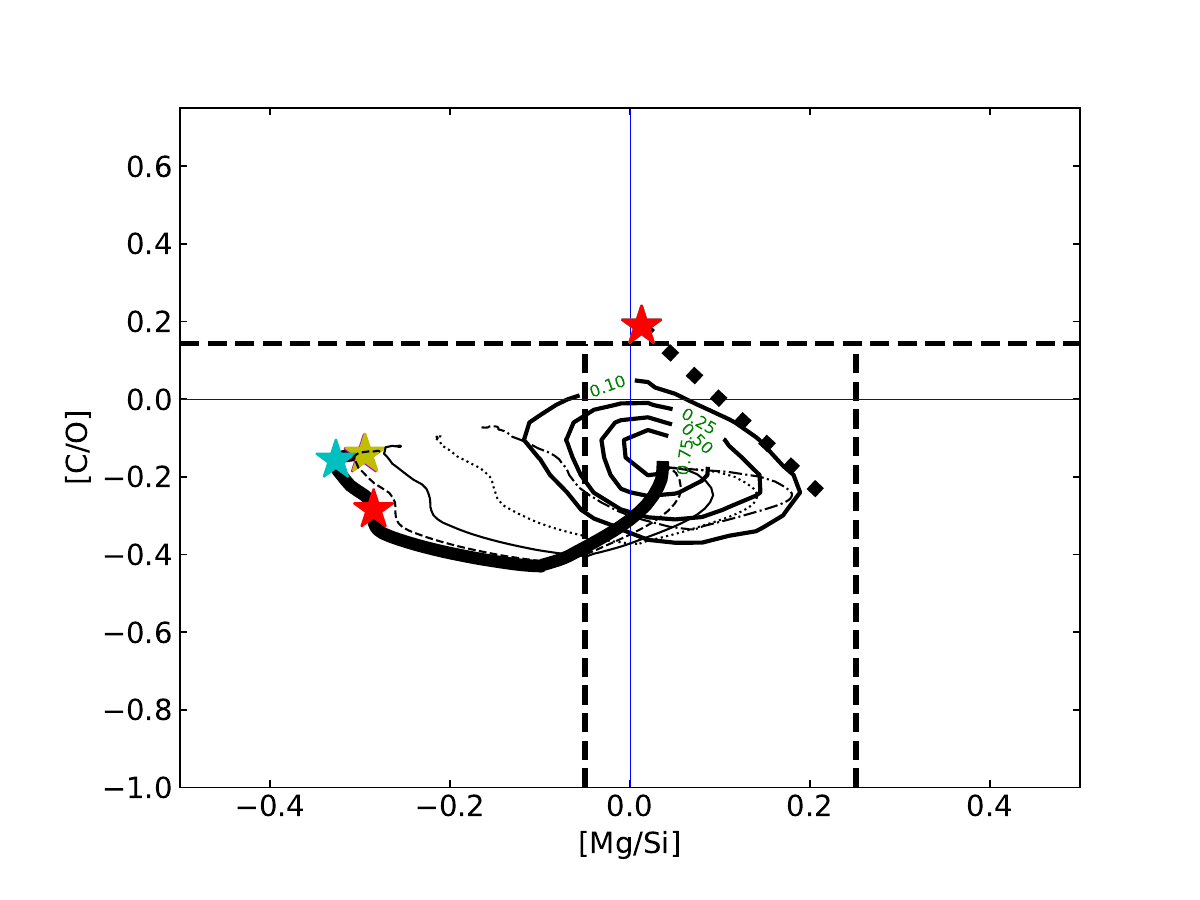}\par
    \end{multicols}
\begin{multicols}{2}
    \includegraphics[width=0.8\linewidth]{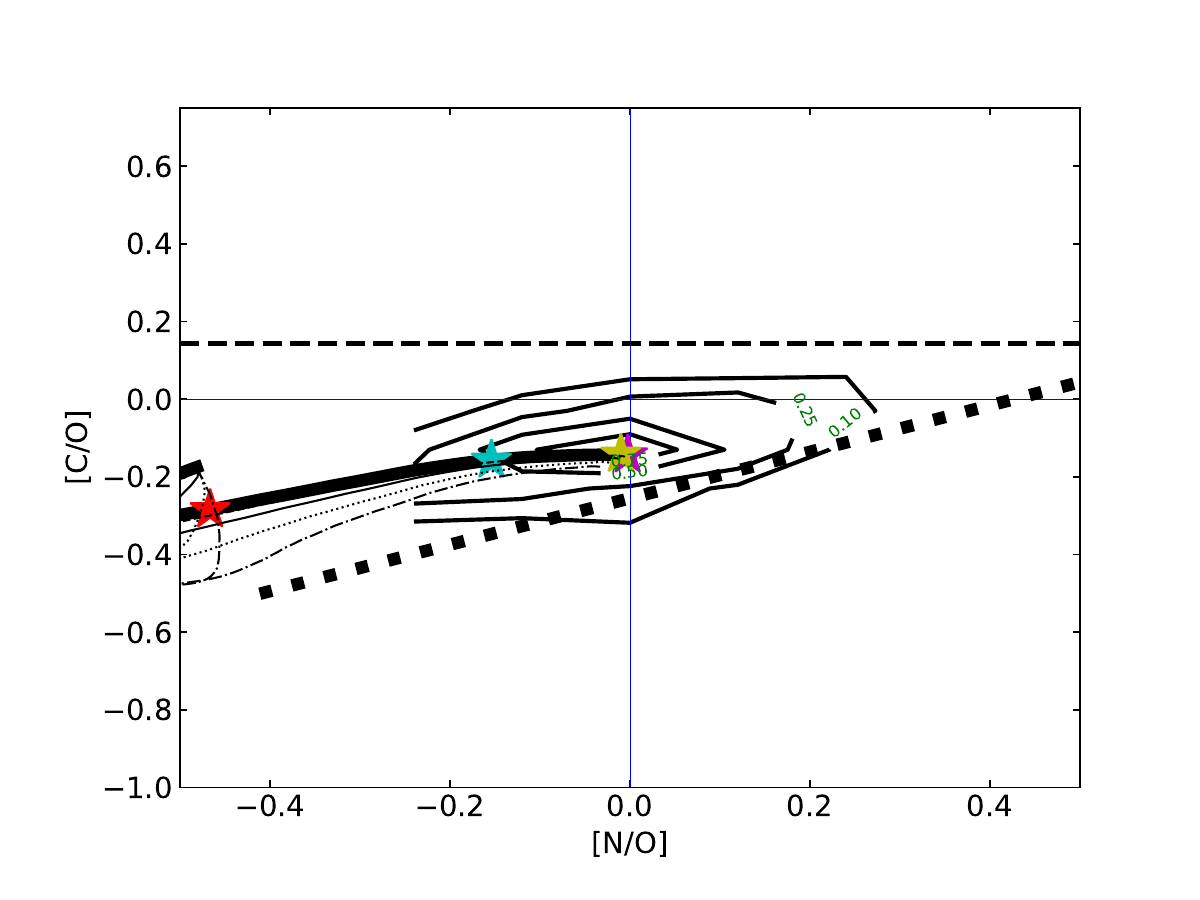}\par
    \includegraphics[width=0.8\linewidth]{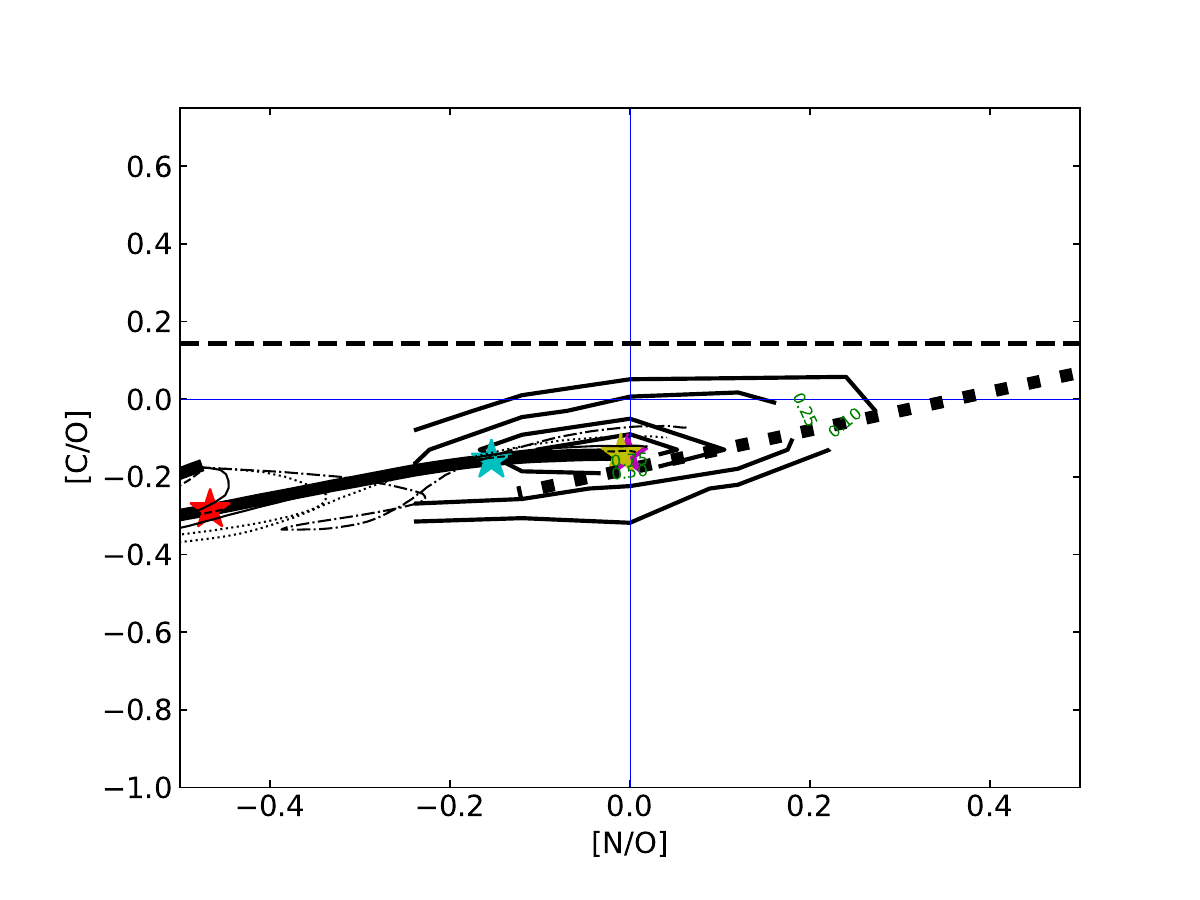}\par
\end{multicols}
\begin{multicols}{2}
    \includegraphics[width=0.8\linewidth]{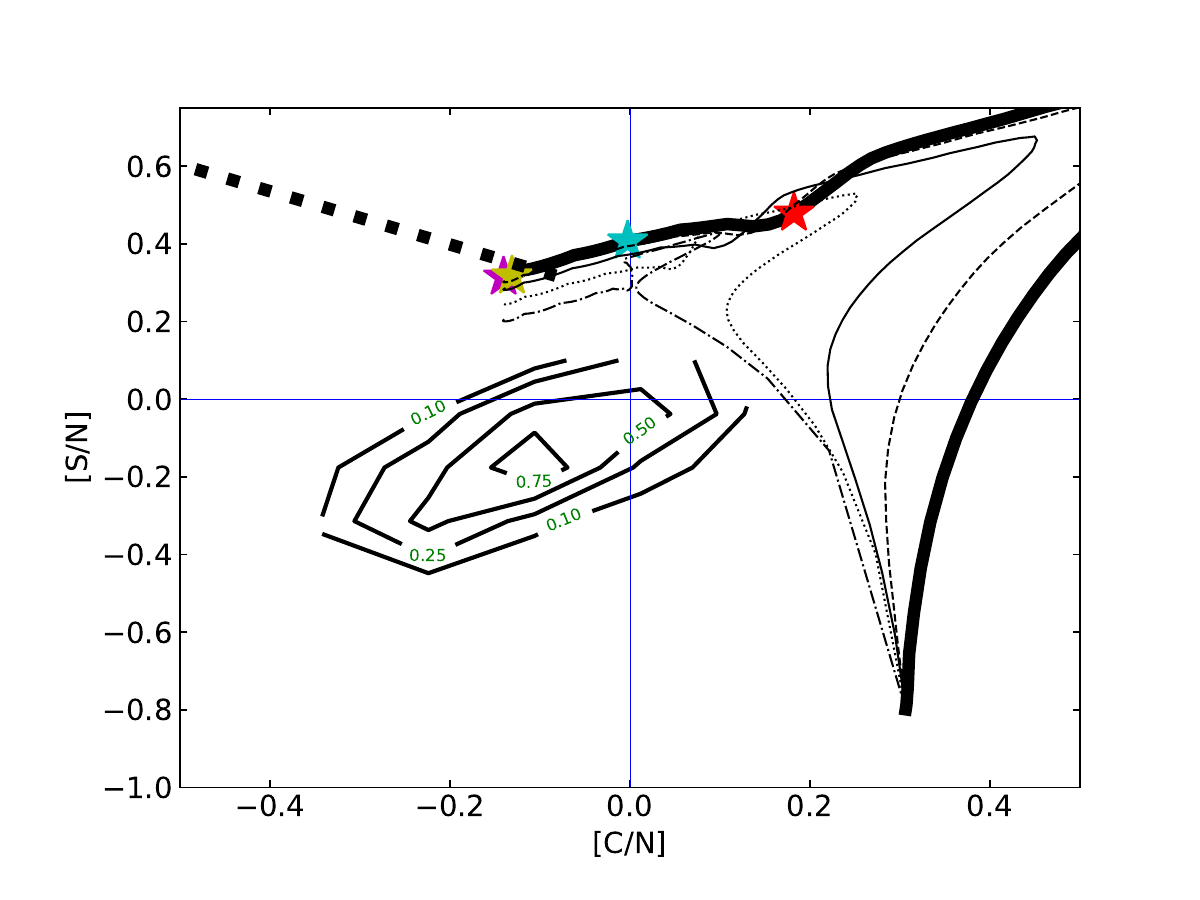}\par
    \includegraphics[width=0.8\linewidth]{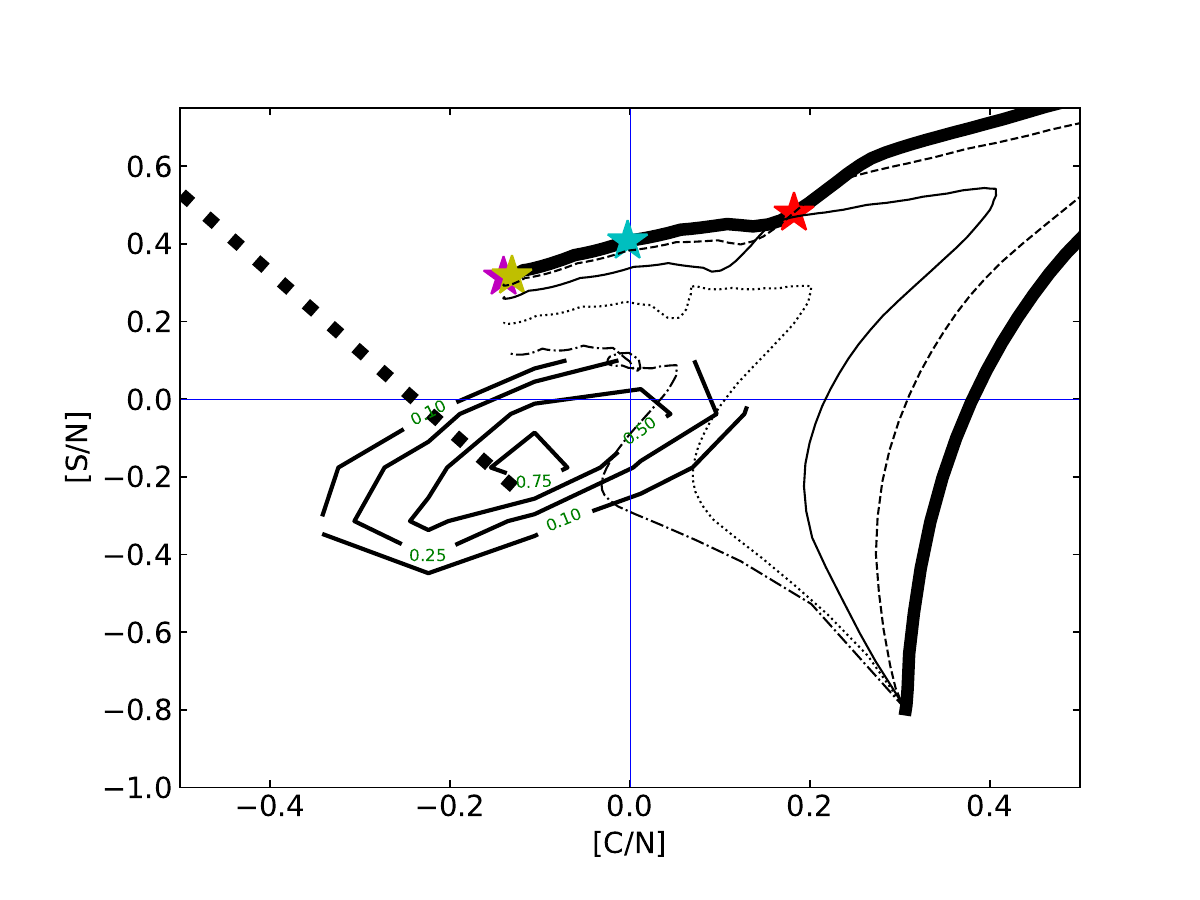}\par
\end{multicols}
\begin{multicols}{2}
    \includegraphics[width=0.8\linewidth]{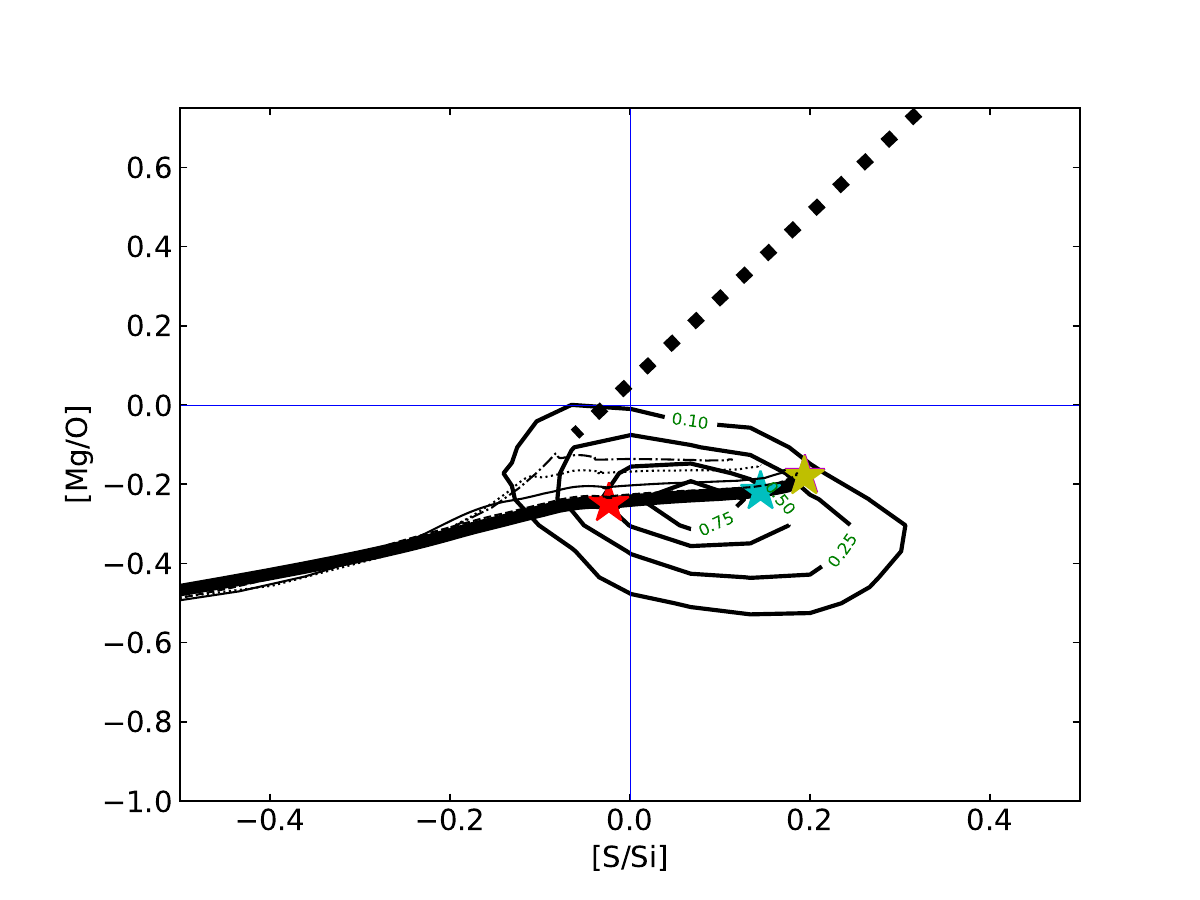}\par
    \includegraphics[width=0.8\linewidth]{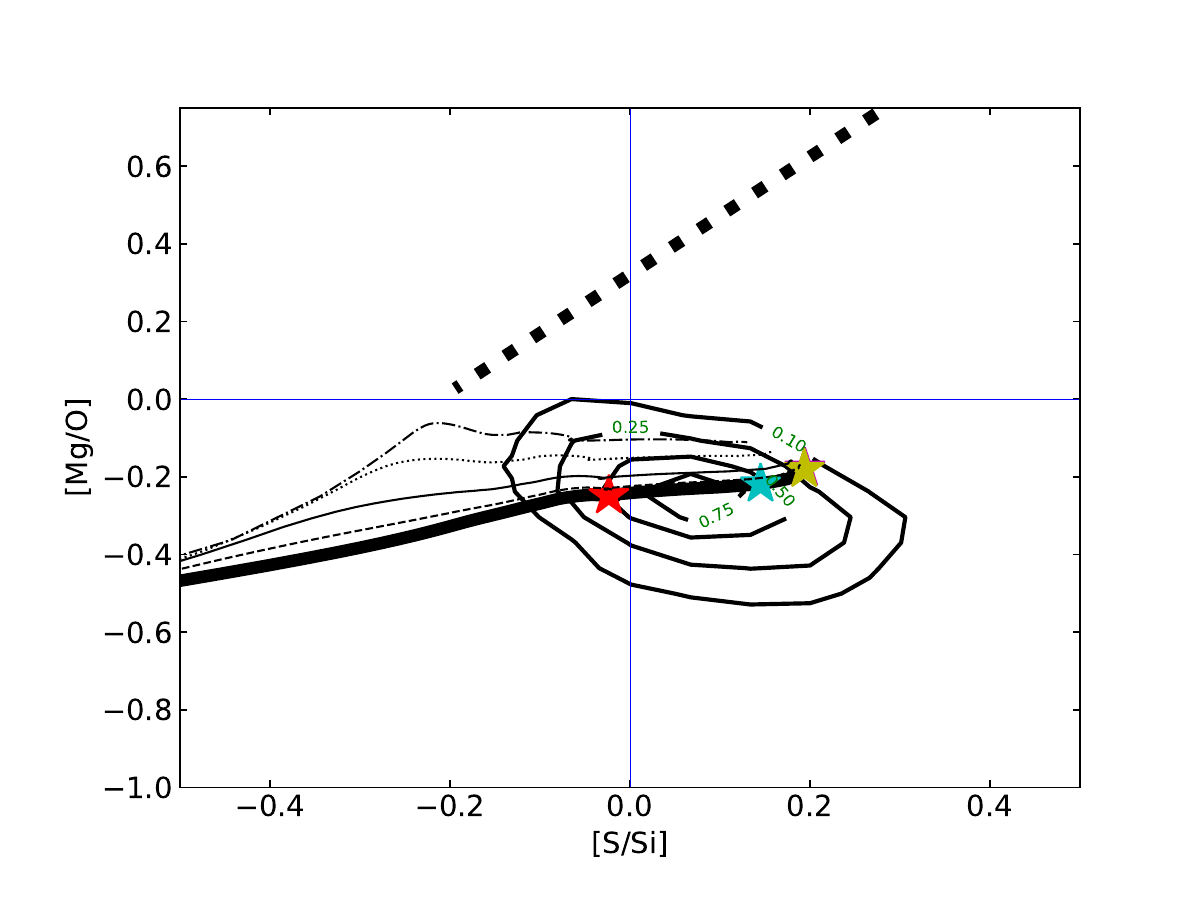}\par
\end{multicols}
    \caption{As in figure~\ref{fig: tk_plots/ratios_oR18h_m40}, but models are shown with CCSN supernovae contribution up to M$_{\rm up}$ = 100 M$_{\odot}$.
    }
    \label{fig: tk_plots/ratios_oR18h_m100}
\end{figure*}

\begin{figure*}
\begin{multicols}{2}
    \includegraphics[width=0.8\linewidth]{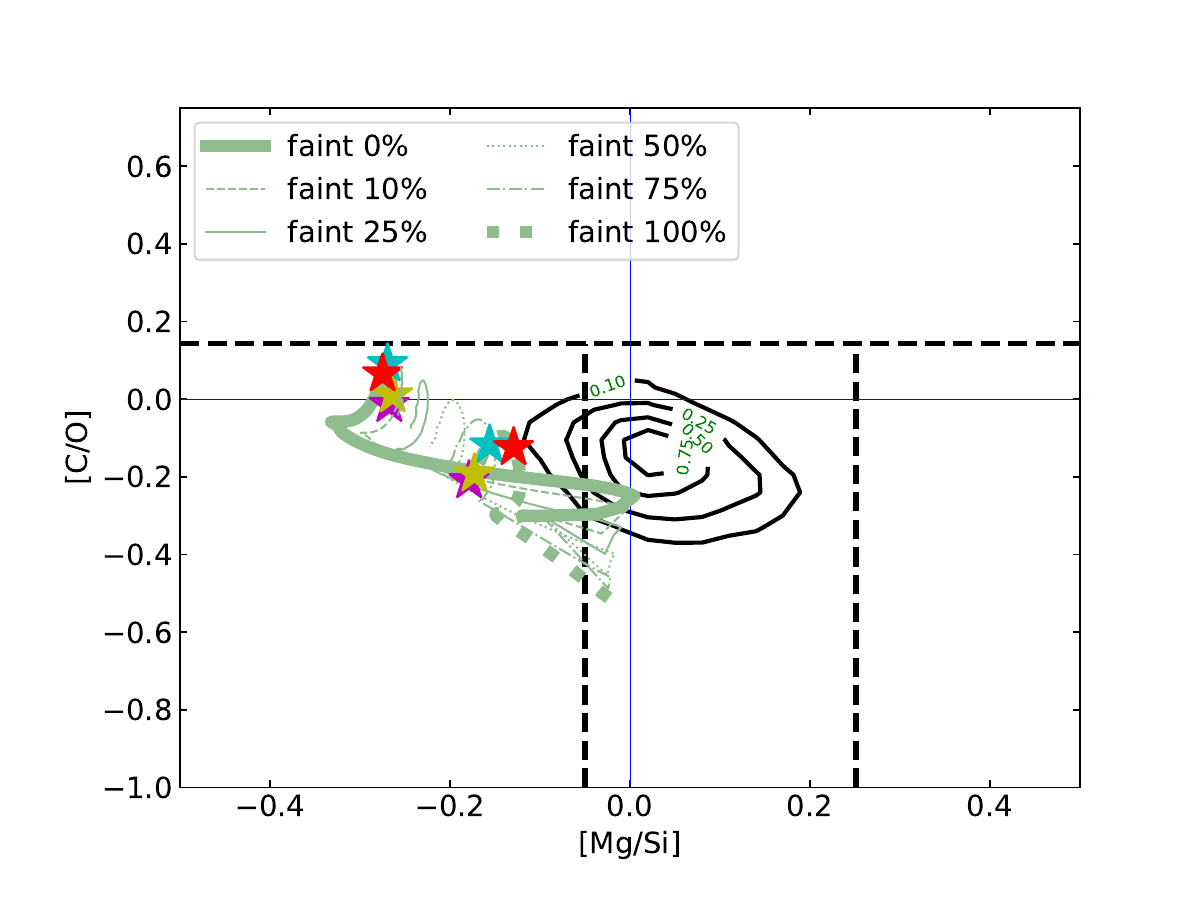}\par
    \includegraphics[width=0.8\linewidth]{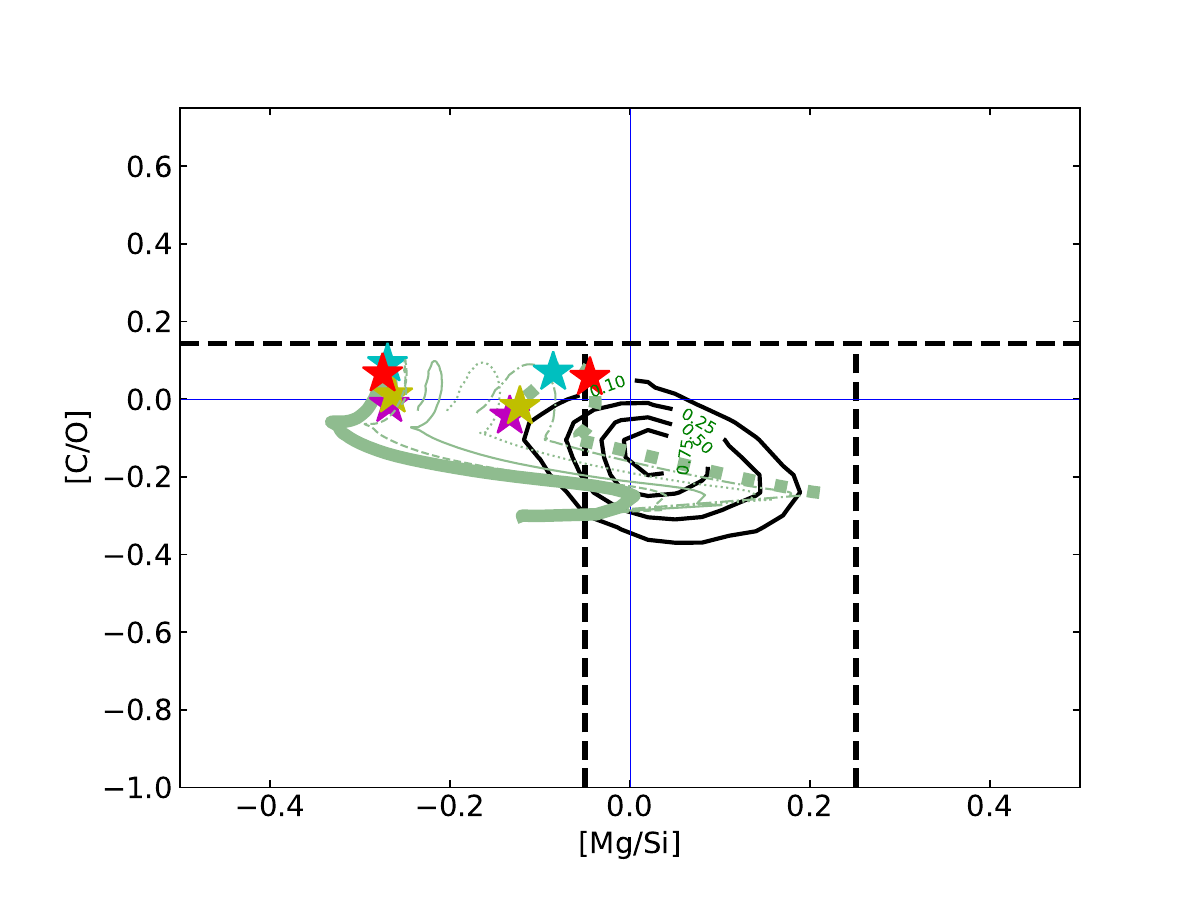}\par
    \end{multicols}
\begin{multicols}{2}
    \includegraphics[width=0.8\linewidth]{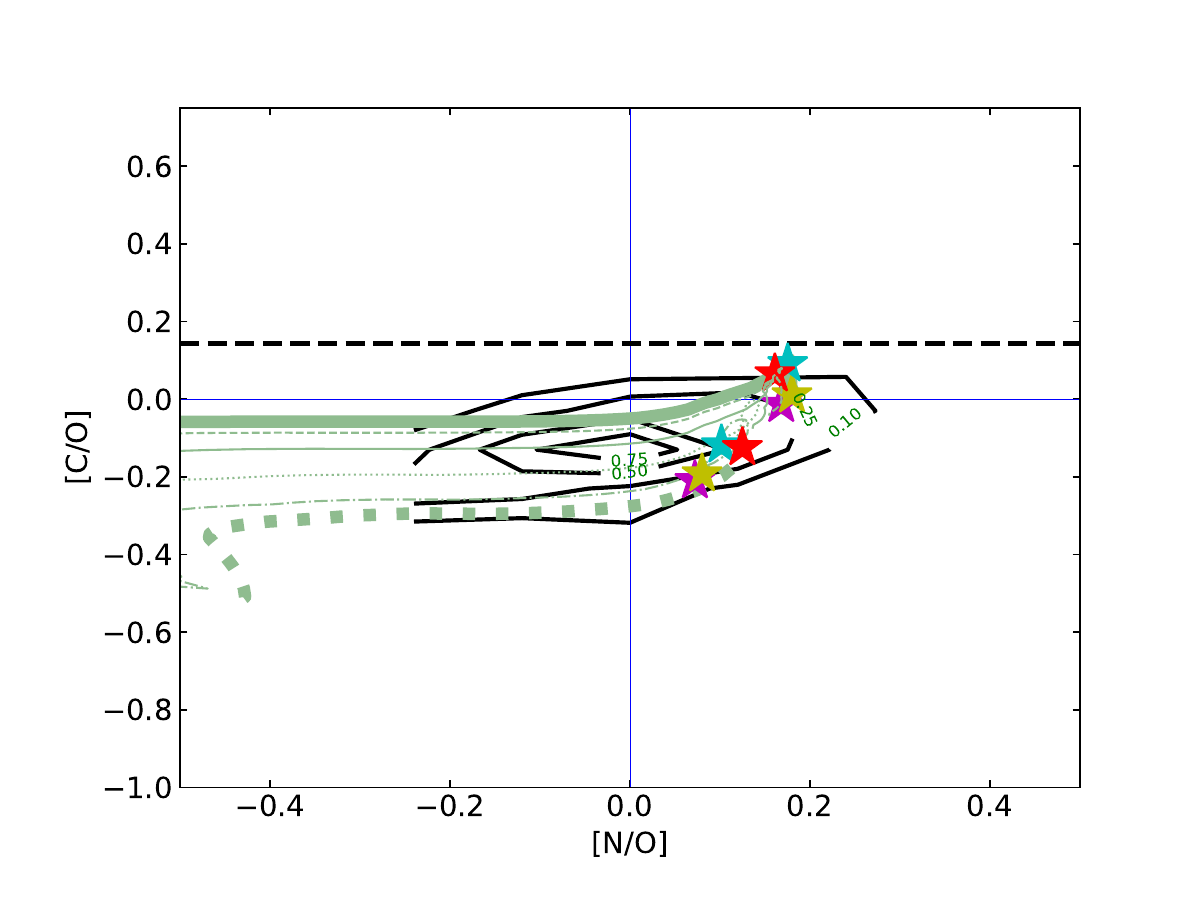}\par
    \includegraphics[width=0.8\linewidth]{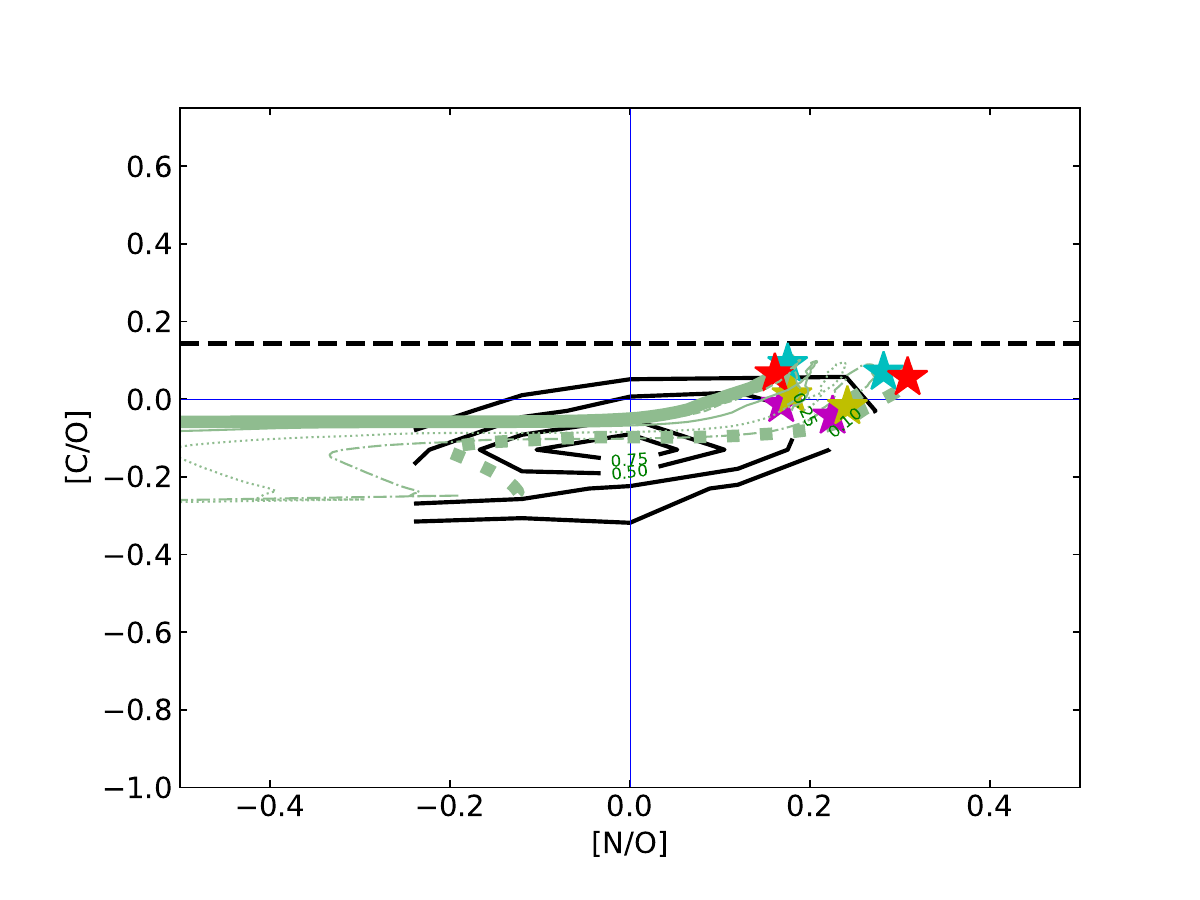}\par
\end{multicols}
\begin{multicols}{2}
    \includegraphics[width=0.8\linewidth]{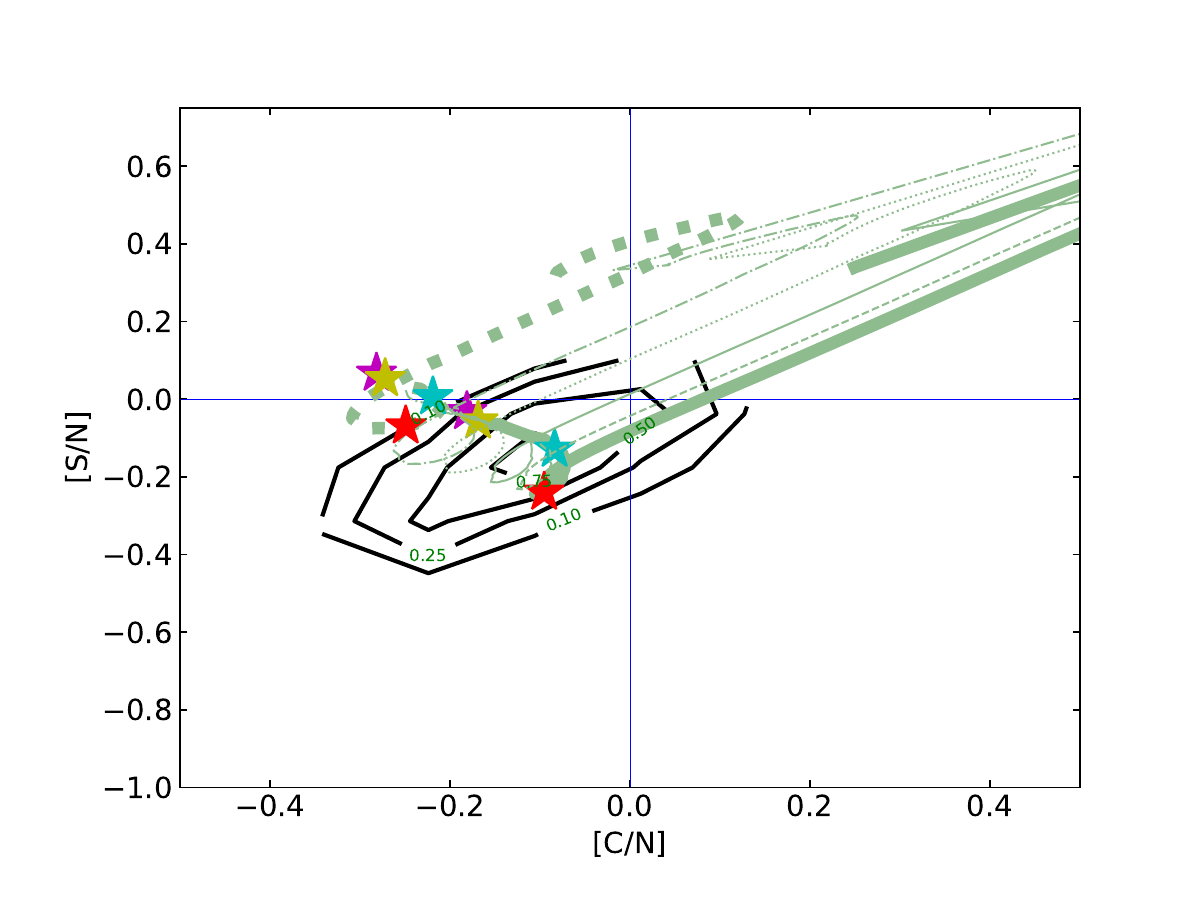}\par
    \includegraphics[width=0.8\linewidth]{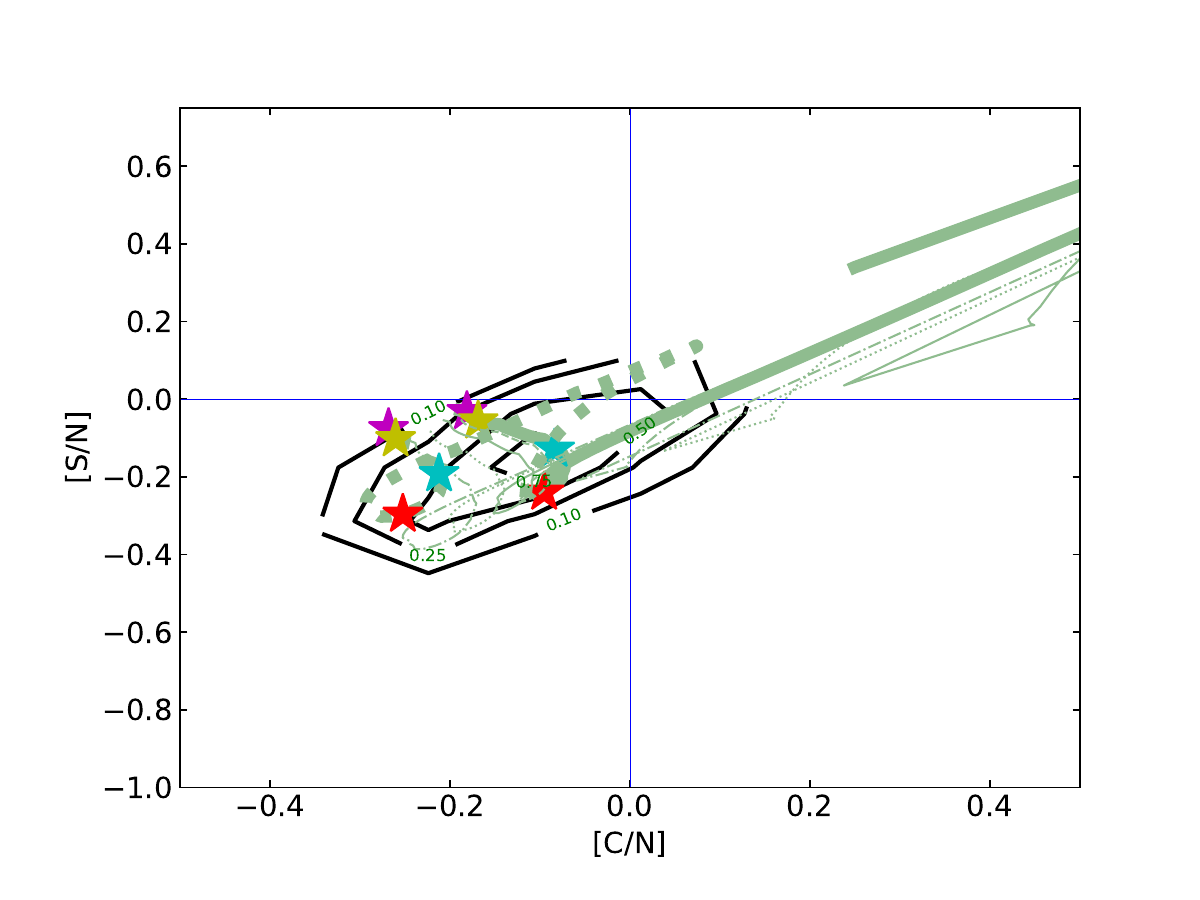}\par
\end{multicols}
\begin{multicols}{2}
    \includegraphics[width=0.8\linewidth]{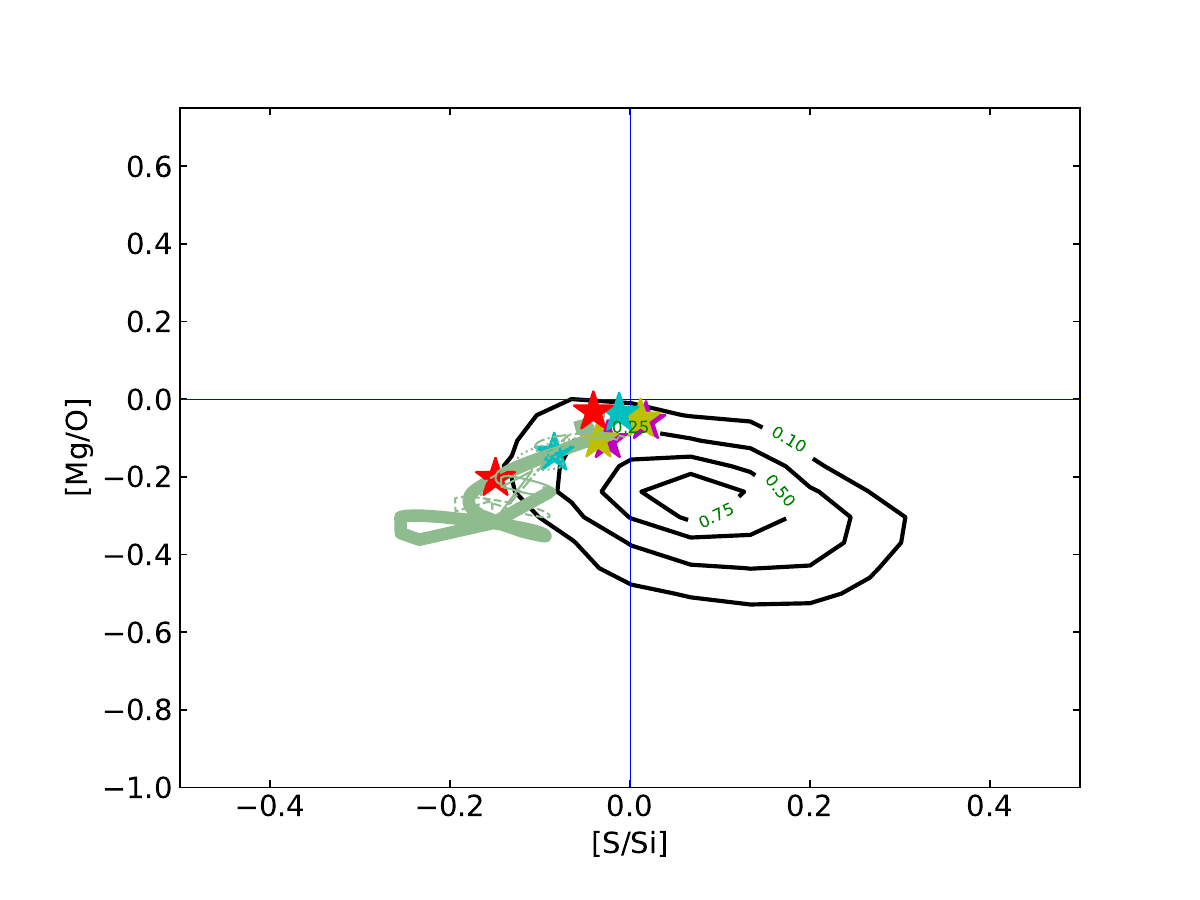}\par
    \includegraphics[width=0.8\linewidth]{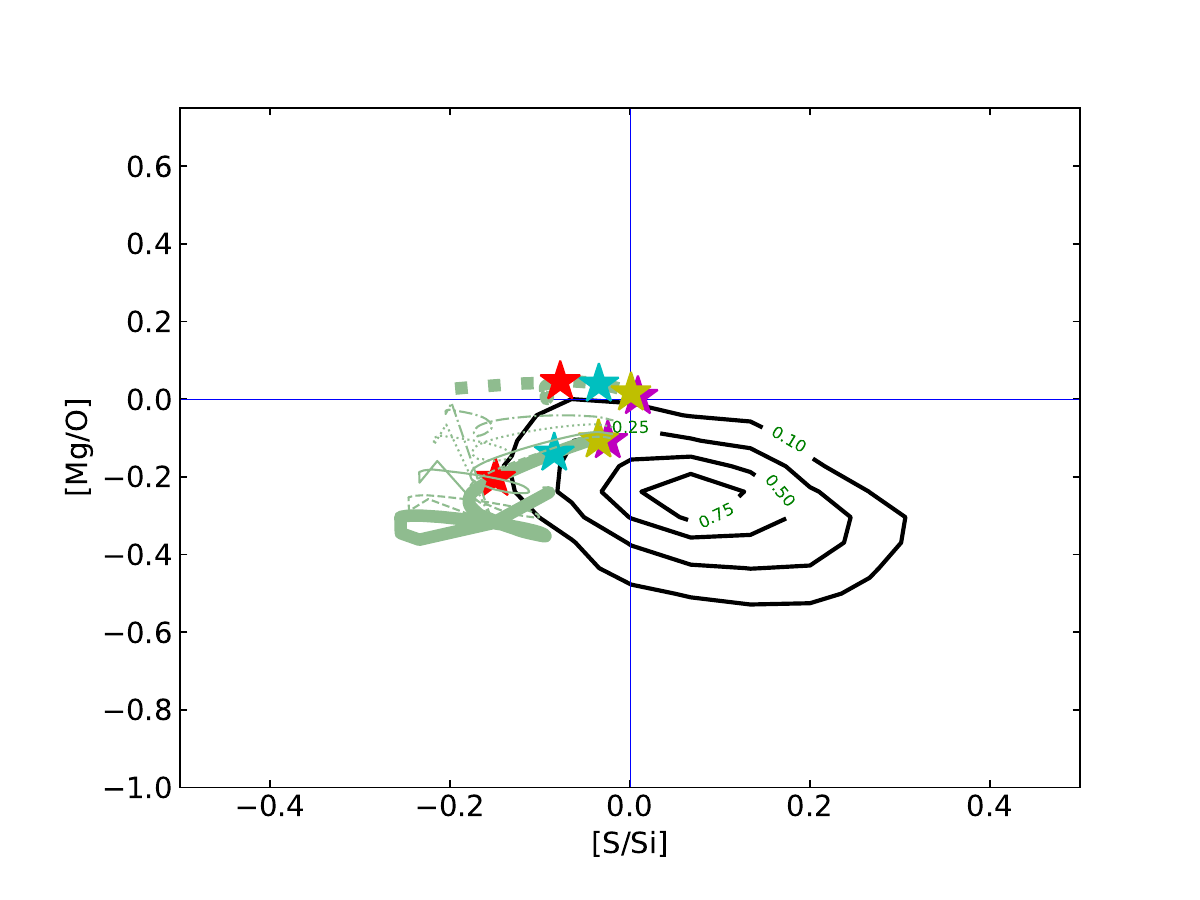}\par
\end{multicols}
    \caption{Same as in Figure~\ref{fig: tk_plots/ratios_gce_oK10m40}, but for the GCE model set oL18.
    }
    \label{fig: tk_plots/ratios_oL18_m40}
\end{figure*}

\begin{figure*}
\begin{multicols}{2}
    \includegraphics[width=0.8\linewidth]{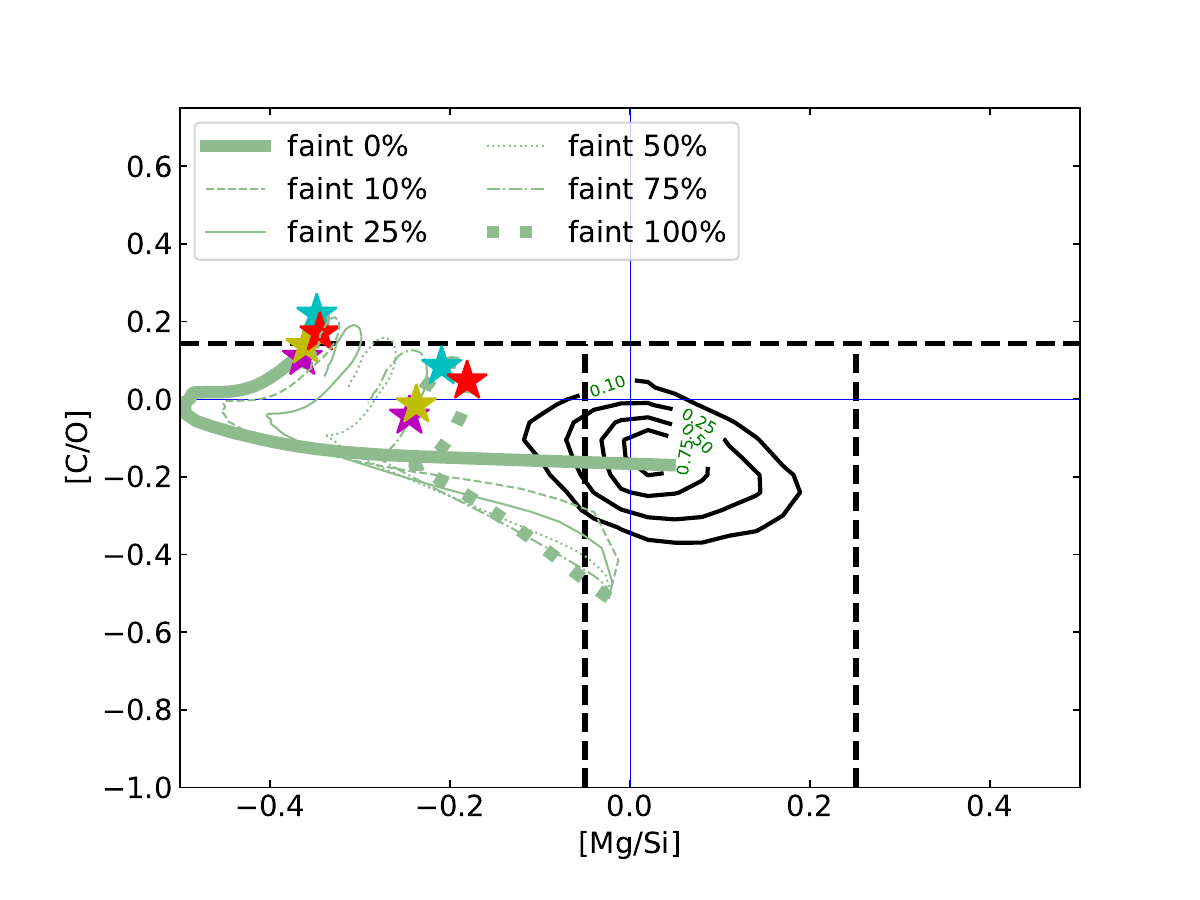}\par
    \includegraphics[width=0.8\linewidth]{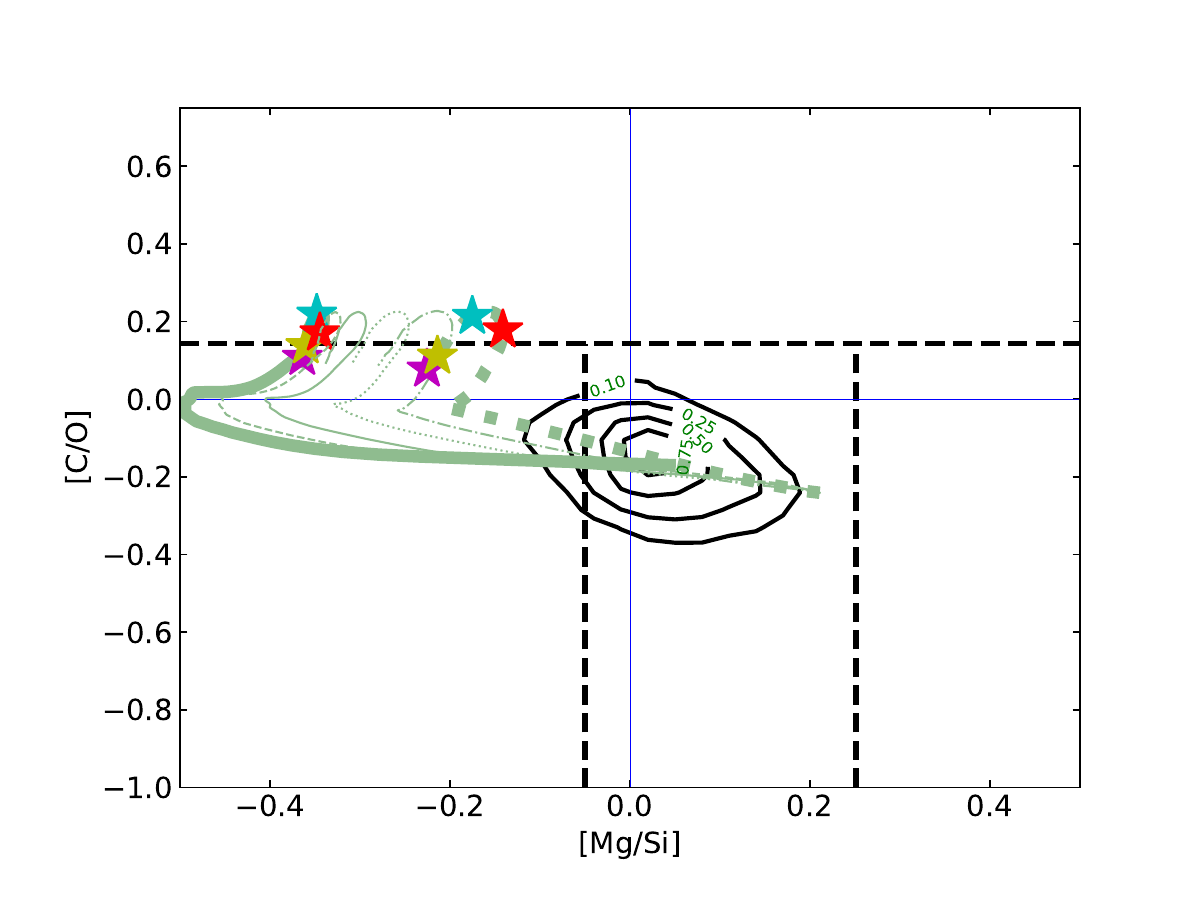}\par
    \end{multicols}
\begin{multicols}{2}
    \includegraphics[width=0.8\linewidth]{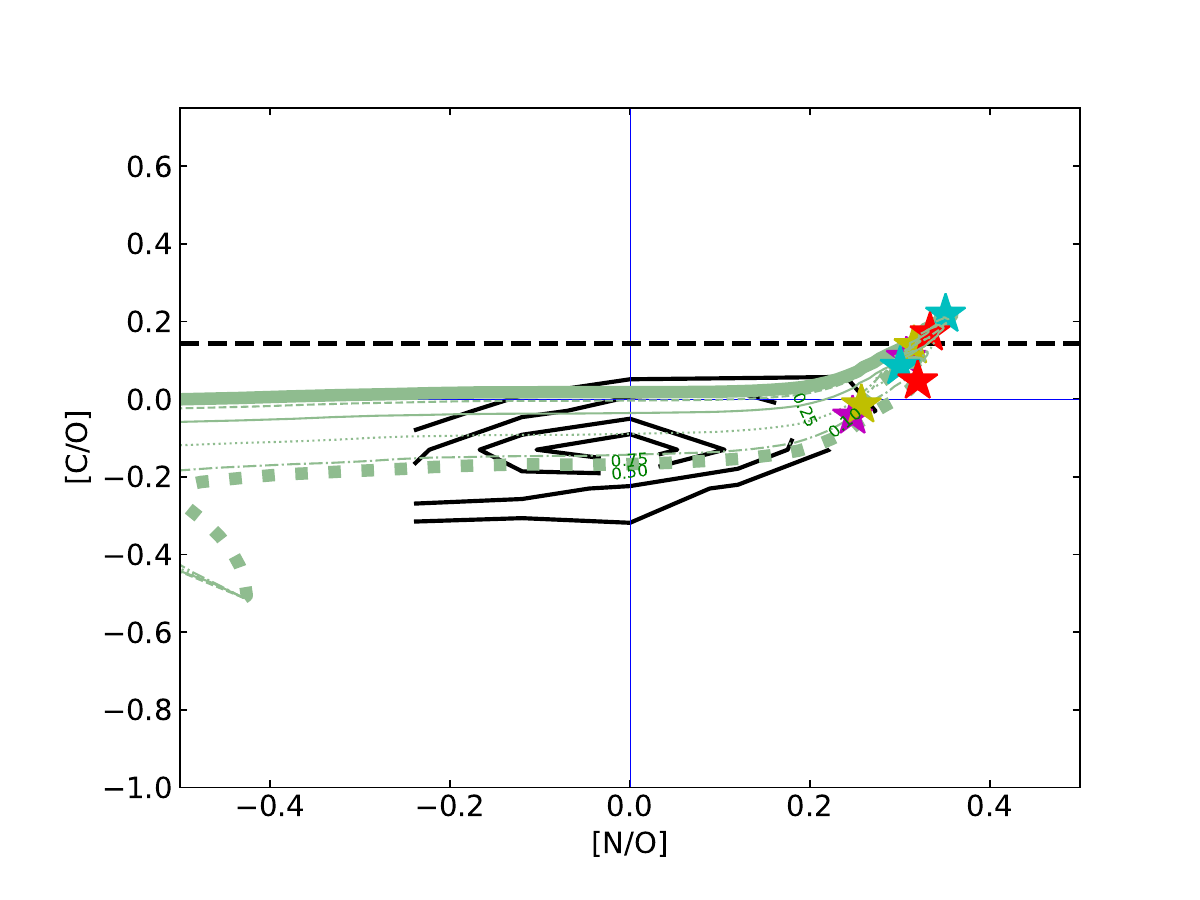}\par
    \includegraphics[width=0.8\linewidth]{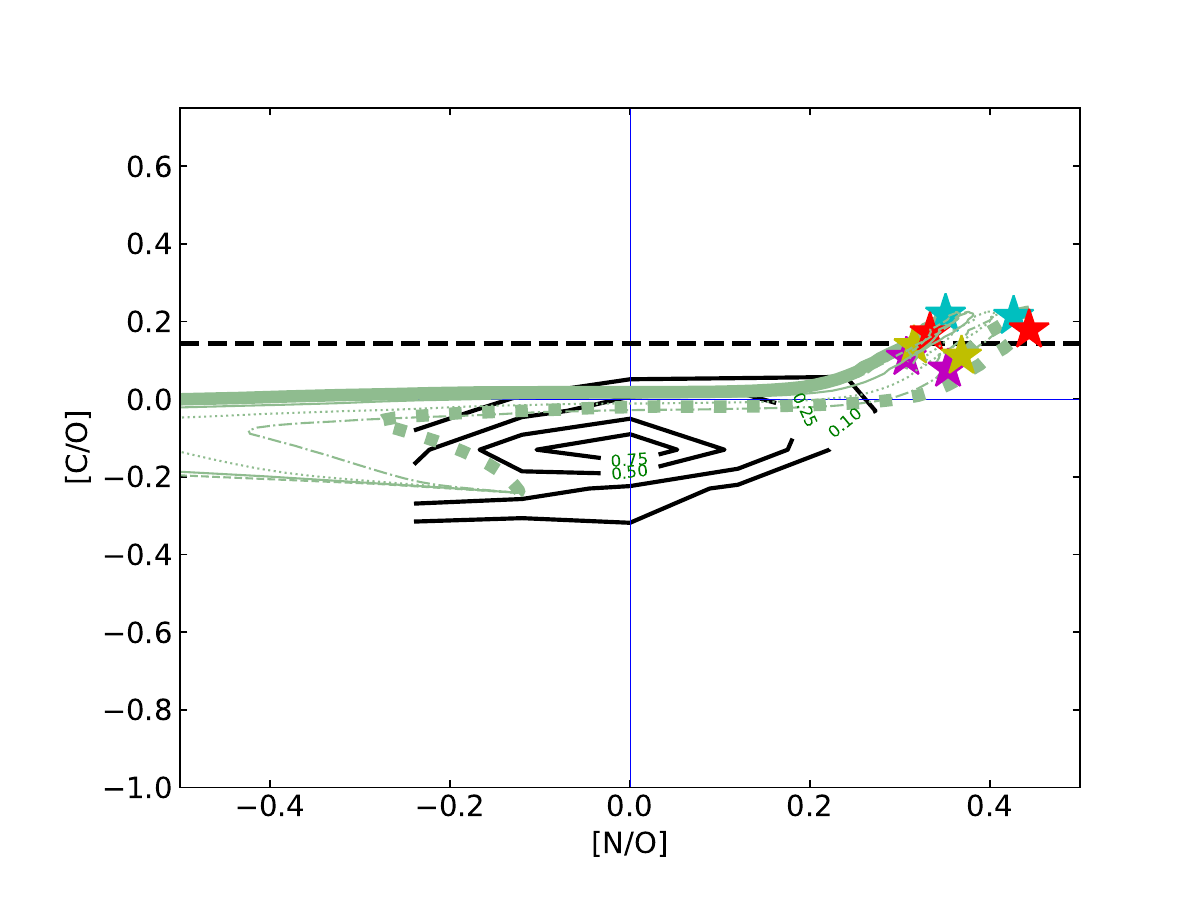}\par
\end{multicols}
\begin{multicols}{2}
    \includegraphics[width=0.8\linewidth]{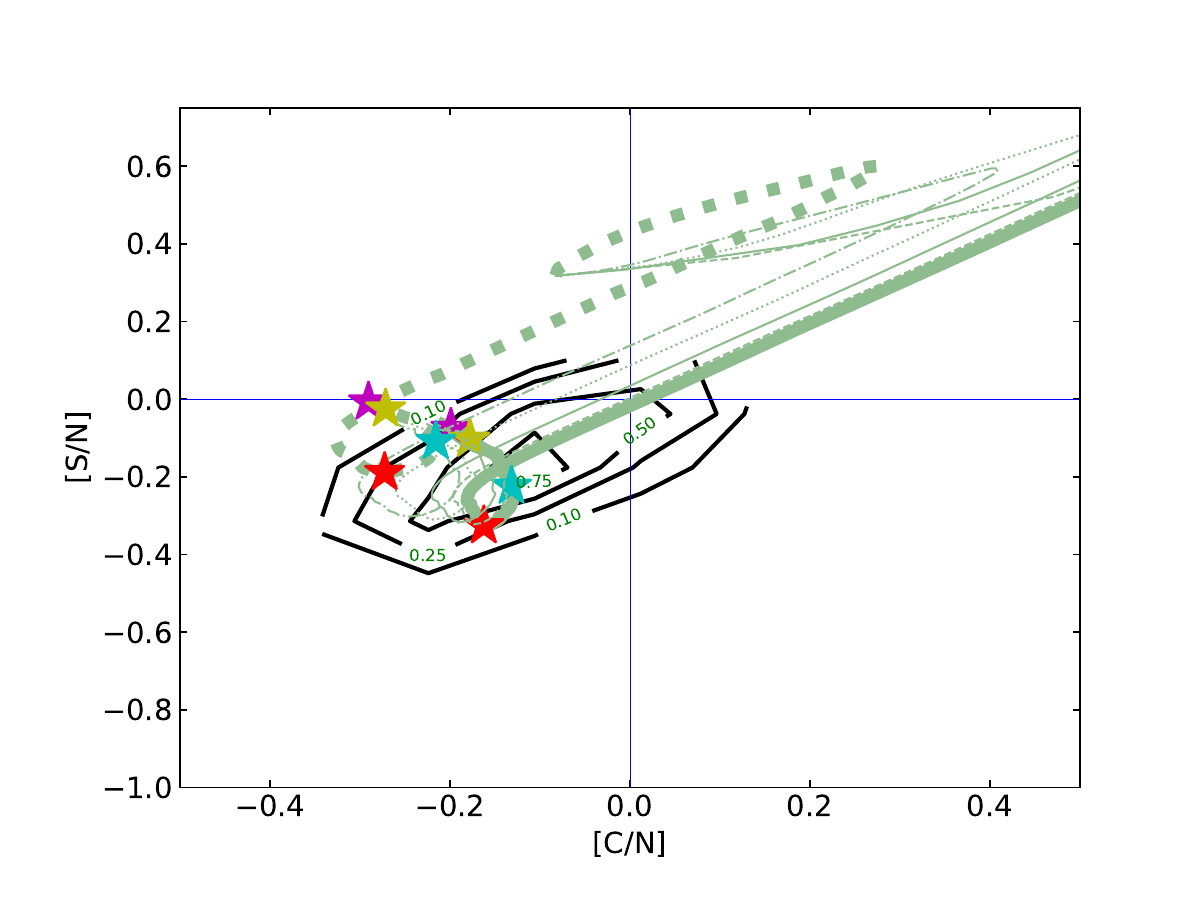}\par
    \includegraphics[width=0.8\linewidth]{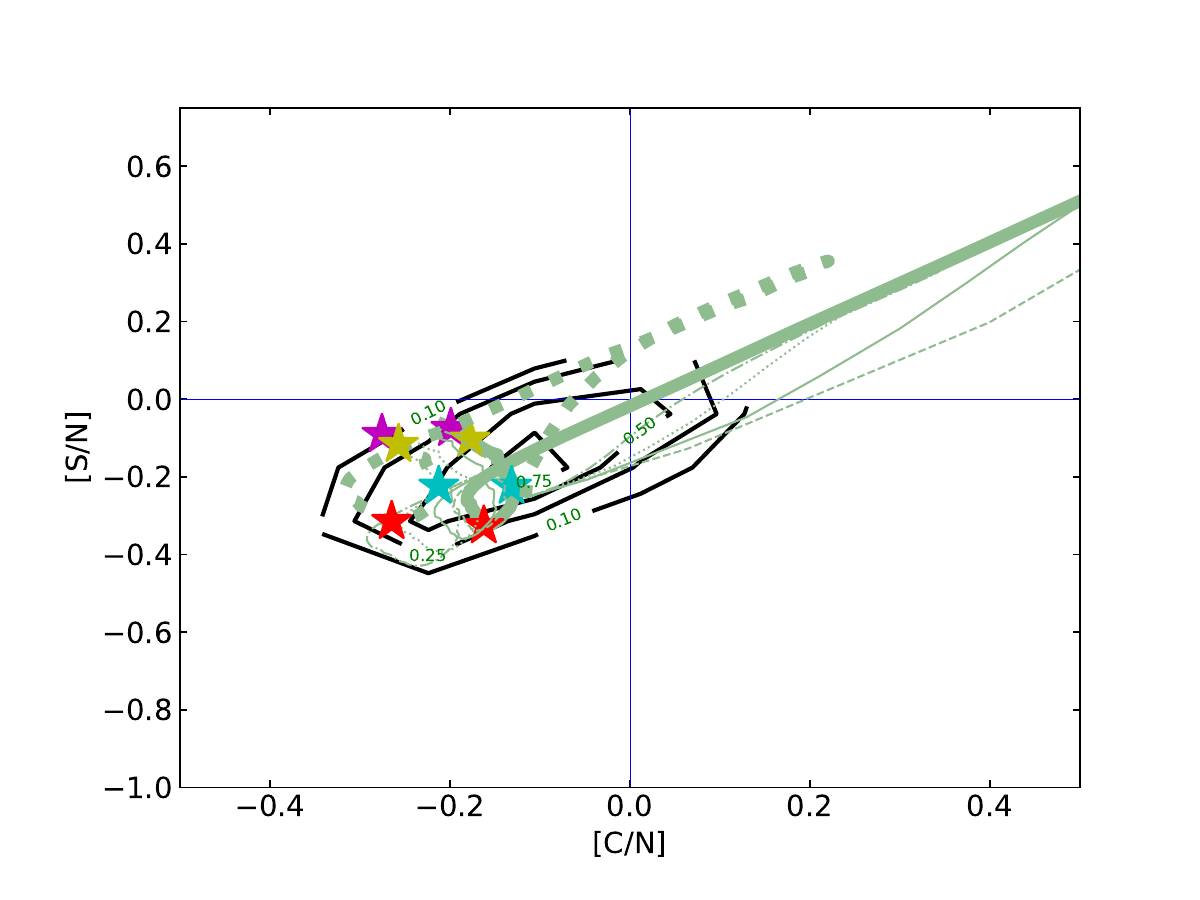}\par
\end{multicols}
\begin{multicols}{2}
    \includegraphics[width=0.8\linewidth]{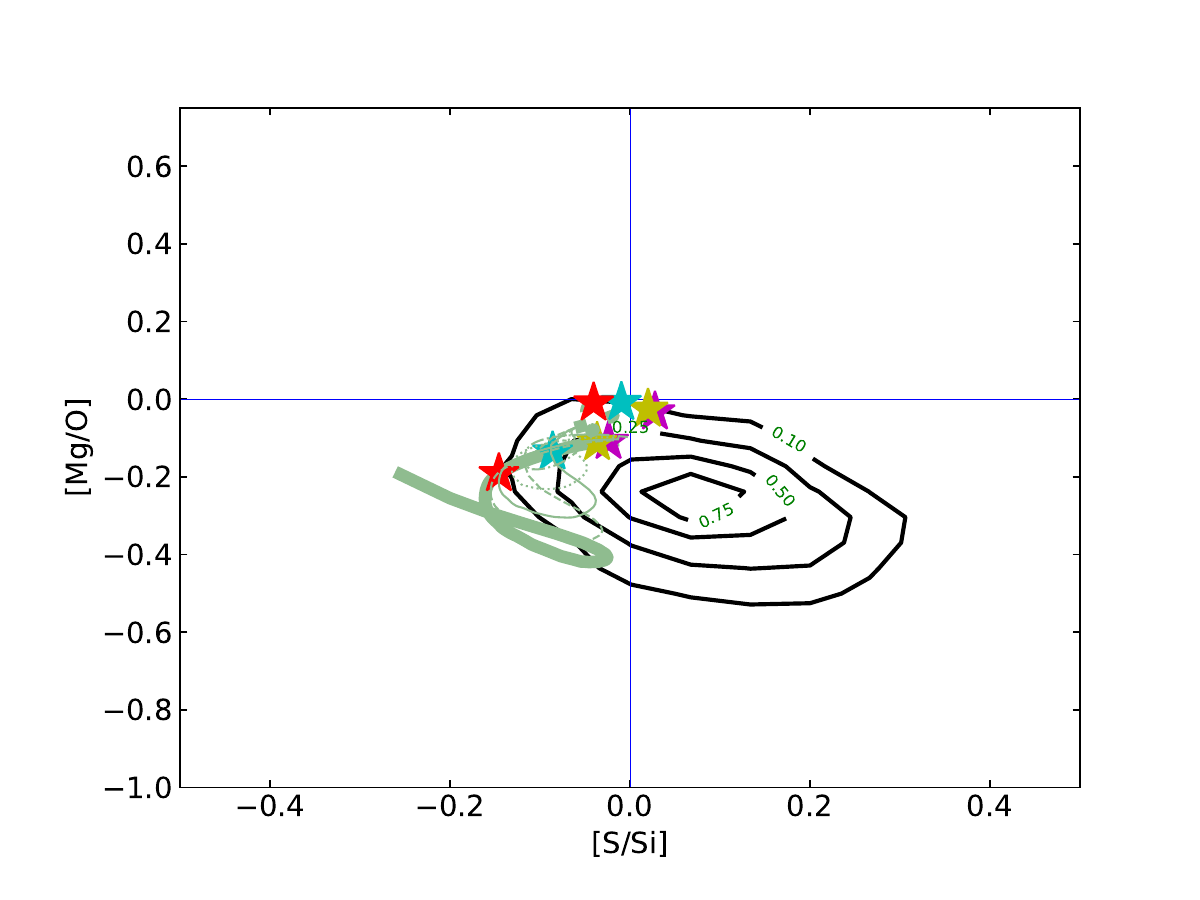}\par
    \includegraphics[width=0.8\linewidth]{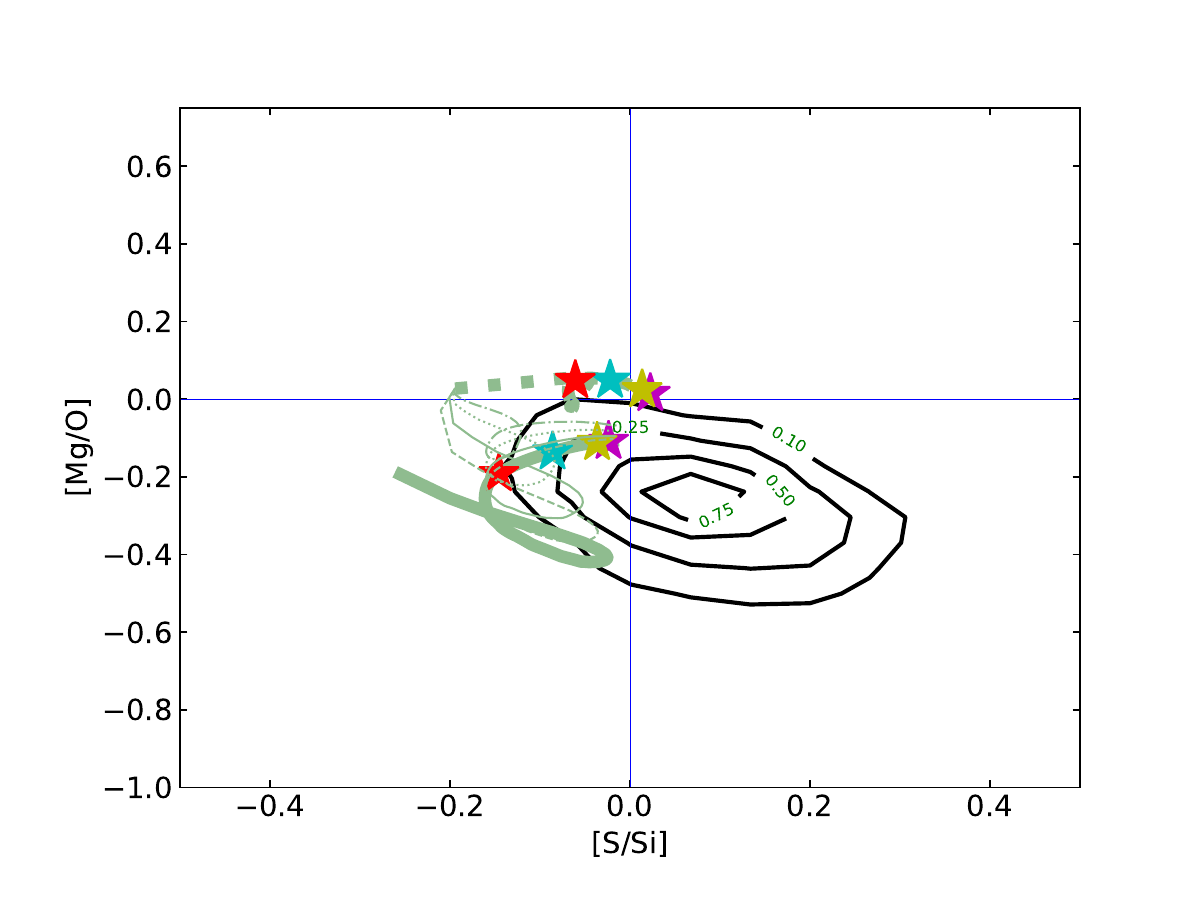}\par
\end{multicols}
    \caption{As in figure~\ref{fig: tk_plots/ratios_oL18_m40}, but models are shown with CCSN supernovae contribution up to M$_{\rm up}$ = 20 M$_{\odot}$.
    }
    \label{fig: tk_plots/ratios_oL18_m20}
\end{figure*}

\begin{figure*}
\begin{multicols}{2}
    \includegraphics[width=0.8\linewidth]{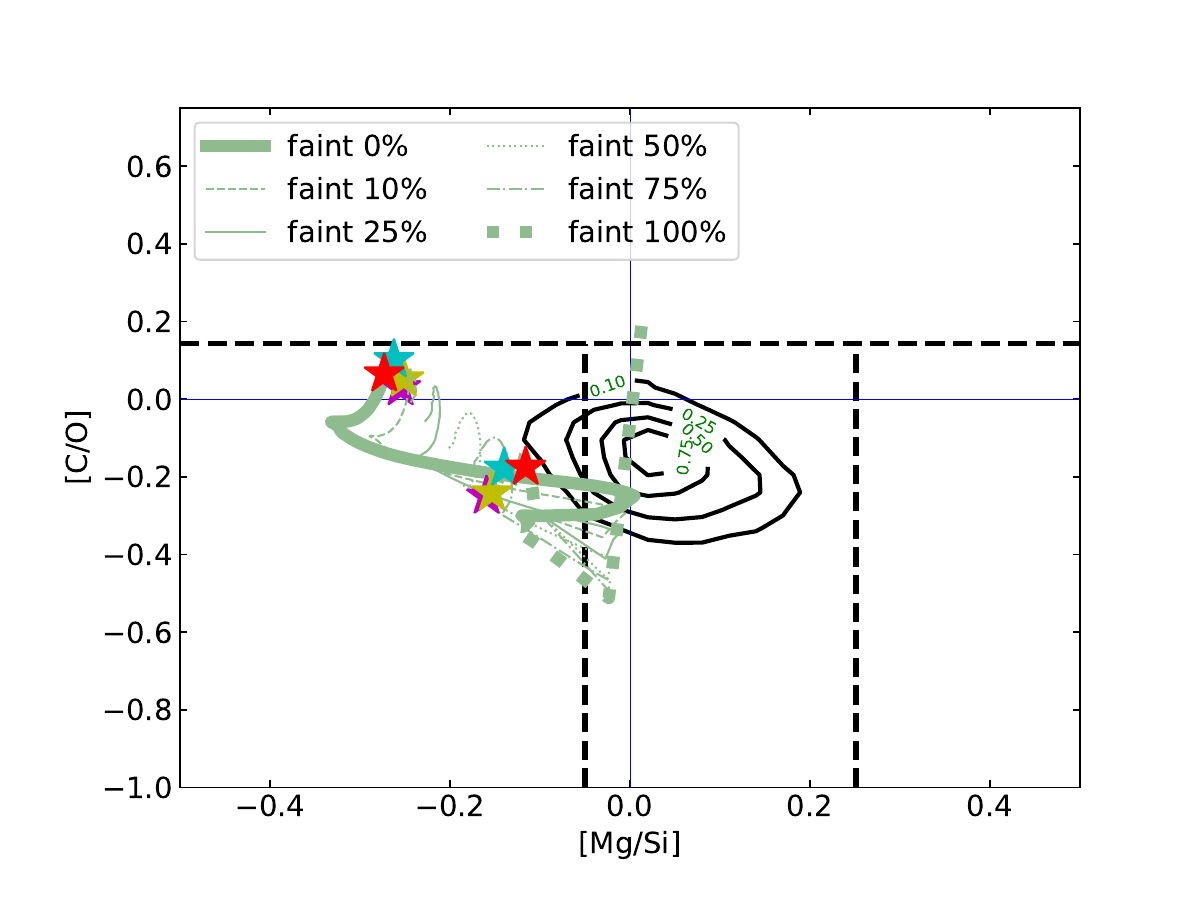}\par
    \includegraphics[width=0.8\linewidth]{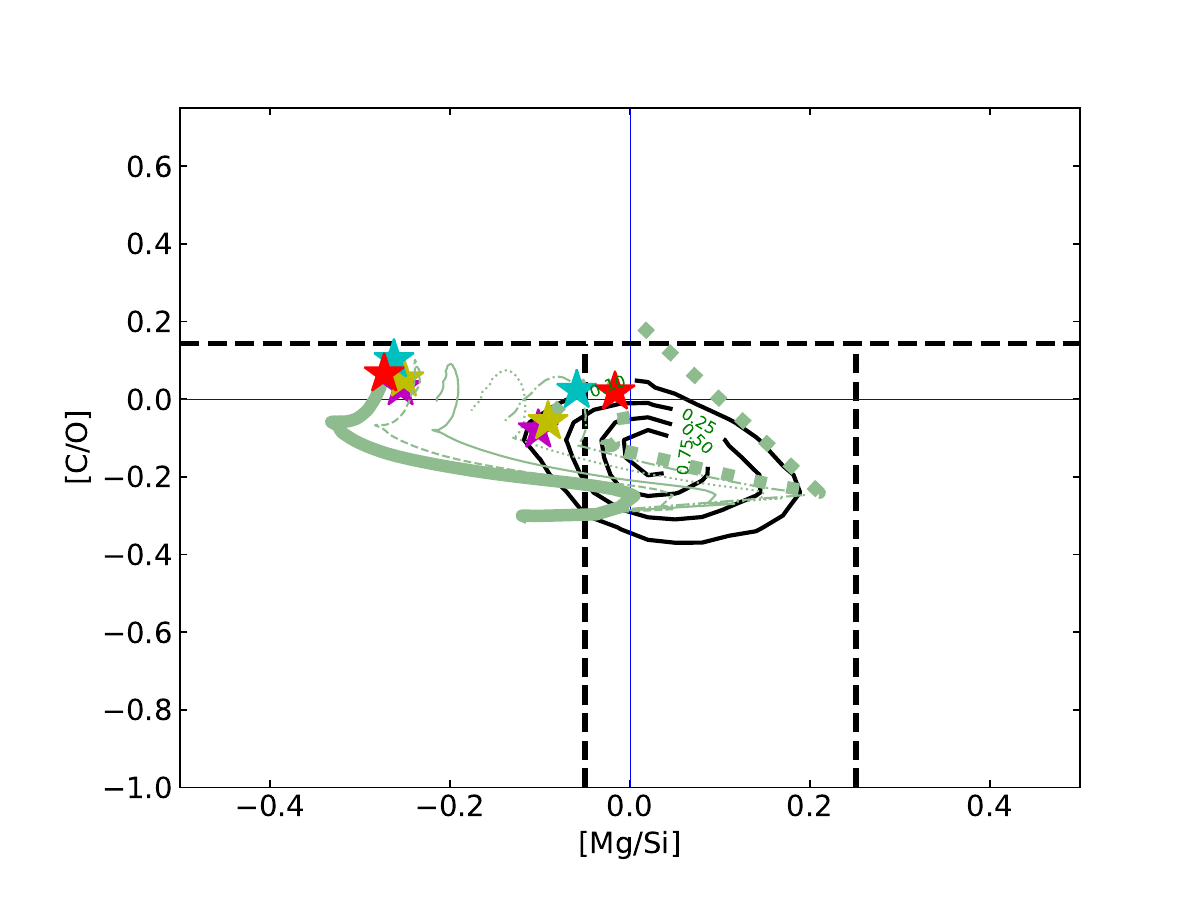}\par
    \end{multicols}
\begin{multicols}{2}
    \includegraphics[width=0.8\linewidth]{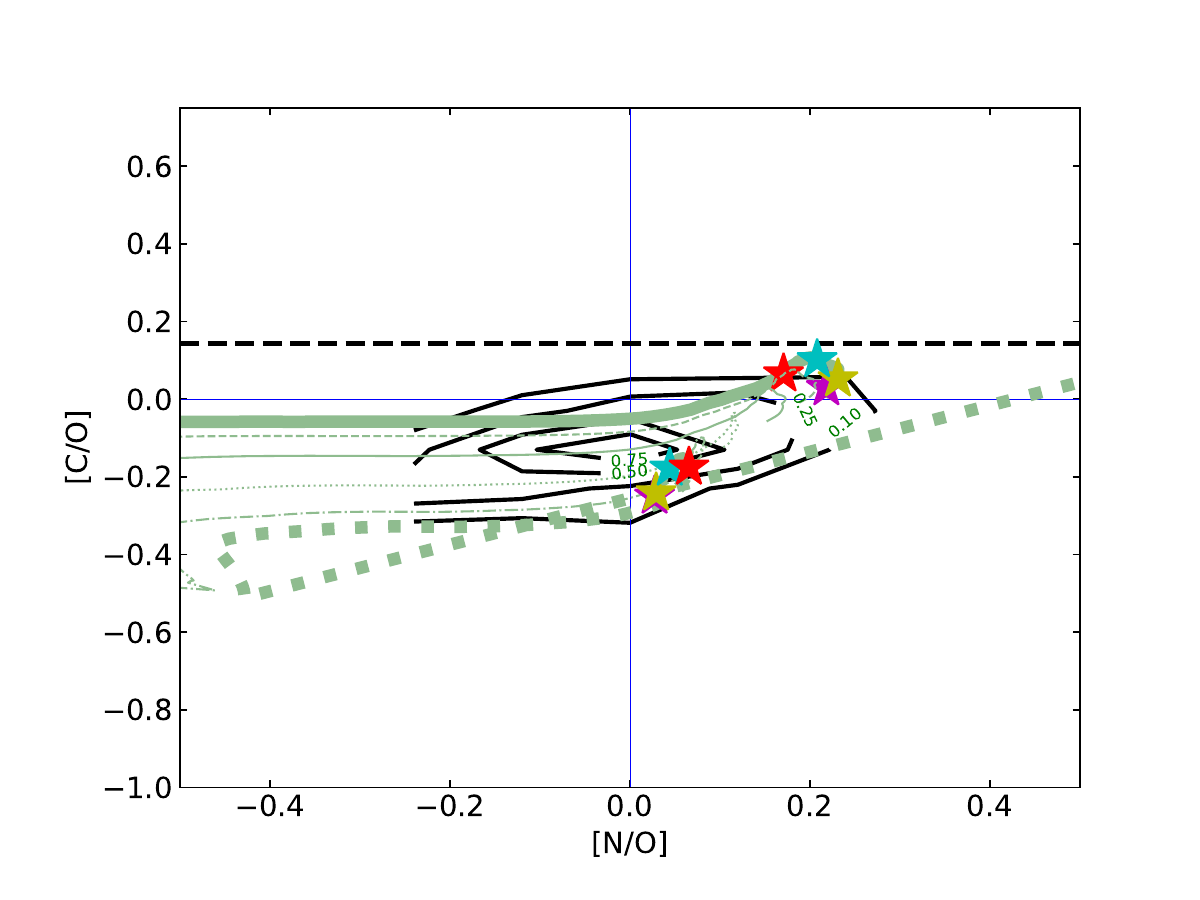}\par
    \includegraphics[width=0.8\linewidth]{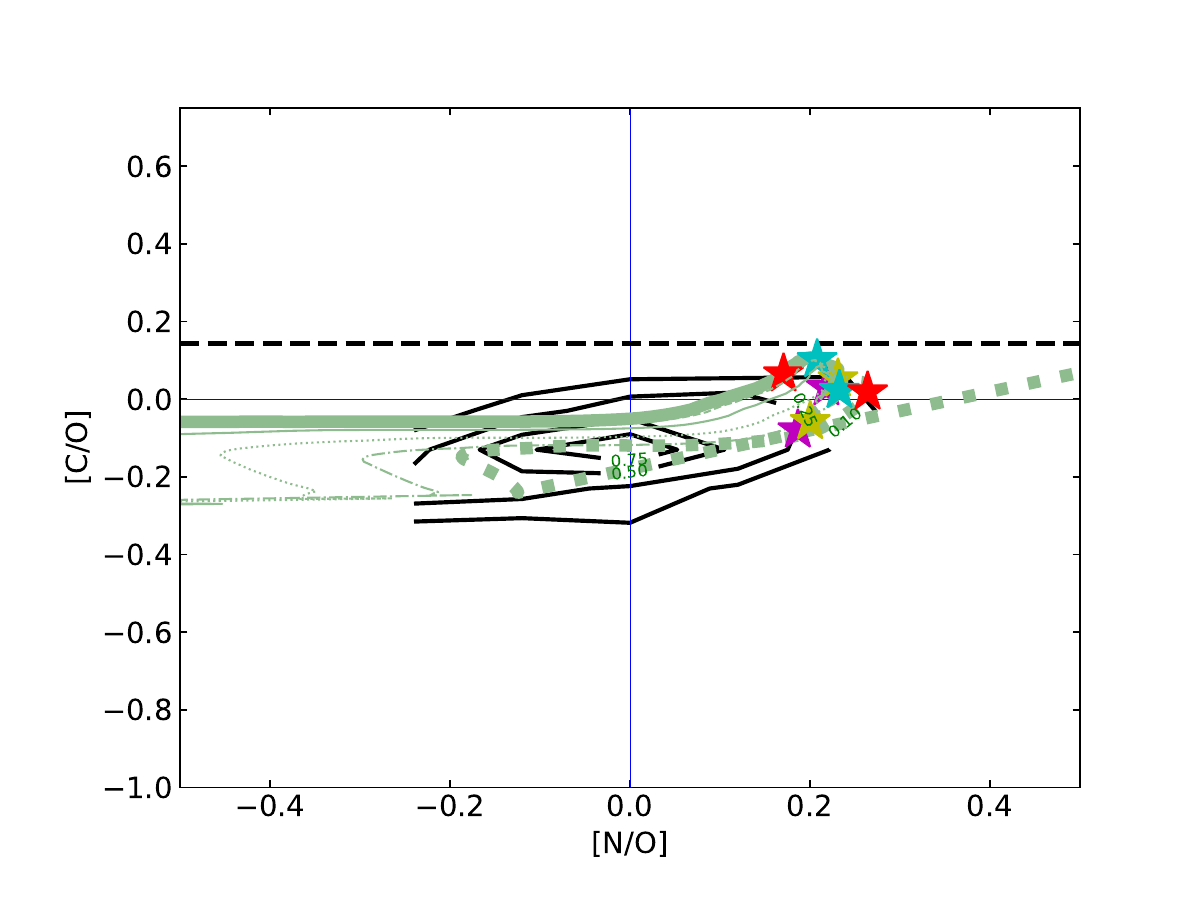}\par
\end{multicols}
\begin{multicols}{2}
    \includegraphics[width=0.8\linewidth]{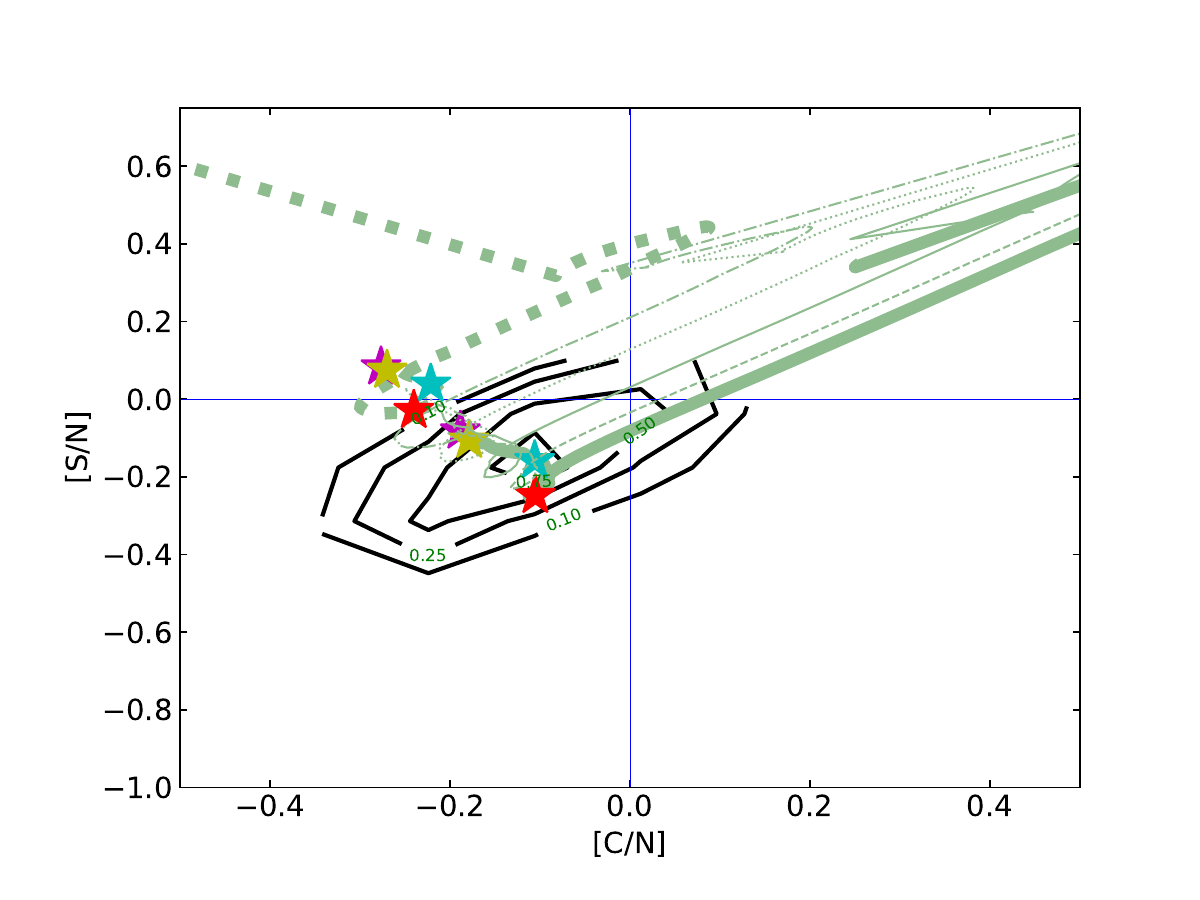}\par
    \includegraphics[width=0.8\linewidth]{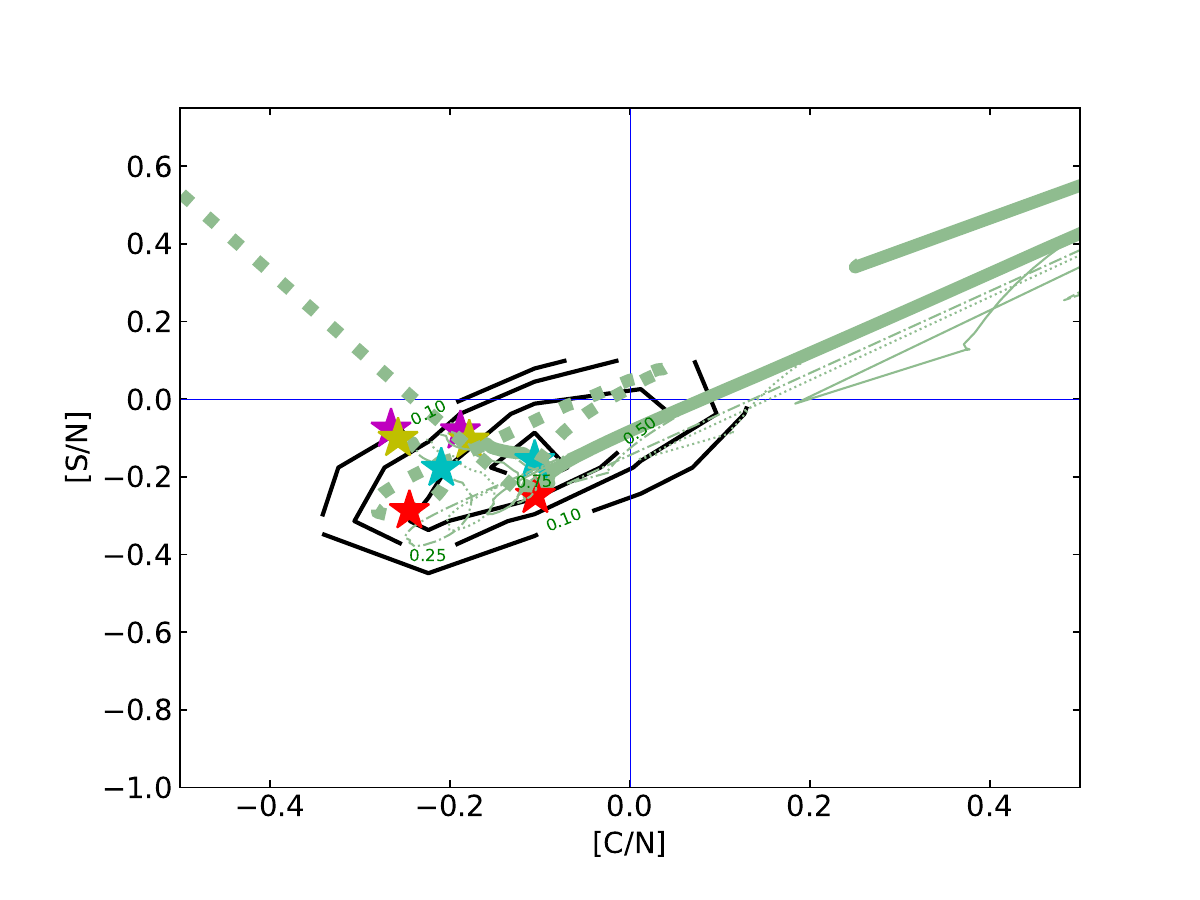}\par
\end{multicols}
\begin{multicols}{2}
    \includegraphics[width=0.8\linewidth]{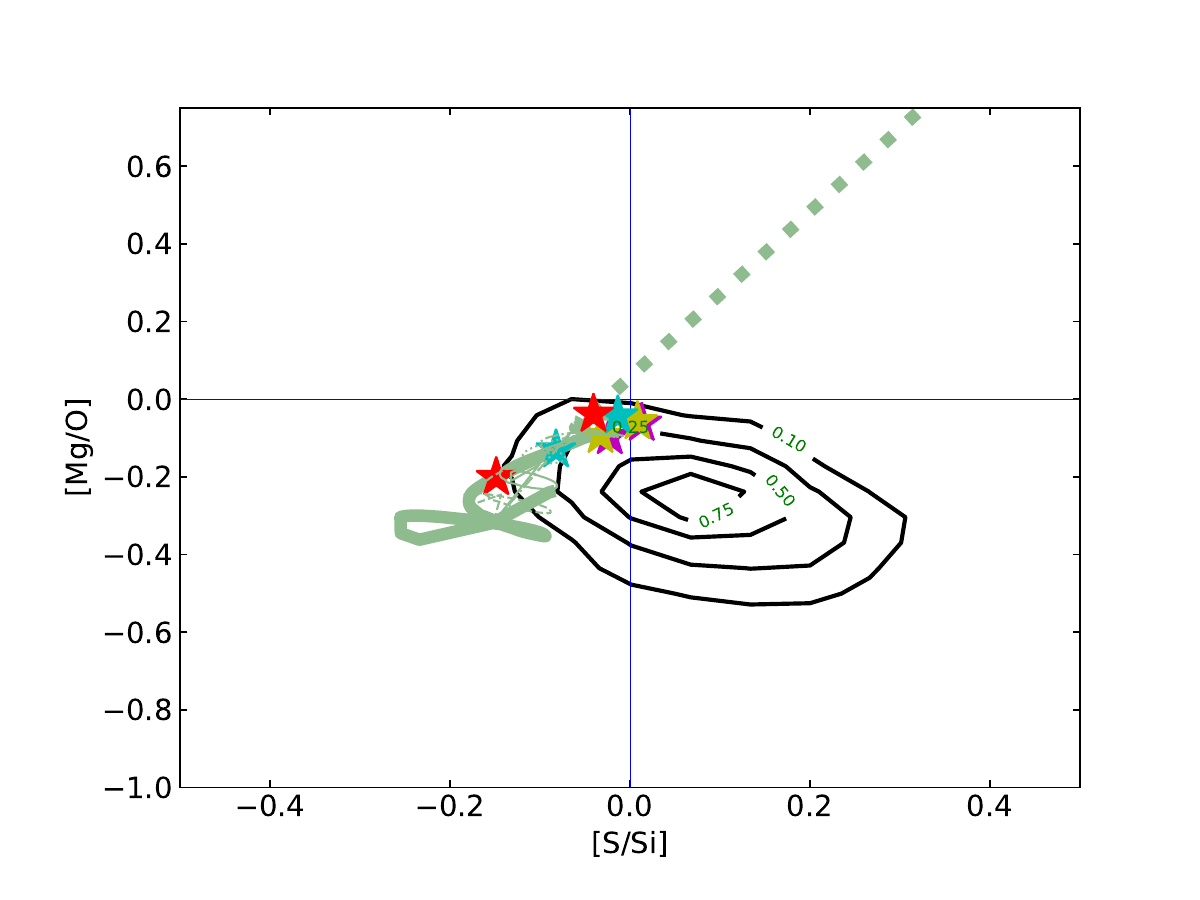}\par
    \includegraphics[width=0.8\linewidth]{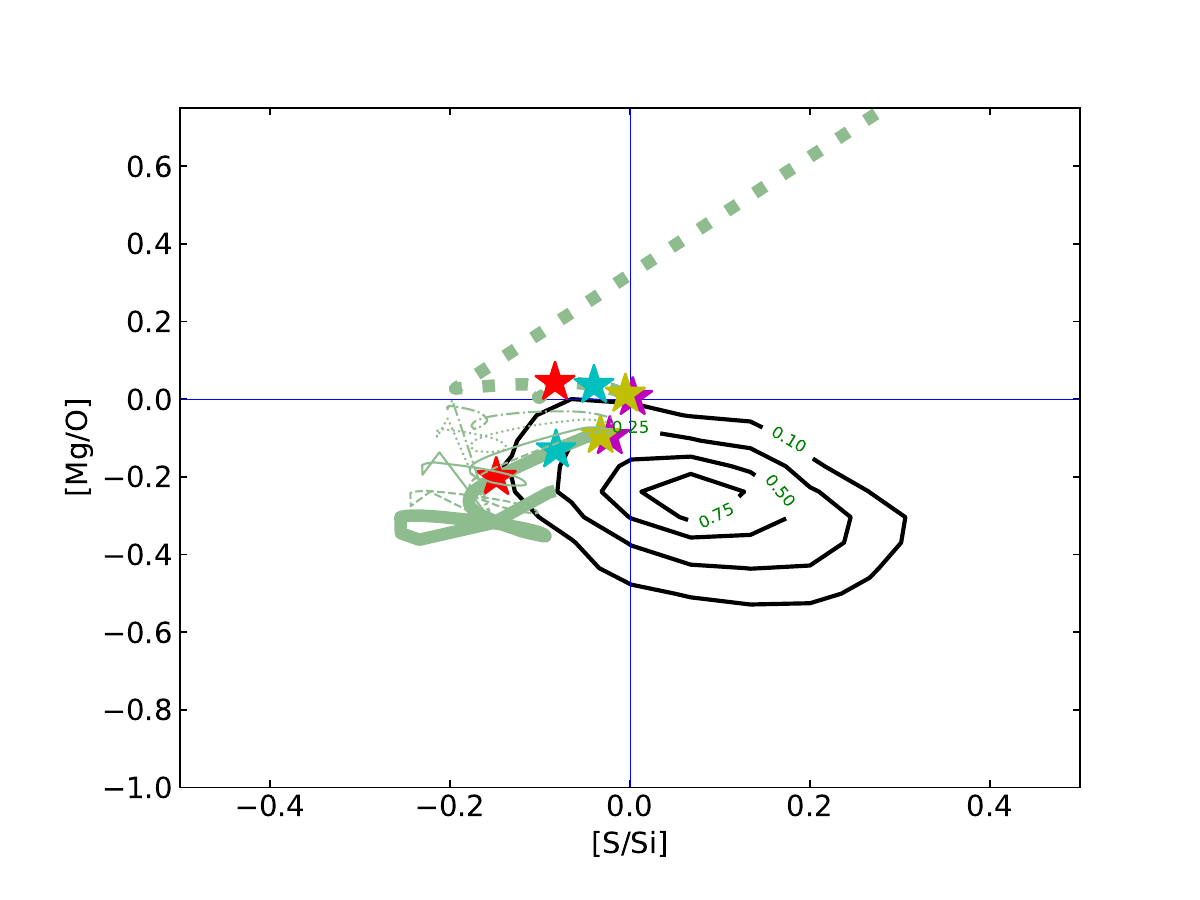}\par
\end{multicols}
    \caption{As in figure~\ref{fig: tk_plots/ratios_oL18_m40}, but models are shown with CCSN supernovae contribution up to M$_{\rm up}$ = 100 M$_{\odot}$.
    }
    \label{fig: tk_plots/ratios_oL18_m100}
\end{figure*}

For models oR18d, the results using M$_{\rm up}$=40M$_{\odot}$, 20M$_{\odot}$ and 100M$_{\odot}$ for different faint supernovae are given in Figures \ref{fig: tk_plots/ratios_oR18d_m40}, \ref{fig: tk_plots/ratios_oR18d_m20} and \ref{fig: tk_plots/ratios_oR18d_m100}, respectively. For models oR18h the results results using M$_{\rm up}$=40M$_{\odot}$, 20M$_{\odot}$ and 100M$_{\odot}$ for different faint supernovae are given in Figures \ref{fig: tk_plots/ratios_oR18h_m40}, \ref{fig: tk_plots/ratios_oR18h_m20} and \ref{fig: tk_plots/ratios_oR18h_m100}, respectively. Finally, the same is reported for models oL18 in Figures \ref{fig: tk_plots/ratios_oL18_m40}, \ref{fig: tk_plots/ratios_oL18_m20} and \ref{fig: tk_plots/ratios_oL18_m100}, respectively.

%


\bsp    
\label{lastpage}
\end{document}